\RequirePackage{lineno}\newdimen\linenumbersep\linenumbersep=2pt
\documentclass[preprint,3p,number,sort&compress,longtitle,times]{elsarticle}

\usepackage{subfigure}
\usepackage{mathtools}
\usepackage{tikz}
\usepackage{rotating}
\usepackage[compat=1.1.0]{tikz-feynman}
\usetikzlibrary{shapes,arrows,positioning,automata,backgrounds,calc,er,patterns}

\usepackage{array, booktabs, color, eurosym, graphicx, longtable, pbox, pdfpages, siunitx, url, verbatim}

\usepackage[colorlinks=true,citecolor=blue,filecolor=blue,linkcolor=blue,urlcolor=blue,pdftex]{hyperref}
\usepackage{xspace}

\begin{document}


\title{The Present and Future Status of Heavy Neutral Leptons}

%
%

\author[fermilab,ippp]{Asli M. Abdullahi}
\author[cern,ciemat]{Pablo Barham~Alzás}
\author[pittsburgh]{Brian Batell}
\author[leiden]{Alexey Boyarsky}
\author[lima]{Saneli Carbajal}
\author[pittsburgh]{Animesh Chatterjee}
\author[ciemat]{Jos\'e I. Crespo-Anad\'on}
\author[ucl]{Frank F. Deppisch}
\author[cern]{Albert De~Roeck}
\author[cp3]{Marco Drewes}
\author[lima]{Alberto Martin Gago}
\author[uppsala]{Rebeca Gonzalez Suarez}
\author[birmingham]{Evgueni Goudzovski}
\author[sjsu]{Athanasios Hatzikoutelis}
\author[ulb]{Marco Hufnagel}
\author[cincinnati]{Philip Ilten}
\author[inrras]{Alexander Izmaylov}
\author[cern]{Kevin J. Kelly}
\author[cp3]{Juraj Klarić}
\author[cern,mainz]{Joachim Kopp}
\author[graz]{Suchita Kulkarni}
\author[infnpd]{Mathieu Lamoureux}
\author[infn]{Gaia Lanfranchi}
\author[ific]{Jacobo L\'opez-Pav\'on}
\author[leiden]{Oleksii Mikulenko}
\author[csu]{Michael Mooney}
\author[ijs]{Miha Nemev\v{s}ek}
\author[leiden]{Maksym Ovchynnikov}
\author[bologna,ippp]{Silvia Pascoli}
\author[kentucky,fermilab]{Ryan Plestid}
\author[Antwerp]{Mohamed Rashad Darwish}
\author[cern]{Federico Leo Redi}
\author[nbi]{Oleg Ruchayskiy}
\author[kracow]{Richard Ruiz}
\author[epfl]{Mikhail Shaposhnikov}
\author[vt]{Ian M.~Shoemaker}
\author[sunysb]{Robert Shrock}
\author[cincinnati]{Alex Sousa}
\author[Antwerp]{Nick Van Remortel}
\author[leiden]{Vsevolod Syvolap}
\author[ipmu]{Volodymyr Takhistov}
\author[ift]{Jean-Loup Tastet}
\author[nbi]{Inar Timiryasov}
\author[queens]{Aaron C.~Vincent}
\author[cern,uta]{Jaehoon Yu}

\address[fermilab]{Theoretical Physics Department, Fermilab, P.O. Box 500, Batavia, IL 60510, USA}

\address[ippp]{Institute for Particle Physics Phenomenology, Department of Physics, Durham University, Durham, DH1 3LF, United Kingdom}

\address[cern]{European Organization for Nuclear Research (CERN), 1211 Geneva 23, Switzerland}

\address[ciemat]{Centro de Investigaciones Energ\'{e}ticas, Medioambientales y Tecnol\'{o}gicas (CIEMAT), Madrid E-28040, Spain}

\address[pittsburgh]{Department of Physics and Astronomy, University of Pittsburgh, 3941 O'Hara St, Pittsburgh, PA 15260, USA}

\address[leiden]{Instituut-Lorentz for Theoretical Physics, Universiteit Leiden, Niels Bohrweg 2, 2333 CA Leiden, Netherlands}

\address[lima]{Secci\'on F\'isica, Departamento de Ciencias, Pontificia Universidad Cat\'olica del Per\'u, Apartado 1761, Lima, Per\'u}

\address[ucl]{Department of Physics and Astronomy, University College London, Gower Street, London WC1E 6BT, UK}

\address[cp3]{Centre for Cosmology, Particle Physics and Phenomenology - CP3 Université catholique de Louvain 2, Chemin du Cyclotron - Box L7.01.05 B-1348, Louvain-la-Neuve, Belgium}

\address[uppsala]{Department of Physics and Astronomy, University of Uppsala, Uppsala, Sweden}

\address[birmingham]{School of Physics and Astronomy, University of Birmingham, Edgbaston, Birmingham, B15 2TT, United Kingdom}

\address[sjsu]{San Jos\'{e} State University, One Washington St., San Jos\'{e}, California,USA}

\address[ulb]{Service de Physique Théorique, Université Libre de Bruxelles, Boulevard du Triomphe, CP225, B-1050 Brussels, Belgium}

\address[cincinnati]{Department of Physics, University of Cincinnati, Cincinnati, OH 45221, USA}

\address[inrras]{Institute  for  Nuclear  Research  of  the  Russian  Academy  of  Sciences, INR RAS, 60-letiya Oktyabrya prospekt 7a, Moscow 117312, Russia}

\address[mainz]{PRISMA Cluster of Excellence, Johannes Gutenberg University, 55128 Mainz, Germany}

\address[graz]{Institute of Physics, NAWI Graz, University of Graz,Universitätsplatz 5, A-8010 Graz, Austria}

\address[infnpd]{INFN Sezione di Padova, 35131 Padova, Italy}

\address[infn]{INFN - Laboratori Nazionali di Frascati, via E. Fermi 40, 00044 Frascati (Rome), Italy}

\address[ific]{Instituto de Física Corpuscular, Universidad de Valencia \& CSIC, Edificio Institutos de Investigaci\'on, Calle Catedr\'atico Jos\'e Beltr\'an 2, 46980 Paterna, Spain}

\address[csu]{Colorado State University, Fort Collins, CO, 80523, USA}

\address[ijs]{Jo\v zef Stefan Institute, Jamova 39, 1000, Ljubljana, Slovenia}

\address[bologna]{Dipartimento di Fisica e Astronomia, Università di Bologna e INFN, Sezione di Bologna, via
Irnerio 46, I-40126 Bologna, Italy}

\address[kentucky]{Department of Physics and Astronomy, University of Kentucky, Lexington, KY 40506, USA}

\address[Antwerp]{Universiteit Antwerpen, Prinsstraat 13, 2000 Antwerpen, Belgium}

\address[nbi]{Niels Bohr Institute, University of Copenhagen,
  Blegdamsvej 17, DK-2010, Copenhagen, Denmark}

\address[kracow]{Institute of Nuclear Physics, Polish Academy of Sciences, ul. Radzikowskiego, Kracow 31-342, Poland}

\address[epfl]{Institute of Physics
Laboratory of Particle Physics and Cosmology
École Polytechnique Fédérale de Lausanne (EPFL)
CH-1015, Lausanne, Switzerland}


\address[vt]{Center for Neutrino Physics, Department of Physics, Virginia Tech University, Blacksburg, VA 24601, USA}

\address[sunysb]{C.  N. Yang Institute for Theoretical Physics and Department of Physics and Astronomy, Stony Brook University, Stony Brook, NY 11794}

\address[ipmu]{Kavli Institute for the Physics and Mathematics of the Universe (WPI), UTIAS, The University of Tokyo, Kashiwa, Chiba 277-8583, Japan}

\address[ift]{Instituto de Física Teórica UAM-CSIC, c/ Nicolás Cabrera 13-15, 28049 Madrid}

\address[queens]{Department of Physics, Engineering Physics and Astronomy, Queen's University, Kingston ON K7L 3N6, Canada}

\address[uta]{ Department of Physics, University of Texas, Arlington, TX 76019, USA}

\date{\today}

\begin{abstract}
The existence of non-zero neutrino masses points to the likely existence of multiple SM neutral fermions. When such states are heavy enough that they cannot be produced in oscillations, they are referred to as Heavy Neutral Leptons (HNLs). In this white paper we discuss the present experimental status of HNLs including colliders, beta decay, accelerators, as well as astrophysical and cosmological impacts. We discuss the importance of continuing to search for HNLs, and its potential impact on our understanding on key fundamental questions, and additionally we outline the future prospects for next-generation future experiments or upcoming accelerator run scenarios.
\end{abstract}

\maketitle

\tableofcontents

\section{Introduction}
\par Neutrinos have proven valuable in elucidating the structure of the Standard Model (SM) of particle physics. Though the SM provides the framework describing how neutrinos interact with leptons and quarks through weak interactions, the SM does not answer fundamental questions about neutrinos.   For many years, neutrinos were thought to be massless, and 
this property is built into the Standard Model.  The earliest evidence for neutrino masses appeared in the observed deficit of solar neutrinos in the Davis radiochemical neutrino experiment \cite{Davis:1968cp}.  The solar neutrino deficit was confirmed by the SAGE \cite{Abazov:1991rx} and GALLEX \cite{GALLEX:1992gcp} gallium experiments and the Kamiokande water Cherenkov experiment \cite{Kamiokande:1996qmi}.  Compelling evidence of neutrino oscillations was obtained by the Super-Kamiokande experiment in 1998 \cite{Super-Kamiokande:1998kpq}, followed by the evidence for transmutation of neutrino flavors by the SNO experiment \cite{SNO:2002tuh}.  Together, these experimental results have produced a very far-reaching transformation in our understanding of particle physics, demonstrating the existence of neutrino masses and lepton mixing.  This is the first confirmed physics beyond the original Standard Model in the laboratory.  Further knowledge has been gained by several generations of accelerator and reactor experiments, as well as further data from deep underground detectors.  We now have reasonably accurate measurements of $\Delta m_{21}^2$ and $|\Delta m_{32}^2|$, where $\Delta m_{ij}^2 = m_{\nu_i}^2 - m_{\nu_j}^2$, and the three Euler angles parametrizing the PMNS lepton mixing matrix, $\theta_{23}$, $\theta_{12}$, and $\theta_{13}$.  Current data yield preferred regions for the CP phase $\delta_{CP}$.  However, there are many fundamental questions concerning neutrinos that are still unanswered.  Perhaps most basic is the question of why their masses, while nonzero, are so small.  An appealing possible explanation for this is the seesaw mechanism \cite{Minkowski:1977sc,Gell-Mann:1979vob,Mohapatra:1979ia,Yanagida:1979as,Schechter:1980gr}, which posits the existence of a number of electroweak-singlet (``sterile") neutrino interaction eigenstates. However, this mechanism does not specify the number of electroweak-singlet neutrinos and, importantly, both high-scale and low-scale seesaw mechanisms are viable.  In the latter case, there can be additional neutrino mass eigenstates that might be observed in current and near future experiments.  Searches for these additional neutrinos thus probe one of the most fundamental and important questions in particle physics.  Indeed, neutrinos provide not only 
the first direct evidence for physics beyond the Standard Model, but also a pathway to search for new physics. 

HNLs with masses in the MeV to GeV scale are a compelling present and future target of opportunity. In \cite{Shrock:1980vy,Shrock:1980ct} it was proposed to search for emission of HNLs via the signature of kinks in the Kurie plots in nuclear beta decays and via anomalous peaks in the energy spectra of charged leptons in two-body leptonic decays of pseudoscalar mesons. These HNLs would also lead to an apparent deviation of the ratio $BR(\pi^+ \to e^+ \nu_e)/BR(\pi^+ \to \mu^+ \nu_\mu)$ and the analogous ratio for $K^+$ from their SM values, as well as an apparent deviation of the spectral parameters in $\mu$ and leptonic $\tau$ decay from their SM values.  In \cite{Shrock:1980vy,Shrock:1980ct,Shrock:1981wq}  analyses were carried out retroactively on existing data to probe for these effects and set upper bounds on HNL emission.  Early peak search experiments were carried out at SIN/PSI \cite{Abela:1981nf,Minehart:1981fv,Daum:1987bg}, TRIUMF \cite{Bryman:1983cja,Azuelos:1986eg,Britton:1992xv}, and KEK \cite{Asano:1981he,Hayano:1982wu}. Recent peak search experiments have been conducted by the E949 experiment at BNL \cite{E949:2014gsn}, the PIENU experiment at TRIUMF \cite{PIENU:2017wbj,PIENU:2019usb}, and the NA62 experiment at CERN  \cite{NA62:2017ynf,NA62:2017qcd,NA62:2020mcv,NA62:2021bji}. Other  accelerator experiments searching for HNLs include PS-191 \cite{Bernardi:1985ny,Bernardi:1987ek}, CHARM \cite{CHARMII:1994jjr,Boiarska:2021yho}, NuTEV \cite{NuTeV:1999kej}, DELPHI \cite{DELPHI:1996qcc}, T2K \cite{T2K:2019jwa}, and a recent reinterpretation of BOREXINO data \cite{Plestid:2020vqf}. The peak search experiments looking for anomalous monochromatic peaks in the charged lepton spectra of decaying $\pi^+$ and $K^+$ appear to place more stringent upper bounds on mixing angles of HNLs than any other terrestrial experiments over a large range of HNL masses from $\sim 40$ MeV to $\sim 2$ GeV. In the range from a few MeV to 40 MeV, the agreement of the ratio $BR(\pi^+ \to e^+\nu_e)/BR(\pi^+ \to \mu^+\nu_\mu)$ with the SM prediction also places stringent upper bounds on $|U_{e4}|^2$ \cite{Bryman:2019ssi,Bryman:2019bjg}.



\section{Theory of Heavy Neutral Leptons}
\label{sec:theory} 



\subsection{Neutrinos masses: evidence of new physics beyond the Standard Model}
\label{sec:numasses}

The discovery of neutrino oscillations just over 20 years ago has established that neutrino have masses and mix. A wide experimental programme has measured with precision most of the oscillation parameters. Two mass-squared differences have been determined, one controlling atmospheric, short-baseline-reactor, and accelerator-neutrino oscillations, and one long-baseline-reactor and solar-neutrino experiments. 
The $\Delta m^2_{21}$ mass-squared splitting is positive and is measured with very good accuracy to be $7.42 \ \times 10^{-5}~\mathrm{eV}^2$ with a 3$\sigma$ range of $6.82$--$8.04 \ \times 10^{-5}~\mathrm{eV}^2$~\cite{Esteban:2020cvm}. $\Delta m^2_{31}$ is known slightly less precisely and its sign is not yet known. The measured values are $\Delta m^2_{31}= 2.515 \  (2.431 $--$ 2.599)\times 10^{-3}~\mathrm{eV}^2$, for the best fit (3$\sigma$ range), which prefers the normal mass ordering (NO) and similarly for the inverted mass ordering (IO) $\Delta m^2_{32}= -2.498 \  (-2.584 $--$ -2.413)\times 10^{-3}~\mathrm{eV}^2$~\cite{Esteban:2020cvm}. 

There are three mixing angles and they control the flavour content of the three mass eigenstates, $|U_{\alpha i}|^2$.
 Their values are given by~\cite{Esteban:2020cvm}: 

\begin{eqnarray}
\theta_{12}= 33.44 \ (31.27--35.86) \phantom{-}& &\text{for \, both \, mass \, orderings},\\
\theta_{23}= 49.2 \ (39.5--52.0) \ \text{(NO)} & &
\theta_{23}= 49.5 \ (39.8--52.1) \  \text{(IO)},\\
\theta_{13}=8.57 \ (8.20-8.97) \ \text{(NO)}  &  &\theta_{13}= 8.60 \ (8.24-8.98) \ \text{(IO)},
\end{eqnarray} 
in degrees. Differently from the quark sector, all three angles are sizable and $\theta_{23}$ could even be maximal ($45^\circ$). The first hints of CP-violation, with $\delta$ close to $-90^\circ$, had been reported but recent data has pushed the preferred value towards CP-conservation at $\delta \sim 180^\circ$. In the coming years, data from the long baseline experiments T2K and NOvA will provide further information, and later DUNE and T2HK will be able to discover leptonic CP violation if $\delta$ is not very close to 0 or $180^\circ$. 


The measurement of two mass-squared differences implies that there are (at least) two massive neutrinos. They can be ordered in two ways:
\begin{itemize}
 \item normal ordering (NO): $m_1 < m_2 < m_3$, i.e. $\Delta m^2_{31} >0$, 
 \item inverted ordering (IO): $m_3 < m_1 < m_2$, i.e. $\Delta m^2_{32}< 0$. 
 \end{itemize}
  The three neutrino masses can be expressed in term of just one unknown parameter, the 
lightest neutrino mass, $m_{\mbox{}_\mathrm{MIN}}$, and the ordering, as
\begin{eqnarray}
m_1 = m_{\mbox{}_\mathrm{MIN}}, &  m_2 = \sqrt{ m_{\mbox{}_\mathrm{MIN}}^2 + \Delta m^2_{21}}, & m_3 = \sqrt{ m_{\mbox{}_\mathrm{MIN}}^2 + \Delta m^2_{31}} ,   \quad \text{for~NO};~~~\\
m_3 = m_{\mbox{}_\mathrm{MIN}}, &  m_1 = \sqrt{ m_{\mbox{}_\mathrm{MIN}}^2 + |\Delta m^2_{32}| - \Delta m^2_{21}}, & m_2 = \sqrt{ m_{\mbox{}_\mathrm{MIN}}^2 + |\Delta m^2_{32}|} , \quad \text{for~IO}. ~~~
\end{eqnarray}
Given these relations and the measured values of $\Delta m_{21}^2$ and $\Delta m_{31}^2$, the sum of the neutrinos can be constrained to be above
\begin{eqnarray}
\sum_{j = 1}^{3} m_j &\gtrsim 60\ \mathrm{meV},\quad \mathrm{(NO)}, \\
&\gtrsim 100\ \mathrm{meV},\quad \mathrm{(IO)}.
\end{eqnarray}
These provide useful benchmarks for cosmological and laboratory searches for the absolute scale of neutrino masses, to which oscillation experiments are not sensitive.

Presently, these searches for the absolute scale of neutrino masses can only place upper limits. The most stringent laboratory search to date, coming from KATRIN, constrains the quantity $m_{\beta} \equiv \sqrt{\sum_{j=1,2,3} \left|U_{ej}\right|^2 m_j^2} < 0.8$ eV~\cite{KATRIN:2021uub}, where $U_{ej}$ are elements of the neutrino mixing matrix (the electron row is well-measured by oscillation experiments). In contrast, cosmological searches constrain $\sum_{j=1,2,3} m_j < 0.12$ eV~\cite{eBOSS:2020yzd,Planck:2018vyg}.

Despite the wealth of information obtained so far, key open questions remain open:
\begin{itemize}
    \item What is the nature of neutrinos? Are neutrinos Dirac or Majorana particles?
    \item What are the absolute values of the masses? In order to answer this question it is necessary to establish the mass ordering and the overall mass scale.
    \item Is there leptonic CP violation? And if so, what is the precise value of the $\delta$ phase?
    \item What are the precise values of the mixing angles? Do they point towards an underlying flavour principle?
    \item Is the standard three-neutrino picture correct or are there other effects, such as sterile neutrinos, non-standard interactions or even more exotic
     ones, e.g. Lorentz-violation?
\end{itemize}
A wide programme of experiments running and under construction will address these questions and we can expect to have a complete picture in the next decade. This information is crucial in order to hunt for the origin of neutrino masses and leptonic mixing.

\subsection{General neutrino mass models and HNLs}
\label{sec:HNLnumasses}

Neutrino masses constitute so far the only particle physics evidence that the SM is incomplete. Not only are they nonzero, they are also much smaller than those of all other fermions and leptonic mixing presents a very different structure compared to that of the quarks. All these fact seem to point towards a different origin of neutrino masses whose unveiling will play a key part in extending the SM to a full theory of particles and interactions. In many cases, a key ingredient to explain neutrino masses are right handed neutrinos, which, for masses $\gg \mathrm{eV}$ we will refer to as HNLs. They enter in Yukawa couplings with the leptonic doublet and the Higgs field, they are predicted in left-right SU(2) gauge models and their embedding in GUT theories, such as SO(10). Their presence is also instrumental in explaining the baryon asymmetry of the Universe via leptogenesis~\cite{Davidson:2008bu}.

\subsubsection{Dirac and Majorana mass terms}
As neutrinos are neutral, they can be of two different types, Dirac or Majorana particles, corresponding to the conservation or not of the leptonic number symmetry. Differently from the case of Dirac particles, common to all other SM fermions, Majorana neutrinos satisfy the Majorana condition
\begin{equation}
    \nu = \nu^c \equiv C \bar{\nu}^T ~,
\end{equation}
where $C$ is the charge-conjugation matrix and $\nu^c$ is the charge-conjugate of $\nu$.
Effectively neutrinos and antineutrinos are not distinguishable. 
Neutrino masses can be of different types, leading to Dirac or Majorana neutrinos.

A Dirac mass term,  $\overline{\nu}  m_\mathrm{D} \nu = \overline{\nu}_L m_\mathrm{D} \nu_R + \mathrm{h.c.}$,  requires the introduction of $\nu_R$ and is analogous to  that of all the SM charged fermions. This term conserves lepton number: giving both chiral components the same lepton number so that under a $U(1)_\mathcal{L}$ transformation $\nu_{L,R} \rightarrow e^{i \eta } \nu_{L,R}$,
    the mass term remains invariant.
%

Majorana masses require only one Weyl spinor $\nu_L$ as  $- \mathcal{L}_{Majorana}= \frac{1}{2}  \overline{\nu^c} m_{\mathrm{M}} \nu = -\frac{1}{2}   \nu_L^T C^\dagger m_{\mathrm{M}} \nu_L + \mathrm{h.c.}$
    This term breaks lepton number by two units. Notice that analogously it is also possible to have Majorana masses for $\nu_R$. 
 If the theory contains also $\nu_R$, it is possible to have  Dirac mass term and both Majorana ones for $\nu_L$ and $\nu_R$. In this case, the resulting massive neutrinos are of Majorana type, as can be understood by the fact that overall the Lagrangian does not conserve lepton number. In a full theory with 3 $\nu_L$ and $N$ $\nu_R$, there will be 3 light Majorana neutrinos and a number $N$ of massive neutrinos that we will denote as HNLs, in the case in which that their masses are much larger than the eV scale. Typically, the latter will be nearly-sterile as only a small admixture of active neutrinos is allowed.

\subsubsection{Neutrino masses beyond the SM}

The mass terms discussed above are forbidden in the SM as there are no right-handed neutrinos and a Majorana mass term using only the $\nu_L$ fields violates the SM gauge invariance. Consequently, the SM, in its minimal form, does not allow for neutrino masses. How can one extend the SM in order to account for neutrino masses in a consistent framework? A vast number of models have been proposed and most them require sterile neutrinos and the corresponding HNL. We briefly review the key features and we discuss more in detail the Seesaw type I mechanism and HNLs in Sec.~\ref{sub:type-i_seesaw}. 

The SM simplest extension can be constructed by adding SM gauge singlets $\nu_R$, typically called sterile neutrinos. The SM gauge group allows for the Yukawa interaction between $\nu_R$, the leptonic doublet $L\equiv (\nu_L^T, \ell^T)^T$ and the Higgs doublet $H$:
\begin{equation}
    - \mathcal{L}_y = \overline{L} F \cdot \widetilde{\Phi} N + \mathrm{h.c.} ~,
\end{equation}
where $\widetilde{\Phi} = i \sigma_2 \Phi^\ast$. After electroweak symmetry breaking, $\langle \tilde{H} \rangle = (v_H/\sqrt{2}, 0)^T$, a Dirac mass term emerges
\begin{equation}
    - \mathcal{L}_y \xrightarrow{\langle \widetilde{H} \rangle\neq0} - \mathcal{L}_{Dirac} = \frac{v_H}{\sqrt{2}} \overline{\nu}_L y_\nu \nu_R + \mathrm{h.c.} 
\end{equation}
This Yukawa interaction and the resulting Dirac mass conserve lepton number. Indeed, a Majorana mass term for $\nu_R$ is not forbidden by gauge invariance and its absence must be imposed by requiring lepton number conservation. In this case, this symmetry goes from being an accidental symmetry of the SM to a fundamental ingredient of the theory of particle interactions. In this sense, this is a major departure from the Standard Model.

Although a viable explanation for neutrino masses, alternatives have been sought as this has several shortcomings: apart from the issue of lepton number discussed above, the Yukawa coupling needs to be exceedingly small, many orders of magnitude smaller than the charged-lepton ones, and moreover one would expect the masses and mixing angles to have a similar hierarchy as the quark ones, a fact that is contradicted by experimental results.

Going beyond dimension-four operators, the SM admits a dimension-five one which is gauge invariant. The so-called Weinberg operator~\cite{Weinberg:1979sa}
\begin{equation}
    \mathcal{L}_{M,BSM}= \frac{c^{[5]}}{\Lambda} L^T \cdot \widetilde{H}^\ast C^\dagger \widetilde{H}^\dagger \cdot L + \mathrm{h.c.} ~.
\end{equation}
indeed leads to Majorana masses
\begin{equation}
    \mathcal{L}_{M,BSM} \xrightarrow{\langle \widetilde{H} \rangle\neq0}  \frac{c^{[5]} v_H^2}{2 \Lambda} \nu_L^T C^\dagger \nu_L + \mathrm{h.c.} ~.
\end{equation}
after electroweak symmetry breaking. It violates lepton number by two units and consequently the massive neutrinos are of Majorana type. This operator suggests the existence of a new theory at a scale $\Lambda$. The hunt for the new particles and interactions involved is at the centre of much of current research in theoretical neutrino physics.

\begin{figure}
    \centering
    \includegraphics[width=1\textwidth]{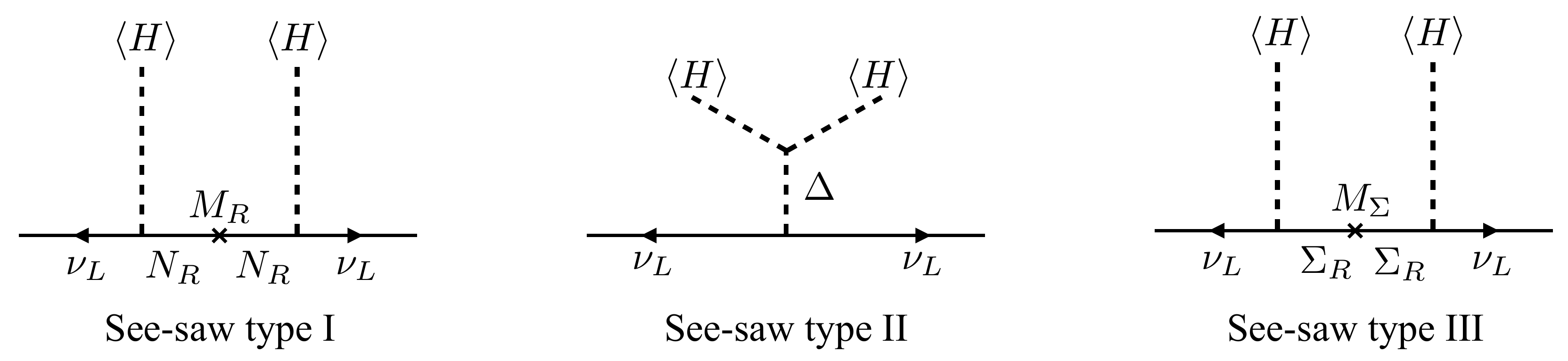}
    \caption{Diagrams contributing to neutrino masses in the three Seesaw scenarios at tree-level. $\langle H\rangle$ indicates the vev of the neutral component of the Higgs field. $\Delta$ is a scalar triplet and $\Sigma$ is the neutral component of a fermion triplet, with mass $M_\Sigma$. }
    \label{fig:see-saws}
\end{figure}

As the Weinberg operator is a low-energy effective term, the key question concerns the full theory which is completed above $\Lambda$, including new particles with mass $M \approx \Lambda$. Conserving Lorentz and gauge symmetries, at tree level, there are three possible options for the origin of this operator:
\begin{itemize}
    \item Seesaw type I~\cite{Minkowski:1977sc,Gell-Mann:1979vob, Mohapatra:1979ia,Yanagida:1979as,Schechter:1980gr} for a singlet fermion; 
    \item Seesaw type II~\cite{Magg:1980ut,Schechter:1980gr} using heavy triplet scalars;
    \item Seesaw type III~\cite{Foot:1988aq,Ma:1998dn} for triplet fermions,
\end{itemize}
as schematically shown in Fig.~\ref{fig:see-saws}.
These operators can be further embedded in a full theory such as GUT models, in which right-handed neutrinos emerge as a completion of the fermionic representations. These terms are not the only way to generate neutrino masses. Models in which they emerge at loop-level have been extensively studied~\cite{Zee:1980ai,Cheng:1980qt,Hall:1983id,Zee:1985id,Chang:1986bp,Babu:1988ki,Dev:2012sg,Zhang:2013ama,Lopez-Pavon:2012yda,Ballett:2019cqp}. 

A typical, albeit by no means ubiquitous, feature of most models behind neutrino masses is the presence of some new neutral fermionic states. Unless a new symmetry forbids a Yukawa interaction between these fermions and the leptonic doublet, generically mixing will arise with the active neutrinos, and potentially other neutral fermions as well. For mass scales in the MeV to multi-TeV, these states will be characterised as HNL with a distinct phenomenology and potentially the possibility to test them.

\subsubsection{The scale of new physics and HNLs}

The smallness of neutrino masses relates to the scale of the new physics but can be ascribed to different approaches (for broad reviews and summaries, see~\cite{Nath:2006ut,Abada:2007ux,Deppisch:2015qwa,Cai:2017jrq,Cai:2017mow,Agrawal:2021dbo,Han:2022qgg}):
\begin{itemize}
    \item Large $\Lambda$. From the Weinberg operator, we see that neutrino masses are suppressed (relative to the charged-fermion ones) by $v/\Lambda$. In many extensions, $\Lambda$ is taken to be very large, even close to the Grand Unification scale, $\sim 10^{14}$~GeV. In this case, couplings are not required to be small and the suppression of neutrino masses is achieved thanks to the heavy masses of the mediator. In this formulation, the Seesaw mechanisms can be embedded in GUT theories: for instance SO(10) models include right handed neutrinos required to implement a Seesaw type I mechanism.  Leptogenesis can be also be realised in this framework to explain the baryon asymmetry of the Universe~\cite{Fukugita:1986hr,Harvey:1990qw,Luty:1992un,Flanz:1994yx,Plumacher:1996kc,Covi:1996wh,Flanz:1996fb}. Despite these many advantages, a significant drawback is that these models cannot be directly tested. Some indirect hints could be present in the case of GUT models, such as proton decay and a possible link to gravitational waves due to cosmic strings~\cite{Dror:2019syi,King:2020hyd,Buchmuller:2019gfy,King:2021gmj}. Moreover, if a stabilising mechanism for the electroweak symmetry breaking scale, e.g. supersymmetry, is absent, the Higgs mass receives corrections which are too large unless $\Lambda < 10^7$~GeV~\cite{Vissani:1997ys,Casas:2004gh,Clarke:2015gwa,Brivio:2017dfq} for Seesaw type I or smaller than the TeV scale for Seesaw type II models.
\item Small couplings. This approach advocates $c^{[5]}\ll 1$. Although generically the naive expectation is for couplings to be of order 1, the SM itself presents many examples of small Yukawa couplings, as small as $10^{-6}$. It is therefore possible that the couplings entering in $c^{[5]}$ make it a small parameter, and the scale $\Lambda$ is much lower. The  advantage is that the new particles are in principle accessible directly at experiments as far as they couple sufficiently strongly to the SM. A lot of interest has been devoted recently to the TeV model as a prime target for searches at the LHC searches. For instance, TeV HNL would induce same-sign dileptons plus jets with no missing energy and lepton flavour violating signals, e.g.~\cite{delAguila:2008cj,Atre:2009rg,Dev:2013wba,Cai:2017mow,Pascoli:2018heg}, see Sec.~\ref{sec:collider}.
The fermion mass scale can be lowered even further for smaller couplings, with  HNL with GeV, MeV and even eV masses. These low energy Seesaw models have distinct signatures which depend on their mass and flavour mixing, as discussed in the other sections. Key signatures are energy peaks in meson decays, decays into visible SM particles, kinks and possible contributions to neutrinoless double beta decays and other lepton number violating processes, e.g. Refs.~\cite{Shrock:1980vy}, as we will discuss in Sec.~\ref{sec:nuclear} and \ref{sec:fixed}. The baryon asymmetry of the Universe could be explained also in this context via the ARS mechanism and resonant leptogenesis at low scales, see Sec.~\ref{sec:cosmo} and references therein including Refs.~\cite{Akhmedov:1998qx,Asaka:2005pn,Hernandez:2016kel,Hambye:2017elz,Drewes:2017zyw, Granelli:2020ysj,Klaric:2021cpi,Drewes:2021nqr}.
\item Quasi-conserved symmetries. As discussed earlier, if a lepton symmetry is preserved, Majorana neutrino masses are zero. Therefore, a small breaking of such symmetries would explain the smallness of masses. This is the principle the inverse~\cite{Mohapatra:1986aw,Mohapatra:1986bd,Bernabeu:1987gr,Gavela:2009cd} and linear~\cite{Akhmedov:1995ip,Akhmedov:1995vm}, as well as extended, Seesaws. This approach is considered ``natural" as setting the small parameters to zero restores the lepton symmetry. The smallness of neutrino masses derives from the partial cancellation of the contributions of multiple HNL to it. As the lepton number violating parameters are small, the heavy neutral lepton states are pseudo-Dirac, i.e. are Majorana particles with nearly degenerate masses and opposite CP-parity~\cite{Wolfenstein:1981kw,Petcov:1982ya}. They can have low scales, including the MeV-TeV one, and large mixing, making them accessible to collider and other terrestrial experiments. As the lepton symmetry is midly broken, lepton number violating processes in this type of framework are typically suppressed for Majorana masses above the electroweak scale while they remain potentially observable for long lives particles.
\end{itemize}

As neutrino experiments do not provide any indication of the scale of the new physics and theoretical considerations do not provide a unique preference for a given scale, both high energy, $\Lambda \gg \mathrm{TeV}$, and low energy $\Lambda < \mathrm{multi \ TeV}$ models can be considered. Above we discussed the advantages of high scale Seesaw models, mainly their simple explanation of the smallness of neutrino masses, their possible embedding in GUT theories and successful leptogenesis, but at the same time their drawbacks, namely their untestability and the possible destabilisation of the Higgs mass.

Low scale Seesaw models, and consequently the existence of HNLs, have strong support from various points of view (for an in-depth discussion see~\cite{Agrawal:2021dbo}). First of all, they are directly testable at experiments and therefore should be a prime choice for experimental searches in order to unveil the origin of neutrino masses. Secondly, in certain mass ranges, they also offer a framework to explain the baryon asymmetry and they can have connections with dark sectors, such as dark photons, dark matter, dark scalars, as discussed in Section~\ref{sec:beyondMinTheory}. In the minimal formulation of the $\nu$MSM, they can even explain evidence beyond the Standard Model, namely neutrino masses, dark matter and the baryon asymmetry, with a very small number of new particles and parameters. They are technically natural, meaning that small parameters are justified if the symmetry of the system is augmented by setting them to zero. In this sense, both options of low $\Lambda$ and quasi-preserved symmetries are natural as they related to the small breaking of a lepton symmetry. Finally, it is non trivial that the SM might be valid up to the Planck scale~\cite{Bezrukov:2012sa}. In this scenario, there is no intermediate scale above the electroweak one and neutrino masses must be generated by a low-energy mechanism, calling for $\Lambda$ below the electroweak scale. The current lack of hints of new physics at the LHC might provide hints in this direction~\cite{Dev:2013wba,Cai:2017mow}. 

\subsection{Type I Seesaw}
\label{sub:type-i_seesaw}

In this section we review the Type I Seesaw mechanism~\cite{Minkowski:1977sc,Gell-Mann:1979vob,Mohapatra:1979ia,Yanagida:1979as,Schechter:1980gr} in greater detail.
The Lagrangian of the model reads
\begin{equation}
	\mathcal{L} = \mathcal{L}_{SM} + i \bar{N}_{I} \gamma^\mu \partial_\mu 
	N_{I}
	- F_{\alpha I} \bar{L}_\alpha \tilde{\Phi} N_{I} - \frac{M_{I J}}{2} 
	\bar{N}_{I}^c N_{J} + h.c.,
	\label{eq:seesaw_I_Lagrangian}
\end{equation}
where $\mathcal{L}_{SM}$ is the Lagrangian of the SM, $N_{I}$ are right-handed (RH) neutrinos
labelled with the generation indices $I, J = 1, 2, 3$, $F_{\alpha I}$ is the matrix of 
Yukawa couplings, $L_\alpha$ are the left-handed lepton doublets labelled with
the flavour index $\alpha = e, \mu, \tau$ and $\tilde{\Phi} = i\sigma_2 \Phi^*$,
 $\Phi$ is the Higgs doublet.
The Majorana mass term introduces a New Physics (NP) scale, given by the mass of the new particles, which is related to the light neutrino masses via the Seesaw mechanism $m_\nu\sim F^2v^2/M$. It is intriguing that this naive Seesaw scaling leads to neutrino masses in the right ball park for $F=\mathcal{O}(1)$ and a NP scale of the order of a Grand
Unified scale, $10^{15}$ GeV. This is, however, far beyond the reach of our present and future colliders. Alternatively, the NP scale could very well be close to the electroweak scale and at the reach of our near future experiments. Indeed, for these low scales, the light neutrino
masses can be explained, in a technically natural way, with the additional suppression from an approximated symmetry \cite{Shaposhnikov:2006nn,Kersten:2007vk}
as in the inverse and linear Seesaw realizations~\cite{Mohapatra:1986aw,Mohapatra:1986bd,Bernabeu:1987gr,Branco:1988ex,Malinsky:2005bi,Gavela:2009cd}.
Moreover, the RH neutrino Majorana mass term violates lepton number. This source of lepton number violation could seed the excess of matter over antimatter in the Universe via Leptogenesis~\cite{Fukugita:1986hr}, as we will see in the next subsection and Sec.~\ref{sub:leptogenesis}.

After the electroweak symmetry breaking, a Majorana mass term is generated for the active neutrinos via the Seesaw mechanism:

\begin{equation}
m_\nu = \frac{v^2}{2}F M^{-1}F^T=\theta M \theta^T = U m U^T,
\label{eq:mnu}
\end{equation}
where $\theta=\frac{vF}{\sqrt{2}}M^{-1}$ is the mixing among active neutrinos and heavy mass eigenstates (the so called HNLs), $m$ are the light neutrino masses and $U$ is the PMNS matrix.
Notice that the matrix $\theta$ is essential for the type I Seesaw phenomenology since it connects the HNLs to the visible sector. In order to derive the above equation, we are only assuming the Seesaw hypothesis $Fv\ll M$. Upon diagonalization of the complete Majorana mass matrix, that includes active and $N$ components, a spectrum of HNLs (composed mainly by $N$ with a small active neutrino contribution) and light neutrinos (a superposition of active neutrinos with a small component of $N$) is obtained.

\paragraph{Symmetry protected scenarios}

Naively, the Seesaw relation~\eqref{eq:mnu} implies mixing angles too small to allow direct detection of the HNLs $|\theta|^2\sim m_\nu/M_h$.
Large mixing angles are instead naturally realized in so-called symmetry protected scenarios, which are associated to the approximate conservation of a generalized lepton number~\cite{Shaposhnikov:2006nn,Kersten:2007vk,Gavela:2009cd,Moffat:2017feq}.
To illustrate how small light neutrino masses are realized in such a scenario, let us consider the case with an even number of HNLs, in which
the full mass matrix of the neutrino sector then takes the form:
\begin{align}
    \begin{pmatrix}
    0 & v Y & v \epsilon Y^\prime\\
    v Y^T & \mu_1 M & M\\
    \epsilon v {Y^\prime}^T & M & \mu_2 M\\
    \end{pmatrix}\,.
\end{align}
The matrices $Y$ and $Y^\prime$ are Yukawa couplings, $M$ is a matrix that determines the scale of the HNL masses, and $\mu_i$ and $\epsilon$ are small lepton number breaking parameters.
The size of the light neutrino masses is directly proportional to these symmetry breaking parameters:
\begin{align}
    m_\nu = v^2 \epsilon \left(Y^\prime M^{-1} Y^T + Y M^{-1} {Y^\prime}^T\right) -v^2 Y \mu_2 M^{-1} Y^T\,,
\end{align}
with the limiting cases: i)  $\mu_2 \neq 0$ and $\mu_1=\epsilon=0$, which correspond to the inverse Seesaw~\cite{Wyler:1982dd,Mohapatra:1986bd,Mohapatra:1986aw,Bernabeu:1987gr,Gonzalez-Garcia:1988okv} (ISS) and ii) for $\epsilon  \neq 0$ and $\mu_1=\mu_2=0$ corresponding to the linear Seesaw~\cite{Akhmedov:1995vm,Akhmedov:1995ip,Barr:2003nn,Malinsky:2005bi,Gavela:2009cd} (LSS)-like scenarios. The suppression with $\mu_i$ and $\epsilon$ allows to generate light neutrino masses compatible with large Yukawa couplings $Y$. Thus, in this scenarios, the active-heavy neutrino mixing can be much larger than the naive Seesaw scaling since is it not suppressed by the small lepton number breaking parameters, $\theta\sim YvM^{-1}$. Another important feature of these scenarios is that the same parameters responsible for the smallness of the light neutrino masses can also suppress LNV observables~\cite{Kersten:2007vk,Ibarra:2010xw,Moffat:2017feq}.
This parametric suppression might not be sufficient to prevent observable LNV effects in HNL decays:
if HNL oscillations are fast compared to their lifetime,
LNV is generally no longer suppressed - therefore the ratio of the physical mass splitting to their decay width is the decisive parameter for LNV~\cite{Kersten:2007vk,Deppisch:2015qwa,Anamiati:2016uxp,Boyanovsky:2014una,Antusch:2020pnn}
(also see~\cite{Cvetic:2015naa,Dib:2016wge,Arbelaez:2017zqq,Balantekin:2018ukw,Hernandez:2018cgc,Boyanovsky:2014una,Cvetic:2015ura,Antusch:2017ebe,Cvetic:2018elt,Tastet:2019nqj} for examples of such processes at specific experiments).
In the minimal scenario with
two HNLs, the measured light neutrino masses impose a lower bound on the LNV parameters, and therefore indirectly on the physical mass splitting between the two HNLs, which implies that LNV effects are expected to be generically observable for long-lived HNLs (with $100$ MeV $ \lesssim M_h \lesssim 10$ GeV) in the absence of fine tuning~\cite{Drewes:2019byd}.

\paragraph{Connection between the HNL and light neutrino sectors: complementary among different observables}
The generation of the light neutrino masses and mixing measured in neutrino oscillation experiments establishes the a link between the HNL and active sectors through the Seesaw formula~\eqref{eq:mnu}. This is translated into the following constraint on the mixing between active neutrinos and HNLs~\cite{Casas:2001sr}

\begin{equation}
\theta\simeq iU m^{1/2}R^\dagger M^{-1/2},
\label{eq:theta}
\end{equation}
where $R$ is a $3\times n$ complex orthogonal matrix associated to the HNL sector and $n$ is the number of $N$ states.
Thus, in addition to the dependence on the parameters associated to the RH neutrino sector (encoded in $R$ and $M$), the mixing $\theta$ depends crucially on the active neutrino parameters: the mixing angles and CP phases included in the PMNS matrix $U$ and the light neutrino masses present in $m$.
As a consequence, a very interesting complementarity among different observables arises in the context of this model. 

As a paradigmatic example, let us start considering the most minimal model in which the number of $RH$ neutrinos is $n=2$. This is the simplest model since it comprises the minimum number of extra degrees of freedom required to accommodate the two squared mass differences observed in neutrino oscillations.
In this model the lightest neutrino mass, for which we currently have only upper bounds from the experiments, is zero and, consequently, the PMNS matrix contains only one Majorana phase (in addition to the Dirac phase $\delta$).

\begin{figure}[t!]
\begin{center}
\includegraphics[width=0.45\textwidth]{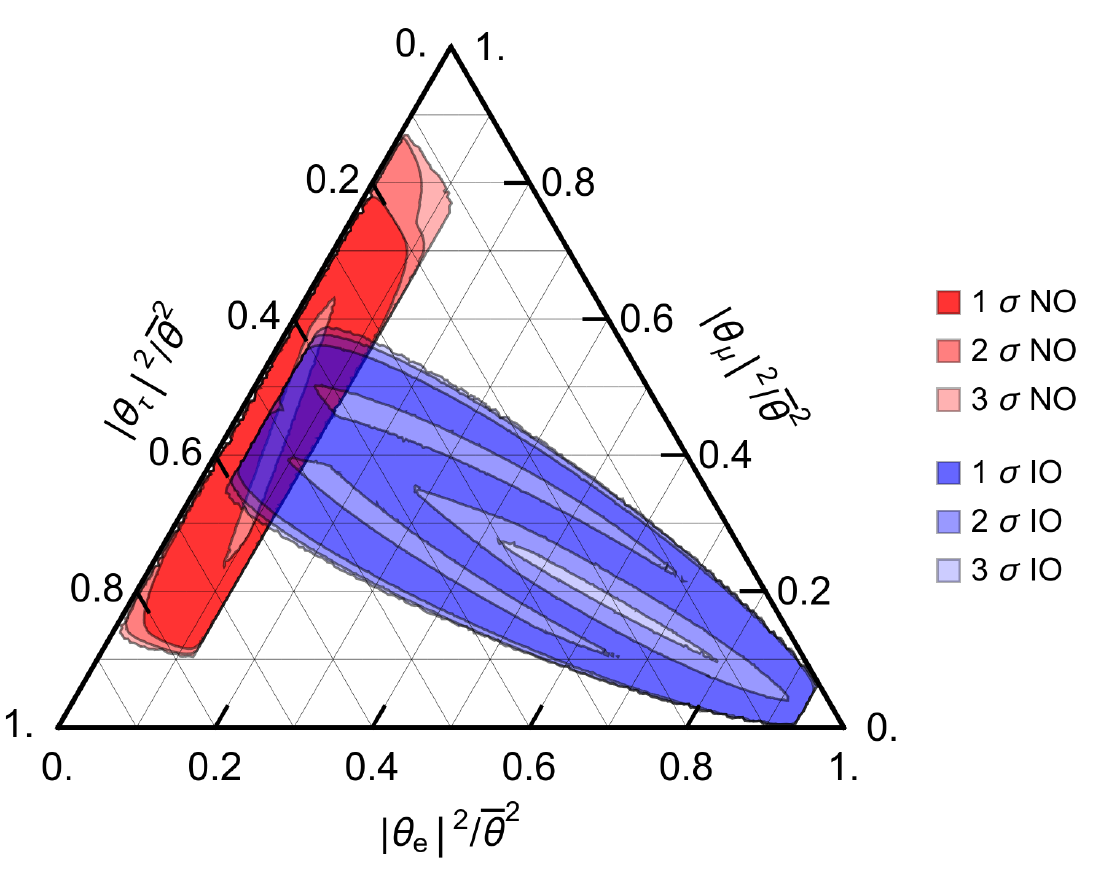}\includegraphics[width=0.5\textwidth]{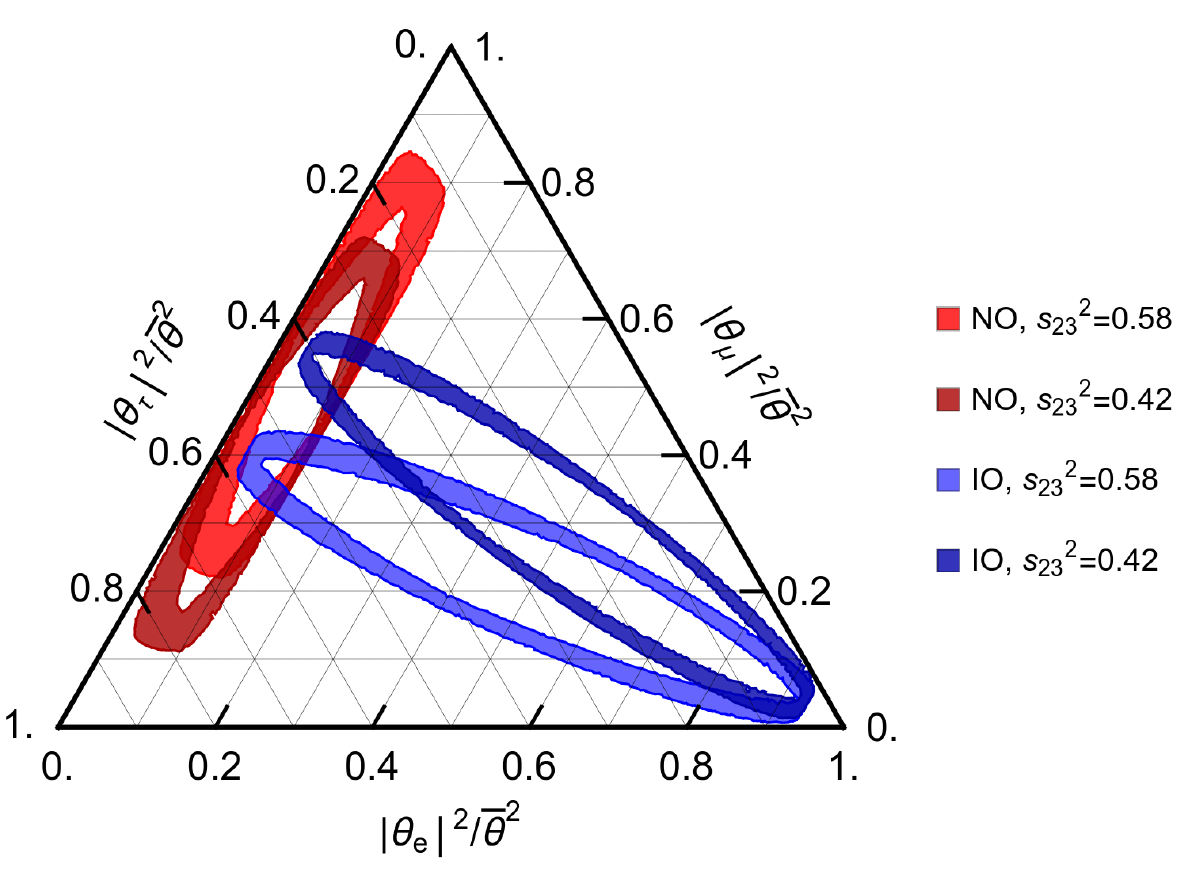}
\caption{\label{fig:triangle}Ternary diagrams  for the normalized flavour mixings $|\theta_{\alpha h}|^2/\bar{\theta}^2$
(in the large mixing regime) with $\bar{\theta}^2\equiv\sum_{\alpha} |\theta_{\alpha h}|^2$. \textit{Left panel}: The range of normalized flavour mixings consistent with the current neutrino oscillation data in the $n=2$ case (see e.g.~\cite{Drewes:2016jae,Caputo:2017pit,Antusch:2017pkq,Drewes:2018gkc}). The different contours correspond to the allowed $\Delta \chi^2$ range from~\cite{Esteban:2020cvm} for the case of NH (red) and IH (blue).
\textit{Right panel}: The projected $90\%$ CL for the mixing ratios in the case of $n=2$, assuming 14 years of neutrino oscillation measurements at DUNE~\cite{DUNE:2020jqi}.
For the projection we assume maximal CP violation $\delta=-\pi/2$ and two benchmark values of the PMNS angle $\theta_{23}$, used in the DUNE TDR~\cite{DUNE:2020ypp}, indicated in the legend.
}
\end{center}
\end{figure}

In the left panel of Fig.~\ref{fig:triangle} we show the allowed flavor structure of the HNLs mixing $\theta$ as dictated by current neutrino oscillation data extracted from~\cite{Esteban:2020cvm}.
The results shown in the figure apply to the large mixing regime $|\theta_{\alpha h}|^2>\mathcal{O}\left(m_\nu/M_h\right)$, which corresponds to the region of the parameter space that can be experimentally probed in the near future for scales in the $\mathcal{O}\left(0.1-100\,\rm{GeV}\right)$ range.
Remarkably, this large mixing regime is automatically realized in symmetry protected scenarios as the linear or inverse Seesaws.
The current allowed regions for the $\theta$ flavor structure will be further reduced thanks to neutrino oscillation measurements in experiments as DUNE~\cite{DUNE:2020jqi} or T2HK~\cite{Hyper-KamiokandeProto-:2015xww,Hyper-Kamiokande:2018ofw}.
Particularly relevant will be the information about the Dirac CP phase $\delta$ and the precise measurement of $\theta_{23}$ (the mixing angle that currently holds the largest uncertainty).
This is illustrated in the right panel of Fig.~\ref{fig:triangle} where we show the future projection from DUNE assuming maximal CP violation for the Dirac CP phase ($\delta=-\pi/2$, pointing to the direction of current hints) and two true values for $\theta_{23}$ at the extremes of its current allowed region.

Furthermore, it should be remarked that future neutrino oscillation data will surely provide the measurement of the light neutrino ordering (NO or IO in Fig.~\ref{fig:triangle}) further improving the predictivity of the minimal model.
Fig.~\ref{fig:triangle} clearly shows the potential of direct HNL searches to falsify this minimal neutrino mass mechanism. A measurement falling inside the colored regions would be a strong indication that the discovered new particles are involved in the light neutrino mass generation.
Moreover, since $|\theta_\alpha|/|\theta_\beta|$ depends mainly on the CP
phases of the PMNS matrix, a precise experimental determination of the HNL mixing with the different flavors would also allow to indirectly determine the leptonic CP phases~\cite{Hernandez:2016kel,Caputo:2016ojx, Drewes:2016jae}, including the Majorana phase whose measurement is extremely challenging. This would allow for a complete determination of the PMNS matrix $U$. However, to fully test this minimal model we would also need to measure all the remaining parameters encoded in $R$ and $M$. While the measurement of all the mixing matrix elements $|\theta_{\alpha I}|^2$ can in principle allow for a full reconstruction of the parameters~\cite{Drewes:2016jae}, this is extremely challenging in practice for the CP phase associated to $R$, which would require a experimental precision $\Delta \theta^2 \lesssim m_\nu/M_h$. Sensitivity to this CP phase can be potentially obtained via the effective interference between the active and HNL contributions to the neutrinoless double beta decay rate~\cite{Hernandez:2016kel}. Complementary information on the value of the HNLs mass splitting can be provided by LNV observables, such as the CP properties of the HNLs~\cite{Bray:2007ru,Cvetic:2015naa}, the correlations among their decay products~\cite{Dib:2016wge,Arbelaez:2017zqq,Balantekin:2018ukw,Hernandez:2018cgc}, or even through direct measurements of coherent HNL oscillations~\cite{Boyanovsky:2014una,Cvetic:2015ura,Anamiati:2016uxp,Antusch:2017ebe,Cvetic:2018elt,Tastet:2019nqj}.

The flavor structure of the HNL mixing $\theta$ is less constrained for the $n=3$ scenario since the number of free parameters is larger than in the $n=2$ case. However, the predicitivity of the model is deeply linked to the magnitude of the lightest neutrino mass (see Fig.~11 from~\cite{Chrzaszcz:2019inj}), another observable that can be probed in the near future by CMB observations~\cite{Abazajian:2019eic}, galaxy surveys such as DESI~\cite{DESI:2016fyo} or Euclid~\cite{Audren:2012vy} or laboratory experiments such as KATRIN~\cite{KATRIN:2001ttj} and Project 8~\cite{Project8:2017nal}. Cosmological observables as CMB and BBN also provide crucial information regarding the scale of the HNLs~\cite{Dolgov:2000pj,Dolgov:2000jw,Dolgov:2003sg,Cirelli:2004cz,Melchiorri:2008gq}. For instance, in~\cite{Hernandez:2014fha} it was found that: (i) masses below $100$ MeV are excluded for the three HNLs as long as $m_{\rm{lightest}}\gtrsim 10^{-3}$ eV; (ii) if $m_{\rm{lightest}}\lesssim10^{-3}$ eV, the same constraint still applies for two of the heavy states but the third HNL can still have a mass lighter than $100$ MeV (depending on the value of its corresponding mixing). A more detailed discussion regarding the cosmological bounds can be found in Sec.~\ref{sec:cosmo}.

The information from cosmology is highly complementary to that from neutrinoless double beta decay ($0\nu\beta\beta$) and neutrino oscillation searches. First of all, the latter leads to a prediction for the magnitude of the active neutrino contribution to the $0\nu\beta\beta$ decay rate.
Additionally, the HNLs can also mediate this lepton number violating process. A sizable contribution (analogous to the long range active neutrino one) is expected for HNL masses lighter than the typical momentum exchange of the process ($\lesssim 100$ MeV), a region that is however subject to stringent constraints from cosmology.
There is still room for a HNL contribution comparable or larger to the active neutrino one for masses in the range $100\, \rm{MeV}\lesssim M_j \lesssim 10 \,\rm{GeV}$, and even for heavier scales if a fine tuned cancellation between the tree and 1 loop induced contributions to the light neutrino masses takes place~\cite{Mitra:2011qr,Lopez-Pavon:2012yda,Lopez-Pavon:2015cga,Bolton:2019pcu}.

In summary, the light neutrino mass generation mechanism introduces a connection between the HNL and active neutrino sectors that can be considered as guidance regarding the experimental
searches.
Complementarity among different observables is a key issue.
The physics reach of the combination of all complementary cosmological and particle physics data is significantly higher than the compilation of independent results.
A good example is that, 
in the minimal Seesaw model, future measurements from neutrino oscillations, $0\nu\beta\beta$ decay and direct search experiments, can provide sufficient information to severely constrain the HNL parameters and falsify~\cite{Asaka:2010kk,Canetti:2012vf,Canetti:2012kh,Hernandez:2015wna,Abada:2015rta,Drewes:2016jae,Drewes:2016gmt,Drewes:2016lqo,Asaka:2016zib,Antusch:2017pkq,Eijima:2018qke,Klaric:2020phc,Klaric:2021cpi}
or even directly test~\cite{Hernandez:2016kel,Drewes:2016jae} the hypothesis that the matter-antimatter asymmetry of the universe has been generated via low scale leptogenesis, see~\ref{sub:leptogenesis}.

\subsection{Neutrino Minimal Standard Model (\texorpdfstring{$\nu$MSM}{nuMSM})\label{sec:nuMSM}}


It is not only neutrino masses that the SM fails to address.
The well established empirical signs of particle physics beyond the Standard Model include the baryon asymmetry of the universe (BAU), and Dark Matter (DM).
It is intriguing that all above-mentioned problems can be solved by the Neutrino Minimal Standard Model ($\nu$MSM)~\cite{Asaka:2005an,Asaka:2005pn}. The $\nu$MSM  extends the particle content of the SM by three right-handed neutrinos $N_{1, 2, 3}$.
Two heavier particles $N_{2, 3}$ generate the masses of active neutrinos via the type I Seesaw mechanism (see section~\ref{sub:type-i_seesaw}).
The same two right-handed neutrinos are also responsible for generating the BAU provided that their masses are close to each other~\cite{Akhmedov:1998qx,Asaka:2005pn}.
The lightest sterile neutrino $N_1$ is the DM candidate~\cite{Dodelson:1993je,Shi:1998km,Abazajian:2001nj,Asaka:2006nq,Laine:2008pg}.
The requirement to be a viable DM candidate forces the Yukawa couplings of $N_1$ to be tiny, leaving the lightest active neutrino almost massless~\cite{Asaka:2005an,Boyarsky:2006jm}.\footnote{This prediction can potentially be tested by the Euclid space mission~\cite{Audren:2012vy} or by directly measuring the mass of the lightest neutrino in an experiment like KATRIN~\cite{KATRIN:2001ttj}.}
Interestingly, the mass degeneracy of $N_{2,\,3}$ along with the tiny couplings of $N_1$ can be a consequence of a slightly broken global symmetry~\cite{Wyler:1982dd,Mohapatra:1986bd,Branco:1988ex,Gonzalez-Garcia:1988okv,Shaposhnikov:2006nn,Kersten:2007vk}.
The measured values of the Higgs and top quark masses are such that the $\nu$MSM is a consistent effective theory to very high scales, possibly all the way up to the Planck scale~\cite{Shaposhnikov:2007nj,Bezrukov:2012sa,Buttazzo:2013uya,Bezrukov:2014ina}.


\paragraph{Leptogenesis in the $\nu$MSM}

In the $\nu$MSM, the baryon asymmetry is generated through
combined action of anomalous processes with fermion number non-conservation \cite{Kuzmin:1985mm} and lepton number and flavor violating reactions involving heavy neutrinos.
This mechanism is known as \emph{low-scale leptogenesis},  where ``low'' refers to the mass scale of $N_{2,3}$ which is much lower than $\sim 10^{9}$~GeV.\footnote{The scale $\sim 10^{9}$~GeV is known as the Davidson-Ibarra bound~\cite{Davidson:2002qv}. Above this scale the usual ``vanilla'' leptogenesis~\cite{Fukugita:1986hr} with hierarchical HNL mass spectrum is possible.}
. Depending on the parameters, the majority of the asymmetry can be produced during \emph{freeze-in} or \emph{freeze-out} of HNLs~\cite{Klaric:2020phc,Klaric:2021cpi}. \emph{Leptogenesis via neutrino oscillations}~\cite{Akhmedov:1998qx,Asaka:2005pn} is usually associated with freeze-in, while \emph{resonant leptogenesis}~\cite{Liu:1993tg,Flanz:1994yx,Flanz:1996fb,Covi:1996wh,Covi:1996fm,Pilaftsis:1997jf,Pilaftsis:1997dr,Pilaftsis:1998pd,Buchmuller:1997yu,Pilaftsis:2003gt,Pilaftsis:2005rv} is typically freeze-out dominated.
In the recent years these last scenarios have received significant attention, see e.g.~\cite{Shaposhnikov:2006nn,Shaposhnikov:2008pf,Canetti:2010aw,Asaka:2010kk,Anisimov:2010gy,Asaka:2011wq,Besak:2012qm,Canetti:2012vf,Drewes:2012ma,Canetti:2012kh,Shuve:2014zua,Bodeker:2014hqa,Abada:2015rta,Hernandez:2015wna,Ghiglieri:2016xye,Hambye:2016sby,Hambye:2017elz,Drewes:2016lqo,Asaka:2016zib,Drewes:2016gmt,Hernandez:2016kel,Drewes:2016jae,Asaka:2017rdj,Eijima:2017anv,Ghiglieri:2017gjz,Eijima:2017cxr,Antusch:2017pkq,Ghiglieri:2017csp,Eijima:2018qke,Ghiglieri:2018wbs,Ghiglieri:2019kbw,Bodeker:2019rvr,Ghiglieri:2020ulj,Klaric:2020phc,Klaric:2021cpi,Domcke:2020ety,Eijima:2020shs,DeSimone:2007edo,DeSimone:2007gkc,Garny:2009qn,Garny:2011hg,Iso:2013lba,Garbrecht:2011aw,BhupalDev:2014pfm,BhupalDev:2014sxe,Garbrecht:2014aga,Dev:2015wpa,Hambye:2016sby,Jiang:2020kbt}).
The parameter space of the model has been thoroughly studied (see section~\ref{sub:leptogenesis}). Existing studies of the parameter space still contain $\mathcal{O}(1)$ uncertainties and a precision study of the full parameter space is highly desirable.

\paragraph{Sterile neutrino dark matter production}
Assuming vanishing initial abundance after inflation~\cite{Bezrukov:2008ut}, DM production in the $\nu$MSM can only occur through mixing with active neutrinos~\cite{Dodelson:1993je,Shi:1998km,Abazajian:2001nj,Asaka:2006rw,Asaka:2006nq,Laine:2008pg}.\footnote{In the $\nu$MSM augmented with Higgs inflation~\cite{Bezrukov:2007ep} HNLs can be produced from higher-dimensional operators~\cite{Bezrukov:2011sz}. $N_1$ can also be produced by a universal four-fermion interaction which is inevitably present in the Einstein-Cartan formulation of gravity~\cite{Shaposhnikov:2020aen}.} The production is efficient in the presence of large lepton asymmetry at temperatures $\mathcal{O}(200)$~MeV~\cite{Shi:1998km,Abazajian:2001nj,Laine:2008pg,Venumadhav:2015pla,Ghiglieri:2015jua,Ghiglieri:2019kbw,Ghiglieri:2020ulj,Bodeker:2020hbo}.
Such an asymmetry is generated in the $\nu$MSM provided that the mass splitting between $N_{2,\,3}$ is tiny~\cite{Shaposhnikov:2008pf}.
The requirement of successful DM production in the $\nu$MSM is the most limiting one~\cite{Roy:2010xq,Canetti:2012kh,Canetti:2012vf,Ghiglieri:2019kbw} and
a comprehensive study of the parameter space accounting for the recent theoretical progress is necessary.

\subsection{Beyond minimal HNL models}\label{sec:beyondMinTheory}

HNL can be considered in isolation, as in minimal extensions of the SM discussed in the previous sections, or they could be part of an extended hidden sector, with new particles and interactions. Indeed, the latter extensions are necessary to explain the origin of their masses via new Yukawa interactions. Hidden sectors could have new gauge symmetries, as in dark sectors, in which a secluded $U(1)_X$ extends the SM gauge group, or in Left-Right models in which parity is restored via a $SU(2)_L\times SU(2)_R$ symmetry at high energies.

%
Historically, the introduction of right handed neutrinos and the consequent Seesaw mechanism was developed in the context of larger gauge groups,
driven by general principles, such as grand unified theories (GUTs) of $SU(5)$~\cite{Glashow:1979nm}, 
$SO(10)$~\cite{Gell-Mann:1979vob}, family symmetries~\cite{Yanagida:1980xy} and parity restoration~\cite{Minkowski:1977sc, Mohapatra:1979ia}.
In the case of GUTs, perhaps the most appealing might be $SO(10)$, where the entire generation
of the SM fermions and the heavy RH Majorana neutrino reside in the spinorial $16_F$
representation.
Despite its mathematical appeal in the matter sector, the Higgs sector of minimal renormalizable models
is rather complicated and the scale of $m_N$ typically is high, from 
$10^{10} - 10^{14}$ GeV, due to proton decay constraints.
This is certainly out of reach of colliders and other direct searches.
Partial unification, such as Pati-Salam~\cite{Pati:1974yy} group $SU(4)_c \times SU(2)_L \times SU(2)_R$ are also
subject to flavor constraints and are inaccessible to present day colliders.
However, parity restoration and Left-Right symmetric models, discussed below, may 
be within reach of the LHC and the upcoming near-future experiments.
Here, we focus on mass scales for the HNL below the multi-TeV one, leaving aside the discussion of GUT-inspired models.

\subsubsection{Left-right models}

The minimal Left-Right (LR) symmetric model based on the $SU(3)_c \times SU(2)_L \times SU(2)_R 
\times U(1)_{B-L}$ gauge group~\cite{Mohapatra:1974gc} may explain parity violation
of weak interaction by restoring it and high energy scales and breaking $SU(2)_R 
\times U(1)_{B-L}$ to $U(1)_Y$ spontaneously~\cite{Senjanovic:1975rk}.
In the minimal model~\cite{Minkowski:1977sc, Mohapatra:1979ia} with scalar triplets
$\Delta_{LR}$, the masses of Majorana neutrinos $N$ is tied to the LR breaking scale.
Moreover, one has two sources of neutrino mass
\begin{align}
    M_\nu = -M_D^T M_N^{-1} M_D + \frac{v_L}{v_R} M_N \, ,
\end{align}
where the first term is usually referred to as type I and the second as type II 
contribution~\cite{Mohapatra:1980yp}.
In contrast to the mostly sterile HNLs in type I Seesaw, the HNLs now also carry gauge
interactions, mediated by heavy $W_R, Z_{LR}$ gauge bosons.
The Majorana masses $m_N$ and LNV originate from a Yukawa interaction with the triplet Higgses, 
and thus $N$'s may also couple to the SM sector via such interactions and scalar mixing.
This leads to couplings such as $hNN$ and $gg\Delta_R$, which makes it possible to produce
$N$s at colliders in the Higgs sector as well.

The desire to restore parity leads to imposition of near-hermiticity or symmetricity on
all the Yukawa matrices, effectively leading to $V_R = V_L = V_{\text{ckm}}$.
With charged gauge currents flavor structure fixed, flavor constraints coming from $K$ and
$B$ meson mixing come about.
Early studies~\cite{Beall:1981ze} had the lower limit $M_{W_R} > 2 \text{ TeV}$, which 
has been revisited~\cite{Zhang:2007da, Maiezza:2010ic} and updated to the current lower limit 
of about a few TeV, depending on the choice of LR parity~\cite{Maiezza:2014ala}, 
see~\cite{Bertolini:2019out} for recent reappraisal.
While the early bound did not motivate the $W_R$ searches at the Tevatron, the LHC has indeed
surpassed the flavor constraints and is in the position to search for $N$s even in the
minimal LR model, as discussed in section~\ref{sec:collider} below.
Finally, the Dirac mass matrices $M_D$ are not arbitrary in the LR model, once parity
is imposed.
This in contrast to the type I Seesaw, where there is significant freedom,
encoded in the Casas-Ibarra parametrization~\cite{Casas:2001sr}, in the couplings that control 
the $\nu-N$ mixing angle.
This connection between Majorana and Dirac mass matrices is particularly clean in the case of 
$\mathcal C$-parity~\cite{Nemevsek:2012iq} and a bit more involved in the case of 
$\mathcal P$~\cite{Senjanovic:2016vxw}.

Apart from the collider signatures, low scale LR symmetry with $M_{W_R} \sim \text{ TeV}$ can manifest
itself in many other phenomenological observables, related to lepton number and flavor violation.
To focus on the most fundamental question of lepton number violation, one should point out the 
interplay between the LHC and neutrinoless double beta decay searches~\cite{Maiezza:2010ic,Tello:2010am,Cai:2017mow}.
As pointed out time ago~\cite{Mohapatra:1980yp}, LNV can manifest itself also in $0\nu2\beta$ not
only through the exchange of light neutrinos, but also via parity inverted diagrams, where $W_L, \nu_L$ are
swapped for their right-handed counterparts.
This interplay was exemplified in the case of type II dominance~\cite{Tello:2010am} and in the presence 
of additional interactions~\cite{Barry:2013xxa}.
The upcoming $0\nu2\beta$ decay searches will thus complement the LNV searches at
the LHC, as seen on the left of Fig.~\ref{fig:LHC_WR_outlook}.

\subsection{Dark sector models}

Dark sector models advocate new particles and interactions at scales below the electroweak one and have gathered much interest in recent years. The key idea is that they contain new particles with feeble couplings to the SM, hence the name of ``dark", ``hidden" or ``secluded" sectors. They can include different type of extensions, with a new gauge sector, new scalar and/or fermions, or a combination of them, as in three-portal models [41]. Each of these option comes with its own connection to the SM, named vector, scalar and neutrino portals, respectively. Their popularity stems from their ability to induce new properties for neutrinos, to embed dark matter beyond the WIMP paradigm, to explain several low energy anomalies and to a blooming programme of searches with a variety of experimental strategies. HNLs have been considered in the context of dark sector models, specifically in fermionic extensions, and can connect to the SM via mixing between sterile neutrinos, dark fermions charged under the new interactions and the standard neutrinos. 

Studies have considered new interactions based on gauging lepton number, such as $B - L$, $L_\mu-L_\tau$, and other combinations, or a completely secluded new gauge symmetry under which no SM fields are charged~\cite{Okada:2014nsa, Diaz:2017edh, Nomura:2018ibs, Hagedorn:2018spx, Shakya:2018qzg, Pospelov:2011ha, Batell:2016zod}. 
Models of this type have been invoked to generate large
neutrino nonstandard interactions~\cite{Farzan:2015doa,Farzan:2016wym,Heeck:2018nzc,Han:2022qgg}, generate new
signatures in DM experiments~\cite{Pospelov:2011ha, Pospelov:2012gm, Pospelov:2013rha, Harnik:2012ni, McKeen:2018pbb}, weaken cosmological
and terrestrial bounds on eV-scale sterile neutrinos for short-baseline neutrino anomalies~\cite{Hannestad:2013ana,Dasgupta:2013zpn, Mirizzi:2014ama,Chu:2015ipa,Cherry:2016jol, Chu:2018gxk, Denton:2018dqq, Esmaili:2018qzu},
and as a potential novel explanation of the MiniBooNE~\cite{MiniBooNE:2007uho, MiniBooNE:2008yuf, Polly:2011zz, MiniBooNE:2013uba, MiniBooNE:2018esg, MiniBooNE:2020pnu} and/or LSND~\cite{LSND:1996ubh,LSND:1997vun} anomalies exploiting new degrees of freedom at the MeV/GeV scale~\cite{Gninenko:2009ks,Gninenko:2010pr,Masip:2012ke,Ballett:2018ynz,Bertuzzo:2018itn,Abdullahi:2020nyr}. A strong connection can also be present with dark matter, especially in the case of non-WIMP candidates.
Dark matter may interact with the SM through several kinds of neutrino portals, with a host of novel cosmological and phenomenological implications~
\cite{Pospelov:2007mp,Falkowski:2009yz,Cherry:2014xra,Bertoni:2014mva,GonzalezMacias:2015rxl,Gonzalez-Macias:2016vxy,Escudero:2016tzx,Escudero:2016ksa,Tang:2015coo,Tang:2016sib,Campos:2017odj,Batell:2017rol,Batell:2017cmf,Schmaltz:2017oov,Berlin:2018ztp,Kelly:2019wow,Blennow:2019fhy,Ballett:2019pyw,Cosme:2020mck,Hurtado:2020vlj,Kelly:2021mcd,Coito:2022kif}.

Due to the new interactions and multiple portals to the SM, such an extension can leave
imprints not just in neutrino experiments but also in
e.g. dark photon and dark scalar searches. Interestingly,
they have been advocated to explain current anomalies, specifically the long-standing $g_\mu-2$ anomaly~\cite{Muong-2:2004fok, Bennett:2006fi, Muong-2:2021ojo}, as a positive contribution to $\Delta a_\mu$ can emerge via  the existence of a kinetically-mixed dark vector boson, the $Z^\prime$. Bounds, such as those from $e^+e^-$ colliders and beam dumps can be avoided in some cases due to the rapid, semi-visible decays of the HNLs~\cite{Mohlabeng:2019vrz}. Indeed, the phenomenology of these models can be significantly different and requires a careful reevaluation of existing limits~\cite{Ballett:2019pyw,Abdullahi:2020nyr}.


\subsubsection{Dark sector HNL phenomenology}

\begin{figure}
   \centering
   \includegraphics[width=0.5\textwidth]{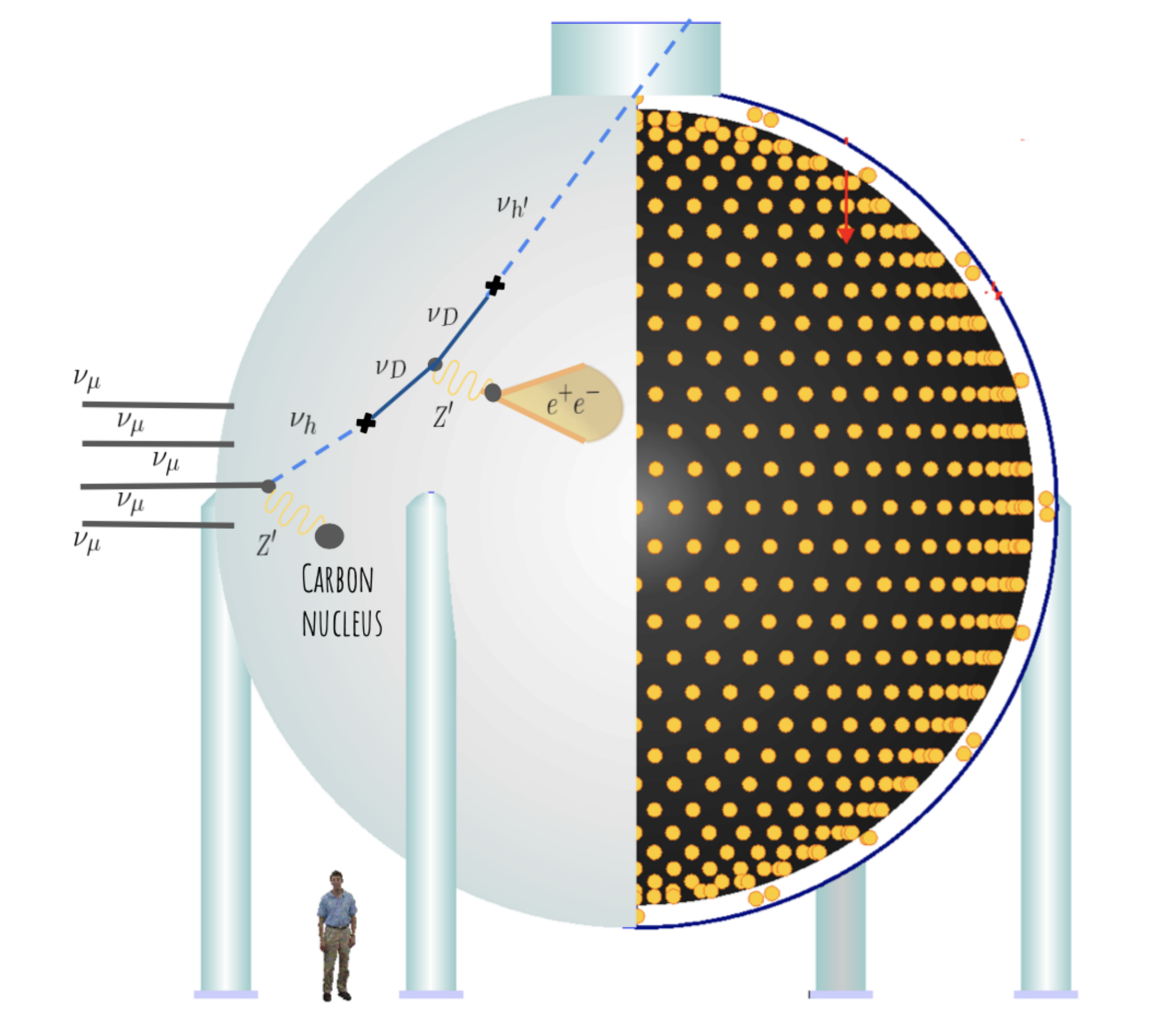}
\caption{HNL production and decay inside the MiniBooNE detector.}
\label{fig:MB_darkHNL}
\end{figure}

In minimal models, the HNL production and decay are controlled by SM interactions and the mixing between HNLs and the active neutrino and typically result in relatively long lifetimes if the masses are in the MeV-GeV range. This is the basis of searches, e.g. at colliders and  beam dump experiments. In models with additional interactions, HNL have new sources of production and decay channels. In the dark sector, interactions can even be strong and lead to enhanced scattering cross sections and fast decays, changing radically the HNL phenomenology. For instance, fast decays into neutrinos or other ``invisible" particles can weaken collider and beam dump limits, as the HNL would have not reached the detector or their  signals are suppressed by the branching ratio into visible channels. Moreover, new signatures can emerge and have been advocated to explain the low-energy excess of electron-like (LEE) events at the MiniBooNE experiment~\cite{MiniBooNE:2007uho, MiniBooNE:2008yuf, Polly:2011zz, MiniBooNE:2013uba, MiniBooNE:2018esg, MiniBooNE:2020pnu}, see Fig.~\ref{fig:MB_darkHNL}. 

In this picture, HNLs, produced in the neutrino beam~\cite{Fischer:2019fbw, Chang:2021myh} or directly inside the MiniBooNE detector via scattering~\cite{Ballett:2018ynz, Ballett:2019pyw,Abdullahi:2020nyr, Datta:2020auq,Dutta:2020scq,Abdallah:2020biq,Abdallah:2020vgg,Hammad:2021mpl, Dutta:2021cip}, subsequently decay to missing energy and an $e^+e^-$ pair, e.g. $\nu_h \to \nu_{h^\prime} \, e^+e^-$. If the $e^+e^-$ pair is sufficiently collinear or the energy of one of the particles falls below the detector energy threshold, the electromagnetic (EM) shower is reconstructed as a single electron or photon. In view of the recent results from the MicroBooNE experiment showing that the leading SM explanation of the excess is strongly disfavoured~\cite{MicroBooNE:2021zai}, and the tension of the most popular BSM explanation of the excess, $3+1$ neutrino oscillations, with other oscillation searches and cosmology~\cite{Planck:2018vyg, Hamann:2011ge,Hagstotz:2020ukm, Dentler:2018sju,Diaz:2019fwt,Dasgupta:2021ies,Berryman:2021yan}, models of $e^+e^-$ have become all the more intriguing. 

\begin{figure}
   \centering
   \includegraphics[width=0.16\linewidth]{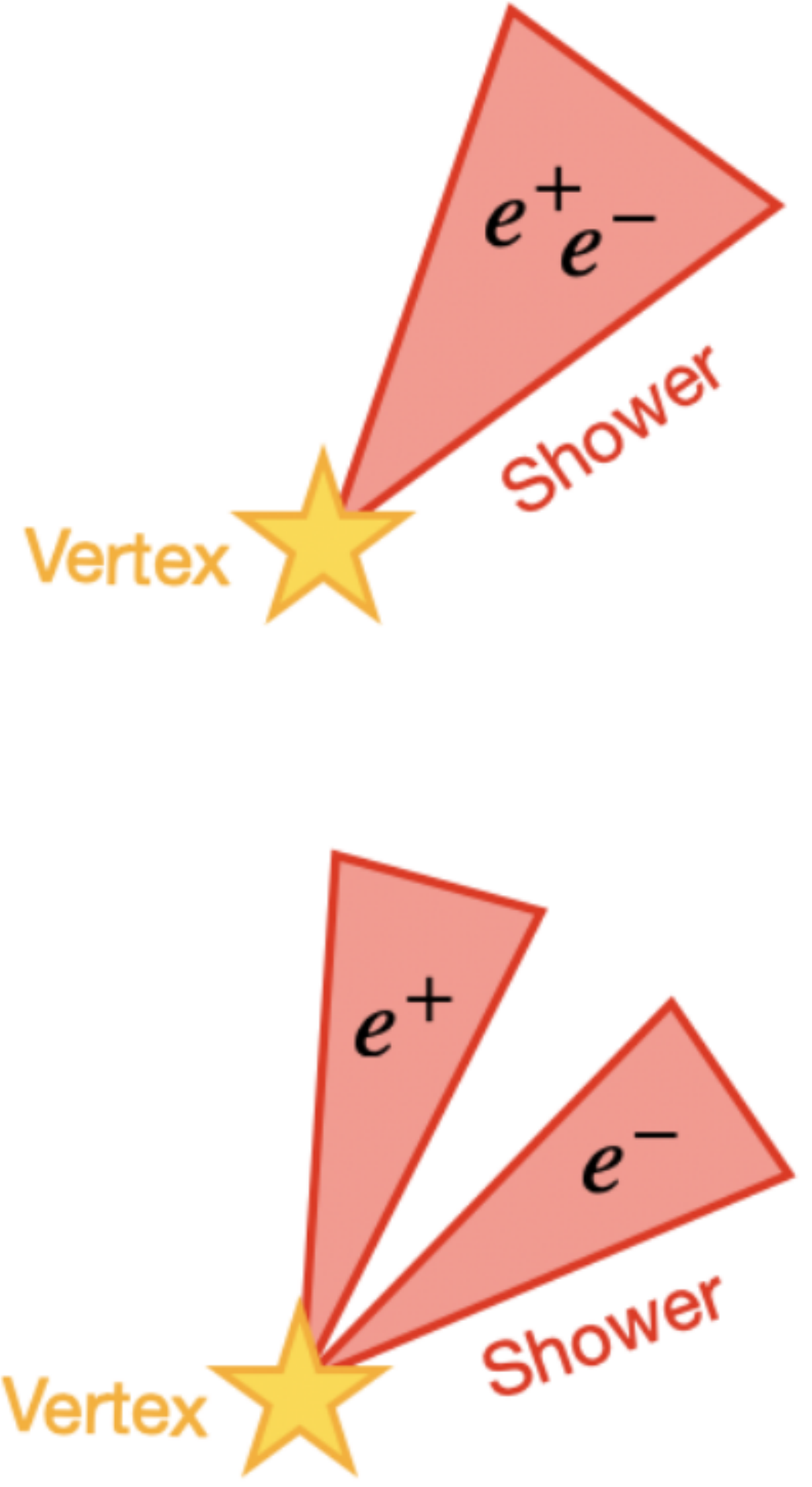}
   \hspace{2cm}
   \includegraphics[width=0.19\linewidth]{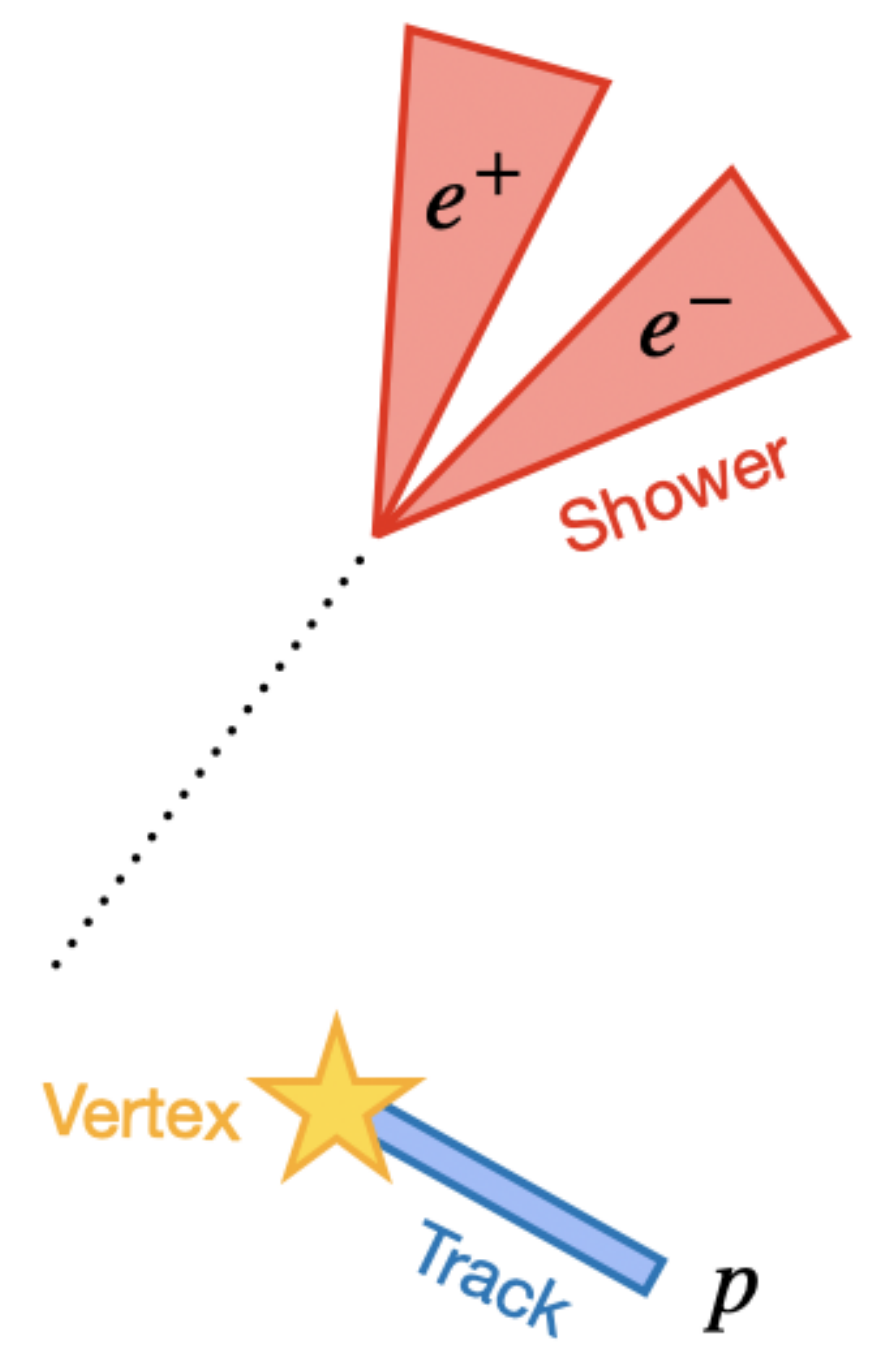}
   \hspace{1.8cm}
   \includegraphics[width=0.18\linewidth]{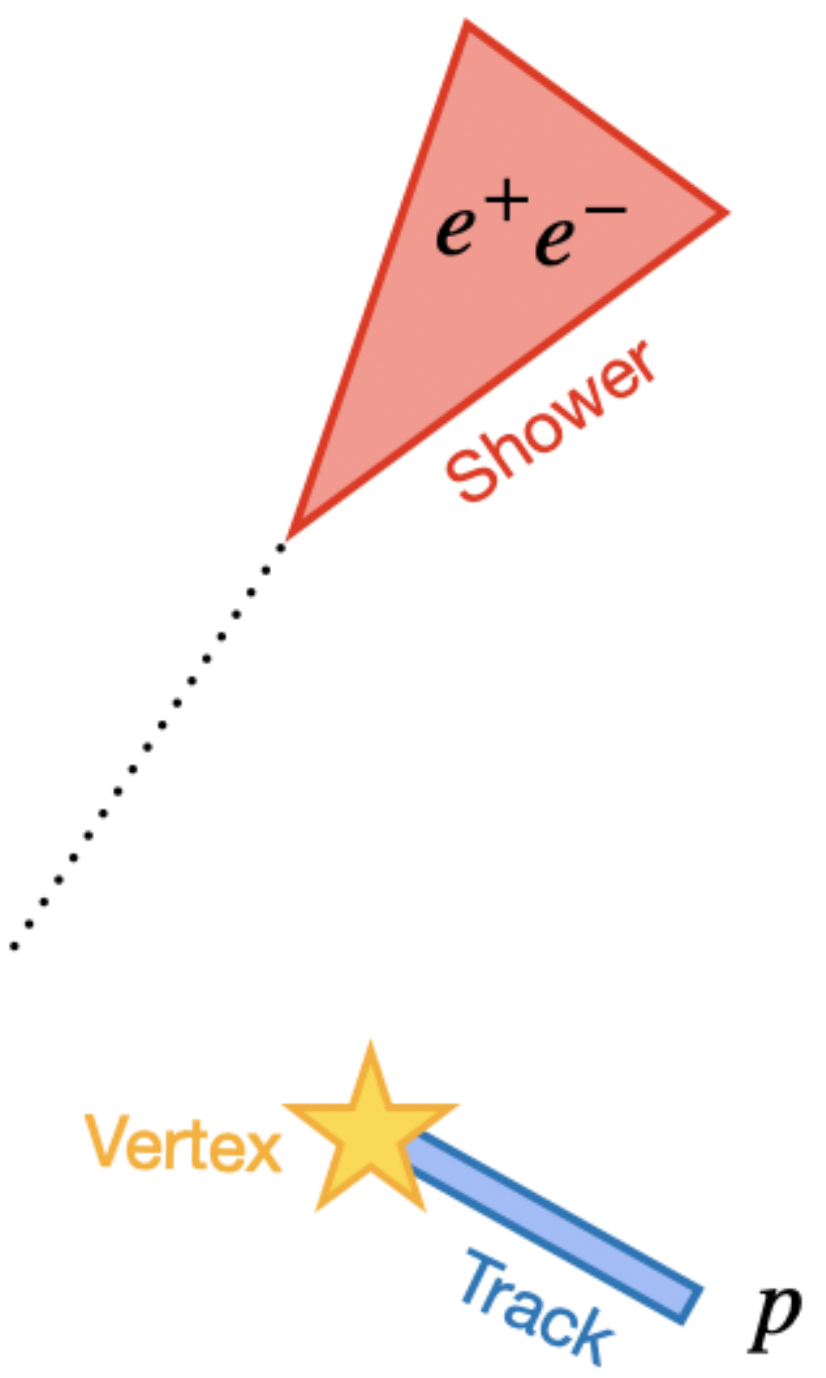}
\caption{The four main topologies of $e^+e^-$ LEE models at MicroBooNE: coherent and incoherent scattering with well-separated or overlapping $e^+e^-$ pairs.}
\label{fig:dark_nu_topologies}
\end{figure}

New opportunities for their search are being offered by neutrino facilities and in particular by short-baseline accelerator neutrino experiments and the near detector complex of long baseline ones, such T2HK and DUNE. A host of new generation liquid argon (LAr) experiments within the Fermilab Short-baseline Neutrino (SBN) program~\cite{Tufanli:2017mwt}, including the MicroBooNE experiment~\cite{MicroBooNE:2016pwy} and the currently running ICARUS one~\cite{}, are ideally suited for these searches. The hodoscopic nature of LAr detectors allows for unprecedented calorimetric and spatial resolution, potentially distinguishing the $e^-, \, e^+e^-, \,  \gamma, \, \gamma\gamma$ EM final states, as well as final state hadron multiplicity. For models in which the HNLs are produced in scattering, typical signatures include a primary neutrino interaction and a HNL decay vertex that can be displaced from it due to the finite HNL lifetimes, and EM showers, in some cases reconstructed as two separate electrons, that do not point back to the original vertex, due to missing energy. For incoherent scattering, there is also a proton track at the initial vertex.  There are  four distinct topologies in LAr that can be searched for
- see Fig.~\ref{fig:dark_nu_topologies}. 

Beyond the short-baseline neutrino facilities, these models can leave signatures at $e^+e^-$ colliders, e.g. BaBar~\cite{BaBar:2001yhh, BaBar:2017tiz}, Belle~\cite{Belle:2000cnh}, and the next generation experiments Belle II~\cite{Kou:2018nap,Duerr:2019dmv} and BESIII~\cite{Zhang:2019wnz}. In particular, monophoton searches at these experiments can test the semi-visible dark photon explanation of the muon $g_\mu -2$~\cite{}. Fixed-target experiments are another testbed for these models. HNLs can be produced in kaon decays at NA62~\cite{NA62:2017rwk}, leading to cascade decays with displaced vertices. Searches for inelastic dark matter produced in electron bremsstrahlung at NA64 are also sensitive to models of this kind~\cite{Banerjee:2019pds, NA64:2020qwq,Cazzaniga:2021hkw}.




\section{Collider Searches}

\label{sec:collider}
\subsection{ATLAS, CMS, and LHCb}

The LHC, the world's most powerful proton-proton collider, presently has nine approved experiments around the ring, distributed over four interaction points. The most general experiments that hunt for new physics are the ATLAS and CMS experiments, both of which are nearly hermetic detectors around the interaction point, and are able to deal with the high proton-proton collision rate produced by the LHC. These collisions have been produced so far 
at center of mass energies of $\sqrt{s} =$ 7, 8, and 13 TeV, but for future runs, the first one starting in 2022, it is expected that the energy will be slightly higher, namely 13.6 TeV. 

The ATLAS and  CMS experiments have a huge search program for physics beyond the Standard Model, covering, e.g., searches for supersymmetry as well as for
other exotic new particles and interactions, 
including searches for heavy neutral leptons (or heavy neutrinos).
In the case of heavy neutrinos, the different detector experiments at the LHC possess a broad coverage of the possible scenarios~\cite{delAguila:2008cj,Atre:2009rg,Tello:2010am,Deppisch:2015qwa,Cai:2017mow,Pascoli:2018heg}, and complements searches at other frontiers~\cite{Abada:2007ux,Deppisch:2015qwa,Cai:2017mow,Agrawal:2021dbo,Han:2022qgg}. Being general-purpose experiments, these detectors can explore a wide range of different signatures, but they are not particularly tailored to the cover the full spectrum of possibilities that needs be covered for a completely comprehensive long lived particle (LLP) search.

 A different experimental set-up at the LHC is the LHCb experiment that is designed to search in particular for new physics in decays of heavy-flavor hadrons. LHCb 
 does not provide hermetic detector coverage
 like ATLAS and CMS but looks instead at the emerging particles from the interaction region in a particular (forward) region with respect to the incoming beams. Tantalizing hints for possible Flavor Non-Universality have been reported by this experiment~\cite{LHCb:2017avl,LHCb:2021lvy}. These observations, however, need to be confirmed and established with stronger significance with more data, and 
 preferably also with independent measurements. 
 
 The ATLAS, CMS and LHCb experiments have been searching for  HNLs in the data samples recorded so far in complementary regions of mass, active-sterile neutrino mixing, and lifetimes.
 This includes searches for HNL in both minimal and non-minimal scenarios.
The status of these searches is now summarized.
 
 
\subsubsection{Current ATLAS and CMS Results}

\begin{figure}[!t]
\centering
\includegraphics[width=0.65\linewidth]{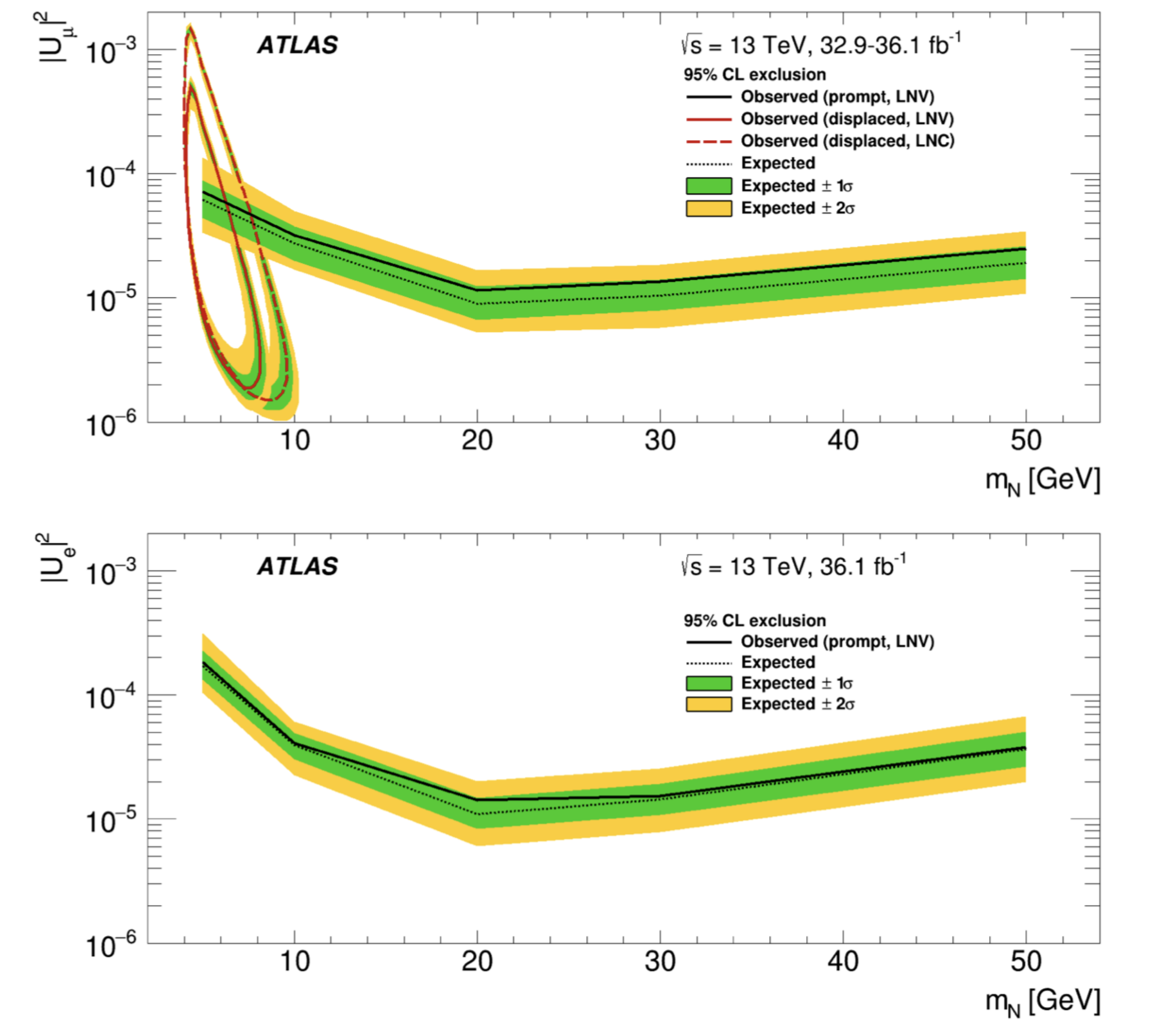}
\caption{Observed 95\% confidence level (CL) exclusion by the ATLAS experiment in $|U_{\mu4} |^2$ (top) and $|U_{e4} |^2$ (bottom),
labeled $|U_{\mu} |^2$ and $|U_e |^2$ respectively,
versus the HNL mass $m_N$ for the prompt signature (the region above the black line is excluded) and the displaced signature (the region enclosed by the red line is excluded). The solid lines show limits assuming lepton number violation (LNV) for 50\% of the decays and the long-dashed line shows the limit in the case of lepton number conservation (LNC). The dotted lines show expected limits and the bands indicate the ranges of expected limits obtained within $1\sigma$ and 
$2\sigma$  of the median limit, reflecting uncertainties in signal and background yields~\cite{ATLAS:2019kpx}.}
\label{fig:ATLAS}
\end{figure}

 \paragraph{{ \it Phenomenological Type I Seesaw}}
At the LHC, HNLs can be produced in the GeV mass range through the decays of heavy meson, $\tau$ leptons, $W$ bosons, $H$ bosons, and even top quarks. If new interactions or gauge symmetries exist in nature, or if masses are above the electro-weak (EW) scale, then HNLs can be produced in many more ways.
For the minimal model, high-mass channels include the Drell-Yan process~\cite{Keung:1983uu,Petcov:1984nf}, gluon fusion~\cite{Willenbrock:1985tj,Hessler:2014ssa,Ruiz:2017yyf}, $W\gamma$ fusion~\cite{Datta:1993nm,Dev:2013wba,Alva:2014gxa,Degrande:2016aje}, and same-sign $WW$ fusion~\cite{Dicus:1991fk,Fuks:2020att}. In the case that HNLs are Majorana particles, decays are allowed into both lepton number conserving and lepton number violating channels; if they are Dirac, only lepton number conserving processes are allowed. Both can mediate lepton flavor violating processes. Moreover, depending on the HNL's mass and mixing angle, the decays may be either prompt (short lifetime) or displaced (large lifetime) from the production vertex.

\begin{figure}[t!]
    \centering
    \includegraphics[width = \textwidth]{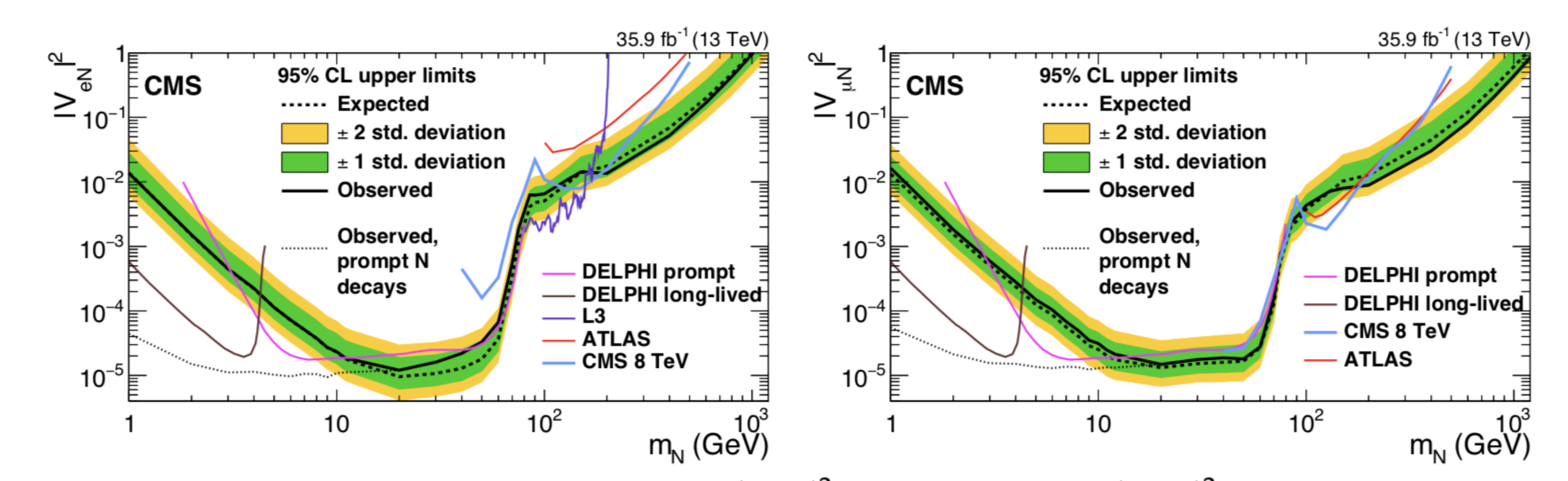}
    \caption{HNL exclusion region at 95\% CL by the CMS experiment in the 
$|V_{eN}|^2$ vs. $m_N$ (left) and $|V_{\mu N}|^2$ vs. $m_N$ (right) planes. The dashed black curve is the expected upper limit, with one and two standard-deviation
bands shown in dark green and light yellow, respectively. The solid black curve is the observed upper limit, while the dotted black curve is the observed limit in the approximation of prompt N decays~\cite{CMS:2018iaf}. Also shown are the best upper limits at 95\% CL from other collider searches in L3, DELPHI, ATLAS (8 TeV), and CMS (8 TeV).}
    \label{fig:CMS2}
\end{figure}

We start with current results from ATLAS and CMS with 
the case for displaced and prompt decays of HNLs in the minimal extension of the SM.
Results that have been released so far at 13 TeV include an ATLAS displaced signature search, that  sets  constraints for mixing with muon-neutrinos at the level $|U_{\mu 4}|^{2} \sim 10^{-6}$ for HNL masses between 4-10 GeV based on a data sample of  33-36 ~${\rm fb}^{-1}$~\cite{ATLAS:2019kpx}. 
The results are shown in Fig.~\ref{fig:ATLAS}. Reinterpretation of these results in terms of models with more than one HNL was performed by Ref.~\cite{Tastet:2021vwp}.
Similarly, CMS has a search with 35.9~${\rm fb}^{-1}$ of integrated luminosity at 13 TeV, which obtained the first direct bounds on HNLs up to and above 500 GeV, and giving $|U_{e4}|^{2}, |U_{\mu 4}|^{2} < 1$ up to TeV HNL masses~\cite{CMS:2018iaf}. 
The results are shown in Fig.~\ref{fig:CMS2}.

\begin{figure}[t!]
    \centering
\includegraphics[width=0.85\linewidth]{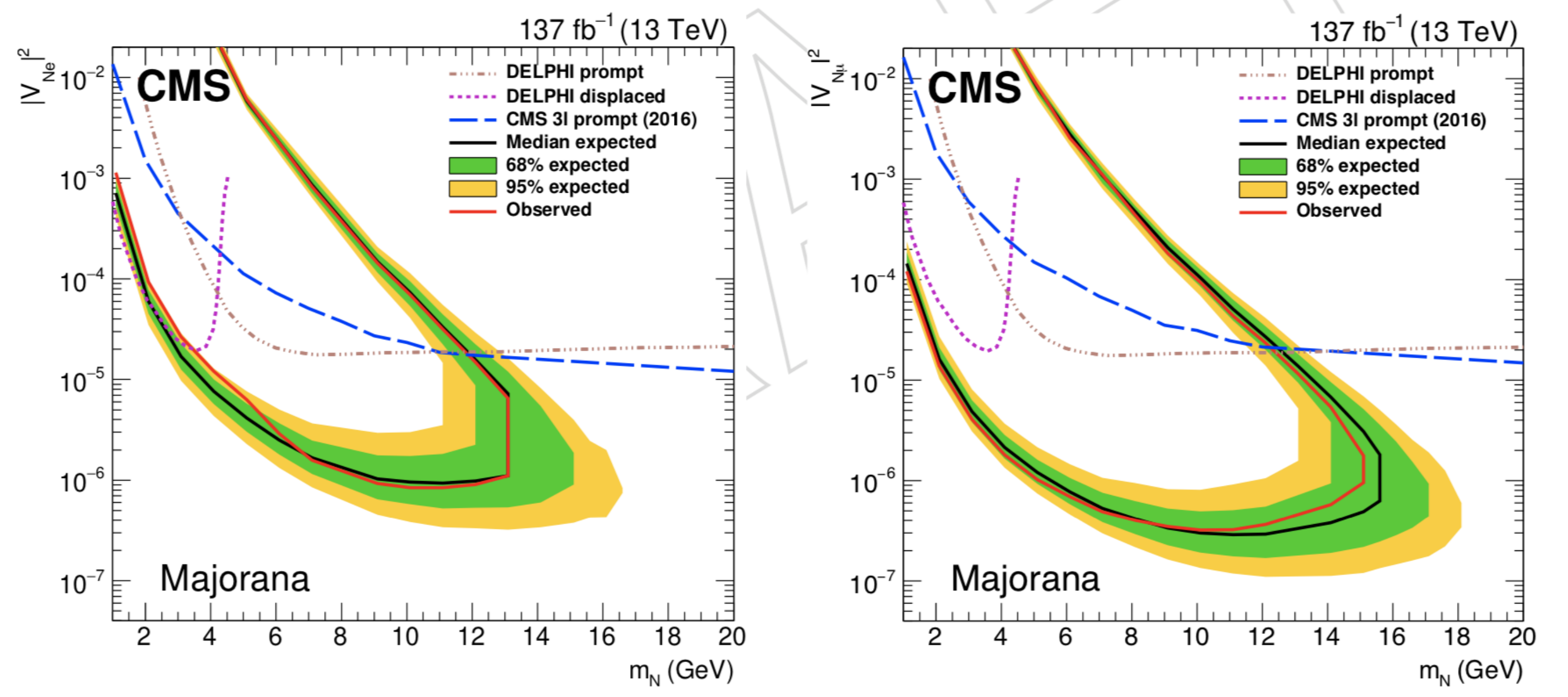}
\includegraphics[width=0.85\linewidth]{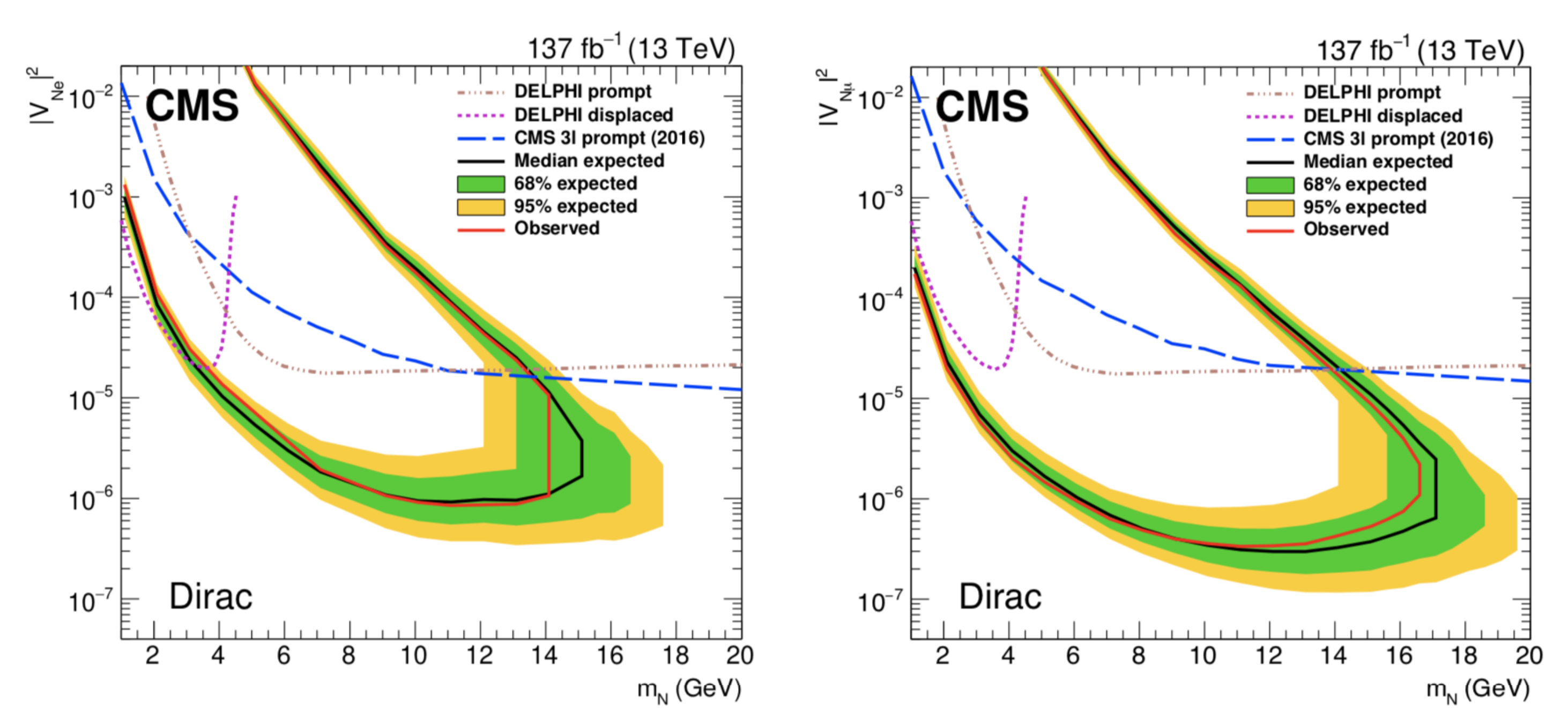}
    \caption{The 95\% CL limits by the CMS experiment on $|V_{eN} |^2$ (left) and $|V_{\mu N} |^2$ (right) as functions of mass $m_N$ for a Majorana (top) and Dirac (bottom) HNL. The area inside the red (black) curve indicates the observed (expected) exclusion region~\cite{CMS:2022fut}. Previous results from the DELPHI and the CMS  Collaborations (2016) are shown for reference.}
    \label{fig:CMS}
\end{figure}

A newly released CMS result of a low mass HNL search~\cite{CMS:2022fut} uses the full data set
of Run 2 corresponding to 138 fb$^{-1}$, using the final state where the HNL decays in  opposite charged
leptons and a neutrino, hence the final state signature is three charged leptons of which two are displaced.
The results are shown in Fig.~\ref{fig:CMS}.
Both the Majorana (top) and Dirac (bottom) HNL decay channels are probed,  using the fact that the HNL production cross section is the same but the HNL lifetime is twice as large for a Dirac HNL due to the missing LNV decay channels.
The results of this study set  constraints for mixing with muon-neutrinos 
 at the level $|U_{\mu4}|^{2} \sim  4\cdot10^{-7}$ for HNL masses between 8-14 GeV.
 The sensitivity in the electron-neutrino channels
 is around $|U_{e4}|^{2} \sim  10^{-6}$ for HNL masses 
 in the same region.
 
 Both ATLAS and CMS have reported results at lower center-of-mass energies~\cite{ATLAS:2011izm,CMS:2012wqj,CMS:2015qur,ATLAS:2015gtp}.

\paragraph{{ \it Extended gauge symmetries and effective field theories}}
HNLs may arise from an extended gauge symmetry, such as $U(1)_{B-L}$, which promotes baryon and lepton number conservation to a gauge symmetry, or Left-Right (LR) symmetric
theories, which restore parity at high-energy scales.
For details on the construction of such models, see Sec.~\ref{sec:beyondMinTheory}.
In contrast to HNLs in minimal Type I models, heavy Majorana neutrinos $N$ in extended gauge scenarios couple directly to new gauge bosons.
Gauge bosons may have couplings that are naturally un-suppressed, as in LR and $B-L$ theories,  or masses that are naturally light, as in $B-L$ theories. 
In these cases, HNLs can be produced copiously at the LHC, as long as the mediators are not too heavy and couplings are not too small.

The paradigmatic example of HNL production through new gauge bosons is the Keung-Senjanovi\'c process~\cite{Keung:1983uu} within
the minimal LR model, where a right-handed (RH) $W_R$ can be produced in the $s$-channel and
decays into a heavy Majorana $N$ and a RH charged lepton of various flavors.
The $N$, in turn, decays into another charged lepton and two partons (quarks and antiquarks).
Depending on the relative sizes of the $W_R$ and $N$ masses, one can obtain four qualitatively distinct signatures:
(i) When $M_{W_R}\gtrsim m_N$, all final state particles are well separated and one can search for the classic collider signature $pp\to W_R \to \ell_i \ell_j j j$~\cite{Keung:1983uu}. The leptons $\ell_i$ and $\ell_j$ may have the same or different electric charges and flavors, and all partons typically have momenta that scale like $M_{W_R}$.
(ii) When $M_{W_R}\gg m_N$, $N$ is boosted and its decay products are collimated. The resulting signature is $pp\to W_R \to \ell_i J$, where $J$ is a multi-prong jet and can be identified as the boosted $N$~\cite{Ferrari:2000sp,Mitra:2016kov,Mattelaer:2016ynf}. The momentum of $\ell$ and $J$ are comparable to $M_{W_R}$.
Importantly, $J$ can be displaced~\cite{Nemevsek:2018bbt}.
(iii) When $m_N$ is within the LHC's kinematic reach but $M_{W_R}$ is not, then  the signature $pp\to \ell_i \ell_j j j$ is still possible through a far off-shell $W_R$~\cite{Ruiz:2017nip}.
In this case, the kinematics of final-state particles are comparable to $m_N$ and strongly resemble the minimal Type I scenario but result in different angular distributions~\cite{Ruiz:2017nip,Nemevsek:2018bbt}.
This limit also has connections to the so-called $\nu$SMEFT framework~\cite{delAguila:2008ir,Aparici:2009fh,Elgaard-Clausen:2017xkq}.
(iv) When $m_N \gtrsim W_R$, the kinematics and search methodology are analogous to a high-mass version of searches for HNLs in the minimal HNL scenario.
The production HNLs through $Z_R$ gauge bosons can similarly be categorized. 
Due to the heaviness of right-handed gauge bosons, vector boson fusion and associated production channels typically feature cross sections too small to be 
produced at the LHC.

Beyond production mechanisms through gauge bosons, HNLs in extended gauge theories can also be produced via Higgs sector couplings and Yukawa interactions.
For example: the SM-like Higgs state at 125 GeV can mix with another scalar that breaks the extended
gauge symmetry and is the source of heavy neutrino mass origin (such as the neutral component of
an SU$(2)_R$ triplet $\Delta_R^0$ in the minimal LR model).
In the mass basis one then has an exotic Higgs decay $h \to N N$~\cite{Maiezza:2015lza} that again
results in two same sign leptons and four jets, all possibly displaced if $m_N$ is low enough.
Likewise, one can produce the extra scalar resonance $\Delta_R^0$ through gluon fusion or associated
$W/Z$ production~\cite{Nemevsek:2016enw}, which leads to an even larger potential signal, as long
as the $WW, ZZ$ final states are suppressed, say for $m_\Delta \lesssim 160$ GeV.
The final states are again LNV with soft $\ell^\pm \ell^\pm$ and additional jets and may be displaced,
thus forming one or more displaced vertices.
It is also possible for $h \to \Delta_R^0 \Delta_R^0$ to occur, which would subsequently produce
4 $N$s and a striking $\Delta L = 4$ final state with essentially zero background.
It is important to stress that the production of $N N$ pairs via a Seesaw Higgs occurs not only in LR scenarios but also in pure $U(1)_{B-L}$~\cite{Accomando:2016sge, Accomando:2017qcs,Deppisch:2018eth}.

\begin{figure}[t!]
  \centering
  \includegraphics[width=0.50\textwidth]{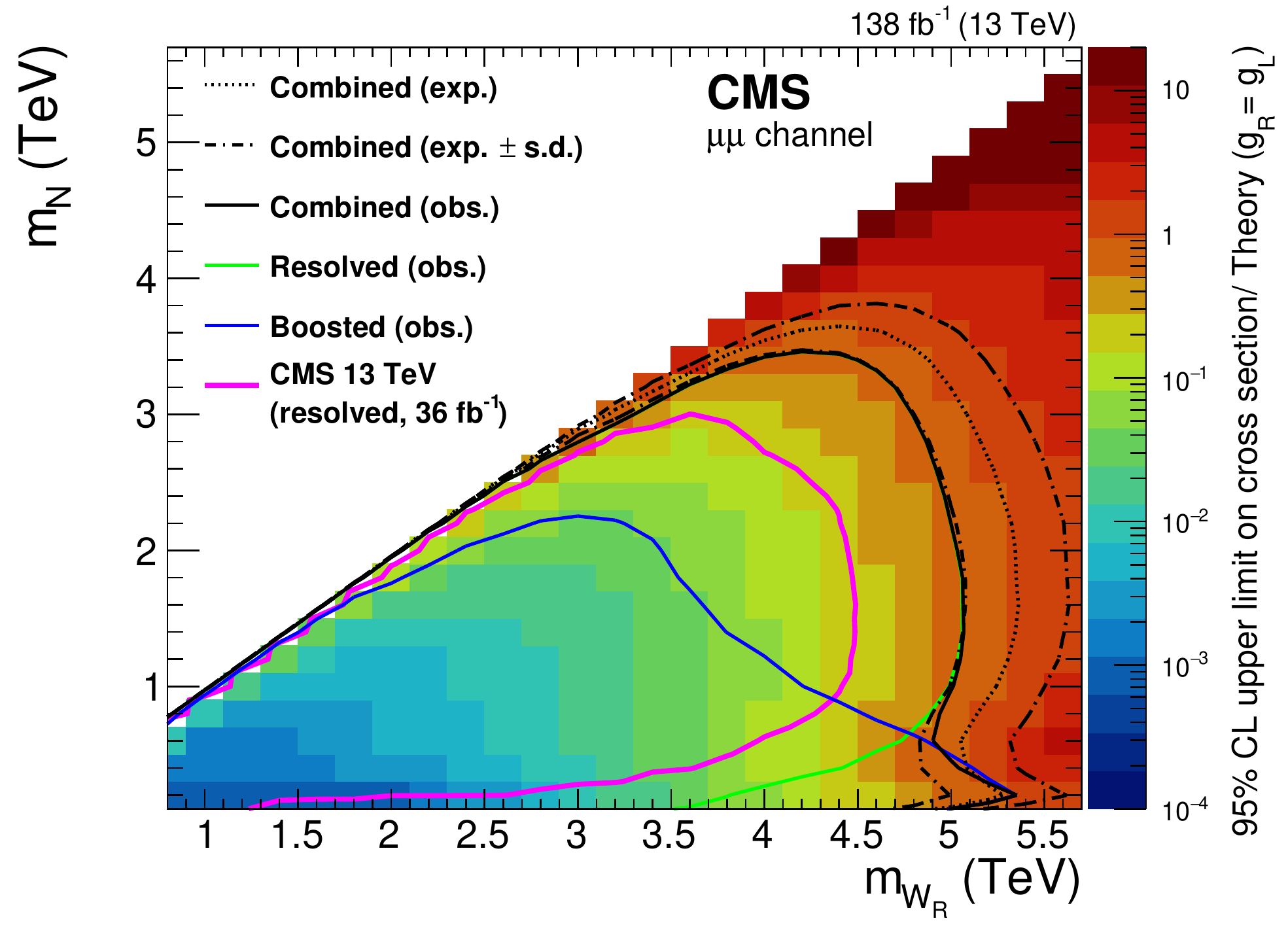}
  \includegraphics[width=0.46\textwidth]{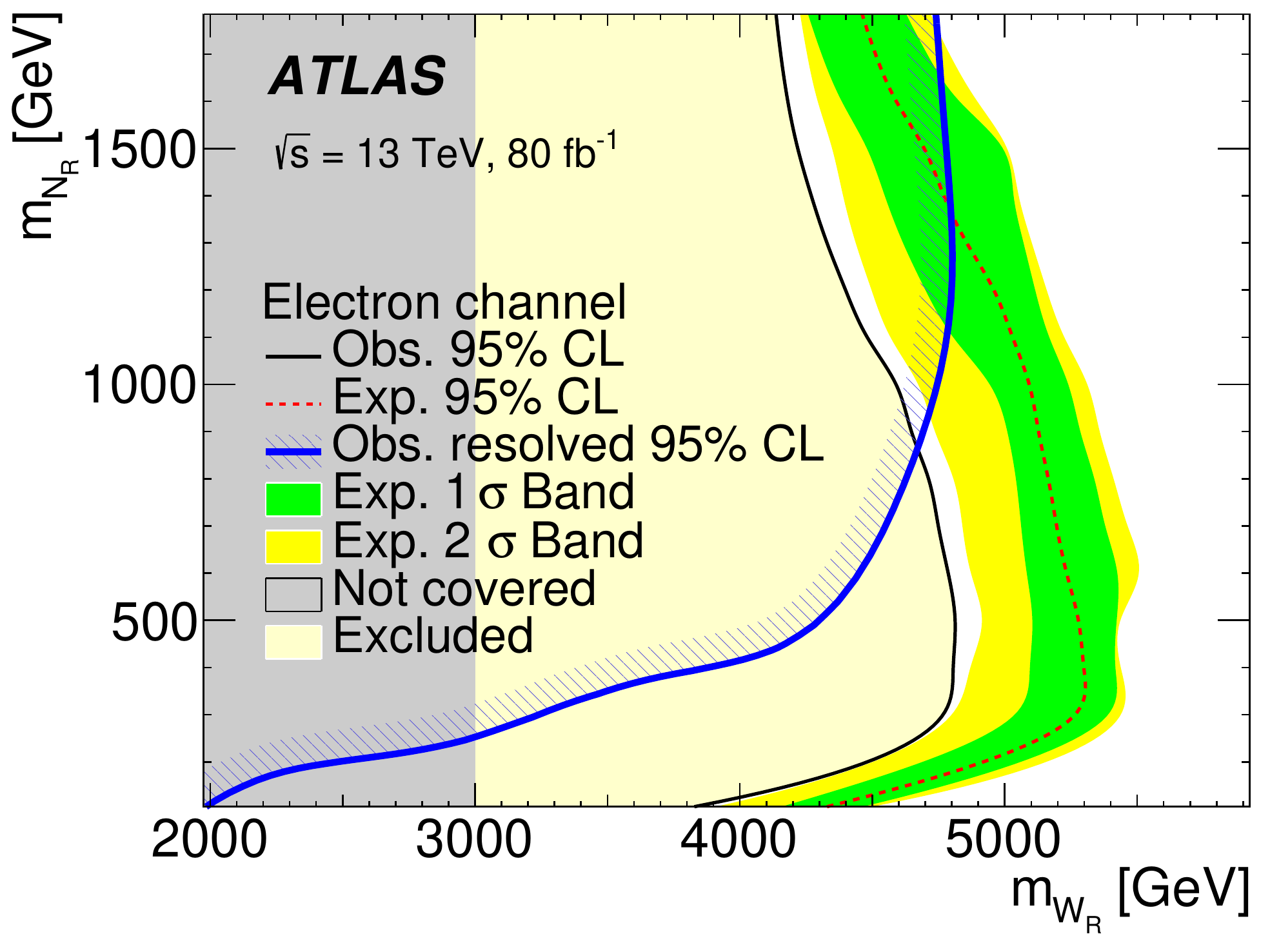}
 \caption{Left: The 95\% CL limit with 138 fb$^{-1}$ of data from CMS~\cite{CMS:2021dzb} on $M_{W_R}$ and $m_N$ in the $\mu-\mu$ channels (resolved and boosted); similar bounds apply to the electron channel, see~\cite{ATLAS:2018dcj} for the ATLAS search.
Right: The 95\% CL limit with 80 fb$^{-1}$ of data from ATLAS~\cite{ATLAS:2019isd} on $M_{W_R}$ and $m_N$ in the $e-e$ channel (boosted); similar bounds apply to the muon channel
}
\label{fig:LHC_WR} 
\end{figure}

For HNLs with $M_{W_R} > m_N > \mathcal O(100)$ GeV, the final state products can be resolved and one gets a $\Delta L = 2$ final state with $\ell^{\pm} \ell^{\pm} j j$.
Current results at 13 TeV 
in the resolved channel exclude $W_R$ masses up to about $M_{W_R} \approx 5$ TeV~\cite{CMS:2018agk,ATLAS:2018dcj,CMS:2021dzb}.
The precise bound depends on the amount of data recorded, with limits using  138 fb$^{-1}$ summarized by the green line in Fig.~\ref{fig:LHC_WR} (left).
%
%
Weaker limits are reported by the two experiments at lower center-of-mass energies~\cite{ATLAS:2012ak,CMS:2012zv,CMS:2014nrz,ATLAS:2015gtp}.
For the boosted channel, experimental searches by CMS~\cite{CMS:2021dzb} and ATLAS~\cite{ATLAS:2019isd} are more sensitive  to  
the parameter space $m_N \lesssim 100$ GeV and $M_{W_R}\gtrsim 1$ TeV, and exclude 
$M_{W_R} < 5.3$ TeV. This is illustrated in  the blue line in Fig.~\ref{fig:LHC_WR} (left) for CMS
and Fig.~\ref{fig:LHC_WR} (right) for ATLAS.
%
For even lighter masses with $m_N \lesssim 50$ GeV, a significant number of $N$'s decay
outside of the detector and one can recast~\cite{Nemevsek:2018bbt} the prompt search for 
$W' \to \ell + \text{MET}$~\cite{ATLAS:2017jbq} (superseded by~\cite{ATLAS:2019lsy}),
which results in a stringent bound of about 5-6 TeV in the electron and muon
channels, as shown in the orange contour on the right of Fig.~\ref{fig:LHC_WR}.
Using a reinterpretation of Run 1 data,  searches for $N$ via far off-shell $W_R$ only exclude $M_{W_R}<1$ TeV for $m_N < 500$ GeV~\cite{Ruiz:2017nip}.
 
\paragraph{ {\it Type III Seesaw}}
 
 It may be the case that HNLs exist but that they are not singlets of the SM's gauge symmetries. 
 In particular, they may appear as the neutral component of a fermionic EW triplet. 
 In this case, not only do HNLs exist but also electrically charged partners $E^\pm$.
 (These states are sometimes also labeled $T^0,T^\pm$, $\Sigma^0,\Sigma^\pm$, or $L^0,L^\pm$.)
 Furthermore, $N$ and $E^\pm$ carry SU$(2)_L$ gauge charges, which allows them to be produced in pairs \textit{without} active-sterile mixing suppression. 
 At the LHC, $N E^\pm$ pairs can be produced through the Drell-Yan and vector boson fusion processes. 
 $E^+E^-$ and $E^\pm\ell^\mp$ pairs  can be produced through the Drell-Yan, photon fusion, gluon fusion, and vector boson fusion; the channels have a non-trivial interplay between mixing, PDF enhancements, and coupling enhancement/suppression~\cite{Ruiz:2015zca,Hessler:2014ssa,Cai:2017mow}.
Decays of heavy, vector leptons typically proceed to EW gauge bosons. 
In principle, $N$ and $E^{\pm}$ have the same masses due to gauge symmetry (they belong to the same multiplet). 
However, small, QED-included mass splittings between $E^\pm$ and $N$ open the $E^\pm \to N\pi^\pm$ decay mode~\cite{Bajc:2007zf}. 
Notably, production of $N E^\pm$ and $E^+E^-$ pairs can result in
final states with multiple leptons and high-$p_T$ jets as well as final states with only  many lepton~\cite{Bajc:2007zf,Franceschini:2008pz,delAguila:2008cj,Arhrib:2009mz}.
In these cases, all high-$p_T$ objects carry momentum comparable to the masses of $N$ and $E^\pm$.
For particularly heavy triplets, one also anticipates final-states with boosted topologies~\cite{Ashanujjaman:2021zrh}.
 
 \begin{figure}[h!]
  \centering
  \includegraphics[width=0.44\textwidth]{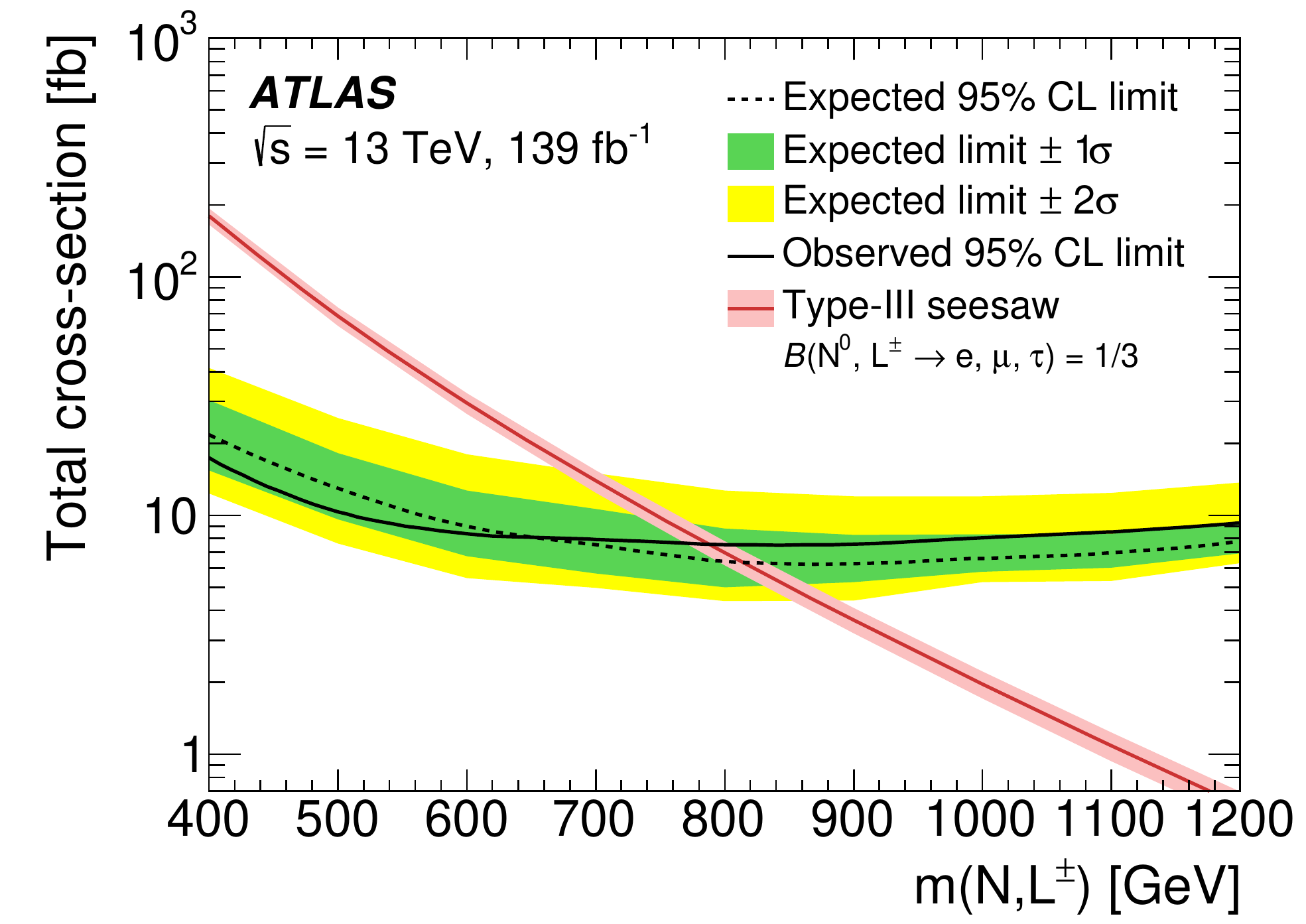}
  \includegraphics[width=0.54\textwidth]{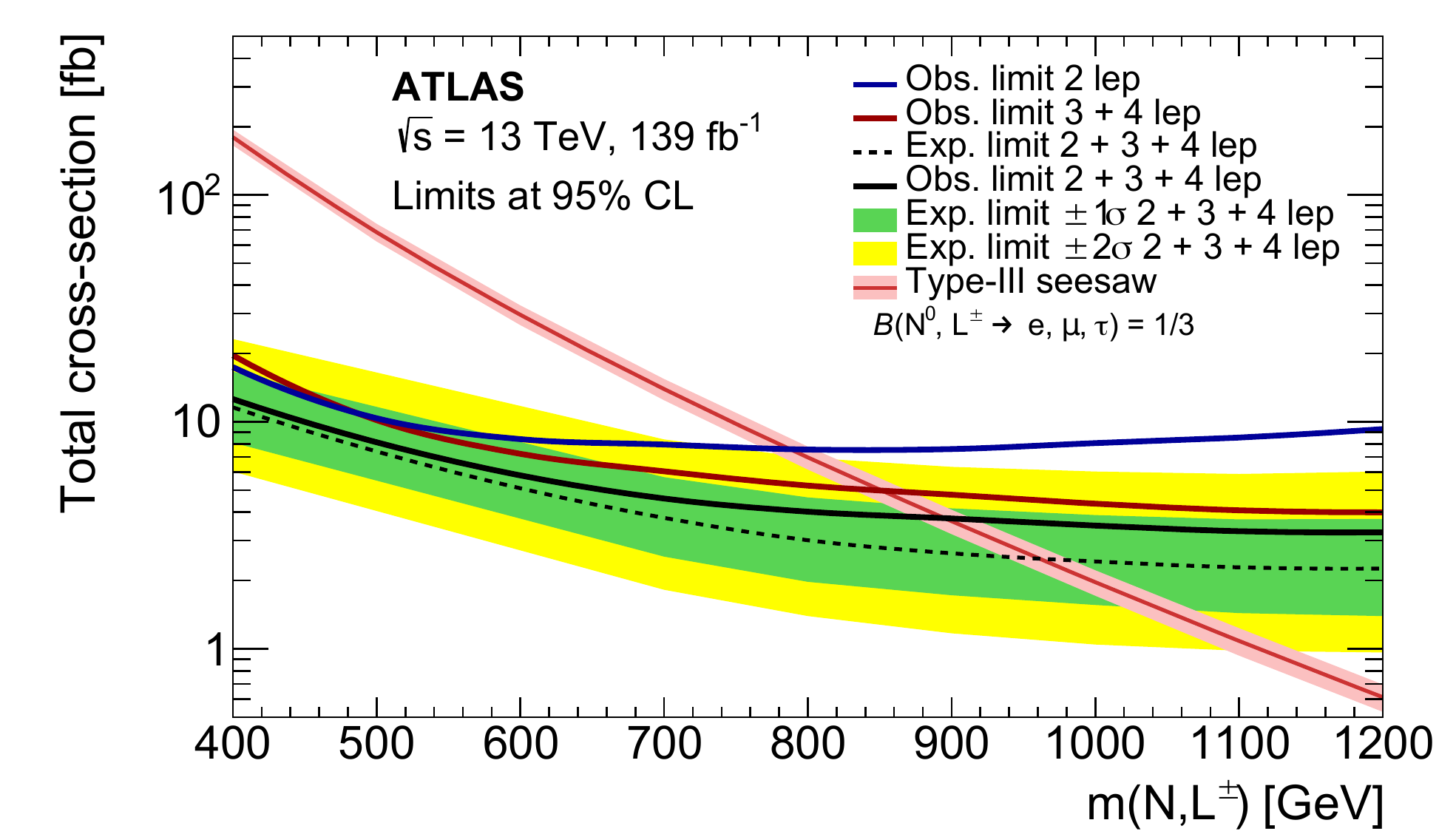}
 \caption{Left: The 95\% CL limit with 139 fb$^{-1}$ of data from ATLAS on
 triplet lepton masses $m_N, m_L$ in the dilepton with MET and jets channel~\cite{ATLAS:2020wop}.
 Right: The same but for the multilepton and MET channel~\cite{ATLAS:2022yhd}.
Similar bounds have been reached by the CMS~\cite{CMS:2019hsm,CMS:2019lwf} 
}
\label{fig:LHC_typeIII} 
\end{figure}

 Presently, searches by the ATLAS and CMS experiments for $N E^\pm$ pairs in few- and many-lepton final states stringently constrain Type III leptons. 
 More specifically, with 139 fb$^{-1}$, the ATLAS experiment 
 has excluded masses below $m_L < 790$ GeV at 95\% CL in the same- and opposite-sign dilepton, MET, and jets configuration~\cite{ATLAS:2020wop}.
 The experiment has also excluded cross sections larger than $\sigma = 22~(7.5)$ fb for representative benchmarks $m_L=400~(700)$ GeV in multilepton  channels~\cite{ATLAS:2021wob,ATLAS:2022yhd}.
 These results are summarized in Fig.~\ref{fig:LHC_typeIII}.
 In events with at least three charged leptons, the CMS experiment has excluded $N E^{\pm}$ with masses below $m_L = 880$ GeV with 137 fb$^{-1}$ at 13 TeV under specific mixing assumptions~\cite{CMS:2019hsm,CMS:2019lwf}.
 At 13 TeV with 36 fb$^{-1}$ of data, CMS has excluded triplet leptons with masses in the range $m_L = 370-900$ GeV in the multi-lepton final state under various mixing and branching rate assumptions, including the contribution from $\tau$ leptons~\cite{CMS:2017ybg}.
 At 8 TeV with 20 fb$^{-1}$ of data, ATLAS has excluded $NE^{\pm}$ pairs with masses below $m_L \lesssim 100 - 468$ GeV in the 3- and 4-lepton channels~\cite{ATLAS:2015qoy}. 
  Less constraining results at 7 TeV have also been reported~\cite{CMS:2012ra}.
 The precise bound depends on the specific lepton flavor configuration assumed. 
 It is important to stress that while some of these searches have included signal processes with $\tau$ leptons, the flavor category remains understudied. 

\begin{figure}[t!]
    \centering
    \begin{minipage}[b][][b]{0.5\textwidth}
        \centering
        \includegraphics[width=0.89\textwidth]{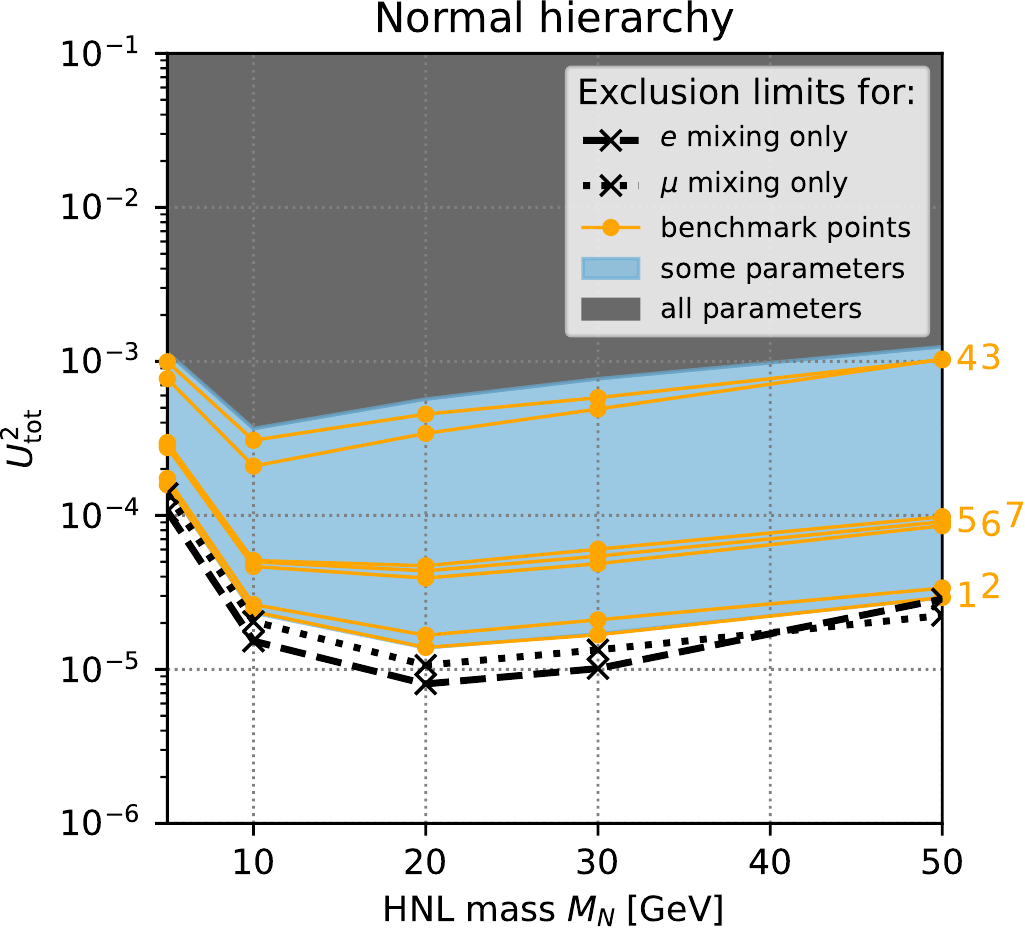}
    \end{minipage}%
    \begin{minipage}[b][][b]{0.5\textwidth}
        \centering
        \includegraphics[width=0.91\textwidth]{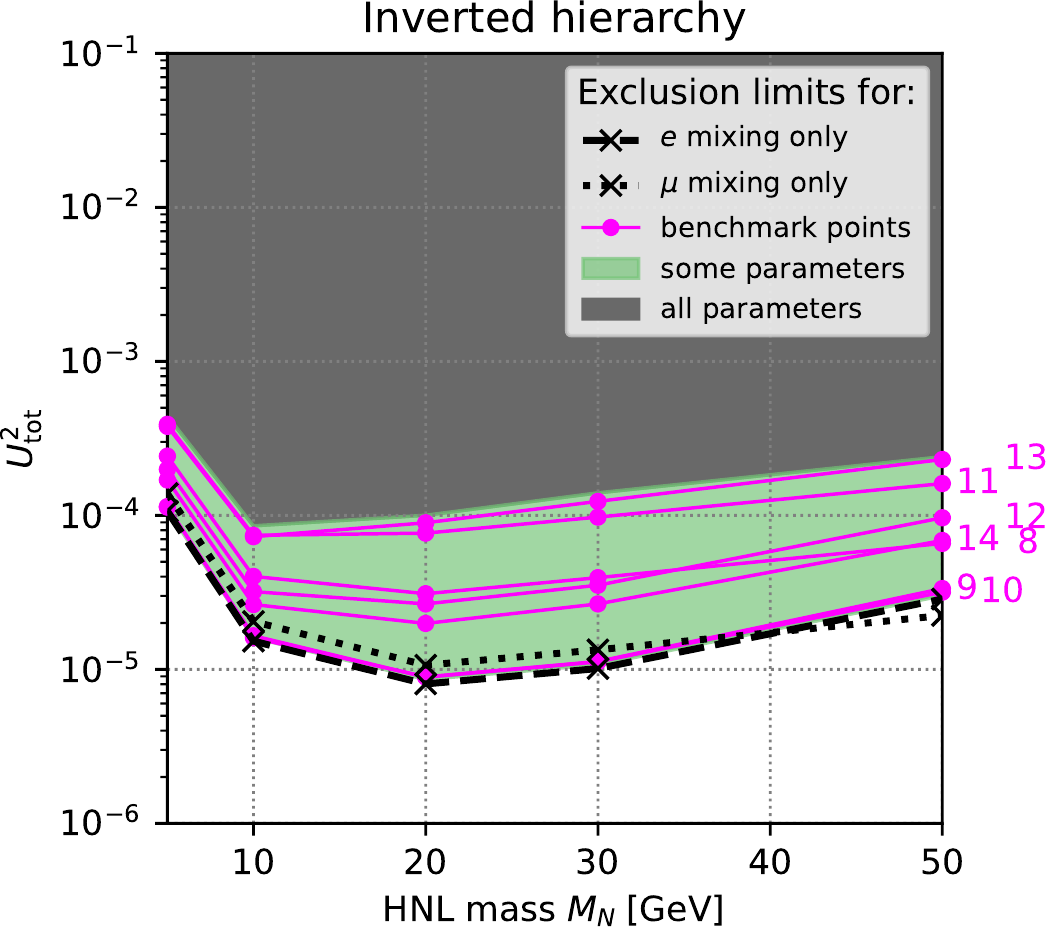}
    \end{minipage}
 \caption{Original (black lines) and reinterpreted (colored lines) $95\%$ exclusion limits on the total mixing angle $U_{\text{tot}}^2 = \sum_{\alpha=e,\mu,\tau} \sum_{I=1,2} |\Theta_{\alpha I}|^2$ for the normal (left) and inverted (right) mass orderings, for Majorana HNLs.
    The black lines are limits obtained under the single-flavor assumption. The solid colored lines denote those obtained for various benchmark points (defined in fig.~2 from Ref.~\cite{Tastet:2021vwp}) spanning the allowed regions shown in figure~\ref{fig:triangle}.
    When scanning over all ratios of mixing angles allowed by neutrino oscillation data, the exclusion limits span the blue (green) shaded regions.
    Correspondingly, the gray filled area is excluded at $\mathrm{CL} > 95\%$ for all possible ratios of mixing angles, and thus constitutes an exclusion limit independent of the specific choice of mixing angles, valid as long as we consider the two HNL model explaining neutrino oscillations.}
 \label{fig:seesaw_reinterpretation}
\end{figure}

 
\paragraph{Reinterpretation of LHC results}
Experiments searching for HNLs typically report their results within simplified models consisting of a single HNL coupled to a single lepton flavor, as in figures \ref{fig:ATLAS} and \ref{fig:CMS2}.  However, the Type~I seesaw---which aims to describe neutrino oscillations---necessarily features at least two HNLs coupled to several flavors (cf.\ section \ref{sub:type-i_seesaw}). 
Ref.~\cite{Tastet:2021vwp} has performed a detailed reinterpretation of the latest ATLAS search~\cite{ATLAS:2019kpx} for prompt HNLs in $W$ decays within a minimal Type~I seesaw with two HNLs. It has shown that the exclusion limits can differ by several orders of magnitude from the limits obtained under the single HNL assumption. Hence comparing the mixing angles from a realistic model to the reported limits could lead to wrongly excluding large regions of the parameter space. Some results of~\cite{Tastet:2021vwp} are shown in figure~\ref{fig:seesaw_reinterpretation}.
The colored regions contain all the possible exclusion limits when scanning over the allowed combinations of mixing angles (cf.\ figure~\ref{fig:triangle}).
As one can see, many allowed parameter sets lead to exclusion limits that are much weaker than the single-flavor ones.
This drastic change in the experimental sensitivity is  related to the opening of new channels (to which the search may not be sensitive) if more than one mixing angle is allowed.

To facilitate reinterpretation, Ref.~\cite{Tastet:2021vwp} proposed a reweighting method which allows---using only a handful of constants that can be easily computed and published by experiments---to exactly extrapolate the expected signal to any combination of mixing angles.
This method is applicable to other models of feebly interacting particles.
In order to perform an accurate reinterpretation, modeling the background and the systematic uncertainties is also critical. To this end, Ref.~\cite{Tastet:2021vwp} recommends that experiments release (preferably) their likelihood or (at least) the covariance matrix between the background counts in all bins from all channels.
This suggestion is in line with the recommendations from the LHC Reinterpretation Forum~\cite{LHCReinterpretationForum:2020xtr}.
 
 \subsubsection{Expected future ATLAS and CMS Results}
 \paragraph{{\it Phenomenological Type I Seesaw}}
New results will be released by both collaborations 
in 2022 based on the full Run 2 data and 
including more final states, 
for example the final states where the HNL decays in a charged lepton and hadrons. These final states have 
larger branching ratios than the full leptonic ones, but tougher background 
conditions to fight. Nevertheless with the help 
machine learning techniques one expects that the resulting limits can 
be competitive with, potentially even by-pass, those from the  3-lepton final states analyses. 
Combining the overall ATLAS+CMS results one can 
anticipate to probe the coupling 
region down to $|U_{\mu N}|^{2} \sim  10^{-7}$ for HNL masses 
in the range of 5-15 GeV in the muon channel, which is the 
most sensitive channel.

Other channels that could be  explored include ultra light HNLs from decays of 
bottom and charm mesons, as already searched for in the LHCb
experiment. The large heavy meson decay samples that can potentially be collected at the experiments, e.g., via so-called ``data parking'' techniques as
exercised  by the ATLAS and CMS experiments in 2018, may 
allow 
for additional sensitivity in the mass range below 5 GeV.
CMS demonstrated in Run 2 that  B meson enriched data samples containing more than $10^{10}$ B hadrons per year 
can be collected, leading to a potentially large source of  neutrinos from leptonic B decays. 

\begin{figure}[t!]
    \centering
\includegraphics[width=0.32\linewidth]{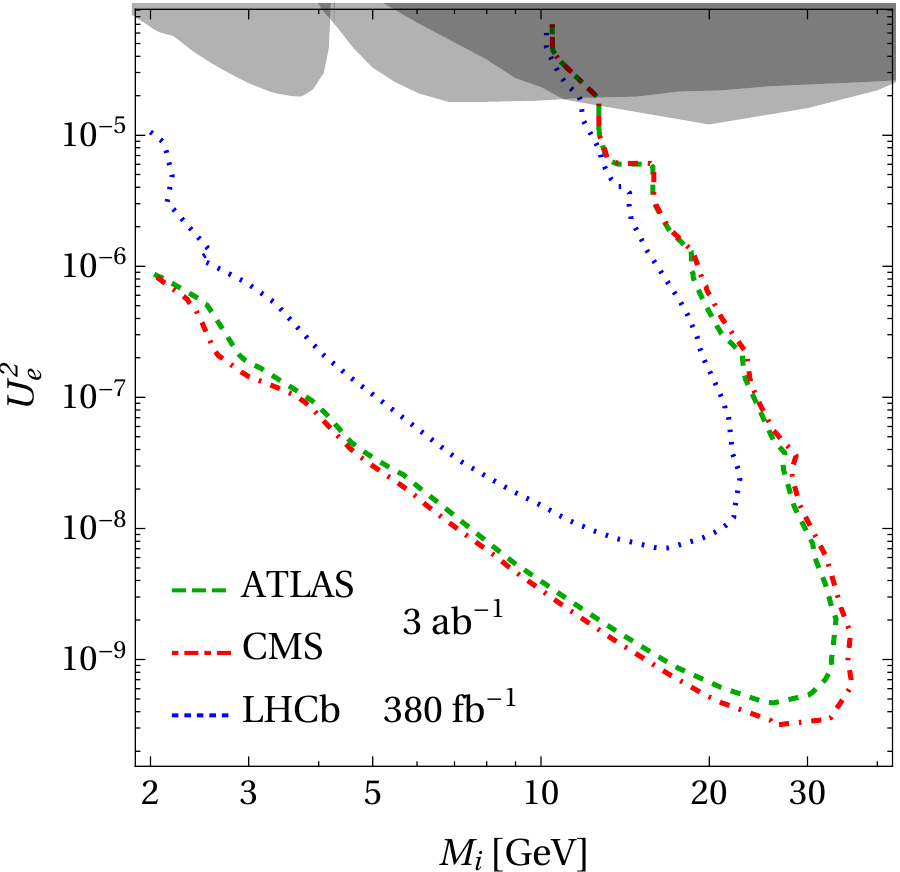}
\includegraphics[width=0.32\linewidth]{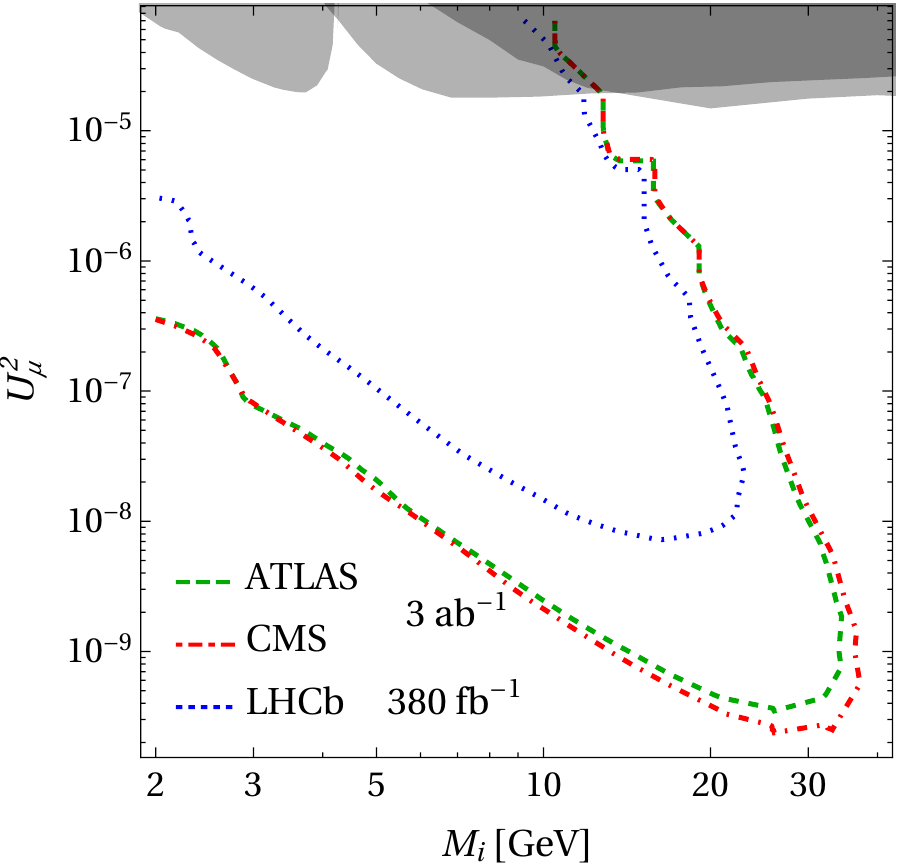}
\includegraphics[width=0.32\linewidth]{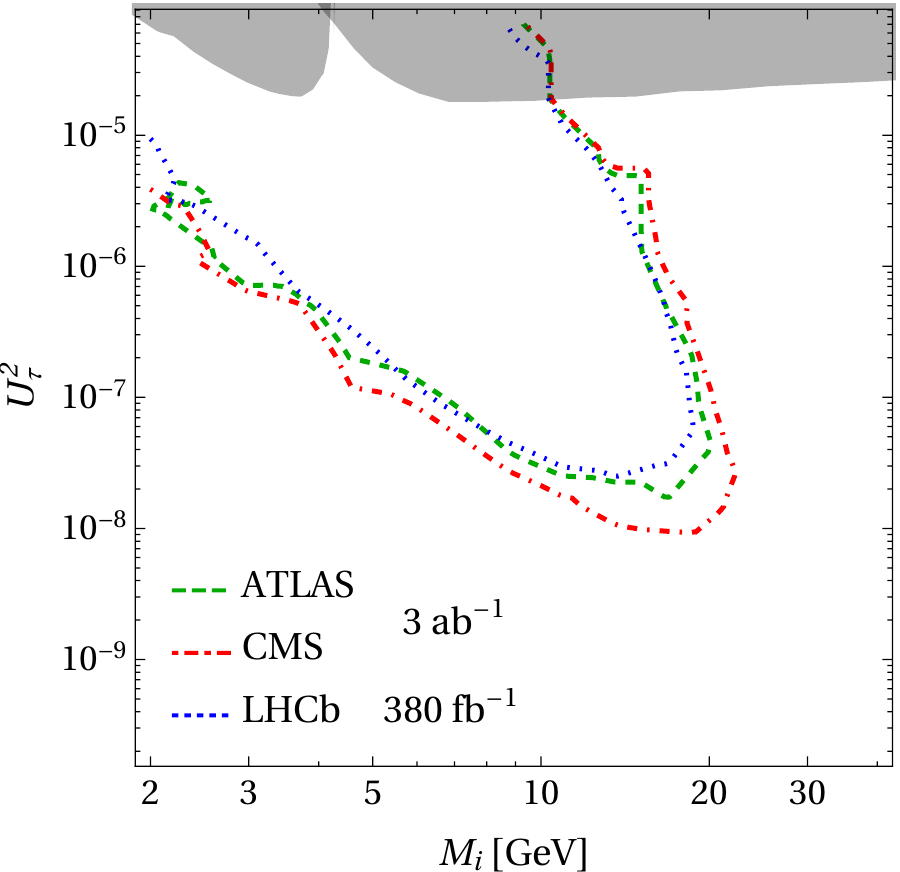}
    \caption{Exclusion reach of ATLAS, CMS with 3 ab$^{-1}$ and LHCb with 380 fb$^{-1}$ for the HL-LHC under ideal conditions for pure electron, muon and tau mixing in panel (a), (b), and (c),  respectively \cite{Drewes:2019fou}.}
    \label{fig:HL-LHC}
\end{figure}

In 2022 the LHC will start a new run, called Run 3, 
which will 
continue till the end of 2025, and should
more than double the luminosity accumulated so far, allowing to 
accumulate a total data sample of as much as perhaps 400-450 fb$^{-1}$. The increase in statistics 
will in particular allow to probe a mass range towards 
about 20
GeV for small couplings.
The next phase  will be the High-Luminosity LHC option, i.e., the HL-LHC. For this era, machine and detector upgrades have been designed and 
are being constructed.  These will replace present detector components that will retire after Run 3. 
The HL-LHC operation should start before the end of this decade, and collect a total data sample 
corresponding to about 3 ab$^{-1}$ per 
experiment in the following 
O(10) years. A simple scaling of the present results 
to the higher luminosity sample compared to
the present one,  and assuming
the experimental upgrades will allow LHC central detectors to collect data samples of the same quality as the present ones despite higher pile-up levels,
these data could allow ATLAS and CMS to probe the parameter space region towards mixing of
$\vert V_{\ell N}\vert^2\approx 10^{-8}$ in the mass range of 10-20 GeV, i.e.,
complementing to the sensitivity by fixed target beam-dump
(and neutrino) experiments in a higher mass range. For lower masses, the increasing lifetime of the HNLs that is associated with smaller mixing 
reduces the potential gain with the present central detectors due to the reduced
acceptance for the HNL decays in the detectors. 
To have more firm predictions of the future potential 
sensitivity of ATLAS and CMS, the full Run 2 analyses 
need to be completed first, and detailed  simulation studies are required by the experimental groups to map the 
sensitivity contours for the HL-LHC's full potential.
We strongly encourage such studies to be conducted in the forthcoming years. 

In addition one can further certainly envisage that the 
future data selection efficiency can be enhanced by tools
that are now being explored, but are still in  
an early phase, such as advanced Machine Learning (ML)
techniques or any new generation of such advanced
analysis algorithms that may become available in some years from now, resulting from the present ongoing wave of 
new developments. 
Early studies on applying ML techniques to HNL searches appear encouraging~\cite{Feng:2021eke}.
Furthermore an important 
improvement of the new upgraded detectors 
that will become available as an additional handle for
long lived particle analyses is the more global precise timing capabilities of
the detectors, with an aimed precision of 30 
pico-seconds. Hence, the hunt for HNL 
signals at ATLAS and CMS will certainly continue
to improve 
over the lifetime of the operation of the 
LHC and should increasingly cover the mass region above 
O(1) GeV. 

Several phenomenological studies have been conducted 
for HNL searches at the LHC.
Focusing first on light HNLs, in
Ref.~\cite{Drewes:2019fou} a phenomenological
study was presented on the potential sensitivity
to low-mass HNLs at the HL-LHC.
The aim of that work was to investigate what HNL
parameter space could be in reach 
in principle, based on 
present  detectors without significant modifications to detector coverage  but  assuming optimizations to obtain 
similar to present 
reconstruction efficiencies, understanding of backgrounds etc.
Specifically it was assumed:
to have a  minimal displacement 5 mm to reduce B meson background; have a displaced vertex invariant mass of 2 GeV to suppress backgrounds from interactions with the material; a displaced vertex has to have at least two tracks with $\Delta$R $\ge $  0.1; a 
primary vertex lepton $p_T$ 
($e, \mu, \tau$ ) = 30, 25, 140 GeV for ATLAS and CMS; a primary vertex lepton $p_T$ ($e, \mu, \tau$ ) = 10, 15, 50 GeV for LHCb; all spatial momenta $|p| >$ 5 GeV to make it out of the magnetic field; muon chambers at CMS can be used for displaced vertex reconstruction.
SM backgrounds were 
included in the study. These, somewhat optimistic,
results shown in Fig.\ref{fig:HL-LHC}
give a clear target for the experiments for the 
parameter phase space that could in principle be covered under ideal conditions by future LHC data.

As mentioned before, dedicated analyses at even lower masses based on searches 
in charm and bottom decays is also  an interesting 
channel to pursue for these experiments. Such
searches will be discussed in the next section
for LHCb; for ATLAS and CMS no results have to date been released that allow to evaluate the power of these 
analyses for the future.

\begin{figure}[!t]
\centering
\includegraphics[width=0.43\linewidth]{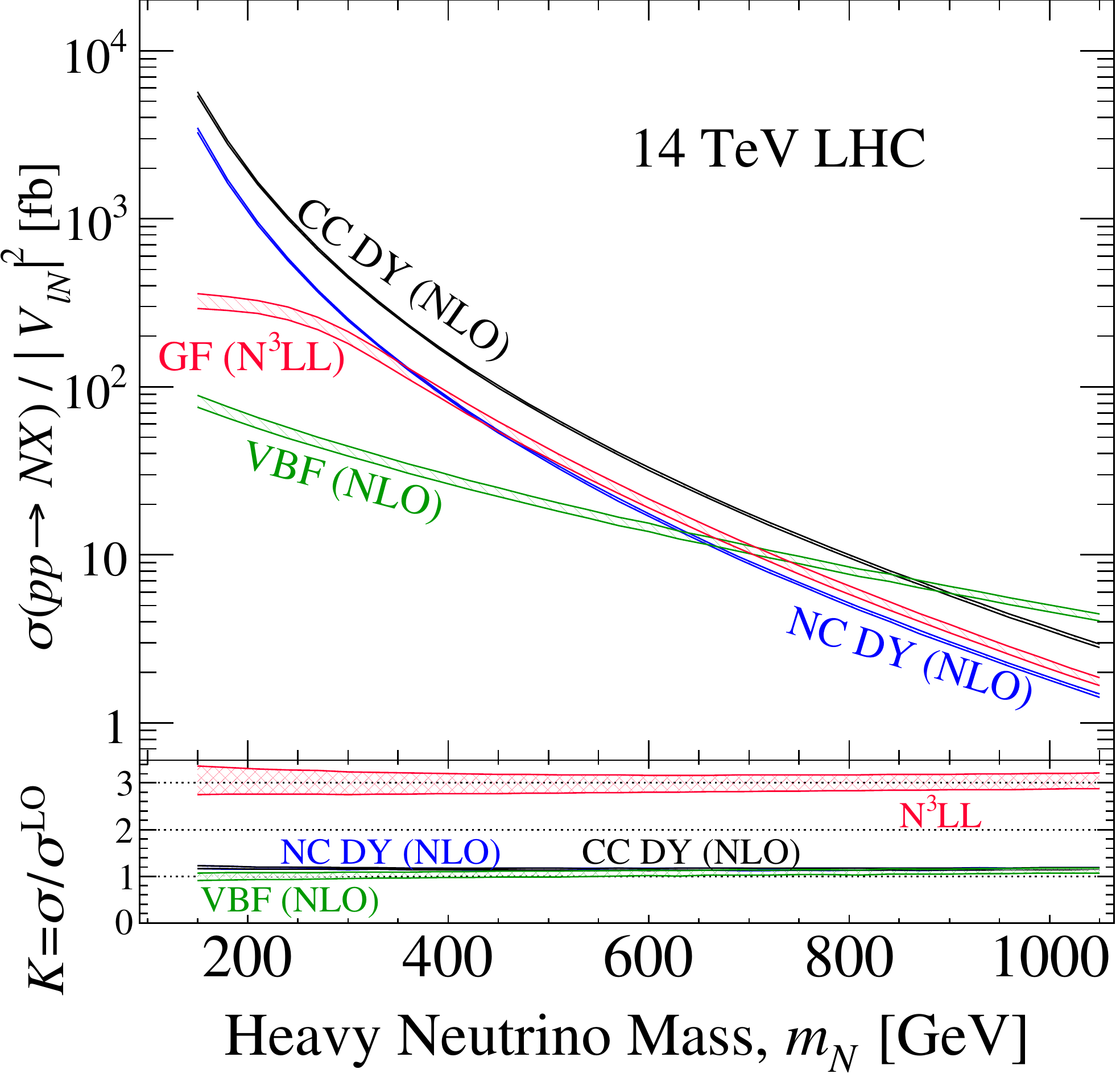}
\includegraphics[width=0.43\linewidth]{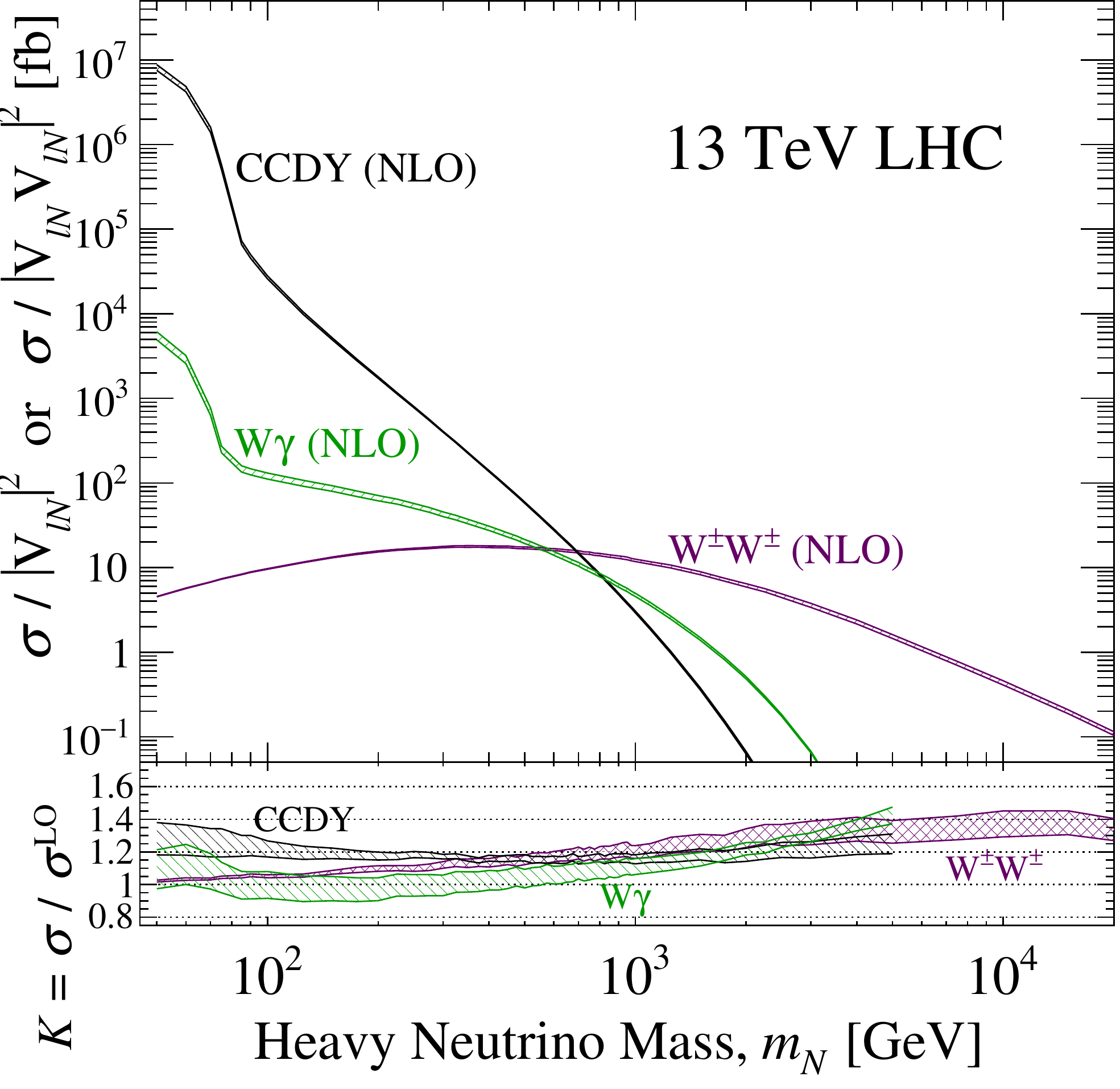}

\caption{
Left: LHC cross sections for HNL production, with active-sterile mixing normalized to unity, as a function of HNL mass for the charged current Drell-Yan, neutral current Drell-Yan, gluon fusion, and $W\gamma$ fusion channels. The lower panel shows the QCD $K$-factors relative to the LO cross section~\cite{Degrande:2016aje,Pascoli:2018heg}.
Right: Same but for the charged current Drell-Yan, $W\gamma$ fusion, and same-sign $WW$ fusion channels~\cite{Fuks:2020att}.
}
\label{fig:hllhc_TypeI_xsec}
\end{figure}

\begin{figure}[t!]
    \centering
\includegraphics[width=0.49\linewidth]{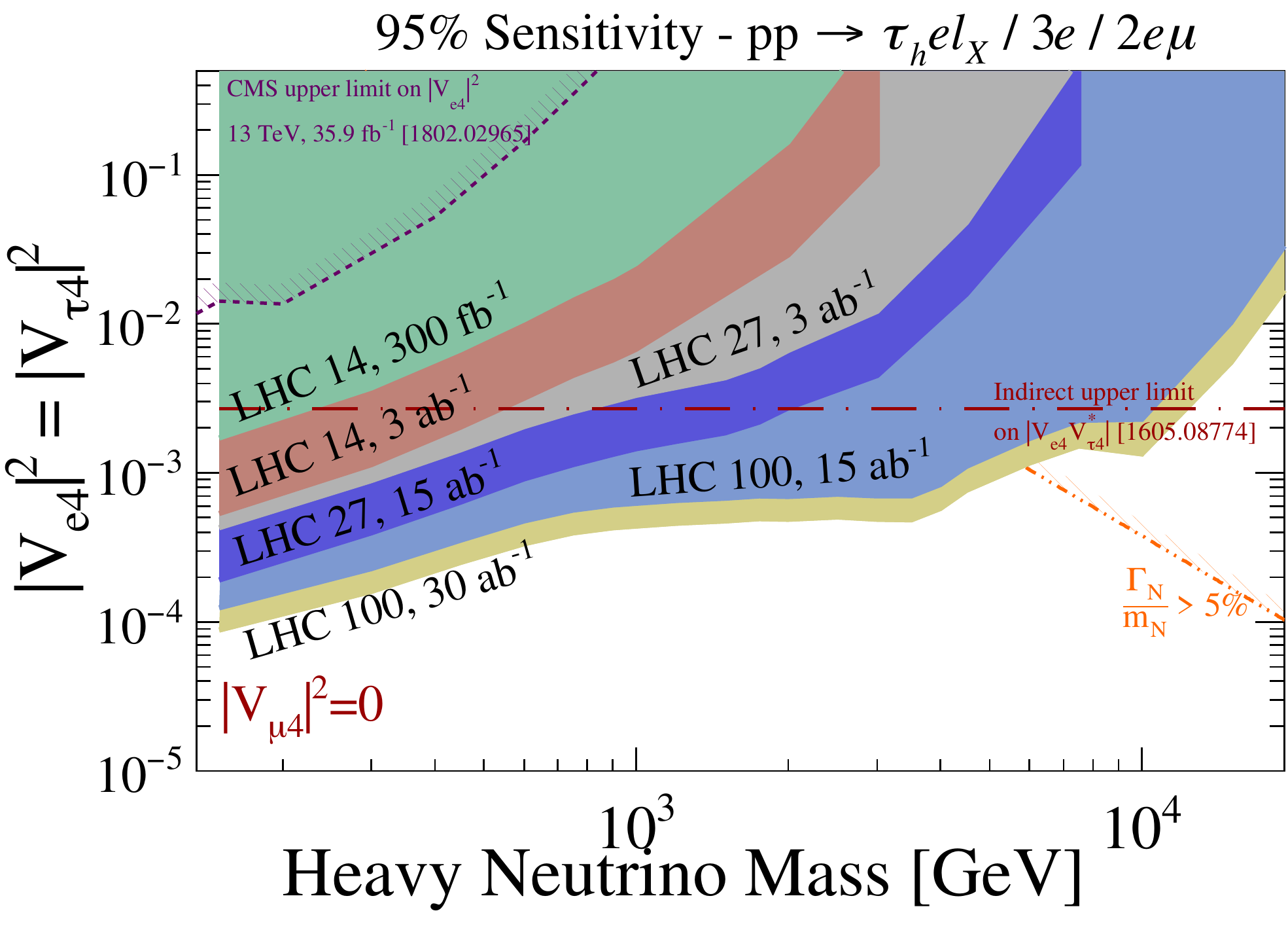}
\includegraphics[width=0.49\linewidth]{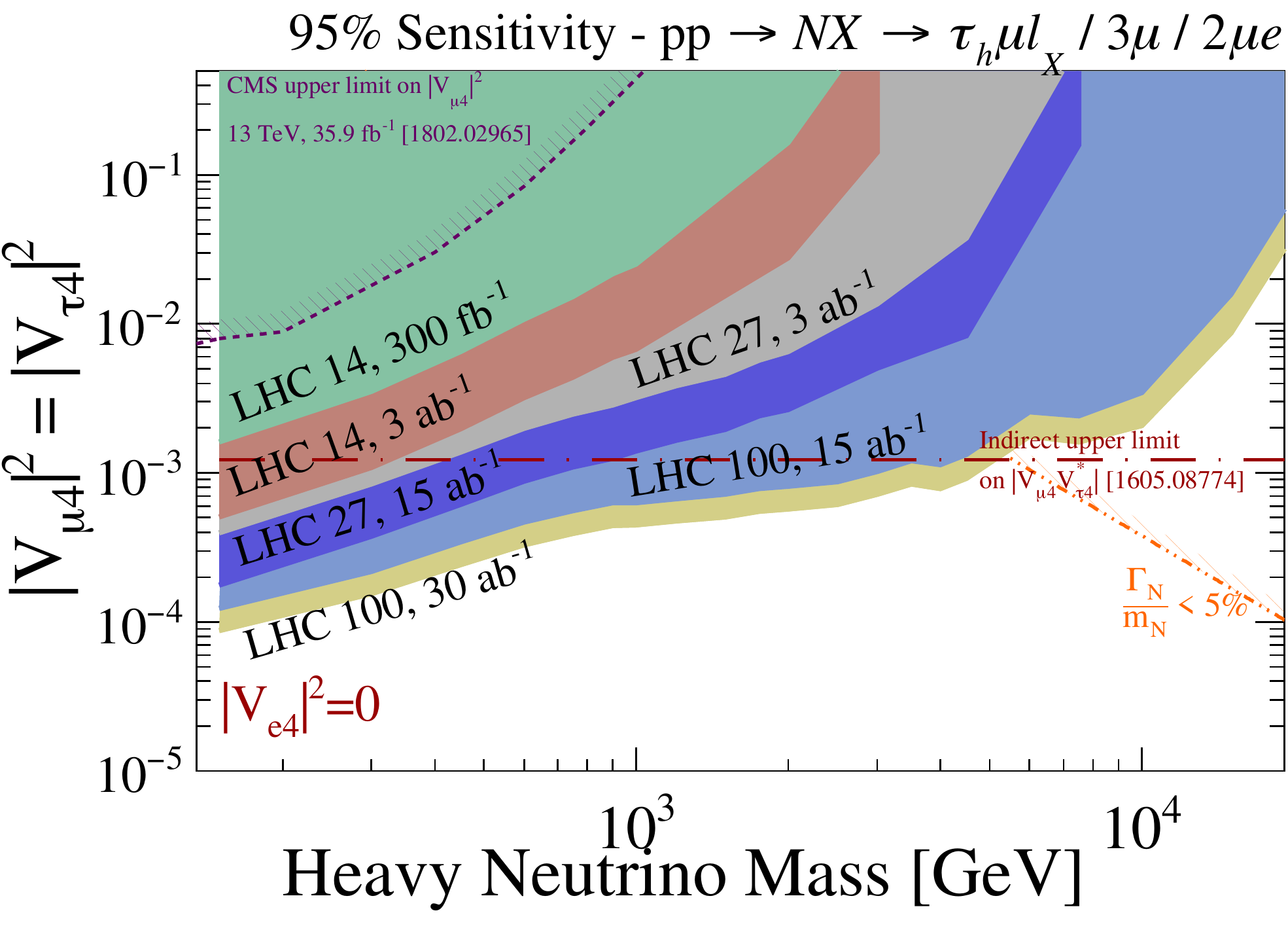}
    \caption{Expected sensitivity at 95\% at the HL-LHC (and future colliders) to Dirac HNLs with EW- and TeV-scale masses, in the trilepton and MET channel assuming the mixing configuration (left) $\vert V_{e4}\vert=\vert V_{\tau 4}\vert$ but $\vert V_{\mu4}\vert=0$,
    and 
    (right) $\vert V_{\mu4}\vert=\vert V_{\tau 4}\vert$ but $\vert V_{e4}\vert=0$~\cite{Pascoli:2018heg}.}
    \label{fig:hllhc_TypeI_outlook}.
\end{figure}

For masses above a 100 GeV, the prospect of discovering HNLs is enhanced by the opening of several production mechanisms with competitive cross sections, as shown for example in Fig.~\ref{fig:hllhc_TypeI_xsec}. 
Some of these production channels involve vector boson fusion, 
which enhance sensitivity to HNL parameter space and to new models~\cite{Alva:2014gxa,Jones-Perez:2019plk,Pascoli:2018heg,Fuks:2020att,Buarque:2021dji}.
With such large masses, HNLs can decay to on-shell EW bosons that are potentially boosted.
In addition to ML techniques~\cite{Feng:2021eke}, the development of novel jet-veto techniques show promise to greatly improve signal-to-noise ratio in searches for heavy HNLs~\cite{Pascoli:2018rsg,Pascoli:2018heg}. As shown in Fig.~\ref{fig:hllhc_TypeI_outlook}, the multi-lepton and MET channel, active-sterile mixing as small as $\vert V_{\ell N}\vert^{2} \sim 10^{-5}-10^{-2}$ in the electron, muon, and tau sectors can be probed at 95\% CL for heavy Dirac neutrinos masses in the range $m_N = 200-1200$ GeV with 3 ab$^{-1}$ of data at $\sqrt{s}=14$ TeV~\cite{Pascoli:2018rsg,Pascoli:2018heg}.
Results are shown for the Dirac channel but comparable sensitivity is expected for the Majorana channel~\cite{Pascoli:2018rsg,Pascoli:2018heg}.


\paragraph{{\it Extended gauge symmetries and effective field theories}}

\begin{figure}[t!]
  \centering
  \includegraphics[width=0.48\textwidth]{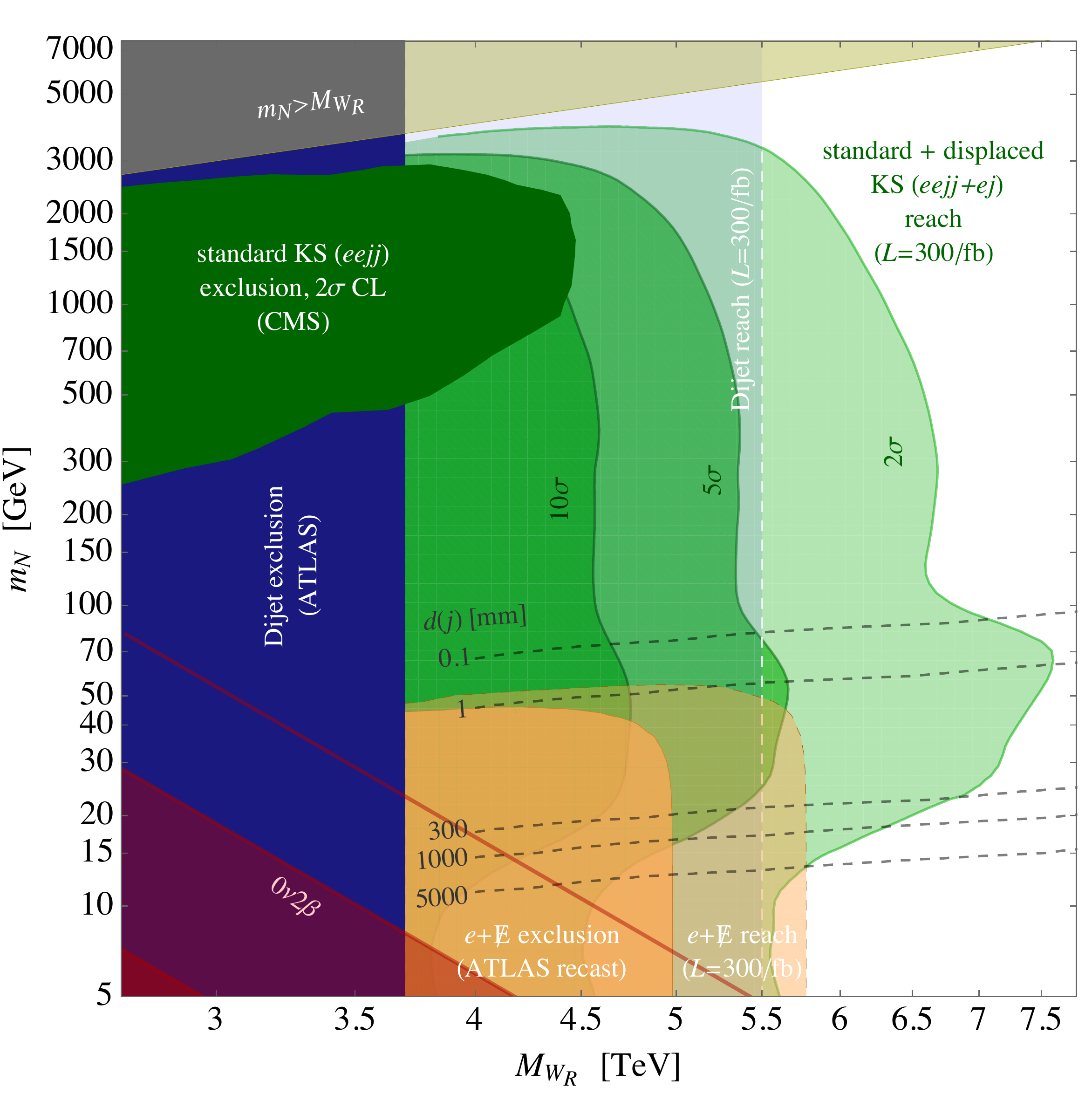}
\includegraphics[width=0.47\textwidth]{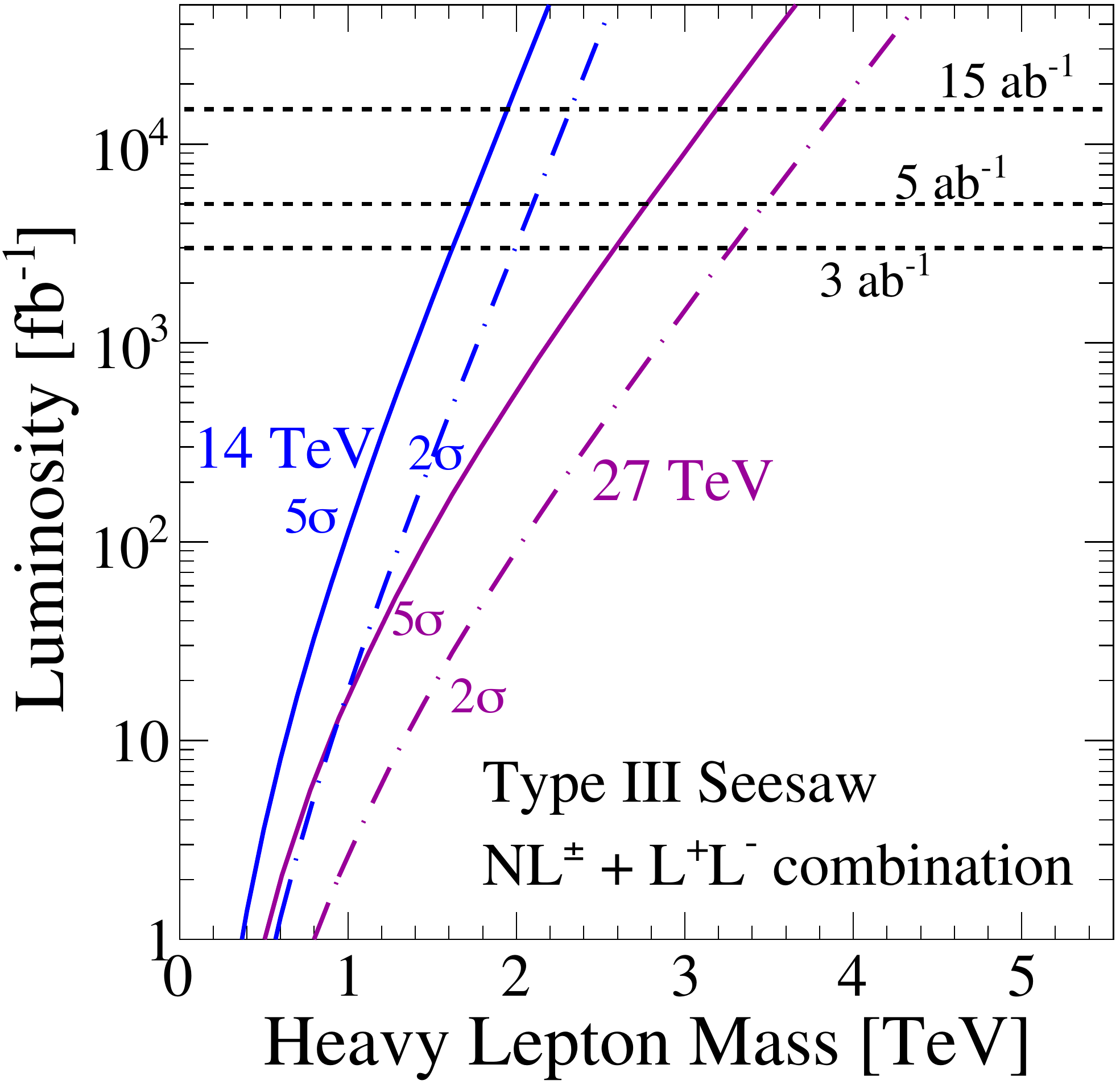}
 \caption{Left: Combined exclusion limits from prompt, displaced (decay length in dashed lines) and missing energy final states in the electron channel~\cite{Nemevsek:2018bbt}.
Also shown is the weaker dijet limit, constraints from neutrinoless double beta
searches and an outlook for 300 fb$^{-1}$.
Right: Expected discovery and exclusion potential of the HL-LHC at 14 TeV to discovering heavy leptons in the Type III Seesaw when the charged current $(NL^\pm)$ and neutral current $(L^+L^-)$ channels are combined~\cite{Ruiz:2015zca,Cai:2017mow}.
}
\label{fig:LHC_WR_outlook} 
\end{figure}

Due to the significantly larger data sets, upgraded detector experiments, and improved Monte Carlo capabilities, the sensitivity to HNLs in extended-gauge models is very promising at the HL-LHC. 
As shown in Fig.~\ref{fig:LHC_WR_outlook} (left), direct searches for HNLs and $W_R$ in the standard and displaced channels are sensitive to $M_{W_R}$ well above 5-6 TeV with the full Run 3 data set~\cite{Nemevsek:2018bbt,Cottin:2019drg}.
Higher masses can be probed with the full HL-LHC data set~\cite{Das:2012ii,Han:2012vk,Mitra:2016kov}.
Searches for $N$ via far-off shell $W_R$ can probe $M_{W_R}\sim \mathcal{O}(10)$ TeV for $m_N < 1$ TeV with 1 ab$^{-1}$ of data~\cite{Ruiz:2017nip}.
More generally, searches for short- and long-lived HNLs with masses $m_N > 100$ GeV via dimension six operators show that EFT scales as large as $\Lambda \sim \mathcal{\mathcal{O}}(10)$ TeV can be probed~\cite{Duarte:2016caz,Ruiz:2017nip}.
For smaller masses, active-sterile mixing below $\vert V_{\ell N}\vert^2 \sim 10^{-10}$ can be probed for a fixed $\Lambda = \mathcal{O}(10)$ TeV~\cite{Chiang:2019ajm,Cottin:2021lzz,Beltran:2021hpq}.
As for the Higgs-mediated channels, in the case of $gg\to \Delta_R^0 \to NN$ production, the LHC is able to probe  significantly higher LR scales with $M_{W_R} \sim \mathcal O(10 \text{ TeV})$, depending on the size of the Higgs-triplet  mixing. 
With such high $M_{W_R}$, the $N$ decay rates are further suppressed and final states
may feature pairs of significantly displaced $N N$ vertices. For further details, see Fig.~8 in Ref.~\cite{Nemevsek:2016enw}.

Finally, in minimal $U(1)_{B-L}$ models, the LHC detector experiments with $\mathcal{O}(100)$ fb$^{-1}$ of data are sensitive to $Z'$ bosons as heavy as 4-5 TeV for effective $B-L$ couplings of $g_{B-L}=0.05$~\cite{Basso:2008iv,Basso:2010pe}. With more data, larger masses and smaller couplings can be probed.
Due to its contribution to the Drell-Yan continuum, it is also possible to constrain very heavy $Z'$ from $B-L$ models in precision measurements of low-$p_T$ dilepton pairs; the precise sensitivity at Run 3 and the HL-LHC depends on the $Z'$ mass and coupling~\cite{Accomando:2017qcs,Amrith:2018yfb,Deppisch:2019ldi}.
A particular novelty of $U(1)_{B-L}$ models is that $NN$ pair production is possible through coupling to the $Z_{B-L}$ gauge boson and $B-L$ Higgs mixing. Among other channels, this implies the potential production of displaced, long-lived $NN$ pairs~\cite{Nemevsek:2016enw,Deppisch:2018eth,Padhan:2022fak}.

\paragraph{{ \it Type III Seesaw}}

If HNLs are the neutral components of a leptonic, EW triplet, as in the Type III Seesaw, then the HL-LHC will greatly improve present sensitivity. In particular, the increase in integrated luminosity will increase sensitivity to heavy lepton pairs up to $m_L=1.8-2$ TeV with 3 ab$^{-1}$ of data~\cite{Ruiz:2015zca,Cai:2017mow}. This assumes the baseline analysis of Ref.~\cite{Arhrib:2009mz}, and varies depending on specific flavor-mixing assumptions.
Such sensitivity also holds for many-lepton final states as well as mult-lepton finale states with fat jets, i.e., boosted, hadronically decay $W/Z/H$ bosons~\cite{Ashanujjaman:2021zrh,Ashanujjaman:2021jhi}. However, further studies are needed to investigate the ultimate sensitivity achievable when fully leptonic, semi-leptonic, and fully hadronic channels are combined.

%


\subsubsection{Current LHCb Results}
The current LHCb results involving HNLs cover a wide mass region. The mass region can be divided into a low mass ($m_{N}<m_{W}$) region and a high mass region ($m_{N}>m_{W}$).
In the low mass region LHCb can be competitive with other experiments. 
For the high mass region this is more challenging.

\begin{figure}[t!]
\centering
\includegraphics[width=0.47\textwidth]{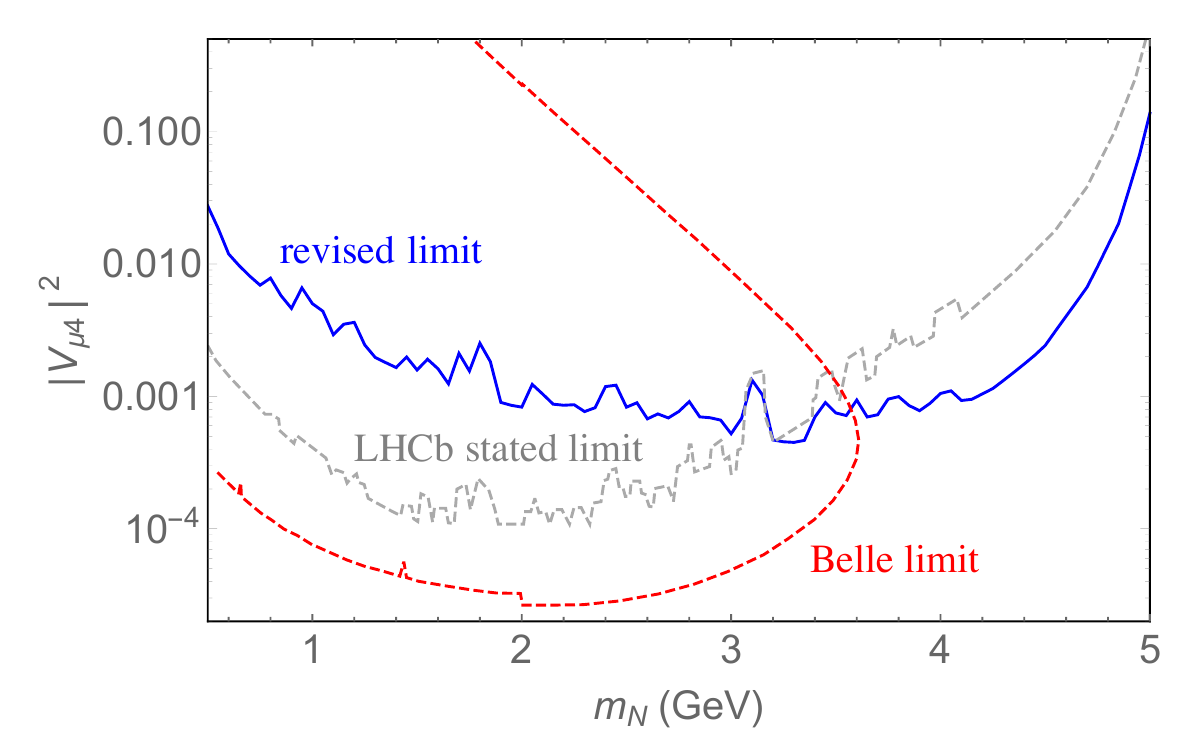}
\includegraphics[width=0.52\textwidth]{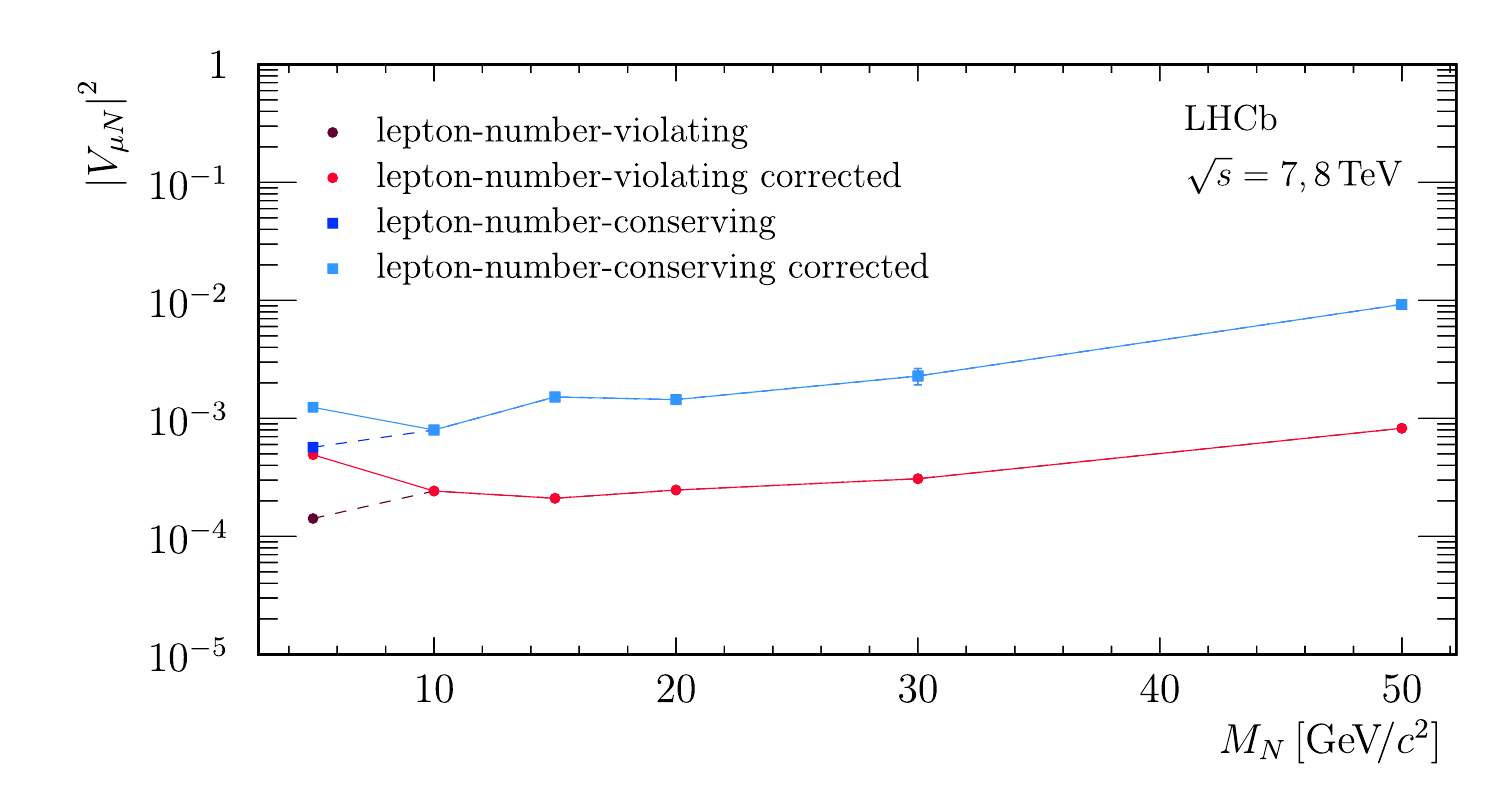}
\caption{Current LHCb reach for the HNL that mixes only with the muon neutrino 
$\nu_\mu$.
Off shell (left) limits are shown here with the theoretically revised reach~\cite{Shuve:2016muy}. On shell results (right) are shown for the lepton number violating and conserving cases~\cite{LHCb:2020wxx}.}
\label{fig:LHCb}
\end{figure}

In the low mass region, the published result by LHCb so far covers the $B^{+} \rightarrow \mu^{+} N \rightarrow (\pi^{+}\mu^{-})$ channel~\cite{LHCb:2014osd} via off-shell $W$ decays and is shown on the left hand side of Fig.~\ref{fig:LHCb}.
The published LHCb result was later corrected 
due to the initial use of incorrect decay rates for $B^\pm$ and $N$~\cite{Shuve:2016muy}. 
In the high mass region the published result covers the $W^{+}\to \mu^{+} \mu^{\pm} \mathrm{jet}$ channel via on-shell $W$ decays and is shown on the right hand side of Fig.~\ref{fig:LHCb}~\cite{LHCb:2020wxx}.
The searches via off-shell $W$ decays suffer from a stringent cut on the $B^{+}$ meson invariant mass. 
This cut has been removed for future studies where contributions from $B^{0}$ and  $B^{0}_{c}$ will also play a central role.

\subsubsection{Expected future LHCb Results}

New results will be released in 2022 based on the full Run 2 data. 
The new results will cover both on- and off-shell $W$ productions of HNLs. 
The first results to be published will most likely be with HNLs decaying in $\pi\mu$ final states. For off-shell $W$ productions of HNLs the decay chain will therefore include, but not be limited to $B^{+} \rightarrow \mu^{+} N \rightarrow (\pi^{+}\mu^{-})$. 
Further extensions, especially using $B_{c}$ initial states should be expected. 
The increase of the delivered luminosity during Run 3 will also be accompanied by an essentially new detector~\cite{LHCb:2012doh}. 
For LHCb this translates to a total integrated luminosity of recorded data for Run 3 of about 50 fb$^{-1}$.
Crucially, the yield for leptonic channels will be increased by a factor 10 and for hadronic channels by a factor 20. 
This is due not only to the increase in luminosity but also to a full detector readout at 30 MHz. This, together with the removal of the hardware trigger stage will allow for an output of 2 to 5 GB/s to storage. 
The LHCb experiment during Run 1 and 2 was able to cover low region of phase space thanks to its low $p_T$  reach.
This will increase even further for Run 3 thanks to a new tracker which will allow good efficiency down to particles with a $p_T$ of about 0.5 GeV.
The greater $p_T$ resolution will be accompanied by a drastic reduction in
ghost rate and a large gain in reconstruction time.

Further development should be expected with the increased in luminosity due to the HL-LHC conditions. A second upgrade has been already proposed and is currently under consideration by the funding agencies~\cite{LHCb:2018roe}. The upgrade, labeled Upgrade II, will potentially make LHCb the only general purpose flavor physics facility in the
world on this timescale and a general purpose experiment in the forward direction. This will be crucial for exploring HNLs coupling to leptons in all generations. Nevertheless LHCb was not built to work under HL-LHC conditions and therefore some of its key elements  will have to be replaced. For example the whole tracking apparatus and some the particle identification. 
If this will happen, LHCb will be able to explore parameter space in regions which will remain unexplored at the LHC. This will be particularly true for the semi-displaced regions just above the charm threshold.

\subsection{Future LHC Experiments: New experiments for Run 3 and proposals for the HL-LHC era}
Given that until now no conclusive signatures 
for new physics have been discovered at the LHC,
it is legitimate to ask the question whether we have been looking so far  at the right place  and to make sure that  no stone is left unturned in searching for it while the LHC will be operated. A systematic study of ideas for searches for LLPs, that 
has culminated in the last years, is a result of this self-assessment in the community. A Community Paper for LLP searches 
with its findings and suggestions has been produced \cite{Alimena:2019zri}
and  sets the direction for future experiments and  measurements. Searches for HNLs feature prominently in this review.

As a result of increasing interest for searches for LLPs in the last five years, many proposals have surfaced for possible new experiments at the LHC.
One of the key benchmark processes used as a metric in comparing the capabilities for these new experimental proposals is the sensitivity to 
searches for long lived HNLs. These searches
have been a key benchmark process for the Physics Beyond Colliders 
study~\cite{Beacham:2019nyx}, and hence the new proposals often included a study for the sensitivity 
of these particles.

The new  proposals include two kind of experiments, so called central/transverse and forward experiments. The latter 
are located generally downstream along the beam-pipe/accelerator direction of central detectors, and are
typically several tens up to hundreds of meters away from the interaction point. Experiments relevant for 
HNL searches include FASER, SND@LHC, FACET, and the proposed Forward Physics Facility (FPF)~\cite{Feng:2022inv}, a new
underground facility for which several potential
experiments are being discussed

The central or transverse detectors search for new particles at central values of pseudo-rapidity ($\eta$) in the pp collision system, and the key
experiments that are or have been discussed 
in this category with 
relevance to possible HNL measurements are MATHUSLA, CODEX-b, ANUBIS, AL3X, and MAPP. 
It is worth noting that among these proposed new experiments FASER and SND@LHC have been approved already, are fully funded and have been installed at the LHC during 2020 and 2021, ready to take data during Run 3.
In the following we give a short description 
of these experiments and discuss the reach for searches for HNLs. 

In the following a few selected example figures are shown 
to give an indication of the 
sensitivities for searches to HNLs, taken from public
results as reported by the various studies for these new 
detector ideas. The figures aim to show the reach for all
detector proposals, sometimes within specific models used for
the study. Also one should go to the original papers of these
studies to appreciate the level of detail for e.g.
the detector response or level of background estimates 
that have been assumed, which may somewhat differs for the 
various studies.

\subsubsection{FASER}
 The FASER experiment was proposed to be installed in the very forward region. FASER, which stands for ForwArd Search ExpeRiment will be in 
 an excellent position to search for low mass and extremely weakly interacting particles at the LHC \cite{FASER:2018bac}. FASER uses the same interaction region as ATLAS and is located in the line-of-sight, behind 100m of rock shielding,
 480 m downstream from the ATLAS interaction point, in the unused service tunnel TI12.  The LHC magnets sweep away charged particle backgrounds hence the experimental area experiences low radiation and low beam backgrounds. The FASER experiment (Phase1) has been approved on March 5th 2019. 
 FASER consists of a tracker system, veto layers, electromagnetic calorimeter modules, a decay volume, and a permanent magnet.  The detector is about 5 m long; the decay volume has a diameter of 20 cm 
 and a length of 1.5 m.
 It makes use
 of spare ATLAS silicon tracking modules and LHCb electromagnetic calorimeter modules. 
 FASER is ready to take data in 2022. An 
 extension of the experimental equipment for 
 neutrino detection based on emulsion technology is the FASER$\nu$ experiment\cite{FASER:2020gpr}.

FASER will be able to place  constraints on GeV-scale HNLs for all flavors of mixing, $|U_{eN}|^{2}$, $|U_{\mu N}|^{2}$, $|U_{\tau N}|^{2}$~\cite{Kling:2018wct}. 
The sensitivity reach for FASER for  HNLs is shown
in Fig.\ref{fig:SND} and Fig.~\ref{fig:CODEXB}, compared 
to other experiments.  
New leading results can be produced on 
$|U_{\tau N}|^{2}$ as seen from the figure.
An upgraded version of FASER, called FASER2
will also be able to produce leading results on the other HNL couplings as well. FASER2, not yet approved, 
is a larger acceptance version of FASER and proposed for the HL-LHC operation of the LHC. 
Recently FASER2 has been included in the FPF facility
discussed below.



\subsubsection{SND@LHC}
The Scattering and Neutrino Detector at the LHC (SND@LHC) experiment is first and foremost a high efficiency neutrino type identification experiment~\cite{SHiP:2020sos}.
It is installed in the TI18 tunnel at the distance of 480 m from the ATLAS interaction (on the 
opposite site of ATLAS compared to FASER) with a pseudorapidity range coverage of $7.2 < \eta < 8.6$.
The installation happens in a symmetrically and opposite location with respect to FASER and therefore SND@LHC will also benefit from the same rock shielding and LHC sweeping magnets.

The target region of the detector is equipped with both offline and online subdetectors. The offline detection consists of emulsion cloud chambers. These chambers are stacked with scintillating fibers; a muon identification system is placed downstream. Different particle identification techniques are used to identify different charged particles.
Micrometric resolutions of the vertexes in combination with a timing resolution of around 100 ps make this target an ideal place to study displaced particles. 

\begin{figure}[t]
\centering
\includegraphics[width=0.95\linewidth]{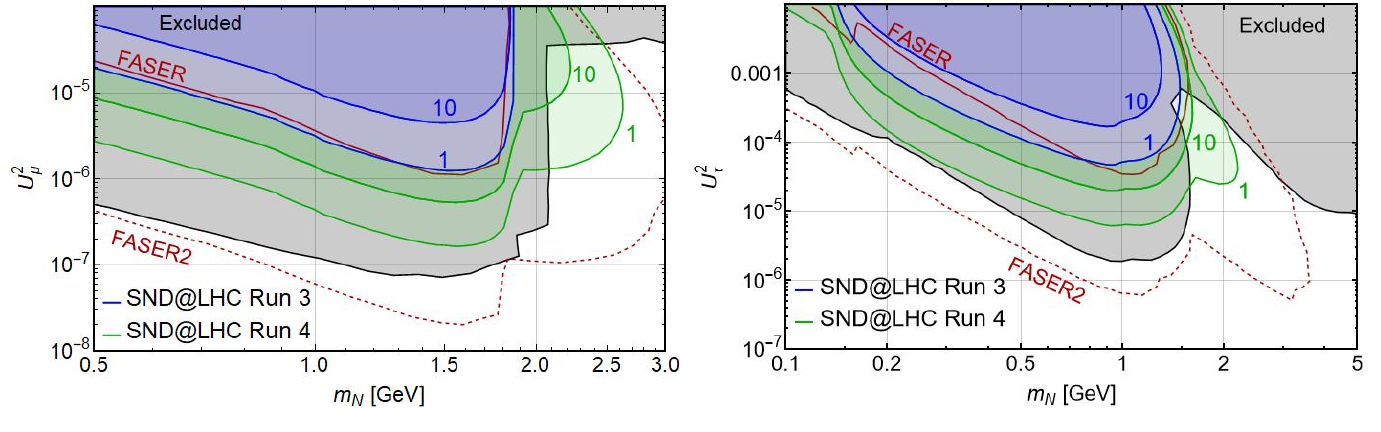}

\caption{FASER and SND@LHC’s reach for the HNL that mixes only with the muon neutrino 
$\nu_\mu$ (left) and only with the tau neutrino
$\nu_\tau$ (right). The gray shaded regions are excluded by current limits. 
Solid blue (green) contours correspond to 1 and 10 events in the (upgraded) SND@LHC target. FASER and FASER2 results are also reproduced. Figure taken from~\cite{Boyarsky:2021moj}}

\label{fig:SND}
\end{figure}

The SND@LHC collaboration has outlined different techniques that will be used to search for physics beyond the SM. Light dark matter can be searched for using scattering off protons with both elastic and NC/CC (inelastic) signatures. When displaced particles decay to a pair of charged particles inside the target region, SND@LHC can also be sensitive to  portals, among which HNLs play an important role. In the case of HNLs mixing with a $\nu_\tau$ the calorimeter property of the target are used to reconstruct the decay $N \to \pi^{0} \nu$. Fig.~\ref{fig:SND}~\cite{Boyarsky:2021moj} shows what was previously reported in the literature. It can be seen that during Run 3 SND@LHC can explore a new  parameter space region if  HNLs  mix exclusively with $\nu_\tau$. The explorable mass region sits below 1 GeV and that should close the existing gap between the CHARM and the DELPHI experiments.
For a potential future upgrade SND@LHC will also be able to probe unexplored parameter space regions  if  HNLs  mix exclusively with $\nu_\mu$.

\subsubsection{MATHUSLA}
The MATHUSLA project \cite{Curtin:2018mvb} proposes a very large surface detector based on an affordable technology to detect decays of ultra long-lived neutral weakly interacting particles, with proper $c\tau$ lifetimes up to $10^7-10^8$ meter. This would allow coverage of lifetimes that range from the start of the early Universe to the Big Bang Nucleosynthesis (BBN) epoch. The presence of remaining heavy long-lived particles in that era could significantly disturb the abundance of the elements, which have strong constraints from data. Hence it means that early Universe long-lived particles should not have lived significantly beyond the start of the BBN phase. MATHUSLA is positioned physically at a significant distance from the LHC detectors interaction point, i.e. more than 100 m, and needs to be huge in size in order to have a sufficient acceptance for observing these centrally produced particles. The proposed size is a detector of 100x100x30m,  build up from individual smaller modules so that a phased installation over years  will be possible. Decays of LLP neutral particles will be detected in 
multiple planes of extruded scintillator, and perhaps complemented with Resistive Plate Chambers.

MATHUSLA will be sensitive down to $10^{-9}$ for mixing matrix elements involving the $e$ and $\mu$ flavors, and down to $10^{-8}$ for $\tau$ flavored mixing~\cite{Curtin:2018mvb}.
The sensitivity curves for MATHUSLA for HNLs
are shown in Fig.\ref{fig:CODEXB} and
Fig.\ref{fig:ANUBIS}.

\subsubsection{CODEX-b}

\begin{figure}[t]
\centering


\includegraphics[width=0.45\linewidth]{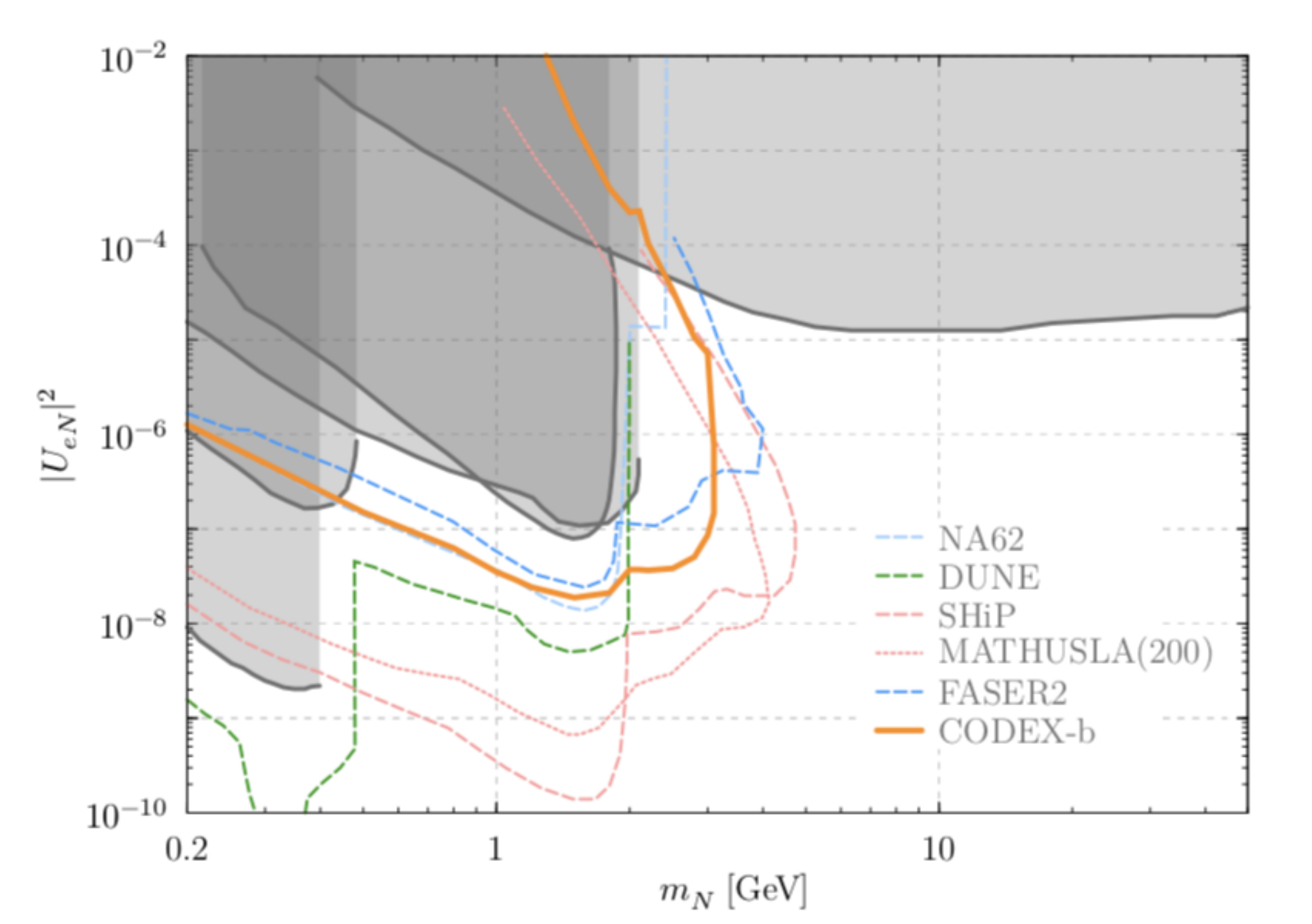}
\includegraphics[width=0.45\linewidth]{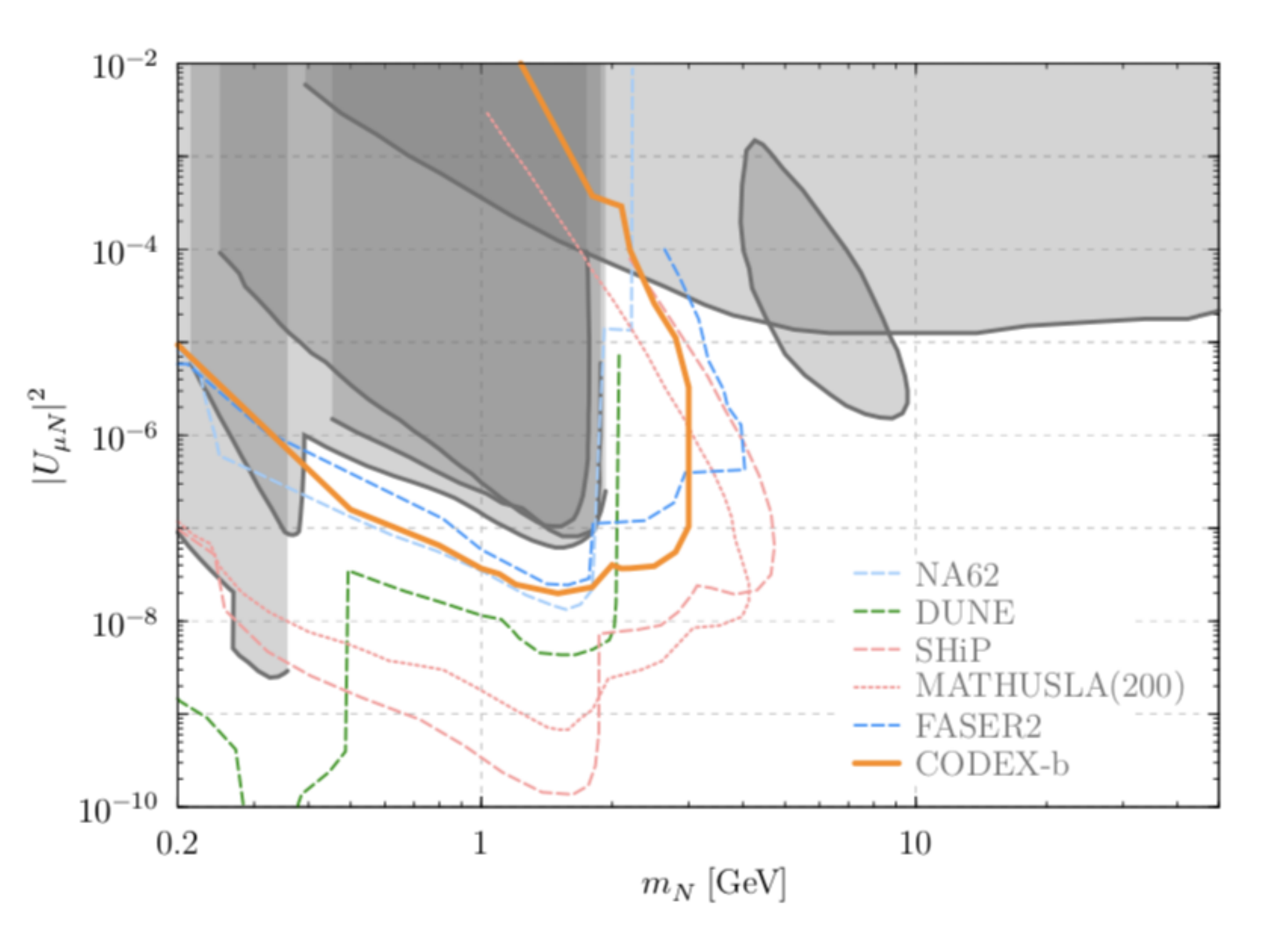}
\includegraphics[width=0.45\linewidth]{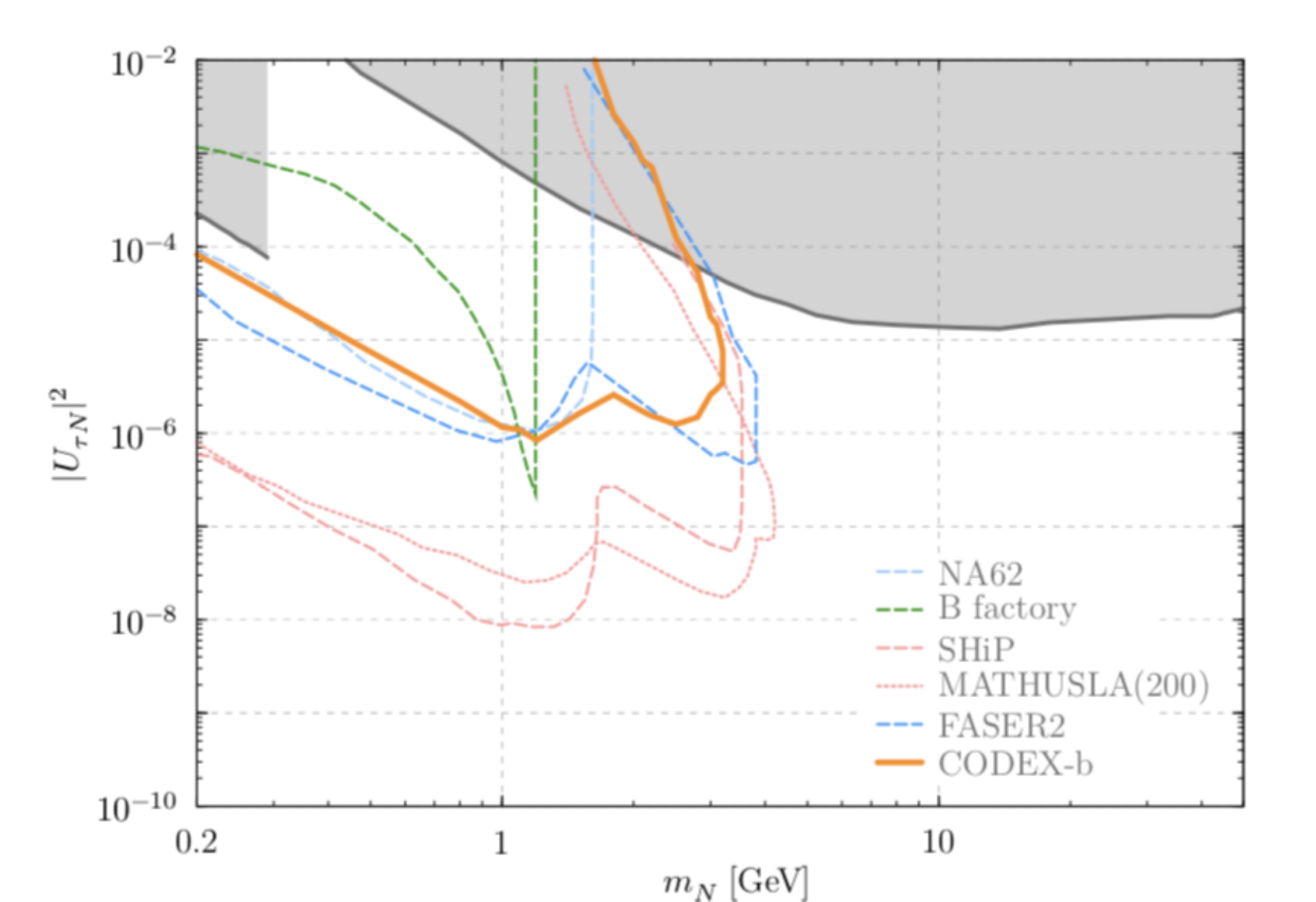}

\caption{Projected sensitivity of CODEX-b/MATHUSLA/FASER2 to Dirac heavy neutral leptons. Comparison are shown with current constraints (gray) and other proposed experiments, for pure electron, muon and tau mixing scenarios. Figures are taken from ~\cite{Aielli:2019ivi}.
}
\label{fig:CODEXB}
\end{figure}

CODEX-b has similar new particle search goals as MATHUSLA but it proposes a detector at a shorter distance to the interaction point, needing a 
smaller detector volume to detect the decays. CODEX-b is proposed to be installed in the LHCb cavern, about 20 meters away from the interaction point in a 10x10x10 m$^3$ area to the side of the LHCb experiment. A lead shield is added between the interaction point and the new detector, in order to stop hadronic activity coming from the interaction point. The detector volume will consist of multi-layered planes of Resistive Plate Chambers, detecting the decays of neutral weakly interacting long-lived particles. Initial test measurements in the anticipated future CODEX-b area have been performed and it was found that the backgrounds were indeed as anticipated.
CODEX-b can be sensitive down to $10^{-8}$ for mixing matrix elements involving the $e$ and $\mu$ flavors, and down to $10^{-6}$ for $\tau$ flavored mixing~\cite{Aielli:2019ivi}.
The expected sensitivity is shown in Fig.~\ref{fig:CODEXB} and
Fig.~\ref{fig:ANUBIS}.

\subsubsection{AL3X}

The AL3X \cite{2019:AL3X} project, an idea which is mentioned for completeness, proposes to use large parts of the present ALICE detector and turn it
into a long-lived particle experiment, based on a similar strategy as for CODEX-b.  The idea is to move the interaction point from the center of ALICE to some 10 meter upstream, put a thick shield with an active veto between the collision point and AL3X, and re-use much of the present ALICE inner tracker, and the 
original L3 magnet that comes with the cavern, as a new long-lived particle detector.  It would require some revision and rebuilding of  accelerator components, and potentially also a more optimized tracking detector for the physics, so this proposal would in any case not be a cheap endeavor. Obviously the ALICE collaboration has ideas and plans of its own for the experiment. No follow up activity on the study of this 
experimental opportunity has been reported 
during the last few years.
The sensitivity to HNLs has been studied for AL3X in \cite{Dercks:2018wum} and is shown in 
Fig.~\ref{fig:ANUBIS}.


\subsubsection{ANUBIS}

Recently in \cite{Bauer:2019vqk} ANUBIS (AN Underground Belayed In-Shaft
experiment)  was
 proposed, an auxiliary detector to be installed in one of the access shafts above the ATLAS or CMS interaction point, as a tool to search for long-lived particles. ANUBIS is proposed to have four (removable) Tracking Stations (TS) spaced 18.5 m apart from each other, providing a uniform coverage of the detector volume. This setup results in a cross-sectional area of 230 m$^2$ per TS. The tracking
stations will be  based on Resistive Plate
Chamber technology. A demonstrator test  with a 
module of 2x1x1m$^{3}$ is proposed to be installed
in the ATLAS shaft during Run 3.
In \cite{Hirsch:2020klk} the sensitivity of this proposal is discussed for long-lived heavy neutral leptons (HNLs) in both minimal and extended scenarios. Both the minimal HNL model where both production and decay of the HNLs are mediated by active-sterile neutrino mixing, and  the case of right-handed neutrinos in a left-right symmetric model are studied.

\begin{figure*}[htbp]
\centering

\subfigure[\label{fig:ANUBIS}]{\includegraphics[width=0.48\textwidth]{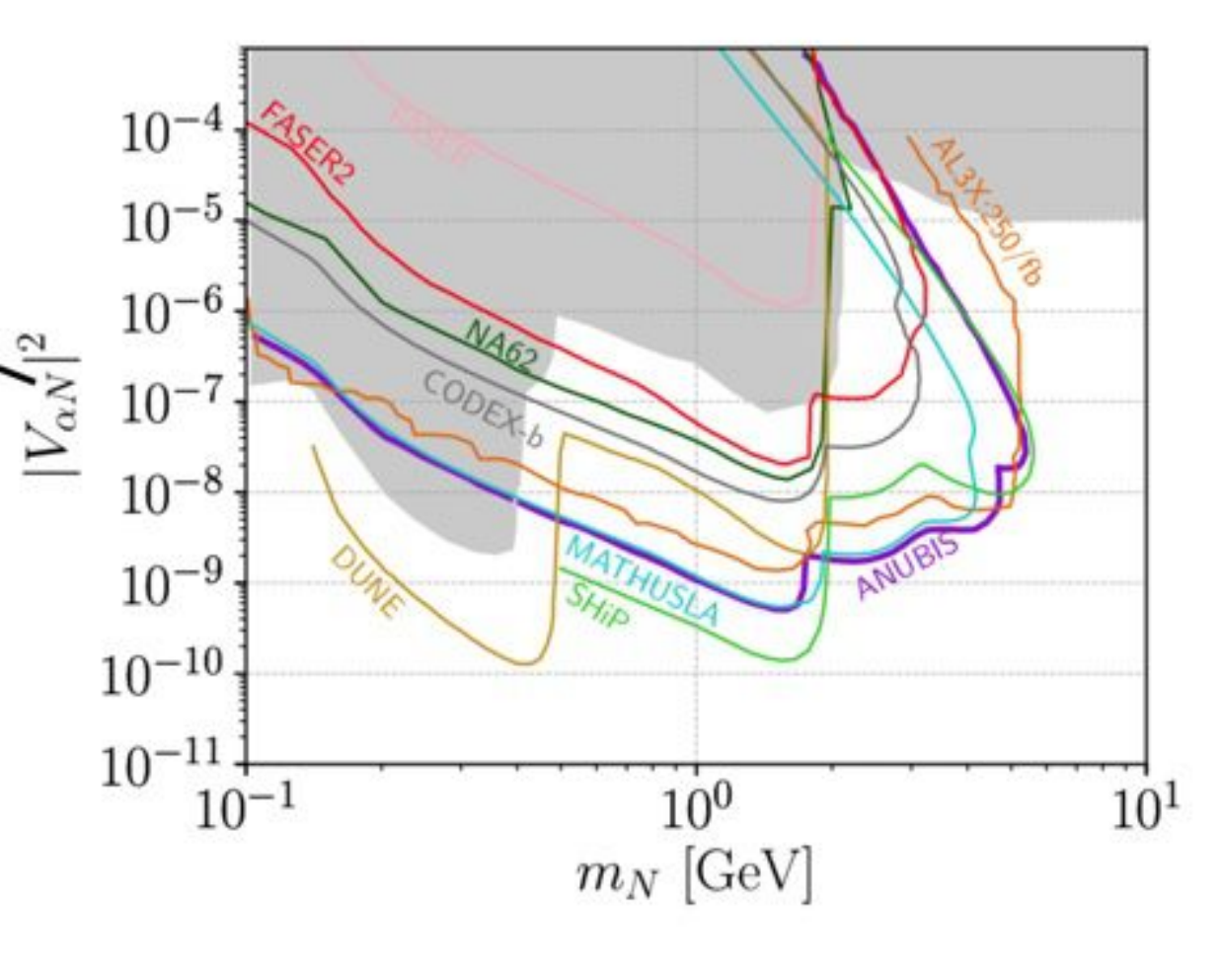}}
\subfigure[\label{fig:FASER2}]{\includegraphics[width=0.42\textwidth]{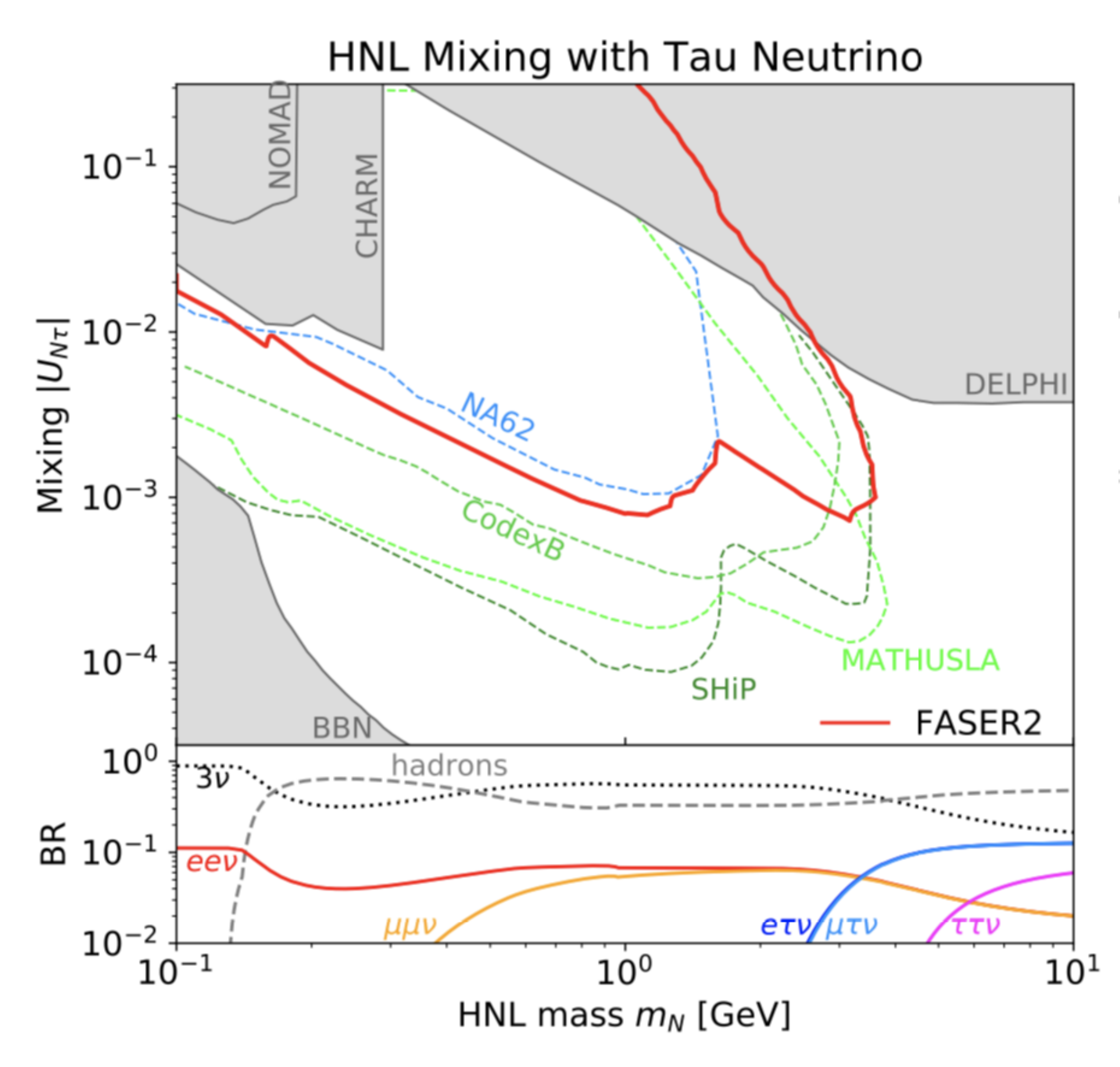}}

\caption{ (left) ANUBIS/AL3X/MATHUSLA reach compared to other future experiments  in the context of the minimal HNL scenario, with one generation of N mixing with either $\nu_e$ or $\nu_\mu$ but 
not both i.e. $  \alpha= e/\mu$. Figure take from 
\cite{Hirsch:2020klk}.
(right) Sensitivities for a HNL mixing with the tau neutrino  in the (mass, coupling) plane. The sensitivity reaches of FASER2 are shown as solid red lines alongside existing constraints (dark gray-shaded regions) and projected sensitivities of other proposed searches and experiments (colorful dashed lines), 
as obtained in Refs. \cite{FASER:2018eoc}
[4, 31, 50, 91] and references therein.
Figure taken from \cite{Anchordoqui:2021ghd,Kling:2021fwx,Beacham:2019nyx,Kling:2020mch}. The
and bottom panels show the LLP’s branching fractions} 
\end{figure*}

Next a  $U(1)_{B-L}$  extension of the Standard Model (SM) is considered. In this model HNLs are produced from  decays of mostly SM-like Higgs boson, via mixing in the scalar sector of the theory. In all cases,  ANUBIS is shown to have a sensitivity reach comparable to the proposed MATHUSLA detector. For the minimal HNL scenario, the contributions from Ws decaying to HNLs are more important at ANUBIS than at MATHUSLA, extending the sensitivity to slightly larger HNL masses for ANUBIS.
The sensitivity curves for ANUBIS for HNLs
are shown in 
Fig.\ref{fig:ANUBIS}.

\subsubsection{MAPP}
MAPP stands for the 
MoEDAL Apparatus for Penetrating 
Particles~\cite{Mitsou:2021vhf}. The
MoEDAL collaboration is proposing MAPP, an upgrade 
detector program for long lived particles in a gallery near IP8 shielded 
from cosmic rays by an overburden of approximately 100 m of limestone. 
A first-stage new detector of MAPP, consisting of scintillator bars and slabs to hunt for milli-charged particles  has
recently been approved and will be installed for  taking data during LHC Run 3. The next stage of the proposal, labeled MAPP-LLP,
targets to  deploy later three nested boxes of scintillator hodoscope detectors, in a ‘Russian doll’ configuration, following as far as possible the contours of the cavern. 
It is designed to be sensitive to long-lived neutral particles from new physics scenarios via their interaction or decay in flight in a decay zone 
of size approximately 5 m (wide) $\times$ 10 m (deep) $\times$ 3 m (high). 

In ~\cite{Deppisch:2019kvs} the pair production of right-handed neutrinos from the decay of an additional neutral $Z'$ boson 
in the gauged $B - L$ model 
is investigated. For this study  the sensitivity for an extended version of 
the MAPP detector proposal covering the whole available tunnel cavern 
area was used. 
The reachable sensitivity  is shown in Fig.~\ref{fig:FACET}.




\subsubsection{The Forward Physics Facility}

The Forward Physics Facility (FPF)~\cite{Feng:2022inv} is a proposal to create a new 
experimental cavern with the space and infrastructure to support a suite of far-forward experiments at the Large Hadron Collider during the High Luminosity era. Located along the beam collision axis and shielded from the interaction point by at least 100 m of concrete and rock, the FPF will house several experiments that will detect particles outside the acceptance of the existing large LHC experiments around the interaction points, and will observe rare and exotic processes in an extremely low-background environment. 
Experiments include a larger acceptance version of FASER, labelled FASER2, a Liquid Argon TPC, and an extended version of 
SND@LHC named AdvSND all of which can be 
used for HNL hunting.
A recent discussion on the status of this project, experiments considered and initial physics potential studies
can be found in \cite{Anchordoqui:2021ghd}.


FASER2  will have larger dimensions than FASER, namely a magnet aperture of 2 m  diameter which will in particular  open the acceptance for decays from $D$ and $B$ mesons, and a decay length volume of 10 m. For the sensitivity studies shown in this
document a decay volume with a length of 5  is used, which 
corresponds to the original FASER2 proposal in pre-FPF study days.
The an angular acceptance of neutral pions will increases from 0.6\% in FASER to 10\% in FASER2. There is a significant improvement in sensitivities to LLPs produced in decays of heavy mesons, due to the additional acceptance of B-meson production. The larger decay volume also improves sensitivity to larger LLP masses and longer LLP lifetimes. The combined effect of all these factors, as well as the increased luminosity expected for the HL-LHC over LHC Run 3, is an improvement in reach of several orders of magnitude
for certain new physics scenarios and will constitute a major step 
forward in sensitivity.

Likewise the  Advanced SND project is meant to extend the physics case of the SND@LHC experiment~\cite{Ahdida:2750060}. 
It will consist of two detectors of which the one placed in the $\eta$ region  i.e.~$7.2 < \eta < 8.4$ can be placed
in the FPF cavern.
The detectors consist of three elements.  Upstream  is the target region for  the vertex reconstruction and the electromagnetic calorimetric energy measurement. The target will be made of thin sensitive layers interleaved with tungsten plates, for a total mass of $\sim$ 5 tons.
Next are a hadronic calorimeter and a muon identification detector. The final most downstream element will be a magnet for measurements of the muon charge and momentum,  allowing for neutrino/anti-neutrino separation for muon neutrinos and for tau neutrinos in the muonic decay channel of the $\tau$ lepton.

FLArE is a LArTPC with transverse dimensions of 1.5x1.5 m$^2$ and a lenght of 7 m. The total LAr mass is
50 tonnes, the fiducial mass will be 10-20 tonnes. New
ideas are being studied to base this detector on liquid krypton. In any case the active material will be dense which 
means that it will not be very favorable for the signal to background ratio for searches for rare decays.
However in Ref.~\cite{Ismail:2021dyp} it is shown that
an interesting sensitivity region can be covered to the search for
neutrino magnetic moments via upscattering of a 
neutrino to a new sterile neutrino state, for heavy neutrino masses below 1 GeV.




At the FPF, HNL decays to hadrons and/or charged leptons will be visible in FASER2.
LHC production of HNLs mixing with all three neutrino flavors have been studied in Refs.
\cite{FASER:2018eoc,Kling:2018wct,Helo:2018qej,Cline:2020mdt,Feng:2022inv}.
Although a HNL giving neutrino masses through the type I seesaw would decay to all flavors, given the mixing angles required by neutrino oscillations, in less minimal models, such as the linear seesaw or inverse seesaw, there is more freedom to fit the neutrino data. In particular, scenarios can be built where the decay of the HNL is predominantly to taus.  This  benchmark was considered in  Fig.\ref{fig:FASER2} showing the expected FASER2 reach. Because the 
$\nu_{\tau}$ has significant production from heavy meson decays, the large rates for forward D and B production at the LHC benefit the FPF, allowing for sensitivity to HNL masses up to several GeV.
In AdvSND the same scattering technique as
discussed for SND@LHC can be used for HNL
detection.

\subsubsection{FACET}

FACET~\cite{Cerci:2021nlb} is  a proposal to add a set of very forward detectors to CMS, integrated within a new purposely-built wide beam-pipe, and
proposed to be deployed during the high-luminosity era of the Large Hadron Collider (LHC) in order to search for beyond the standard model(BSM) long-lived particles (LLPs), such as dark photons, dark Higgs, axion-like particles, as well as
 heavy neutral leptons.  FACET stands for Forward-Aperture CMS
 ExTension, and will be sensitive to  LLPs that can penetrate up to 50 m of magnetised
 iron and decay in an 18 m long, 1 m diameter wide vacuum pipe. The decay products will be
 measured in tracking and calorimeter
 detectors using identical technologies as
 for the the planned upgrades of CMS for the
 high-luminosity LHC running.
 
 FACET is not proposed to become a new stand alone experiment, but 
 rather as a new subsystem or extension
 of the CMS experiment
 that, while overlapping in the parameter space with other searches, will cover a different and
 unique region for LLPs. FACET will be sensitive to particles produced with polar angle 1 $< \theta <$ 4 mrad
 (7.6 $> \eta >$ 6.2). It is closer to the interaction region (IP5) than 
 e.g. FASER2 at IP1; it  has an long decay volume spanning from 101 $< z <$ 119 m from the IP; it is followed by an 8 m long region instrumented with various
 particle detectors. FACET covers a range of (unboosted) lifetimes $c\tau$ from 0.1 to 100 m. A
 unique feature among the proposed new
 LHC experiments is that the decay volume is at LHC vacuum
 quality, thus virtually eliminating any background from secondary interactions 
 with e.g. air in the decay volume.

\begin{figure}[t]
\centering

\includegraphics[width=0.45\linewidth]{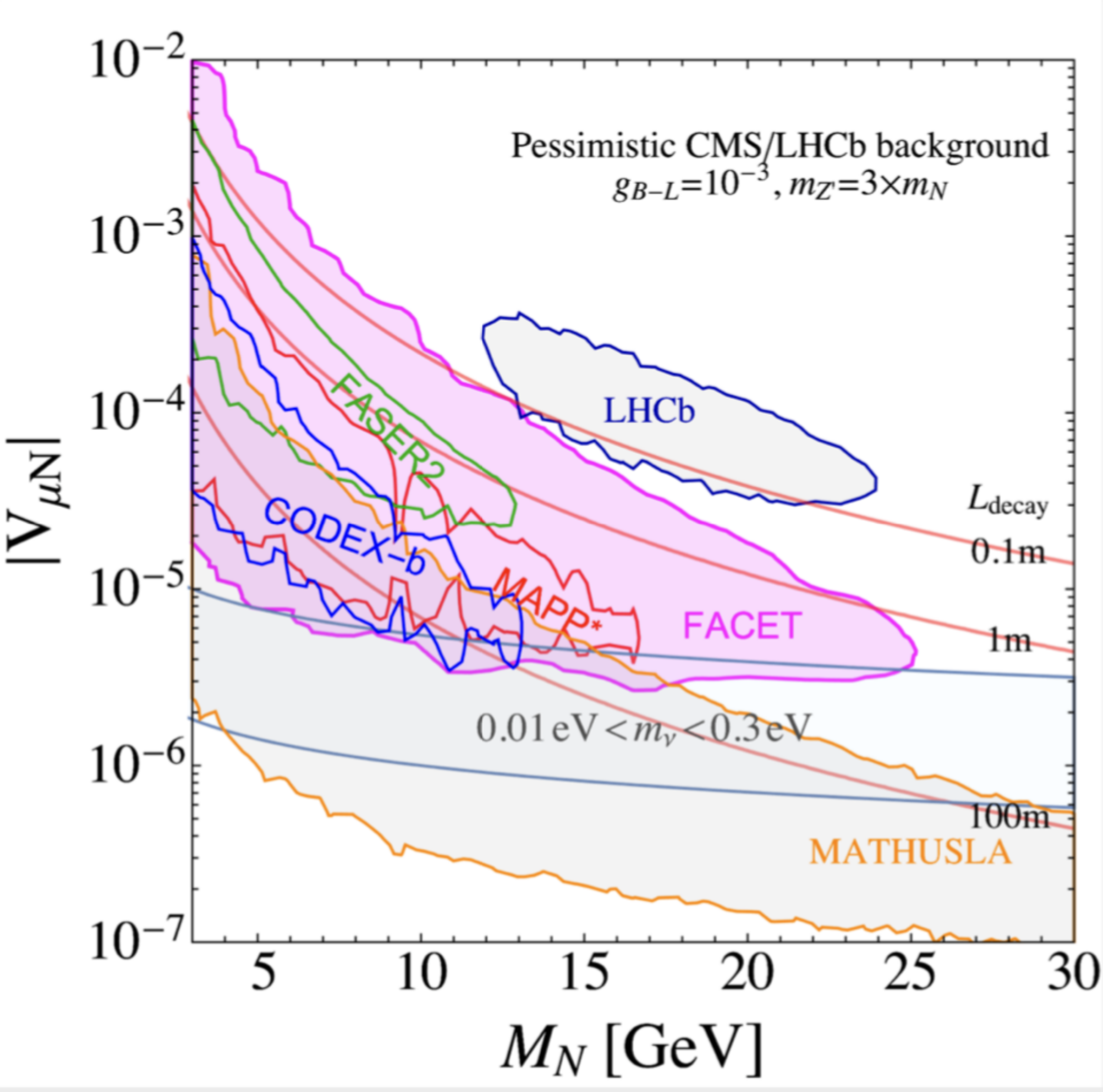}
\includegraphics[width=0.45\linewidth]{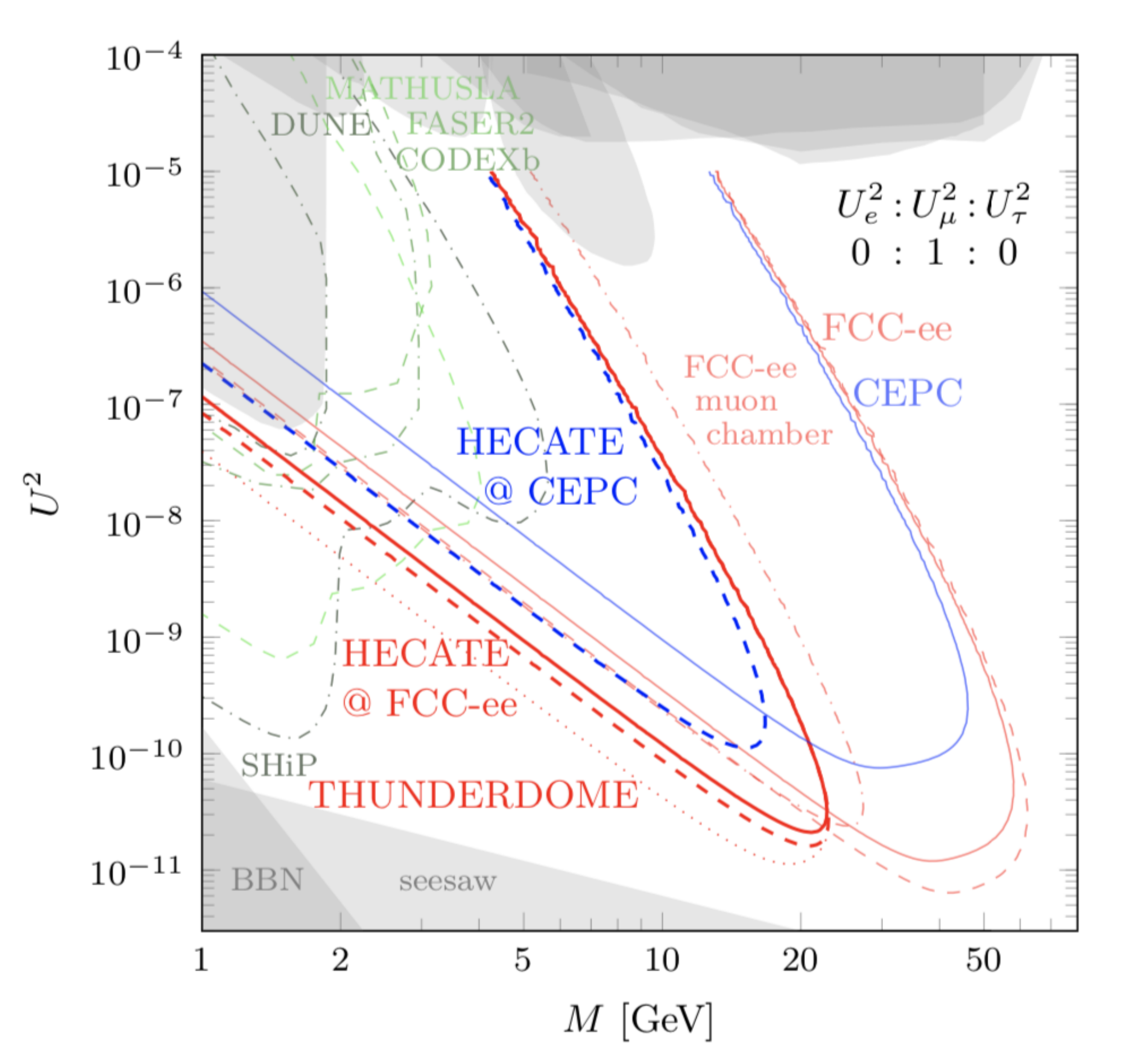}

\caption{
(Left) FACET reach compared to other future experiments  in the context of the $U(1)_{B-L}$ HNL scenario, with one generation of $N$ mixing. Figure take from  
 Ref.~\cite{Cerci:2021nlb}.
(Right) Comparison of the sensitivities for nine signal events that can be achieved at the FCC-ee with 2.5$\cdot$10$^{12}$ Z bosons (red) or CEPC with 3.5$\cdot$10$^{11}$ Z bosons (blue). The faint solid curves show the main detector sensitivity ($l_0$ = 5mm, $l_1$ = 1.22 m). The faint dash-dotted curve indicates the additional gain if the muon chambers are used at the FCC-ee ($l_0$ = 1.22 m, $l_1$ = 4 m).
The thick curves show the sensitivity of HECATE with $l_0$ = 4 m, $l_1$ = 15 m (solid) and 
$l_0$ = 4 m, $l_1$ = 25 m (dashed), respectively. Finally, the faint dashed red line shows the FCC-ee main detector sensitivity with 5$\cdot$10$^{12}$ Z bosons, corresponding to the luminosity at two IPs. 
For comparison we indicate the expected sensitivity of selected other experiments with the different green curves as indicated in the plot [8–10, 45, 48]. The gray areas in the upper part of the plot show the region excluded by past experiments, the gray areas at the bottom mark the regions that are disfavored by BBN and neutrino oscillation data in the $\nu$MSM.
}

\label{fig:FACET}
\end{figure}

The reach for  HNLs within the $U(1)_{B-L}$ model with FACET has been studied\cite{Deppisch:2019kvs}. 
In particular decays of $Z'$ bosons to muon flavored neutrinos were studied. 
 The $Z'$ can be light (e.g. 20 GeV) and yet have escaped detection in earlier experiments due
 to the smallness of the  coupling to SM particles. In this scenario, FACET has a unique sensitivity at high HNL
 masses, above about 10 GeV for lifetimes between 10 cm and 100 m.
 Fig.\ref{fig:FACET} shows the coverage in the coupling $|V_{\mu N} | - M(N)$ plane in the model, taken from 
 Ref.\cite{Cerci:2021nlb}.

\subsection{FCC} 

The Future Circular Collider (FCC) integrated program~\cite{FCC:2018byv} is proposed as a two-stage project, using one 100\,km circular tunnel at CERN. In the first stage, an electron-positron collider (FCC-ee)~\cite{FCC:2018evy} is proposed as a frontier Higgs, Top, Electroweak, and Flavor factory,  as the first step towards $\geq$ 100\,TeV proton-proton collisions in the same infrastructure (FCC-hh)~\cite{FCC:2018vvp}, with heavy ions and lepton-hadron (FCC-eh) options. The FCC program is one of the main options for a future particle collider that follow from the recommendations of the Updated European Strategy for Particle Physics in 2020~\cite{CERN-ESU-015}.
In terms of HNLs, the FCC, in its different, complementary stages, can probe very large areas of the parameter space, that are not constrained by astrophysics or cosmology, and complementary to beam dump and neutrino facilities~\cite{Cai:2017mow}. 

\subsubsection{FCC-ee}
At the electron-positron stage of the FCC, high-precision measurements will be performed; but FCC-ee also offers powerful opportunities for the discovery of new phenomena. Direct and indirect evidence for BSM physics, can be obtained via a combination of high precision measurements and direct searches. Among the later, searches for HNLs  will have an extraordinary potential at the high-luminosity $Z$ pole run~\cite{Chrzaszcz:2021nuk,Verhaaren:2022ify}.

The main production mode will then be ${Z}\rightarrow \nu {N}$, with subsequent decays of the $N$ into off-shell $W$ or $Z$ bosons. The FCC-ee will have good sensitivity above the charm mass all the way up to the $W$ boson mass for different values of the mixing angle with the existing neutrinos. Sensitivity down to heavy-light mixing of $\vert V_{\ell N}\vert^2 \sim 10^{-8}-10^{-12}$ was obtained in Refs.~\cite{Blondel:2014bra,Shen:2022ffi}, covering a large parameter space for heavy neutrino masses between 5 and 80 GeV. This sensitivity includes the regime where $Z\to N$ decays result in boosted HNLs~\cite{Shen:2022ffi}.

\begin{figure*}[ht]
\centering
\includegraphics[width=0.8\textwidth]{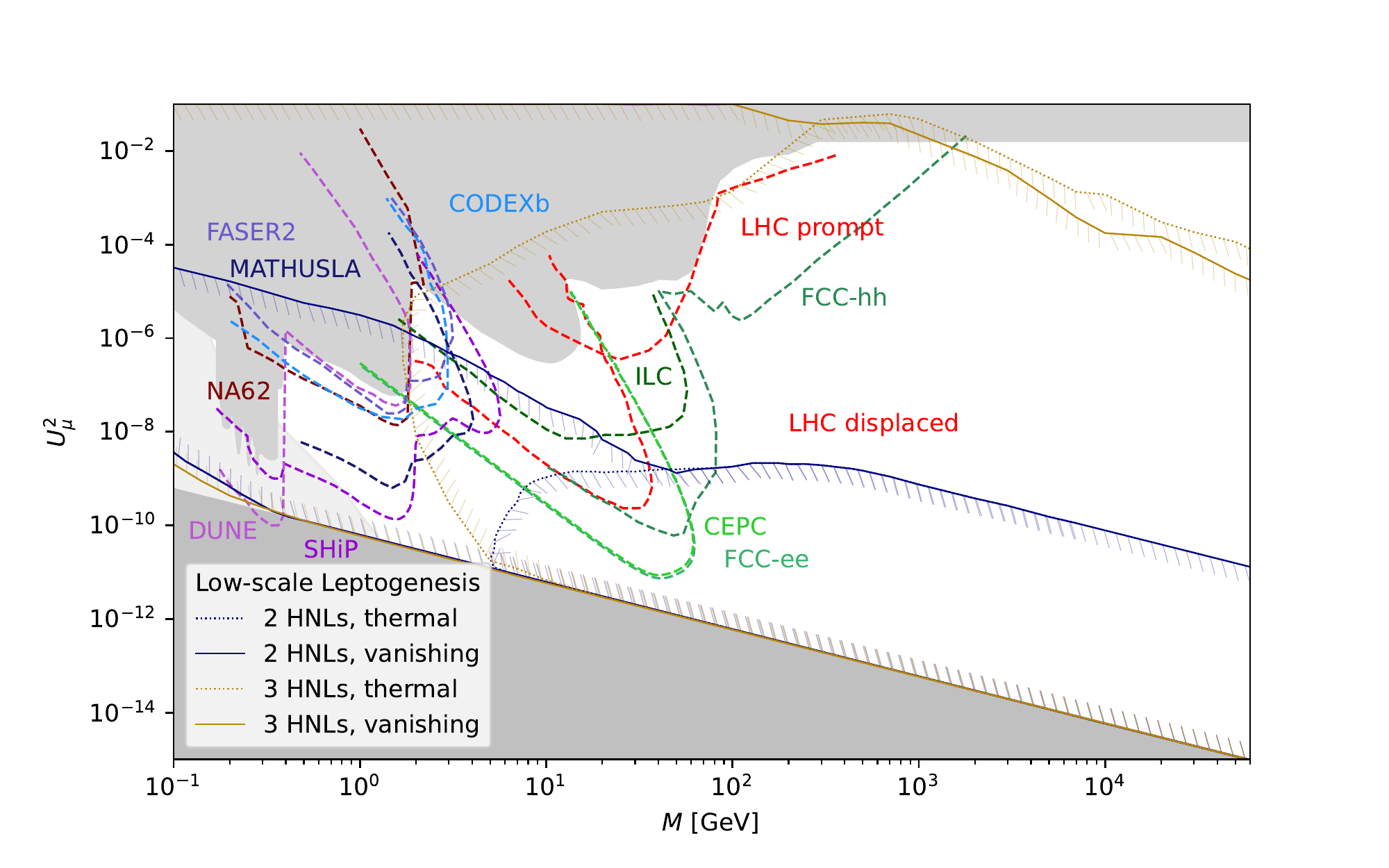}
\caption{
\emph{Bold green line:} Sensitivity of displaced vertex searches at FCC-ee with $5\times 10^{12}$ $Z$ bosons corresponding to 4 observed HNL decays, assuming no background and $75\%$ reconstructed HNL decays with a displacement between $400\mu m$ and $1.22$m. For comparison, we show what CEPC can achieve  with  $4.2\times 10^{12}$ $Z$ bosons for the same parameters.
\emph{Bold turquoise line:} Gain in sensitivity if the maximal observable displacement is increased to $5$m with a HECATE-like detector~\cite{Chrzaszcz:2020emg}.
\emph{Dark gray:} Lower bound on the total HNL mixing from the requirement to explain the light neutrino oscillation data \cite{Esteban:2020cvm}.
\emph{Medium gray:} Constraints on the mixing $|V_{\mu i}|^2$ of HNLs from past experiments~\cite{CHARM:1985nku,Abela:1981nf,Yamazaki:1984sj,E949:2014gsn,Bernardi:1987ek,NuTeV:1999kej,Vaitaitis:2000vc,CMS:2018iaf,DELPHI:1996qcc,ATLAS:2019kpx,CMS:2022fut}, obtained under the assumption $|V_{ \ell N }|^2=\delta_{\ell \mu}U_\mu^2$. 
\emph{Light gray:} Lower bound on $U_\mu^2$
 from BBN~\cite{Sabti:2020yrt,Boyarsky:2020dzc}. 
 \emph{Hashed orange and violet lines:} Regions in which the observed baryon asymmetry of the universe can be explained with two \cite{Klaric:2020phc,Klaric:2021cpi} or three \cite{Drewes:2021nqr} HNL flavours and different initial conditions, as explained in the legend.
 \emph{Other colourful lines:}
Estimated sensitivities of the LHC main detectors (taken from \cite{Izaguirre:2015pga,Drewes:2019fou,Pascoli:2018heg}) and NA62 \cite{Drewes:2018gkc} as well as the sensitivities of selected planned or proposed experiments (DUNE~\cite{Ballett:2019bgd},FASER2~\cite{FASER:2018eoc},		 SHiP~\cite{SHiP:2018xqw,Gorbunov:2020rjx}, MATHUSLA~\cite{Curtin:2018mvb}, Codex-b~\cite{Aielli:2019ivi}) as well as FCC-hh~\cite{Antusch:2016ejd}.
Figure from~\cite{Verhaaren:2022ify}. 
}
\label{fig:HNLsummary}
\end{figure*}

For low values of the neutrino mixing angles, the decay length of the HNLs can be significant, leading to long-lived signatures~\cite{Antusch:2016ejd,Antusch:2017ebe,Verhaaren:2022ify}. At the FCC-ee, HNLs could be detected
for decays up to about 1 m away from the collision point, generating a secondary vertex in the middle of the tracking system. These searches will suffer from very low backgrounds.
In case of an observation and with enough luminosity it could even be possible to study
kinematic properties that differ between Dirac and Majorana neutrinos~\cite{Blondel:2021mss,Verhaaren:2022ify}
using asymmetry observables as defined in Refs.~\cite{Bray:2007ru,Anamiati:2016uxp,Antusch:2017ebe,Ruiz:2020cjx}.
For illustration, Fig.~\ref{fig:HNLsummary} shows an updated estimation of different sensitivities for current and proposed detectors including FCC-ee displaced vertex analysis~\cite{Verhaaren:2022ify}.

Following the plans and ideas to add additional LLP experiments at the HL-LHC it is possible to also envision a similar concept at new future colliders. 
This is the motivation behind e.g. HECATE~\cite{Chrzaszcz:2020emg}, a long-lived particle detector concept for the FCC-ee or CEPC facility. The civil engineering of the FCC-ee will have much bigger detector caverns than needed for a lepton collider experiments since they 
are intended to be used later for a hadron collider facility, which requires larger volumes. This would allow to install extra instrumentation on the large cavern walls to further boost the sensitivity for HNLs.
Studies for long-lived particles with detectors at
displaced locations w.r.t the central detector for
future lepton colliders have also 
been reported in \cite{Wang:2019xvx}.
Fig.\ref{fig:FACET} (right) shows a phenomenological study
on the reach for HNLs 
at the FCC-ee (and CEPC) and compare the 
results for a standard detector (ie central tracker and outer muon system) with 
detector
scenarios where the additional space in the large cavern is equipped with muon 
detector layers.
The quantities 
$l_0$ and $l_1$ denote the minimal and maximal distance from the interaction point where the cavern is instrumented to detect an HNL decay into charged particles. The study 
shown is made for the channels 
with HNLs decaying to muon final states.





Indirect constrains from precision SM measurements are also sensitive
 to HNLs: the potential sensitivity 
of the FCC-ee for the HNL reach
 has been discussed in~\cite{Chrzaszcz:2021nuk}. 
Leptonic colliders are also well suited to look for light $m_N \sim 10 \text{ GeV}$ 
produced in the Higgs sector of extended gauge theories, LR and $U(1)_{B-L}$.
These may be produced in the associated $Z \Delta_R^0$ or $W \Delta_R^0$ or via 
$W W$ fusion.
In both cases, $e e$ colliders have the advantage of reduced fake rates and absence
of triggers and can thus access lighter (and displaced) $m_N$.

If any HNLs with masses at or below the EW scale are discovered, FCC-ee would provide an powerful tool to falsify or support the hypothesis that they are responsible for the neutrino masses and/or baryon asymmetry of the universe, cf.~Sec.~\ref{sub:leptogenesis}.
FCC-ee could measure the flavoru mixing pattern displayed in Fig.~\ref{fig:triangle} at the percent level \cite{Antusch:2017pkq}. 
Leptogenesis imposes further constraints on the flavour mixing pattern in addition to the experimental fits shown in Fig.~\ref{fig:triangle}, which can be tested by comparing flavoured branching ratios in displaced decays. Finally, determing the angular distribution of decay products can help address the question whether HNLs are Dirac or Majorana or the potential existence of additional intermediate states \cite{Han:2012vk,Chen:2013foz,Ruiz:2017nip,Ruiz:2020cjx,Blondel:2021mss,Verhaaren:2022ify}, cf.~Sec.~\ref{sub:type-i_seesaw}, while its dependence on the displacement is sensitive to the HNL mass splitting \cite{Antusch:2017pkq}, a crucial parameter for leptogenesis.


\subsubsection{FCC-hh}

The FCC-hh is the second stage of the FCC and will be sensitive to higher masses and smaller mixing than accessible at the LHC for both prompt and displaced searches. In 100~TeV hadron collision, HNLs can be produced in a variety of channels just as they can be at the LHC.
Notably, the gigantic increase in the gluon parton distribution function has a significant impact on production cross section~\cite{Degrande:2016aje,Pascoli:2018heg}, particularly production channels involving intermediate the $Z$ and Higgs bosons~\cite{Ruiz:2017yyf,Pascoli:2018heg}. 
The signatures and backgrounds at 100 TeV are largely the same as at 13-14 TeV. However, at 100 TeV, the hadronic environment is significantly more energetic and (fake) background rates categorically grow much faster than HNL signal rates~\cite{Pascoli:2018heg}. This means analysis strategies designed for 14 TeV must be adapted for a qualitatively different environment.

In this stage, the environment would be much more complex than FCC-ee, and pile-up, backgrounds characterization, and triggers would be all crucial elements in the success of these searches.
As shown in Fig.~\ref{fig:hllhc_TypeI_outlook}, assuming a comparable acceptance to the ATLAS and CMS detectors, it could be possible to test heavy, Dirac-like neutrinos with masses up to 10~TeV (100 GeV)
to the level of $\vert V_{\ell N}\vert^2 \sim 10^{-3}~(10^{-4})$ in the trilepton and MET 
channel involving final-state $\tau$ leptons (in their leptonic and hadronic decay modes)~\cite{Pascoli:2018heg}.
Comparable sensitivity holds for the Majorana case in the trilepton channel.
Much smaller mixing can potentially be probed at lower masses and with lighter lepton flavors~\cite{Antusch:2016ejd,Pascoli:2018heg}. 
It is possible that even higher mass HNL scales can be probed indirectly with the same-sign $W^\pm W^\pm$ scattering channel~\cite{Fuks:2020att,Fuks:2020zbm}.
Finally, it is important to stress that the FCC-hh conditions also allow for characterization both in flavor and charge of the produced neutrino, thus information of the flavor-sensitive mixing angles and a test of the fermion violating nature of the intermediate (Majorana) particle.

Recently, a study was reported on HNL production 
in the $U(1)_{B-L}$ model discussed earlier, showing a potential 
sensitivity reach at the FCC-hh \cite{Liu:2022kid}. The study, 
which considered final states with muons, shows couplings 
can be probed down to $\vert V_{\ell N}\vert^2 \sim (10^{-11})$
for high HNL mass values around 1 TeV.
Decays of a heavy Z’ boson resulting in boosted decays of light HNLs in 
$\mu j j$ final states, boosted fatjet topologies or track bases can be used. While this is difficult for HL-LHC, for FCC-hh searches involving at least one displaced fatjet in the inner detector or displaced tracks in muon system can probe seesaw region for light neutrino mass as low as 30 GeV~\cite{Padhan:2022fak}

The FCC-hh program would also greatly improve sensitivity to non-minimal HNL models. 
In LR models for instance, direct searches for $W_R$ and $N$ are sensitive to $M_{W_R}\sim35-40$ TeV with 10 ab$^{-1}$ of data~\cite{Mitra:2016kov}. 
Indirect searches can readily exceed this benchmark~\cite{Ruiz:2017nip}.
For HNLs in the Type III Seesaw, masses up to 8 TeV can be probed via $E^\pm N^0$ and $E^+E^-$ pair production and decay with 3 ab$^{-1}$ of data~\cite{Ruiz:2015zca,Cai:2017mow}. However, these analyses can be improved in light of new techniques, e.g., machine learning, that have been developed over recent years.

\subsubsection{FCC-eh}

The clean environment and low pileup at $ep$ colliders makes it also an excellent 
place to search for long-lived particles (LLP) using the long lifetime and displaced 
vertex signature~\cite{Curtin:2017bxr}.    
Electron-proton colliders can be used to search for heavy neutrinos, $N$, e.g.
in the process $ep \rightarrow Nj; Nljj$
~\cite{Antusch:2016ejd}. In particular, in low-scale see-saw models with two sterile neutrinos, as discussed before 
the mixing with SM neutrinos is expected to be very weak, leading to a displaced vertex from the decay of the heavy neutrino. 
Searches ~\cite{Antusch:2019eiz} 
for $ep \rightarrow \mu +3j$ based on BDT optimization, assuming a tracking resolution of 8 $\mu m$ and a sensitivity of 40 $\mu m$ for the displacement of the vertex, shows that the constraint on the mixing angles 
down to a few times $10^{-9}$ can be reached for an HNL mass between approximately 10 and 60 GeV.






\subsection{Other Future Projects}

The FCC is one of the proposed future high energy colliders projects. The CEPC 
project~\cite{CEPCStudyGroup:2018rmc}, which is studied as a possible high 
energy collider for China, has a similar proposed circular set-up as for the FCC-ee. Other projects concern linear colliders,
which include the ILC~\cite{Behnke:2013xla},
for CM energies up to 1 TeV, and CLIC~\cite{linssen2012physics} for CM energies up to 3 TeV scale. In~\cite{Mekala:2022cmm} the producton
of HNLs with displaced vertices is discussed. It shows that sensitivities down to $\vert V_{\ell N}\vert^2 \sim (10^{-6}-10^{-7})$ can be reached
up to the kinematical limits for the HNL mass for the different ILC energy versions and down to 
$\vert V_{\ell N}\vert^2 \sim (10^{-5})$ for CLIC for masses up to a few TeV.
The authors further note that it might be very interesting to perform a similar study at high-energy muon colliders, which due to its higher anticipated energy of 10 TeV or even beyond could reach much higher neutrino masses. The smaller dilution due to   beamstrahlung will also improve the signal-to- background ratio, and given to the incident muon flavor different flavor mixing structures will be probed.

\subsection{Monte Carlo developments for studies at the LHC and beyond}

Monte Carlo event generator programs are an important 
component of the sensitivity studies.
In recent years, an immense effort has gone into modernizing the Monte Carlo tools that are used to simulate minimal and non-minmal neutrino mass models at the LHC experiments~\cite{Cai:2017mow}. While theory predictions for Run 1 analyses were successfully derived at leading order (LO) with parton shower accuracy using the \texttt{ALPGEN}~\cite{delAguila:2007qnc} or \texttt{PYTHIA}~\cite{Sjostrand:2006za} frameworks, this was restricted to only a handful of processes. In conjunction with other advancements, fully differential predictions up to next-to-leading order (NLO) in QCD with parton shower matching for arbitrary partonic processes are now available using a tool chain combining \texttt{FeynRules}~\cite{Alloul:2013bka}, \texttt{MadGraph5\_aMC@NLO}~\cite{Stelzer:1994ta,Alwall:2014hca},
and mainstream parton showers.
In particular, a number of so-called Universal \texttt{FeynRules} Output (UFO) libraries~\cite{Christensen:2009jx,Degrande:2011ua}, which encode the Feynman rules of a model, have been published containing the ultraviolet counter terms needed~\cite{Degrande:2014vpa,Frixione:2019fxg} for numerical, NLO-accurate computations.
The ability to compute NLO-accurate matrix elements depends on the generator. That said, ``NLO'' UFOs can still be imported into popular event generators to support LO computations.

A number of NLO UFOs are already publicly available from the \texttt{FeynRules} database\footnote{Available from the URL: \url{https://feynrules.irmp.ucl.ac.be/wiki/ModelDatabaseMainPage}.}.
These include:
\texttt{HeavyN\_NLO}~\cite{Alva:2014gxa,Degrande:2016aje} and \texttt{HeavyN\_Dirac\_NLO}~\cite{Degrande:2016aje,Pascoli:2018heg}, which implement the Phenomenological Type I Seesaw for Majorana and Dirac-like HNLs, respectively,
as well as a variant that includes dimension $d=6$ operators~\cite{Degrande:2016aje,Cirigliano:2021peb};
\texttt{VPrime\_NLO}~\cite{Fuks:2017vtl}, which implements generic $W'$ and $Z'$ bosons that couple to all fermions;
\texttt{EffLRSM\_NLO}~\cite{Mitra:2016kov,Mattelaer:2016ynf}, which implements a description of $W_R$, $Z_R$, and Majorana HNLs from the Left-Right Symmetryic Model;
\texttt{TypeIISeesaw}~\cite{Fuks:2019clu}, which implements the full Type II Seesaw for the normal and inverse ordering of neutrino masses;
and
\texttt{SMWeinbergNLO}~\cite{Fuks:2020zbm}, which implements the dimension $d=5$ Weinberg operator.
Also available is a full description of the LRSM at LO~\cite{Nemevsek:2016enw,Nemevsek:2018bbt} as well as a description of the Type III Seesaw at LO~\cite{Biggio:2011ja}. A publicly available implementation of the $U(1)_{B-L}$ is also given in 
\cite{Deppisch:2018eth}.
The treatment of helicity propagation and spin correlation of HNLs in LNV and LNC processes in some of these implementations has been checked (and verified) explicitly~\cite{Ruiz:2020cjx}.

Low-mass HNL decays, and HNL production from heavy-flavor decays, have been implemented in a preliminary version of \textsc{Pythia}, which after sufficient validation will be incorporated into the official \textsc{Pythia} release\footnote{Available from the URL: \url{https://gitlab.com/hnls/pythia}.}. This code does not provide relevant branching fractions for either the HNLs or heavy-flavor mesons, as there are a number of publicly available codes which can already perform these calculations~\cite{Bondarenko:2018ptm}. Instead, the focus here is to provide a more accurate description of the relevant decay kinematics, both for the production of the HNL from the heavy-flavor meson, as well as the decay of the HNL itself. However, incorporating such codes in the future might provide a more unified and convenient package for HNL production and decay.

For the production of HNLs from heavy-flavor decays, a detailed study~\cite{Gorkavenko:2021mpj} has been performed which indicates that the parton-level matrix element already available in \textsc{Pythia} for weak decays provides a sufficient approximation for most HNL heavy-flavor production studies. Note that the usage of this matrix element differs from the standard usage for semi-leptonic decays within \textsc{Pythia}. Here, the spin of the decaying heavy-flavor state determines the ordering of the charged lepton and HNL, as passed to the matrix element. A configuration script is available, which given the mass of the HNL and its lepton mixing angles, provides a \textsc{Pythia} decay table configuration of all possible heavy-flavor decays that can produce the specified HNL. The output of this configuration script can be read by \textsc{Pythia}, or directly used to configure \textsc{Pythia} through its \textsc{Python} interface, and ensures the correct ordering of the decay products, as well as the correct matrix element code.

The decays of the HNLs in \textsc{Pythia} include full spin effects and are performed using the full helicity decay framework originally developed to perform $\tau$ decays~\cite{Ilten:2013yed}.The matrix elements typically used for $\tau$ decays have been generalized to also handle the decays of HNLs. Again, the configuration script can be used to produce a \textsc{Pythia} readable decay table which provides all possible decay channels for an HNL, given its mass and lepton mixing angles. These channels can then be individually selected by the user. Using the \textsc{Python} interface to \textsc{Pythia}, a simple example generation script is provided which produces the full HNL production and decay chain from a heavy-flavor meson particle gun. The user simply needs to provide the HNL mass, lepton mixing angles, selected HNL parent particle, selected production channel, and selected decay channel. Generation and decays of HNLs are also possible through the standard mechanisms available in \textsc{Pythia}, using the text output of the configuraiton script.

This current HNL framework in \textsc{Pythia} is a prototype, and suffers from a number of shortcomings. A full hellicity treatment of the heavy-flavor mesons has not yet been included. Recursive helicity decay structure is not yet available; consequently neither correlated $\tau$ decays produced from HNLs, or the full helicity structure for HNLs produced from charged lepton decays are modeled. Additionally, the interference between the neutral and charged current HNL decays into $\ell\ell\nu$ final states has not yet been included, as well as a number of neutral currents. Finally, while some validation of the code has been performed, further testing is required before full-scale production is recommended.

In preparation for the HL-LHC program, Monte Carlo tools for neutrino mass models can be improved in several ways. For instance: there are several UFO libraries that are only accurate at LO. For many models, no UFO has ever been made publicly available. Improving support for Effective Field Theories is also of interest to several communities.  Finally, developing new or extending existing tools to more fully support activities at LHCb, beam dumps, and forward detector experiments is highly desired.

\subsection{Summary}

The LHC and future high energy colliders will offer excellent opportunities
to search for heavy neutral leptons. With the 
full 
high luminosity event statistics the CMS and ATLAS experiments 
can potentially reach values of couplings in the minimal 
HNL model
on $|V_{e N}|^2$ and $|V_{\mu N}|^2$ down to or below $10^{-7}-10^{-8}$, in
the mass region $m_N $ of 5-20 GeV. LHCb will extend the range for lower mass values. One of the issues hindering the reach to 
smaller couplings at small mass hypotheses is the decreasing acceptance of the 
central detectors due to the  correspondingly increasing HNL 
lifetimes. 
Proposed solutions to overcome this deficiency are the new ideas for experiments at the LHC,
with typically reduced space angle  acceptance, but 
 positioned at locations further away
 from the HNL production point. 

Current proposals are grouped in transverse and forward type 
of detectors. The transverse 
detector ideas encompass MATHUSLA, CODEX-b, AL3X, ANIBUS and MAPP-LLP
experiments. These are typically experiments optimized for searches for observing new weakly 
interacting neutral particle decays, and placed at distances of tens to more than a 
hundred meters away from the new particle production point.
Forward detectors, such as FASER, SND@LHC, the forward physics 
facility and FACET, are located along the  
direction of the LHC beam line and are mostly sensitive to 
the production of new neutral particles 
originating in decays of mesons.

These additional detectors will cover an important part of the HNL 
parameter space, mostly 
for masses $m_N$ less than 5 GeV, and will be complementary to planned 
or proposed experiments at high intensity fixed target experiments.
The sensitivity  will be 
reaching values  of
$|V_{e N}|^2$ and $|V_{\mu N}|^2$ roughly 
down to $10^{-8}-10^{-9}$, maybe even below, 
in a mass region between 100 MeV and 5 GeV. This constitutes a large 
newly covered region in  so far unexplored HNL parameter space.

As is clear from e.g. Fig.~\ref{fig:ANUBIS} the 
coverage can be reached 
via different experimental set-ups. Eg for the HNL search the 
sensitivities of ANUBIS and MATHUSLA are very similar, obtained with 
similar method but rather different detector set-ups (and different 
budget requirements). Clearly while experimental coverage provided 
by these central/transverse detectors will be a large gain for 
long-lived particle physics studies and HNL searches in particular, it will be 
necessary to evaluate and compare on par these different 
proposals on their
performance and HNL parameter space coverage. Such comparison should
of course not just consider the potential for HNL searches but look at 
complete picture of the full spectrum of capabilities. It will 
be equally important include in that discussion  the capabilities of these detectors to measure properties of these putative newly
found particles 
when observed, e.g., the Majorana versus Dirac character in case of 
new neutrinos, 
and what can we learn from the new physics related to these
particles. The "Physics Beyond Collider" \cite{CERNPBC} initiative and related working groups, can offer an opportune forum for conducting these discussions.

In a more distant future a facility like
the FCC project could be realized. 
Such a facility, based on a circular
ring collider, could host and $e^+e^-$ 
precision physics facility as well as in a later stage
a hadron-hadron (and electron-hadron), high-energy-frontier collider.
In terms of searches for HNLs, the FCC in its different complementary stages, can probe very large areas of the parameter space towards
in the tens of GeV mass region, that are not constrained by astrophysics or cosmology, and are complementary to beam dump and neutrino facilities. Heavy new neutrinos with masses larger than 
10 TeV can be searched for at the high energy frontier at the 
FCC-hh.

For many if the future options and proposal given in this section 
- both for the near and more distant future - first estimates on the 
sensitivity for HNL discoveries have been made, and demonstrate 
the potential HNL parameter space coverage.
Certainly further studies e.g on detector optimization
 are strongly desirable and needed for this important physics target.
 Such studies need to go in future also beyond the simplest version 
 of the HNL models, covering non-minimal scenarios.



\section{Nuclear Decay Searches}
\label{sec:nuclear}

A novel approach to searching for heavy neutral leptons involves exploiting energy-momentum conservation in nuclear reactions in which an electron neutrino or electron antineutrino is involved.  If a heavy neutral lepton (such as a keV-scale sterile neutrino) mixes with the electron neutrino/antineutrino, by reconstructing the kinematics of the nuclear decay products and looking for irregularities, one can obtain sensitivity to the existence of heavy neutral leptons, setting upper limits on $|U_{e4}|^2$ as a function of $m_{4}$ in the event of a null result.  In this section, two different categories of nuclear decays (beta decay and electron capture) are discussed, emphasizing proposed experiments to probe these decays in the next decade in order to search for heavy neutral leptons.  For more comprehensive discussion of previous measurements and global limits, see~\cite{deGouvea:2015euy,Dragoun:2016,Bolton:2019pcu}. 

Finally, at the end of this section we remind on some opportunities 
at reactors for detectors measuring the inverse beta decay event
rates.

\subsection{Beta Decay Searches}
\label{sec:betadecay}

One type of heavy neutral lepton search involving nuclear decays requires detecting ``kinks'' in beta decay electron energy spectra, providing a handle on $|U_{e4}|^2$~\cite{Shrock:1980vy}.  These features are normally present in beta decay electron energy spectra due to mixing between the three Standard Model neutrinos, but primarily lead to spectral distortions very close to the end point (the $Q$ value of the beta decay).  The existence of a heavy neutral lepton mixing with the electron neutrino or electron antineutrino would lead to a ``kink'' at $Q - m_{4}c^{2}$, much further from the end point for large $m_{4}$.  The impact to the energy spectrum of the final-state electron is broad, not being restricted to the localized ``kink'' signature, such that it can be probed experimentally with detectors of even modest energy resolution.  The values of $m_{4}$ probed in this search method are limited on the high side from the $Q$ value of the beta decay, as well as the detector energy threshold/resolution, and on the low side strictly from the detector energy resolution.

Beta decays of tritium ($\mathrm{{}^{3}H}$) provides access to $m_{4}$ values below the end point of $Q = \SI{18.6}{keV}$.  Given that it is a superallowed beta decay, theoretical uncertainties are especially small for this isotope, enabling sensitive searches for heavy neutral leptons when coupled with a spectrometer of significant precision.  The availability of intense $\mathrm{{}^{3}H}$ sources~\cite{KATRIN:2001ttj,KATRIN:2010} allows for sensitive probes of even small values of $|U_{e4}|^2$ in the relevant range of $m_{4}$ values.

\begin{figure}[t]
  \centering
  \includegraphics[width=.48\textwidth]{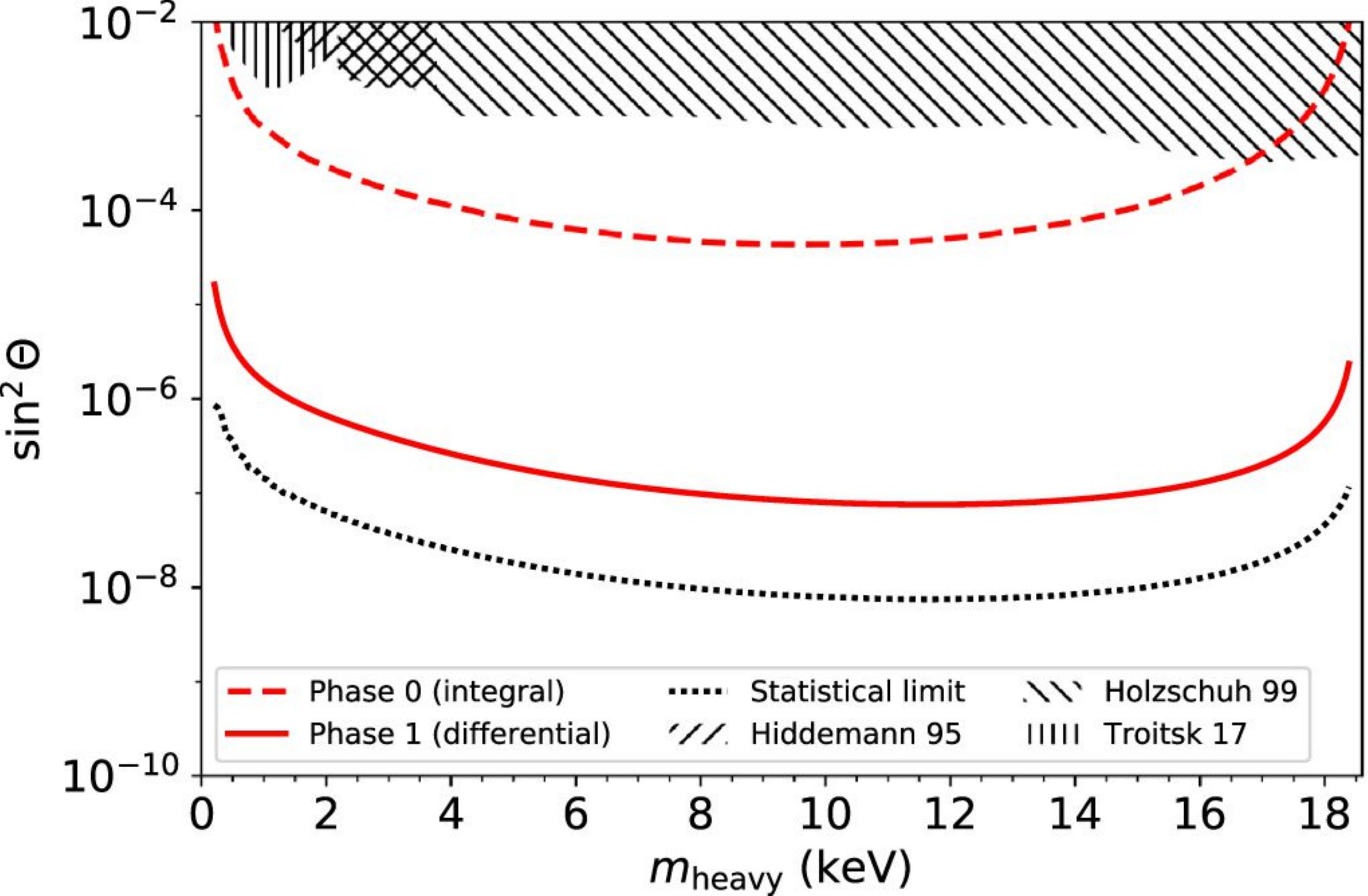}
  \hspace{2mm}
  \includegraphics[width=.49\textwidth]{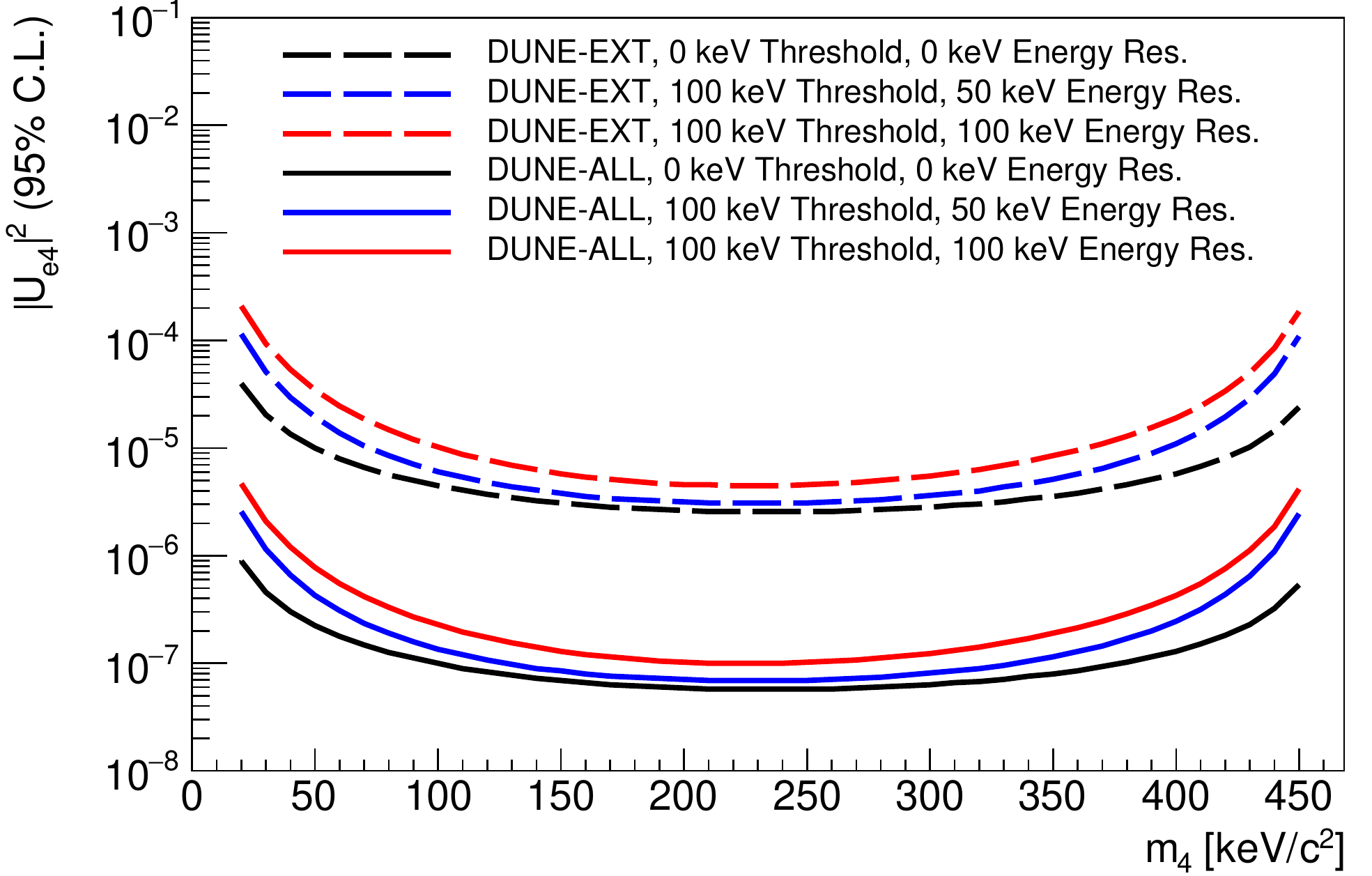}
  \caption{Projected upper limits (95\% C.L.) for $\sin^{2}\theta = |U_{e4}|^2$ as a function of heavy neutral lepton mass $m_{\mathrm{heavy}} = m_{4}$ for KATRIN/TRISTAN (left)~\cite{TRISTAN:2019} and DUNE (right).  In the case of KATRIN/TRISTAN, the dashed red line represents a measurement with $6 \times 10^{11}$ electrons, the solid red line represents a measurement with $1 \times 10^{16}$ electrons, and the dotted black line represents the statistical limit with full $\mathrm{{}^{3}H}$ source strength ($1 \times 10^{18}$ electrons); an energy resolution of \SI{0.3}{keV} (FWHM) is assumed for all calculations.  In the case of DUNE, ``DUNE-ALL'' refers to a limiting dataset containing every single $\mathrm{{}^{39}Ar}$ beta decay occurring during detector operations ($2 \times 10^{16}$ electrons), while ``DUNE-EXT'' refers to a baseline dataset obtained using external triggers and full detector readout without special localized triggering for low-energy activity in the detector ($1 \times 10^{13}$ electrons).  All upper limits shown are purely statistical (no theoretical/experimental systematic uncertainties included) and assume zero background.}
\label{fig:betadecays}
\end{figure}

A heavy neutral lepton search using $\mathrm{{}^{3}H}$ is planned by KATRIN~\cite{KATRIN:2001ttj,KATRIN:2010} and follow-up experiment TRISTAN~\cite{TRISTAN:2019}, which can look for heavy neutral leptons in the \SIrange{1}{18}{keV} mass range by making use of an enriched source of $\mathrm{{}^{3}H}$ and a MAC-E filter.  Figure~\ref{fig:betadecays} (left) shows projected statistical upper limits (95\% C.L.) for $\sin^{2}\theta = |U_{e4}|^2$ at KATRIN/TRISTAN as a function of heavy neutral lepton mass $m_{\mathrm{heavy}} = m_{4}$.  By minimizing theoretical and experimental systematic uncertainties, $|U_{e4}|^2$ values as low as $10^{-7}$ can be probed, with a TRISTAN design goal of $10^{-6}$.  Reaching these levels of sensitivity will require upgrading the detector and readout to handle the high event rates that are expected when probing beta decays further from the $\mathrm{{}^{3}H}$ end point.  The TRISTAN upgrade is planned to take place after the completion of the Standard Model absolute neutrino mass measurement of KATRIN, prospectively in 2025.  Another planned $\mathrm{{}^{3}H}$ beta decay experiment, Project 8~\cite{Project8:2017nal}, is designed to measure the absolute neutrino mass using cyclotron radiation emission spectroscopy (CRES)~\cite{Monreal:2009} and may also have the capability to search for heavy neutral leptons in the same mass range~\cite{Adhikari:2017}.

Large liquid argon detectors making use of atmospheric argon, associated with $\mathrm{{}^{39}Ar}$ beta decay activity of roughly \SI{1}{Bq/kg}~\cite{Benetti:2007}, will enable the probing of $m_{4}$ values below the $\mathrm{{}^{39}Ar}$ beta decay end point of $Q = \SI{565}{keV}$.  By utilizing very large detectors with large volumes of liquid argon, a substantial amount of beta decays can be detected as to enable sensitive measurements of $|U_{e4}|^2$ at larger $m_{4}$ values than for $\mathrm{{}^{3}H}$.  Given that liquid argon detectors (such as liquid argon time projection chambers, or LArTPCs) function as total absorption calorimeters, the liquid argon provides both the source and the detector for such measurements, in contrast to measurements using $\mathrm{{}^{3}H}$.

The Deep Underground Neutrino Experiment (DUNE)~\cite{DUNE:2020ypp} at the Long-Baseline Neutrino Facility will use massive LArTPCs to study accelerator neutrinos (\SI{\sim1}{GeV}) undergoing flavor oscillations over a long baseline (\SI{\sim1300}{km}) in order to probe leptonic CP violation.  A search for heavy neutral leptons can also be carried out using ionization charge measurements of $\mathrm{{}^{39}Ar}$ beta decays in the DUNE far detector, enabling sensitivity to $|U_{e4}|^2$ in the \SIrange{20}{450}{keV} mass range.  Figure~\ref{fig:betadecays} (right) shows projected statistical upper limits (95\% C.L.) for $|U_{e4}|^2$ at DUNE as a function of $m_{4}$.  Recording the full DUNE $\mathrm{{}^{39}Ar}$ beta decay dataset toward maximum sensitivity to $|U_{e4}|^2$ (better than $10^{-6}$) requires substantial development of the trigger system at DUNE, including selectively recording low-energy activity within ``regions of interest'' inside of the detector; this capability is currently under development.  Reconstruction of $\mathrm{{}^{39}Ar}$ beta decays has been previously carried out at MicroBooNE~\cite{MicroBooNE:2017}, a LArTPC neutrino experiment at Fermilab, demonstrating that low thresholds (roughly \SI{100}{keV}) and good energy resolution from low TPC noise levels (roughly \SI{50}{keV}) are achievable in large LArTPC detectors~\cite{MicroBooNE:2018}; more comprehensive studies at ProtoDUNE-SP~\cite{ProtoDUNESP:2017,Abi:2020} are currently in progress.  Further studies on the impact of $\mathrm{{}^{39}Ar}$ beta decay spectrum theoretical uncertainties~\cite{Kostensalo:2017}, experimental systematic uncertainties, and radiological backgrounds in the DUNE far detector are also in progress.  While these additional considerations may lower the sensitivity to $|U_{e4}|^2$ considerably, given the current global limits on $|U_{e4}|^2$ of $10^{-2}$ to $10^{-4}$ in the relevant range of $m_{4}$ values~\cite{deGouvea:2015euy,Dragoun:2016,Bolton:2019pcu}, significant improvement is expected.

\subsection{Electron Capture Searches}
\label{sec:electroncapture}

The second type of heavy neutral lepton search involving nuclear decays requires involves total energy-momentum reconstruction of the non-neutrino final-state particles in electron capture events~\cite{Smith:2019}.  Electron capture provides a pure two-body final state that consists of the recoiling daughter atom and the emitted electron neutrino, both of which (in principle) are mono-energetic.  By making a precision measurement of the low-energy recoiling atom and associated activity from de-excitation of the nucleus (e.g.\ X-rays, Auger electrons), information on energy-momentum conservation with the neutrino can be directly probed.  More specifically, this is done by seeking a separated nonzero missing-mass peak due to the existence of a heavy neutral lepton mixing with the electron neutrino/antineutrino.

The proposed HUNTER experiment~\cite{Smith:2019,HUNTER:2020} will search for heavy neutral leptons in the \SIrange{20}{300}{keV} mass range by fully reconstructing K-capture decays from $\mathrm{{}^{131}Cs}$ using magneto-optical trapping and recoil-ion momentum spectroscopy (RIMS).  HUNTER will perform event-by-event reconstruction of the decays, detecting the recoil $\mathrm{{}^{131}Xe}$, the atomic X-ray, and one or more Auger electrons.  $\mathrm{{}^{131}Cs}$ is commercially available at reasonable cost as a brachytherapy source, and decays 100\% of the time by electron capture, making it an ideal source for this type of heavy neutral lepton search.  HUNTER expects to significantly improve upper limits on $|U_{e4}|^2$ in the relevant mass range.

Another experiment using electron capture decays to search for heavy neutral leptons is the BeEST experiment~\cite{Leach:2021}, which is searching for heavy neutral leptons in the \SIrange{100}{850}{keV} mass range using superconducting quantum sensors embedded with $\mathrm{{}^{7}Be}$.  The measurement is carried out by implanting intense beams of unstable $\mathrm{{}^{7}Be}$ atoms created at the TRIUMF-ISAC rare-isotope beam facility into high-rate superconducting tunnel junction (STJ) quantum sensors, which are used to perform low-energy calorimetry of the decay products following $\mathrm{{}^{7}Be}$ electron capture.  The pure electron-capture-decaying nucleus of $\mathrm{{}^{7}Be}$ is an excellent source for neutrino studies via energy-momentum reconstruction due to its large $Q$ value (\SI{862}{keV}), relatively high recoil energy (\SI{\sim50}{eV}), and simple atomic and nuclear structure.  Previous results from BeEST~\cite{Friedrich:2021} achieve upper limits on $|U_{e4}|^2$ as low as $10^{-4}$ in the relevant range of $m_{4}$ values, with projected upper limits on $|U_{e4}|^2$ as low as $10^{-7}$ after future upgrades to the experiment.





\subsection{Reactor Searches}
\label{sec:reactors}

In this subsection we discuss the opportunity for a search of HNL particles with a mass in the MeV range in short baseline reactor experiments.
Nuclear reactors are a widely used source of electron antineutrinos in the intermediate energy range up to 12 MeV. If the mixing between electron antineutrinos and a heavy neutral counterpart exist, they can produce a measurable rate of HNLs in the mass range
up to 12 MeV. Up to date, no dedicated reactor experiment exists to search for HNLs in this mass range, but there exists, however a variety of short and long baseline reactor neutrino oscillation experiments worldwide, operating at various distances from either commercial nuclear power plants or research reactors. For HNL masses and couplings
relevant to the current best exclusion limits for a decay in flight measurement, the decay probability at distances ranging from meters to kilometers away from the production source can be considered as uniform. Therefore, the total detectable HNL flux for increasing reactor-detector distance can be compensated with a larger detector volume, and especially a larger solid angle coverage of the sensitive detectors. The HNL production rate will scale linearly with reactor power, together with the reactor induced gamma and/or neutron background, thus favoring experiments at high power reactors. Backgrounds can be further reduced by an appropriate detector location providing natural shielding or by the use of dedicated active or passive shielding.

In this mass range the 
 visible decay channel are $N \rightarrow \nu e^+e^-$ 
 and radiative modes
  $N \rightarrow \nu \gamma, \nu\gamma\gamma$
 leading to two electromagnetic signature activities in the detector. 
 In the energy range of the reactor neutrinos the dominant decay into visible particles is the $e^+e^-$ mode. 
 A pioneering analysis in this field is Ref.~\cite{Hagner:1995bn}, reporting limits down to a few times $\vert V_{eN}\vert^2 \sim 10^{-4}$ in the mass range of 2 to 7 MeV.

Very short baseline neutrino experiments combine interesting features of being located at very close proximity to either power or research reactors, and capable of background reduction via pulse shape discrimination and/or detector segmentation. The detection technologies are typically based on liquid and plastic/composite scintillators. 
A number of experiments is installed at a distance roughly between 
  5-25m away from intense reactor cores, at
 O(100) MW research or O(1) GW commercial reactors, and 
 with 1 to 5 m$^3$ volume detectors. These are  
 currently taking data or have recorded a large amount of data in the last few years. 
An overview of examples of  short baseline 
detectors of this type 
is reported in \cite{Lasserre:2014ita}. While the design of these experiments are optimized
to address the long standing reactor anomalies through the  precise measurements of the 
inverse beta decay  process, ie the measurement of a prompt positron and a delayed thermalised neutron that is captured,
 most can likely probe the decay in flight to electron positron pairs induced by HNL decays.
 
The experiments often use  a trigger scheme designed to detect the captured delayed neutron,
 which is a signature that is absent in HNL decays. However, most of the short baseline detectors already
 have an electromagnetic energy threshold trigger in place, used to either calibrate or
 monitor the detector stability. Such a trigger scheme is likely suitable to detect the stopped
 electromagnetic decay products of an HNL in the mass range between 1 and 12 MeV.
 The key challenges for such a study are the control of the large anticipated backgrounds from photons of the reactor, the intrinsic radioactivity in the detector muon and neutron backgrounds which need to be mastered, and depend on the detailed capabilities 
 of the experiments. An exploratory study was recently presented 
 in \cite{Verstraeten:2021bcw}.
 There could however be an opportunity to improve on direct experimental limits in this 
 mass range with dedicated analyses on present and possible future reactor experiments, and we strongly encourage to explore that.

 \subsection{Neutrinoless Double Beta Decay Searches}
\label{sec:0vbb}

The most sensitive probe of the Majorana nature of the light active neutrinos is neutrinoless double beta ($0\nu\beta\beta$) decay, see Ref.~\cite{Agostini:2022zub} for a recent review. Observing this rare nuclear decay in select isotopes such as $^{76}\text{Ge} \to {}^{76}\text{Se} + e^- e^-$, only allowed if total electron number is violated, would require that the light active neutrinos are Majorana fermions \cite{Schechter:1982bd, Duerr:2011zd}, though, possibly, of a quasi-Dirac nature. In addition, $0\nu\beta\beta$ decay is sensitive to other BSM sources of lepton number violation, typically at or below the $\mathcal{O}(10)$~TeV scale \cite{Doi:1981, Cirigliano:2017djv, Graf:2018ozy, Cirigliano:2018yza, Deppisch:2020ztt}. Such scenarios involving HNLs were discussed in Sec.~2, with the exchange of a sterile neutrino $N$ coupling through a SM charged-current interaction as the most prominent example. 

As $0\nu\beta\beta$ decay involves a virtual HNL, it should be considered an indirect probe, and it is not the focus of this white paper. Taken in isolation, $0\nu\beta\beta$ decay is highly sensitive to heavy Majorana neutrinos, with the decay half-life $T_{1/2}^{0\nu}$ approximately given by
\begin{align}
\label{eq:0vbb:half-life-heavy}
	\frac{10^{28}~\text{yr}}{T_{1/2}^{0\nu}} \approx \left(\frac{|U_{eN}|^2}{10^{-9}}\cdot\frac{1~\text{GeV}}{m_N}\right)^2,
\end{align}
for $m_N \gtrsim 100$~MeV and 
\begin{align}
\label{eq:0vbb:half-life-light}
	\frac{10^{28}~\text{yr}}{T_{1/2}^{0\nu}} \approx \left(\frac{|U_{eN}|^2}{10^{-9}}\cdot\frac{m_N}{15~\text{MeV}}\right)^2,
\end{align}
for $m_N \lesssim 100$~MeV, using a nuclear matrix element for the isotope $^{76}$Ge calculated in Ref.~\cite{Deppisch:2020ztt}. The behaviour changes at around $m_N\approx 100$~MeV as this is the nuclear scale of $0\nu\beta\beta$ decay, and at the crossover the momentum dependence must be carefully accounted for \cite{Babic:2018ikc, Dekens:2020ttz}. Current best limits on the $0\nu\beta\beta$ decay half life are set by KamLAND-Zen, $T^{0\nu}_{1/2}(^{136}\text{Xe}) > 1.1\times 10^{26}$~yr \cite{PhysRevLett.117.082503} and GERDA, $T^{0\nu}_{1/2}(^{76}\text{Ge}) > 1.8\times 10^{26}$~yr \cite{GERDA:2020xhi}, at 90\%~CL. There is a host of proposed experiments \cite{Agostini:2022zub} with projected sensitivities reaching $T^{0\nu}_{1/2}(^{76}\text{Ge}) \approx 10^{28}$~yr (LEGEND-1000, \cite{Zsigmond:2020bfx}), driven by the goal to probe the full light neutrino parameter space for an inverted mass ordering. Thus, future $0\nu\beta\beta$ experiments can reach sensitivities $|U_{eN}|^2 \approx 2\times 10^{-10}$ to $10^{-9}$, in the regime $10~\text{MeV}\lesssim m_N \lesssim 1$~GeV. This is close to the value expected in a seesaw scenario of light neutrino mass generation around $m_N = 100$~MeV, and also comparable to future direct searches in this mass window. For much heavier masses, the nominal sensitivity to Majorana HNLs are still strong, but as detailed in Sec.~2.3, the successful generation of light neutrino masses requires the HNLs to become quasi-Dirac states with an associated small mass splitting that suppresses $0\nu\beta\beta$ decay as an LNV process. For example, with a pair of quasi-Dirac HNL states with average mass $m_N$ and relative mass splitting $\delta_N = \Delta m_N/m_N$, Eq.~\eqref{eq:0vbb:half-life-heavy} is modified as 
\begin{align}
	\label{eq:0vbb:half-life-heavy-quasi-dirac}
	\frac{10^{28}~\text{yr}}{T_{1/2}^{0\nu}} \approx \left(\frac{\delta_N}{0.01}\cdot\frac{|U_{eN}|^2}{10^{-7}}\cdot\frac{1~\text{GeV}}{m_N}\right)^2.
\end{align}
Larger masses $m_N \gtrsim 10$~GeV and splittings are in principle possible but require a fine-tuned cancellation of the induced loop contributions to the light neutrino masses \cite{Mitra:2011qr, Lopez-Pavon:2012yda, Lopez-Pavon:2015cga, Bolton:2019pcu}. Direct searches that do not rely on an LNV signal do not suffer from such issues and can probe even Dirac HNLs.

The above limits assume that the HNL contribution saturates the $0\nu\beta\beta$ decay sensitivity, but other contributions may be present. Most importantly, the light active neutrinos (if Majorana) will induce the so called standard mass mechanism which will coherently interfere with any HNL contribution. This makes a model-independent interpretation difficult. Moreover, it is worthwhile to reiterate that $0\nu\beta\beta$ decay only probes HNLs coupling to electron flavour via the active-sterile mixing $U_{eN}$. It does not set model-independent constraints on $U_{\mu N}$ and $U_{\tau N}$. The differences to other searches make $0\nu\beta\beta$ decay highly complementary, and its experimental discovery will have profound consequences for neutrino physics.


\section{Searches at Extracted Beamlines} 
\label{sec:fixed}

\subsection{Introduction}  
\label{subsec:HNL:intro}

Heavy neutral leptons with masses between 1~MeV and 5~GeV can emerge from the decay of charged pions, kaons, and heavy-quark hadrons produced in the interactions of protons with a target or a dump. Since their couplings to SM particles are very suppressed leading to expected production rates of $10^{-10}$ or less, high-intensity proton beams are required to improve over the current results~\cite{Agrawal:2021dbo}. Moreover since the charm and beauty cross-sections steeply increase with the energy of the proton beam~\cite{Lourenco:2006vw, ZEUS:2013fws}, high-energy proton beam lines are preferred with respect to low-energy ones for producing HNLs with masses above the kaon mass. 

The smallness of the HNL couplings implies that the HNLs are also very long-lived (up to $\sim 0.1$~s) compared to the bulk of the SM particles. Therefore the HNL decays to SM particles can optimally be detected only using experiments with decay volumes tens of meters long followed by spectrometers with particle identification capabilities. The typical expected signature is a displaced vertex in the decay volume and nothing else. Depending on which mixing angle is nonzero, decay channels in this HNL mass region include $N \to \ell_\alpha \pi$, $N \to \nu \ell_\alpha^+ \ell_\beta^-$, $N \to \nu \pi^0$ (and other neutral mesons), etc. 
In addition to peak searches, it is also valuable to search for specific decays of HNLs \cite{Gronau:1984ct}. 
Simultaneous searches for different final states can allow for enhanced prospects for HNL discovery and potentially distinguishing this new physics from other similar signatures in experiments at extracted beam lines, such as those from Vector, Pseudo-scalar or Higgs-Portals. 


In the past experiments at extracted beam lines have pioneered the field: After the original proposal to search for HNL emission in two-body leptonic decays of charged pseudoscalar mesons \cite{Shrock:1980vy,Shrock:1980ct}, dedicated peak search experiments have been carried out at SIN/PSI, TRIUMF, J-PARC, BNL, and CERN, as noted in the introduction.
In 1984 the PS191~\cite{Bernardi:1985ny,Bernardi:1987ek} experiment at CERN appears to have been the earliest beam dump to report HNL bounds from the direct production and decay.
Other bounds from past experiments at extracted beam lines come from CHARM~\cite{Orloff:2002de} and NOMAD~\cite{NOMAD:2001eyx}, although these latter limits have now been superseded by a new generation of experiments. CHARM results have been also recently reinterpreted~\cite{Boiarska:2021yho} and their reach extended to include also the coupling to the third lepton generation.

Today, a lively activity to search for HNLs (and in general for feebly-interacting particles) is going on in all leading worldwide laboratories provided of extracted beam lines. Among the most stringent bounds for the decays of heavy neutrinos at extracted beams are recent searches performed at PIENU~\cite{PIENU:2017wbj,PIENU:2019usb}, NA62~\cite{NA62:2020mcv,NA62:2021bji}, T2K~\cite{T2K:2019jwa,Arguelles:2021dqn}, MicroBooNE~\cite{MicroBooNE:2019izn,Kelly:2021xbv}, and ArgoNeuT~\cite{ArgoNeuT:2021clc} experiments, whose details are reported in Sections~\ref{subsubsec:HNL:NA62},~\ref{subsubsec:HNL:T2K},~\ref{subsubsec:HNL:SBN}, and~\ref{subsubsec:HNL:ArgoNeuT}, respectively. 
An even more lively activity is foreseen for the short to long term timescale: Future opportunities at CERN, J-PARC, FNAL, and PSI are discussed in Sections~\ref{subsec:HNL:CERN},~\ref{subsec:HNL:FNAL_KEK}, and~\ref{subsec:HNL:PSI}.  

Despite the fact that sometimes searches have focused on a variety of model scenarios (e.g. different nonzero mixing angles with different weights), in presenting the results, we will consider only one nonzero mixing angle at a time, $|U_{e4}|^2$, $|U_{\mu 4}|^2$, or $|U_{\tau 4}|^2$, as it has become the current standard in the community. In the event of a discovery of an HNL, more thorough, multi-parameter analyses could consider multiple non-zero mixing angles simultaneously. Beyond this, considering additional interactions among the $N$ and the SM allow for stronger probes in fixed-target experiments~\cite{Arguelles:2021dqn} without violating constraints from, for instance, Big Bang Nucleosynthesis (BBN). Moreover, when considering some extended scenarios, predictive phenomenology can connect constraints/observations in these experiments with those from other sectors, allowing for additional handles on discovering or constraining a particular model scenario~\cite{Ballett:2019pyw,Abdullahi:2020nyr}.

After discussing current and future prospects, in Section~\ref{subsec:HNL:PostDiscovery} we discuss further prospects in the event of a HNL discovery. This includes determining whether the HNL is a Dirac or Majorana fermion, and measuring parameters associated with HNL mixing when they are responsible for the observed light neutrino masses from oscillation experiments. Finally, we will conclude by providing in Section~\ref{subsec:HNL:Conclusions} the reader with a summary of what could be the status of the field in 10 and 20 year timescales. 

\subsection{Opportunities at CERN: NA62, SHADOWS, and SHiP}
\label{subsec:HNL:CERN}

Leveraging on a long tradition of successful experiments at extracted beam lines (including CHARM, PS191, NA31, NA48, NA62), the CERN North Area offers compelling opportunities in the future for searching for HNLs in a region of parameters compatible with the origin of neutrino masses and successful leptogenesis (see Section~\ref{sec:theory}). This can be done using 400~GeV/$c$ primary protons beam lines slowly extracted  from the CERN Super-Proton-Synchrotron (SPS). In the near to medium term future, the NA62 experiment~\cite{NA62:2017rwk}, the NA62-upgrade~\cite{NA62:2020upd}, and the SHADOWS project~\cite{Baldini:2021hfw}, both served by the P42/K12 beamline and hosted in the same experimental area ECN3, can search for HNLs from leptonic and semileptonic decays of either kaons (NA62-upgrade) or charm and beauty hadrons (NA62-upgrade and SHADOWS) produced by the interaction of a high-intensity proton beam with a dump. The NA62-upgrade and SHADOWS experiments, running the P42/K12 beam line up to a factor of six higher the current intensity, are aiming to collect each up to $5\times 10^{19}$~protons-on-target (pot) in dump-mode (NA62-upgrade and SHADOWS) and $5\times 10^{19}$~pot in kaon mode (NA62-upgrade only) by the end of the High Luminosity LHC era ($\sim$2040 or so). The SHiP experiment~\cite{SHiP:2015vad} proposed at the Beam Dump Facility (BDF)~\cite{Ahdida:2019ubf} with a dataset of $2\times 10^{20}$ pot can further expand the physics reach for HNLs between the kaon and beauty masses, pushing the sensitivity further down to the seesaw bound. In the following three subsections we discuss the opportunities at NA62, SHADOWS, and SHiP.

\subsubsection{NA62 at CERN} 
\label{subsubsec:HNL:NA62}

The NA62 experiment at CERN~\cite{NA62:2017rwk} is focused on the measurement of the $K^+\to\pi^+\nu\bar\nu$ decay rate to a 10\% precision using the decay in flight technique, and pursues a wide rare kaon decay program. The experiment is served by the P42/K12 beamline from the SPS. Slowly extracted 400~GeV protons, delivered in 4.8-second spills at a nominal intensity of $3.3\times 10^{12}$ protons per pulse (ppp), impinge over a 400~mm long beryllium target to obtain a monochromatic (75~GeV) secondary mixed $\pi^+$, $K^+$ and $p$ beam. The NA62 Run~1 dataset collected in 2016--2018, corresponding to $2.2\times 10^{18}$ protons on target (pot), has led to the first observation of the $K^+\to\pi^+\nu\bar\nu$ decay based on 20~candidates~\cite{NA62:2021zjw}. NA62 Run~2 started in 2021, and is approved until the CERN Long Shutdown 3 (LS3). In the longer term, a plan for a high-intensity $K^+$ decay experiment (NA62-upgrade) to be performed at CERN between possibly LS3 and LS4 is taking shape~\cite{NA62:2020upd}. The primary beam intensity in the high-intensity experiment is expected to be four times that of NA62, and collection of about $5\times 10^{19}$ pot is expected by 2038 or so.

The NA62 experiment has conducted a search for HNL production in the $K^+\to e^+N$ and $K^+\to\mu^+N$ decays with the Run~1 dataset~\cite{NA62:2020mcv,NA62:2021bji}. The former search uses the main trigger line designed for the $K^+\to\pi^+\nu\bar\nu$ decay, while the latter search is based on the control trigger downscaled by a factor of 400. The analysis represents a peak search in the reconstructed squared missing mass, $m^2_{\rm miss} = (P_K-P_{\ell})^2$, where $P_K$ and $P_\ell$ are the measured kaon and lepton 4-momenta. The HNL mass ranges covered are 144--462~MeV in the electron case (complementing PIENU results for $m_N<135$~MeV, see Section~\ref{subsec:HNL:PSI}), and 200--384~MeV in the muon case (complementing the $K^+\to\mu^+N$ peak search in the 178--300~MeV range with stopped kaons at BNL E949~\cite{E949:2014gsn}). No HNL signals have been observed in either decay mode, and the corresponding upper limits of the decay branching fractions and the mixing parameters $|U_{\ell 4}|^2$ have been established. The sensitivity is limited by backgrounds in both cases; in particular, the $K^+\to\mu^+\nu$ decay followed by the $\mu^+\to e^+\nu\bar\nu$ decay in flight represents an irreducible background to the $K^+\to e^+N$ process. The exclusion limits at 90\% CL obtained with NA62 Run~1 data are shown in Figs.~\ref{fig:hnl_exclusion_electron} and \ref{fig:hnl_exclusion_muon} for electron and muon couplings, respectively. The NA62 bounds on HNL production are comparable to those obtained from HNL decay searches at T2K with almost its full dataset~\cite{T2K:2019jwa}.

The sensitivity to the mixing parameters $|U_{\ell 4}|^2$ is expected to improve in future as the inverse square root of the integrated kaon flux. The projected sensitivities of the $K^+\to\ell^+N$ searches at NA62-upgrade~\cite{NA62:2020upd} are displayed in Figs.~\ref{fig:hnl_exclusion_electron} and \ref{fig:hnl_exclusion_muon}. A search for $\pi^+\to e^+N$ decay of the beam pions at NA62-upgrade is expected to improve the PIENU limits on $|U_{e4}|^2$~\cite{PIENU:2017wbj} in the range $m_{N}<m_\pi$; the corresponding sensitivity is yet to be evaluated and is not shown. In the electron case, the gap at $m_N\approx m_\pi$ between searches in $\pi^+$ and $K^+$ decays can be covered by the search for $K^+\to\pi^0e^+N$ decay, although with a limited sensitivity~\cite{Tastet:2020tzh}.

The P42/K12 beamline can also be operated in beam dump mode. A switch to beam dump mode is performed by removing the Beryllium target from the proton beamline. The primary proton beam is then dumped into a pair of 1.6~m thick movable Cu-Fe collimators (TAX) placed between two pairs of bending dipole magnets. During the standard data taking in kaon mode, these bending magnets are used to select the momentum of the secondary beam passing through suitably aligned holes. In beam dump operation, the TAX collimator is used as hadron absorber, while the pair of magnets downstream the TAX is used to provide sweeping of the secondary muons produced in the TAX. A sample of $1.5\times 10^{17}$ pot was collected by NA62 in beam dump mode during a 1-week long run in November 2021 at a typical beam intensity of $6\times 10^{12}$ ppp, using trigger streams for decays of exotic particles into charged and neutral particles. NA62 operation in beam dump mode with the goal of accumulating $10^{18}$~pot by LS3 has been approved by the CERN scientific committees~\cite{Collaboration:2691873}; the corresponding sensitivity curves assuming electron, muon, and tau flavor dominance are displayed in Figs.~\ref{fig:hnl_exclusion_electron},~\ref{fig:hnl_exclusion_muon}, \ref{fig:hnl_exclusion_tau}. In the longer term, a collection of $10^{19}$ pot in beam dump mode by LS4, and of additional $4\times 10^{19}$ pot after LS4, is foreseen~\cite{Baldini:2021hfw}.
However the exact schedule and beam sharing between kaon and dump mode will be a matter of scientific policy, which is why the end-date for collecting the required data samples varies between 2032 and 2038 in Table~\ref{tab:fixed-target-all-proposals}.

\subsubsection{SHADOWS at CERN}
\label{subsubsec:HNL:SHADOWS}

A novel experiment, SHADOWS (Search for Hidden And Dark Objects With the SPS), has been recently proposed within the Physics Beyond Colliders CERN study group\footnote{\url{https://pbc.web.cern.ch/}} and the Expression of Interest submitted to the CERN PS and SPS Committee~\cite{Baldini:2021hfw}. SHADOWS aims to use the 400~GeV primary proton beam (P42/K12 beamline) extracted from the CERN SPS currently serving the NA62 experiment, and will be located in the same experimental hall (ECN3) in the CERN North Area. SHADOWS will take data {\it off-axis} concurrently with NA62-upgrade when the beamline is operated in the dump mode. In fact HNLs (and in general feebly interacting particles) emerging from the decays of charmed and beauty hadrons have a non-negligible transverse momentum and therefore can be detected by an experiment placed off-axis with respect to the direction of the impinging beam on the dump.

The off-axis position allows SHADOWS to be placed very close to the dump, at about 10~m with respect to the proton interaction point, with the twofold advantage of: i) maximizing the signal acceptance due to the larger flux intercepted by the relatively small transverse dimensions; ii) minimizing the background that is mostly concentrated in the forward direction. The SHADOWS current plan, synchronized with that of the NA62-upgrade (Section~\ref{subsubsec:HNL:NA62}), involves running at 4--6 times the nominal P42/K12 intensity (at $2\times 10^{13}$~pot per pulse), and collection of $10^{19}$~pot in a year of data taking between LS3 and LS4, and additional $4\times 10^{19}$~pot in 3--4 years of data taking after LS4.
However, as in the NA62 case, the exact schedule and beam sharing between kaon and dump mode is still to be defined. Therefore in Table~\ref{tab:fixed-target-all-proposals} the end-date for collecting the required data sample ($\sim 5\cdot 10^{19}$~pot) varies also in this case between 2032 and 2038.

The current detector baseline consists of a standard spectrometer of $2.5\times 2.5$~m$^2$ transverse dimensions with a $\sim$20~m long, in-vacuum decay volume placed about 1~m off-axis with respect to the beam. The spectrometer comprises a tracking system with a dipole magnet of 1~Tm bending power, a timing detector, an electromagnetic and (possibly) a hadron calorimeter, and a muon detector. The upstream face of the decay vessel is instrumented with an active veto. A second spectrometer of $3\times 4$~m$^2$ transverse dimensions and 30~m long decay volume following the first one is still compatible with the NA62-upgrade operation, and is being envisaged for a future upgrade.

The proton interactions with the dump, along with feebly-interacting particles, give rise to a copious direct production of short-lived resonances, pions and kaons. While the dump thickness ($22 \, \lambda_{\rm I}$) is sufficient to absorb the hadrons and the electromagnetic radiation produced in the proton interactions, the decays of pions, kaons and short-lived resonances result in a large flux of muons and neutrinos. Muons and neutrinos from the dump are the two major sources of background for FIP searches.
For an off-axis detector, however, the background originated by neutrino interactions with the residual air in the in-vacuum decay volume or the detector material is negligible as the large majority (those  emerging from the decays of pions and kaons) is intrinsically produced in the forward direction. The most relevant backgrounds for SHADOWS are those originated by the beam-induced muon flux. Leveraging on the fact that off-axis muons have a much lower momentum than those on-axis (typically below 15~GeV for off-axis muons against up to 400~GeV for on-axis muons), a magnetized iron block placed in front of the SHADOWS decay vessel should be able to reduce this background to a manageable level. Preliminary studies are very promising.

The projections for the 90\%~CL exclusion limits on HNL couplings corresponding to $5 \times 10^{19}$ pot assuming electron, muon and tau flavor dominance are shown in Figs.~\ref{fig:hnl_exclusion_electron}, \ref{fig:hnl_exclusion_muon}, \ref{fig:hnl_exclusion_tau}. The addition of a second spectrometer is equivalent to the increase of the proton intensity by a factor 2--5, depending on the value and type of coupling and the considered mass range. These results do not include trigger, reconstruction, and selection efficiencies. However, realistic estimates obtained in a similar spectrometer as NA62 show that the inclusion of such efficiencies would reduce the available dataset by at most a factor of two. 

SHADOWS is currently under review by the SPS Committee at CERN. It aims to be installed during the LS3 (2026--2028) and take data in Run~4. It is fully compatible with the NA62-upgrade operation, and is complementary to NA62 in the HNL physics reach: NA62 is mostly sensitive below the kaon mass, while SHADOWS is sensitive mainly between the kaon and the beauty masses. The relatively small dimensions and the use of existing beamline, experimental area and related infrastructure make SHADOWS a relatively cheap, compelling, and competitive project for HNL searches in this decade. 

\subsubsection{SHiP at CERN}
\label{subsubsec:HNL:SHiP}
The Search for Hidden Particles (SHiP)~\cite{SHIP:2021tpn} is an experiment that would be placed at a proposed general-purpose intensity-frontier facility, called the Beam Dump Facility (BDF). SHiP is designed to probe a variety of new physics models containing long-lived particles with masses in $\mathcal{O}(10\text{ GeV})$ range and has unique potential for the HNL searches.

SHiP will be served by the SPS accelerator, exploiting the additional $4\times 10^{19}$ protons at 400 GeV per year that is not exploited by any experiment since the decommissioning of the CNGS facility. The experiment consists of a large density target~\cite{LopezSola:2019xjb,LopezSola:2019sfp,Kershaw:2018pyb}, followed by a hadron stopper, and an active muon shield based on the magnetic deflection~\cite{SHiP:2017wac}. Downstream the muon shield, there is the Scattering and Neutrino Detector (SND), designed for studying scatterings of neutrinos and dark sector particles. Next to the SND there is the Hidden sector decay volume, which is a pyramidal frustum with a length of 50 m and upstream dimensions of $2.2\times 4.5$~m$^2$ located on-axis approximately 45 meters downstream the center of the target. The decay volume ends by the Hidden Sector Decay Spectrometer (HSDS), consisting of straw trackers, spectrometer magnet, timing detector, EM calorimeter and the muon system. HSDS allows studying decays of hidden sector particles. It reconstructs the decay vertex of hidden sector particles, measures their invariant mass, and identifies decay products. 

The background for decays, including the muon combinatorial background, muons and neutrinos DIS events in the detector material, is efficiently suppressed by the combination of the target configuration, the hadron stopper, the muon shield, HSDS surrounding background tagger, and HSDS timing detector. Full Monte Carlo numeric simulations performed with the help of the full MC suite called FairShip (based on the FairRoot software framework~\cite{Al-Turany:2012zfk}), supplemented by the measurements done in test beam on the detector prototypes have shown that the background yield expected during the working time of the SHiP experiment is $<0.1$ for fully reconstructed signals and $6.8$ for partially reconstructed signals. The latter may be easily eliminated by requiring the invariant mass of the reconstructed tracks to exceed 100~MeV~\cite{SHIP:2021tpn}.

The SHiP experiment has a perfect potential to probe HNLs with masses above the kaon mass, being complementary to the experiments optimized to search for HNLs produced in decays of long-lived kaons, such as DUNE and NA62 in the kaon mode. Indeed, first, kaons produced at SHiP would get mostly absorbed by the hadron stopper (see however~\cite{Gorbunov:2020rjx} discussing the sensitivity of SHiP to HNLs from kaons). This allows for background-free searches for partially reconstructed signals from decays of HNLs. Second, the main production channels of HNLs in the mass range $m_{K}<m_{N}<m_{B}$ are decays of $D$ and $B$ mesons. Due to the large intensity of the protons at SHiP, around $3\times 10^{17}$ of $D$ mesons and $5\times 10^{13}$ of $B$ mesons would be produced~\cite{CERN-SHiP-NOTE-2015-002}. Their angular distribution is peaked in the far-forward region, and hence most of them fly within the geometrical acceptance of HSDS, providing the maximal geometric acceptance for HNLs.

The analysis of the SHiP sensitivity to HNLs has been performed in~\cite{SHiP:2018xqw}. The signal, decays of HNLs into at least two charged particles, has been modeled by full MC simulations with the help of FairShip, with the detector response simulated by GEANT4~\cite{GEANT4:2002zbu}. The 90\% CL sensitivity regions for the setup with $2\times 10^{20}$ pot for the case of the pure $e,\mu,\tau$ mixings are shown in Figs.~\ref{fig:hnl_exclusion_electron}, \ref{fig:hnl_exclusion_muon} and \ref{fig:hnl_exclusion_tau}. The sensitivity to the HNLs with arbitrary mixing pattern may be calculated using a script~\cite{Zenodo}. To summarize, SHiP may probe the HNL mixing angles down to $U^{2} \sim 10^{-10}$ in the region $m_{N}<m_{D_{s}}$, and $U^{2}\sim 10^{-8}$ in the region $m_{D_{s}}<m_{N}<m_{B}$.

\subsection{Opportunities at FNAL and J-PARC: Neutrino-Beam Experiments}
\label{subsec:HNL:FNAL_KEK}

Many neutrino experiments, both those currently operating and those in the planning stages, have demonstrated powerful sensitivity to HNLs with masses between 1~MeV and 2~GeV. This comes about from the intense proton beams that are used to produce intense neutrino beams. The neutrino beams are the byproduct of the decays of charged SM mesons, including $\pi^\pm$, $K^\pm$, $D_{(s)}^\pm$, etc (the exact makeup of which mesons are produced depends strongly on the energy of the incident proton beam). For HNLs with small mixing, there is the possibility that one can be produced in rare meson decays, e.g., $K^+\to\mu^+ N$ with a decay width proportional to $|U_{\mu 4}|^2$, in lieu of the process $K^+\to\mu^+\nu_\mu$.

Accelerator neutrino experiments are (typically) equipped with near detectors that are used to monitor the neutrino flux and measure neutrino cross section properties, which are then used for extraction of information when combined with far detector measurements. However, these near detectors, typically located within a few hundred meters of the proton-beam target, can be used for searches for displaced decays of the metastable HNLs.

The main backgrounds to these searches come from SM neutrino–nucleon scattering events in which the hadronic activity at the vertex is below threshold, see e.g.~\cite{Ballett:2019bgd}. Charged-current quasi-elastic events with pion emission from resonances are background to the semi-leptonic decay channels. Moreover, mis-identification of long pion tracks as muons can constitute a background to three-body leptonic decays. Neutral pions are often emitted in neutrino scattering events and can be a challenging background for HNL decays that include a neutral meson or channels with electrons in the final state. However, the possibility of using gaseous detectors, e.g., the one proposed for the DUNE Near Detector complex~\cite{DUNE:2021tad}, has been explored as well~\cite{Berryman:2019dme}. The signal rate of HNL decays scales like the volume of the detector, whereas neutrino-scattering backgrounds scale with its mass -- this implies that for similar-volume gaseous detectors, the signal-to-background ratio will scale inversely with the density. Finally, neutrino-nucleus scattering backgrounds can also be suppressed using timing information.

HNL searches with the current generation of neutrino experiments are led by T2K ND280 at J-PARC and ArgoNeuT at Fermilab, and the results are discussed in Sections~\ref{subsubsec:HNL:T2K} and~\ref{subsubsec:HNL:ArgoNeuT}, respectively. Upcoming experiments have great sensitivity in searches for neutrino-portal particles, much in the same way that the existing searches have operated. These next-generation experiments additionally offer excellent particle identification capabilities, allowing for searches of a variety of final states of $N$ decays simultaneously. Recent sensitivity studies have been carried out for T2K-II/HyperK near detectors at J-PARC, the Fermilab SBN detectors~\cite{Ballett:2016opr} and DUNE Near Detector complex~\cite{Krasnov:2019kdc,Ballett:2019bgd,Berryman:2019dme,Coloma:2020lgy,Breitbach:2021gvv}. They are detailed in Sections~\ref{subsubsec:HNL:HK}, \ref{subsubsec:HNL:SBN} and \ref{subsubsec:HNL:DUNE}, respectively.

\subsubsection{T2K at J-PARC}
\label{subsubsec:HNL:T2K}
T2K (Tokai-to-Kamioka)~\cite{T2K:2011qtm} is a long-baseline neutrino experiment located in Japan with the primary goal of measuring muon (anti-)neutrino oscillations using Super-Kamiokande as its far detector. The T2K neutrino beam is produced at the J-PARC center (Tokai, Ibaraki) by colliding 30 GeV protons on a graphite target. The produced pions and kaons are focused and selected by charge with magnetic horns and subsequently decay in flight to neutrinos. Depending on the polarity of the current in the horns, the experiment can be run either in neutrino or anti-neutrino mode. The near detector ND280, located 280 meters from the proton target, is composed of several sub-detectors, in particular three time projection chambers (TPCs) filled with Argon gas~\cite{T2KND280TPC:2010nnd} and two scintillator-based fine-grained detectors, surrounded by an electromagnetic calorimeter and a 0.2\,T magnet.

The HNL search was performed by considering their production from kaon parents ($K^{\pm}\to \ell^{\pm}_{\alpha}N, \alpha = e,\mu$) and subsequent decay in ND280~\cite{T2K:2019jwa}. T2K data from November 2010 to May 2017 were considered for a total of $1.86 \times 10^{21}$~pot ($\sim$2:1 of neutrino to anti-neutrino mode). All the possible decay modes $N \to \ell^{\pm} \pi^{\mp}$ and $N \to \ell^{\pm} \ell^{\mp} \nu$ were considered, including the neutral current decay modes $N \to e^+ e^- \nu_{\tau}$ and $N \to \mu^+ \mu^- \nu_{\tau}$ that are directly sensitive to the mixing element $U_{\tau}^2$. Production and decay branching ratios were extracted from~\cite{MSM2007}, and effects related to heavy neutrino polarization~\cite{levy2019rates} and delayed arrival time (with respect to light neutrinos) are also taken into account.

In order to significantly improve the signal to background ratio, only events occurring in the TPC gas volume are considered for this analysis. Events are selected by identifying two tracks of opposite charge originating from a vertex in a TPC, and kinematic cuts are applied. Upstream detector activity is also vetoed. Five signal channels are then identified based on particle identification (obtained from TPC energy loss): $\mu^{\pm}\pi^{\mp}$, $e^- \pi^+$, $e^+ \pi^-$, $e^+ e^-$, $\mu^+ \mu^-$. The remaining background events are mainly originating from neutrino-induced coherent pion production on argon nuclei in the TPC gas, as well as misreconstructed interactions outside the gas e.g., photon conversion.

No events are observed in the defined signal regions, which is consistent with the background-only hypothesis. Several approaches are applied to obtain final limits on the $U_{\alpha}$ mixing elements. On the one hand, each production/decay mode is considered independently and the corresponding analysis channel is used to put limits on the associated mixing elements. On the other hand, a combined method has been developed to fit simultaneously all the heavy neutrino production and decay modes and the ten different analysis/signal channels (five for each beam mode). The Bayesian framework is using a Markov Chain Monte Carlo to marginalize over nuisance parameters and flat priors on the mixing elements $U_\alpha^2$ are assumed. The limits on $U_e^2$ and $U_\mu^2$ obtained with profiling (setting $U_\mu^2 = U_\tau^2 = 0$ and $U_e^2 = U_\tau^2 = 0$, respectively) are shown in Figs.~\ref{fig:hnl_exclusion_electron} and \ref{fig:hnl_exclusion_muon}, while Ref.~\cite{T2K:2019jwa} presents as well marginalized limits (including on $U_\tau^2$) and constraints in the plane $U_e^2-U_\mu^2$.

These limits have been recently reinterpreted to extend to lower masses ($20 \lesssim m_N \lesssim 200$\,MeV) in Ref.~\cite{Arguelles:2021dqn} by considering $N \to e^+ e^- \nu_\mu$ decays. The T2K group is currently performing the corresponding analysis including as well using pion production of heavy neutrinos, and searching for other decay modes, as well as taking benefit of the increased statistics (a total of $3.8\times 10^{21}$ POT has been collected as for February 2022).

\subsubsection{ArgoNeuT at FNAL} 
\label{subsubsec:HNL:ArgoNeuT}
ArgoNeuT was a LArTPC that operated in the Neutrinos from the Main Injector (NuMI) beam at Fermilab. The NuMI beam is produced colliding 120~GeV protons on a graphite target, producing mesons that are focused by a two-horn system and allowed to decay over a 675~m long decay pipe. ArgoNeuT collected data between 2009--2010, during which time the NuMI beam was operating in antineutrino mode for a total exposure of $1.25\times 10^{20}$ POT. The detector had a target mass of 0.24 ton and an active volume of $40 \times 47 \times 90$ cm$^3$. It was situated a distance of 1033 m downstream of the proton target and 318 m from the NuMI absorber. Finally, it was located immediately upstream of the MINOS near detector. The MINOS near detector, being significantly larger and also magnetized, can act as a muon spectrometer (and a muon/pion separator) for particles that exit the ArgoNeuT LArTPC before stopping.

The ArgoNeuT collaboration~\cite{ArgoNeuT:2021clc} carried out a search for heavy neutral leptons with nonzero $|U_{\tau 4}|^2$ where the $N$ are produced in a chain of $D$ meson decays -- $D_{(s)}^\pm \to \tau^\pm \nu_\tau$, $\tau^\pm \to N X^\pm$ (where $X^\pm$ is one or more SM particles). These are produced in the NuMI beam and absorber, and then can decay either inside or in front of the ArgoNeuT detector via $N \to \nu_\tau \mu^+ \mu^-$. With no observed signal, Ref.~\cite{ArgoNeuT:2021clc} places a constraint improving on existing searches in this mass range for tau-coupled $N$ from CHARM~\cite{Orloff:2002de} and DELPHI~\cite{Abreu:1996pa}. However, in the time since, Ref.~\cite{Boiarska:2021yho} demonstrated that CHARM has sensitivity beyond the reach of Ref.~\cite{Orloff:2002de}, yielding a more stringent constraint than ArgoNeuT.

\subsubsection{NOvA at FNAL}
\label{subsubsec:HNL:NovA}

The NO$\nu$A Near Detector at FNAL has a potential for HNL studies and is briefly assessed here. There have been comparative studies, before 2018, of the sensitivity to HNL detection with the Near Detector (ND) of the NO$\nu$A experiment hosted by Fermilab and falling under the general concept of Lightweight Dark Matter~\cite{Hatzikoutelis:2015nlf} 
The 990~m distance of the ND to the NuMI target could provide a search for micro-second HNL lifetimes. 
The acceptance efficiency into the 300-ton NO$\nu$A ND has been estimated to be on the order of $10^{-4}$.
The main mass range for the HNL that can be produced by the Fermilab's NuMI beam extends primarily to 500 MeV from the decay of the light mesons that make up the secondary beam of NO$\nu$A. 
There is also a possibility for accessing masses up to 2 GeV from the decay of heavier mesons, like the charmed meson $D_{s}$, which is produced with 1 $\mu b$ within the NuMI beam~\cite{PhysRevLett.77.2388,E769:1996jqf}. 

The preliminary studies with NO$\nu$A ND focused on the first few $10^{20}$ POT of exposure and the di-lepton decay channel for the HNL. The advantage of NO$\nu$A ND to leptons is based on its segmented design, the low-Z plastic, and the advanced PID efficiency used for measuring the NC neutrino interaction signatures. The current upper limit of the mixing between the electron-neutrino and the HNL in the mass range 100--500~MeV to the SM is no more than $10^{-7}$ for an expected HNL yield of O(100) events 
\cite{2016,Alekhin:2007686,Alekhin:2015byh}.
There are much stronger HNL hadronic and semi-leptonic decays, still unexplored, within the NO$\nu$A existing data set. There can be e-showers surrounded by pion-showers or purely multi-pion signatures, or combinations of a muon track of high energy with an e-shower. 
In particular, a muon with a jet combination is believed to be the strongest channel yet. 
However, the preliminary NO$\nu$A ND studies, before 2018, did no show any competitive sensitivity for these channels. 
This is mainly due to the lack of magnetic field in the fiducial volume which compromises any high energy reconstruction efficiency. Also, to a lesser degree, it was due to the hadronic PID analysis still being in the early development stage. 

At the time of  Snowmass 2021, though, the NO$\nu$A analysis-production of training samples for PID has been substantially increased and extended. 
The experience with the use of neural networks along with the maturity of analysis selections, the understanding of the reconstruction resolution and selections, has matured sufficiently to warrant a new investment in the sensitivity studies of HNL detection in the NO$\nu$A ND. 
The non-magnetic fiducial region may still restrict the resolution at high energy signatures, but this is somewhat compensated by the already existing large data set of detected events, several years ahead of any new detector developments.
Analysis of the existing data set of NO$\nu$A can improve the understanding of backgrounds to HNL, further constrain model rates about the coupling to SM, and strengthen the case for such exotic studies of New Physics studies within the neutrino experiments.  

\subsubsection{T2K-II and Hyper-Kamiokande} 
\label{subsubsec:HNL:HK}
The reported T2K results~\cite{T2K:2019jwa} are still statistically limited. The experiment will continue taking data up to 2026 (T2K-II phase), with an upgrade of the near detector ND280~\cite{Abe:ND280Upgrade} complex as well as an increase of the accelerator beam power, with the aim to collect additional $10 \times 10^{21}$ POT. The new ND280 detector will in particular embed two additional TPCs, that will approximately double the active volume available for HNL search. The operation will then continue in the context of the Hyper-Kamiokande (Hyper-K) experiment~\cite{protocollaboration2018hyperkamiokande}. 

Ten years of Hyper-K operation with a usage of (upgraded) ND280 as a near detector, correspond to statistics enhanced by a factor of $\sim 30$ with respect to current T2K results on HNLs. Scaling the latter based on this factor, we obtain a sensitivity shown in Figs.~\ref{fig:hnl_exclusion_electron} and~\ref{fig:hnl_exclusion_muon}. This is assuming the measurement would still be limited by the collected statistics, which is reasonable given the very low background levels reported in Ref.~\cite{T2K:2019jwa}. Furthermore, additional data will also allow one to better constrain the backgrounds by using more populated control regions, and improvements in event reconstruction are also expected that can potentially lead to an increase in the HNL signal selection efficiency and purity.

\subsubsection{The Short-Baseline Neutrino experiments: SBND, MicroBooNE and ICARUS at FNAL} 
\noindent
\label{subsubsec:HNL:SBN}
The Short-Baseline Neutrino Program at Fermilab consists of three liquid argon time projection chamber detectors situated along the Booster Neutrino Beam (BNB), a neutrino beam produced colliding 8~GeV protons on a Be target, resulting in predominantly pions and kaons that are subsequently focused using a single horn and allowed to decay over a 50~m decay pipe. The Short-Baseline Near Detector (SBND) is located at 110 m from the neutrino beam origin and has 112 ton of active mass. The SBND detector is currently being installed, with beam data taking expected in 2023. MicroBooNE is placed at 463 m downstream the neutrino beam origin and has 85 ton of active mass. MicroBooNE started beam data taking on October 2015 and finished on October 2021 after accumulating $\sim 10^{21}$ POT. ICARUS is located at 600~m from the neutrino beam origin and has 476 ton of active mass, and is finalizing commissioning.

The MicroBooNE collaboration published its first results from searching for kaon-decay production of $N$ and its subsequent decay into a charged pion/muon pair in Ref.~\cite{MicroBooNE:2019izn}. Due to large backgrounds from neutrino scattering (CC$1\mu1\pi$), the time structure of the neutrino beam was leveraged to reduce backgrounds and optimize signal searches. This (null) result leads to a constraint which is relatively weak compared to T2K and other existing constraints. In addition, MicroBooNE is exposed to the off-axis component of the NuMI beam. MicroBooNE is located at 680 m and $8^\circ$ from the NuMI target and 100 m and $120^\circ$ from the NuMI absorber. MicroBooNE has also searched for Higgs-Portal Scalars~\cite{MicroBooNE:2021usw} where the new-physics particle is produced in the absorber of the NuMI neutrino beam. Ref.~\cite{Kelly:2021xbv} reinterpreted this result in the neutrino-portal context and found that it improves on existing constraints in the 30--150~MeV region, comparable to the results demonstrated for T2K in Ref.~\cite{Arguelles:2021dqn}.
 
The ICARUS collaboration has recently started to study the physics case and technical feasibility for a dedicated HNL search using the NuMI beam. The ICARUS detector is located off-axis to the NuMI beamline with an angle of ${5.7}^\circ$, about 800 m from the NuMI target. The production of HNL mostly comes from charged mesons (i.\ e.\ pion/kaon) decay, with contributions from both decay-at-rest and decay-in-flight. It is expected that the HNL flux will peak at off-axis angles (due to the mass of the HNLs), whereas the SM neutrino flux drops off significantly at these angles. This allows for HNL searches to benefit from an increased signal-to-background ratio.
 
The SBND collaboration is carrying out a sensitivity study for HNLs in the BNB beam. Due to the closer location of SBND to the beam origin (compared to both MicroBooNE and ICARUS), a search using a delayed trigger window exploiting the longer time of flight of the HNL, as employed by MicroBooNE \cite{MicroBooNE:2019izn}, is expected to be less effective. Different strategies must be applied to search for HNL signatures and reduce the neutrino-scattering background, like the double ionization signature used by ArgoNeuT for dimuon searches \cite{ArgoNeuT:2021clc} or exploit the internal structure of the proton beam as MiniBooNE did \cite{MiniBooNEDM:2018cxm}. In order to obtain a realistic estimation of the signal efficiency and background rejection, SBND is using a full-detector simulation to include the reconstruction effects. SBND is projected to accumulate between $10 \times 10^{20}$ and $16 \times 10^{20}$ POT during its run.

\subsubsection{DUNE near detectors at FNAL}
\label{subsubsec:HNL:DUNE}
The DUNE (Deep Underground Neutrino Experiment) is a long-baseline neutrino experiment using the neutrino beam delivered by the new LNBF accelerator facility under design and construction at FNAL. This facility will use 120~GeV protons and provide $1.1\times 10^{21}$ POT per year, delivering a neutrino beam with 1.2~MW of beam power. After a running period of approximately 6~years an upgrade of the beam facility power to 2.4~MW is planned, doubling the number of POT/year. Both neutrino and antineutrino beams will be provided roughly with equal running times. The neutrinos are sent to a 40~kton fiducial volume far detector located 1.5~km underground, based in the Sanford Underground Research Facility in South Dakota, located 1300~km northwest from FNAL. 

Closer to the neutrino source will be a Near Detector (ND) complex which, among many other tasks, will monitor the neutrino flux produced by the facility. Apart from being a neutrino source, the high-intensity beam interactions with the target constitute also a potential source of HNLs and other new exotic particles, which are produced in particular in decays of light mesons. With the LBNF beam energy and intensity, the DUNE ND will cover an interesting and novel search region for decay-in-flight HNL particles with masses up to 2~GeV. 

The dominant decay channels into visible final states for HNLs in the low-mass region are:  HNL $\rightarrow e\pi, \nu e\mu, \nu e e, \nu \mu\mu, \nu\pi^0,$ and $\mu\pi$.  
The mass range for HNLs up to 2~GeV can be explored in all flavor-mixing channels. Decay channels for masses above 500~MeV include
channels containing heavier mesons and tau leptons (see, e.g., Table~1 in \cite{Ballett:2019bgd}).

The expected main background to this search comes from SM neutrino-–nucleon scattering events in which the hadronic activity at the vertex is below threshold, see 
e.g.~\cite{Ballett:2019bgd,Breitbach:2021gvv}. Charged-current quasi-elastic events with pion emission from resonances are background to the semi-leptonic decay channels. Moreover, misidentification of long pion tracks as muons can constitute a background to three-body leptonic decays. Neutral pions are often emitted in neutrino scattering events and can be a challenging background for HNL decays that include a neutral meson or channels with electrons in the final state. 

The DUNE ND~\cite{DUNE:2021tad} will be built in a shallow underground hall located 574~m downstream from the neutrino beam origin. Currently, it is expected that it will consist of three different components: a 150-tonne LArTPC (ND-LAr); a magnetized, gaseous argon time-projection chamber surrounded by an electromagnetic calorimeter (ND-GAr); and a smaller magnetized beam monitor consisting of straw tube chambers and an electromagnetic calorimeter (SAND). They will serve important individual and overlapping functions in the primary mission of the ND: the precise characterization of the neutrino beam energy and composition, as well as vastly improving knowledge of cross sections and particle yields for neutrino scattering processes in the few-GeV energy region. The two argon TPCs (ND-LAr and ND-GAr) will be moved laterally off the beam axis for some of the running, as to sample the beam at off-axis angles to better constrain flux uncertainties. Details on this DUNE-PRISM 
project are given in 
\cite{DUNE:2021tad}.


\begin{figure}[ht]
\includegraphics[width=0.95\textwidth]{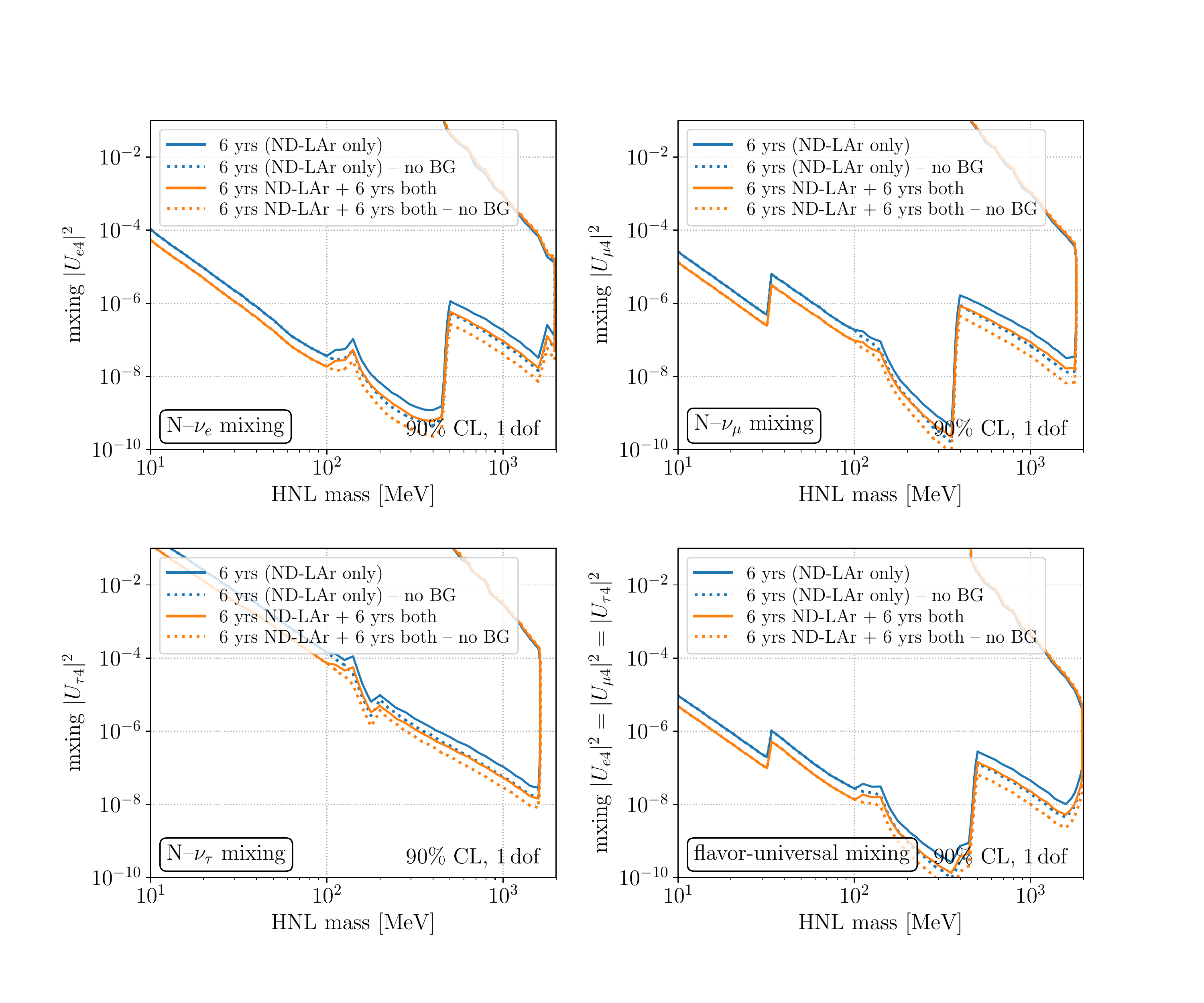}
\vspace{-1cm}
\caption{\normalsize DUNE sensitivity to Heavy Neutral Leptons with exclusive couplings to the first, second and third lepton generation 
and flavor-universal mixing $|U_{e4}|^2 = |U_{\mu 4}|^2 = |U_{\tau 4}|^2 = |U|^2$ in the bottom right panel.
Sensitivities are shown for a total of $6.6 \times 10^{21}$ POT in the first 6 years 
with the ND-LAr near detector alone (blue lines), and when 
another 6 years are added with $1.4 \times 10^{22}$ POT 
and having both ND-LAr and ND-GAr present (orange lines). The detectors 
are assumed to be 50\% of the time on axis and 50\% of axis
(see text). Solid lines show the sensitivity for signal plus  background estimates, while
dotted lines include only signal.}
\label{fig:DUNE_HNLs}
\end{figure}

The optimal detector component in this Near Detector complex for HNL studies is the ND-GAr as it has a larger (decay) volume than SAND and has a gaseous rather than liquid argon medium.~\cite{Berryman:2019dme,Breitbach:2021gvv}. For detectors with a dense medium a much larger background from neutrino interactions with a similar topology as the signal decays is expected, reducing its discovery potential. 
The studies indicate that one can expect about a factor two better sensitivity for HNLs searches in the gas-based TPC.
The LBNF and ND facility are expected to become  operational at the start of the next decade.

Several studies have recently examined the sensitivity for a DUNE-like experimental set-up, all showing the interesting potential for HNL searches below the kaon mass limit, where the anticipated high intensity proton beams may outmatch those of high-energy beam dump experiments~\cite{Krasnov:2019kdc,
Ballett:2019bgd,DUNE:2021tad,Berryman:2019dme,Coloma:2020lgy,Breitbach:2021gvv}.
In Ref.~\cite{Krasnov:2019kdc} only the signal is studied, based
on meson decays generated by PYTHIA. It creates its own neutrino flux from pp collisions. All the other studies 
make use of officially released neutrino fluxes delivered by the FNAL machine-experiment interface group~\cite{LauraField}
-- albeit for different versions, as the proton accelerator and neutrino beam-line have been still further optimized during the last few years.  
These flux files have been modified in~\cite{Berryman:2019dme,Coloma:2020lgy,Breitbach:2021gvv} to replace the quasi-massless SM neutrinos by massive HNL and change correspondingly the kinematics of these new particles. In Ref.~\cite{Coloma:2020lgy} effective Lagrangians are obtained for both the Dirac and Majorana scenarios, and made available as FeynRules models, so that fully differential event distributions can be easily simulated. 


Ref.~\cite{Berryman:2019dme} highlighted the prospects of ND-GAr for such a study, given both the lower detector density (higher signal-to-background ratio) and the magnetization, allowing for charge-identification of the HNL decay products. If an HNL is discovered in a channel where it decays fully visibly, e.g. $N \to \mu^\pm \pi^\mp$, then measurements of the respective rates of these two charges, $\mu^+\pi^-$ and $\mu^-\pi^+$ can allow for discrimination between Majorana- and Dirac-fermion hypotheses for the HNLs and SM neutrinos. With the magnetized ND-GAr and a relatively pure neutrino beam from LBNF, such discrimination is possible for many HNL masses in ten years of data collection~\cite{Berryman:2019dme}. 

For the sensitivity studies in this document we 
follow \cite{Breitbach:2021gvv}, which includes 
background estimates and a simple parametrized detector simulation. 
The studies have been performed for different scenarios of the total amount of protons on target (POT), implementation of the background, and selection of the near detectors (ND-LAr + ND-GAr or ND-LAr only). Compared to the original study in 
\cite{Breitbach:2021gvv} an improved background rejection
method has been used.
The study in Ref~\cite{Breitbach:2021gvv} also includes the  effect of the DUNE-PRISM running scheme, and shows that during the off-axis operation the experiment still has excellent prospects for discovering HNLs.
For that study the detectors were kept on axis for 
50\% of the time and the rest of the time the
detectors were moved laterally and with equal
exposure at 5 different 
positions: 6~m, 12~m, 18~m, 24~m, 30~m.
Fig.~\ref{fig:DUNE_HNLs} shows 
the sensitivity for a scenario of 6 years of LBNF with 1.2~MW
followed by 6 years with 2.4~MW, corresponding  to a total of $2\times 10^{22}$~POT for ND-GAr and $1.4\times 10^{22}$~POT for ND-GAr, which should have been collected by approximately {$\sim$}2040, with the ND-GAr staging scenario.
The results show that DUNE will be very competitive for masses in the region of a few 100~MeV, and can provide interesting limits also in the higher mass region.

For the summary figures 
Fig.~\ref{fig:hnl_exclusion_electron},  Fig.~\ref{fig:hnl_exclusion_muon} and Fig.~\ref{fig:hnl_exclusion_tau}, the scenario of 12 years 
operation, and including signal plus background, was chosen,
corresponding to the orange solid line in Fig.~\ref{fig:DUNE_HNLs}.

A new DUNE study is forthcoming, using the most recent neutrino flux calculations,
including the flux corrections and decays as implemented in Ref.~\cite{Coloma:2020lgy}, for the signal and  backgrounds, and including a parameterized response of the detectors for smearing of the kinematical variables and particle detection efficiency. Also
the neutrino--nucleus background could possibly be further suppressed exploiting the timing
difference between SM neutrinos and HNLs, if e.g.\
very fast timing detectors can be included. The use of Machine Learning is another direction which can be explored, utilizing the very detailed event information 
recorded by the detectors to select HNL decays.
To quantify these background levels, a detailed simulation study is required to estimate the impact on the detector requirements.

At sufficiently small mixing angles, HNLs will be essentially stable on detector length scales, making their direct experimental observation difficult. However a recent proposal~\cite{carbajal2022indirect} suggests  looking for a deficit in the muon or electron neutrino charged current events in a neutrino detector near the production point. This method may in principle allow one to probe mixing angles as low as $\mathcal{O} (10^{-5})$ for low masses (MeV region), 
it relies on 
a precise knowledge of the active neutrino flux coming from the target
which has not yet been demonstrated.

\subsection{Opportunities at FNAL: DarkQuest}
\noindent
The DarkQuest proton beam dump experiment, a proposed upgrade of the existing SeaQuest/SpinQuest experiment at Fermilab, has excellent prospects to search for light, long-lived exotic particles~\cite{Gardner:2015wea,Berlin:2018tvf,Berlin:2018pwi,Blinov:2021say}, including HNLs~\cite{Batell:2020vqn}. The SeaQuest/SpinQuest experiment utilizes part of the Fermilab 120 GeV Main Injector proton beam. The beam impacts a 5m iron dump, which is immediately followed by a $\sim$~20~m spectrometer capable of detecting and precisely measuring $\mu^+ \mu^-$ final states~\cite{SeaQuest:2017kjt}. The proposed DarkQuest upgrade will install an electromagnetic calorimeter, expanding its capability to detect additional visible signatures, including electron, pions, and photons~\cite{PHENIX:2003fvo}. In comparison to other proton-fixed target experiments, one novelty of the DarkQuest setup is its compact detector geometry and shorter baseline. Especially for light BSM states with moderate lifetimes, this results in a larger geometric acceptance and access to otherwise challenging regions of parameter space. The 5m beam dump dramatically reduces the potential SM backgrounds, while the remaining SM backgrounds from penetrating neutral SM particles can be further mitigated by several means, such as additional shielding, vertexing, and topological and kinematical handles. The DarkQuest experiment can produce HNLs through meson or tau lepton decay, and detect their visible decay modes over a wide range of masses below the $B$ meson mass. In Figs.~\ref{fig:hnl_exclusion_muon} and \ref{fig:hnl_exclusion_tau} we display the DarkQuest sensitivity corresponding to 10 HNL signal events for the muon and tau flavor dominance hypotheses, respectively~\cite{Batell:2020vqn}. The reach on the electron dominance scenario is very similar to the reach for the muon dominance case. These estimates assume the detection of all HNL decays to visible final states.  In each case, two sensitivity lines are presented corresponding to a Phase I run scenario for $10^{18}$ POT and 5m-6m fiducial decay region (solid), which can be achieved on the few years time scale, and a Phase II run scenario of $10^{20}$ POT and 7m-12m fiducial decay region (dashed), which is feasible on the 5--10 year time scale. As argued in~\cite{Batell:2020vqn}, it is anticipated that SM backgrounds can be brought down to manageable levels for both phases of running, provided additional shielding is installed between 5-7 m for Phase II. As we can see from the figures, particular with a $10^{20}$ POT run, DarkQuest has the potential to extend the reach beyond previous experiments.

\subsection{Opportunities at PSI: PIONEER} \label{subsec:HNL:PSI}
\noindent
The PIONEER experiment~\cite{Mazza:2021adt,PIONEER:2022yag} will carry out a peak search and plans to measure the ratio $R^{(\pi)}_{e/\mu} = {\cal B}(\pi^+ \to e^+\nu_e)/{\cal B}(\pi^+ \to \mu^+ \nu_\mu)$ to an order of magnitude higher precision than the current most accurate measurement, which is from the PIENU experiment at TRIUMF \cite{PiENu:2015seu}. In this context, it is important to recall that even if an HNL is too heavy to be emitted in $\pi^+_{e2}$ decay, it will, in general, cause a deviation the ratio $R^{(\pi)}_{e/\mu}$ from its SM value~\cite{Shrock:1980vy,Shrock:1980ct}. Indeed, the agreement of $R^{(\pi)}_{e/\mu}$ with the SM prediction has been used to obtain a stringent upper bound on $|U_{e4}|^2$ in the HNL mass range from a few MeV to about 60~MeV~\cite{Bryman:2019ssi,Bryman:2019bjg}. 
The PIENU and PSI PEN\footnote{See D. Počanić et al. (PEN), http://pen.phys.virginia.edu/ (2006).} collaborations expect to make further improvements on $R^\pi_{e/\mu}$. 
The new measurement of $R^{(\pi)}_{e/\mu}$ planned by the PIONEER experiment at PSI~\cite{PIONEER:2022yag} will yield not only an improved constraint on HNL effects, but also a higher-precision test of $e$--$\mu$ universality~\cite{PIONEER:2022alm}.

Limits on HNL mixing parameter $|U_{e4}|^2$ from the PIENU pion decay experiment (a predecessor of PIONEER), including results from $\pi^+\to e^+N$ peak search in the 62--135~MeV range~\cite{PIENU:2017wbj} and a constraint imposed by the $R^{(\pi)}_{e/\mu}$ measurement at lower $m_N$ values~\cite{PiENu:2015seu,Bryman:2019ssi,Bryman:2019bjg}, are displayed in Fig.~\ref{fig:hnl_exclusion_electron}, along with projections based on proposed PIONEER sensitivities.

\subsection{Post-Discovery Potential}\label{subsec:HNL:PostDiscovery}

In the event of discovery of a new neutrino-portal particle, the next step will be to understand its properties and address a large number of questions -- does it have interactions beyond those of mixing with the SM neutrinos, is it connected to the origin of light neutrino masses, do its interactions preserve or violate Lepton Number? Depending on the context in which such a particle is discovered, we may approach these questions in qualitatively different ways.

\textbf{Neutrino-Beam Discovery:} In neutrino-beam environments, it is possible to study the possibility of Lepton-Number Violation (LNV) and whether $N$ is a Dirac or Majorana fermion. This is attainable by measuring the relative rate of fully-visible decays in a detector, e.g., $N \to \mu^+ \pi^-$ and $N \to \mu^- \pi^+$. If $N$ is a Majorana fermion, then these will occur with equal probability, whereas if $N$ is a Dirac fermion, only one process can occur (depending on the polarity of the beam, which must be pure to perform this separation). Ref.~\cite{Berryman:2019dme} demonstrated that for large parts of parameter space, DUNE ND-GAr (thanks to the magnetization of the detector) can not only discover a new particle $N$ but also determine, via its decay rates, whether it is a Dirac or Majorana fermion. If $N$ decays in a partially-invisible channel, then decay kinematics can be used to perform this separation~\cite{BahaBalantekin:2018ppj,Balantekin:2018ukw,deGouvea:2021ual,deGouvea:2021rpa}.

\textbf{Beam-Dump Discovery:} In a complementary way, several well-motivated models may be explored in detail if the new particle is discovered in a beam-dump scenario.
As a specific example, Ref.~\cite{Tastet:2019nqj} explored the possibility of examining a two-quasi-mass-degenerate HNL scenario at SHiP. This could  be studied at other beam-dump experiments as well if they are provided by a
tracking system that guarantees good momentum and vertex resolutions. This is the case for example for NA62 in dump mode, SHADOWS, and DarkQuest. Studies at this point in time assume perfect performance of reconstruction, in identifying decay products and momentum/vertex resolution.

\begin{figure}[ht]
    \centering
    \begin{minipage}[b][][b]{0.5\textwidth}
        \centering
        \includegraphics[width = 0.920\linewidth]{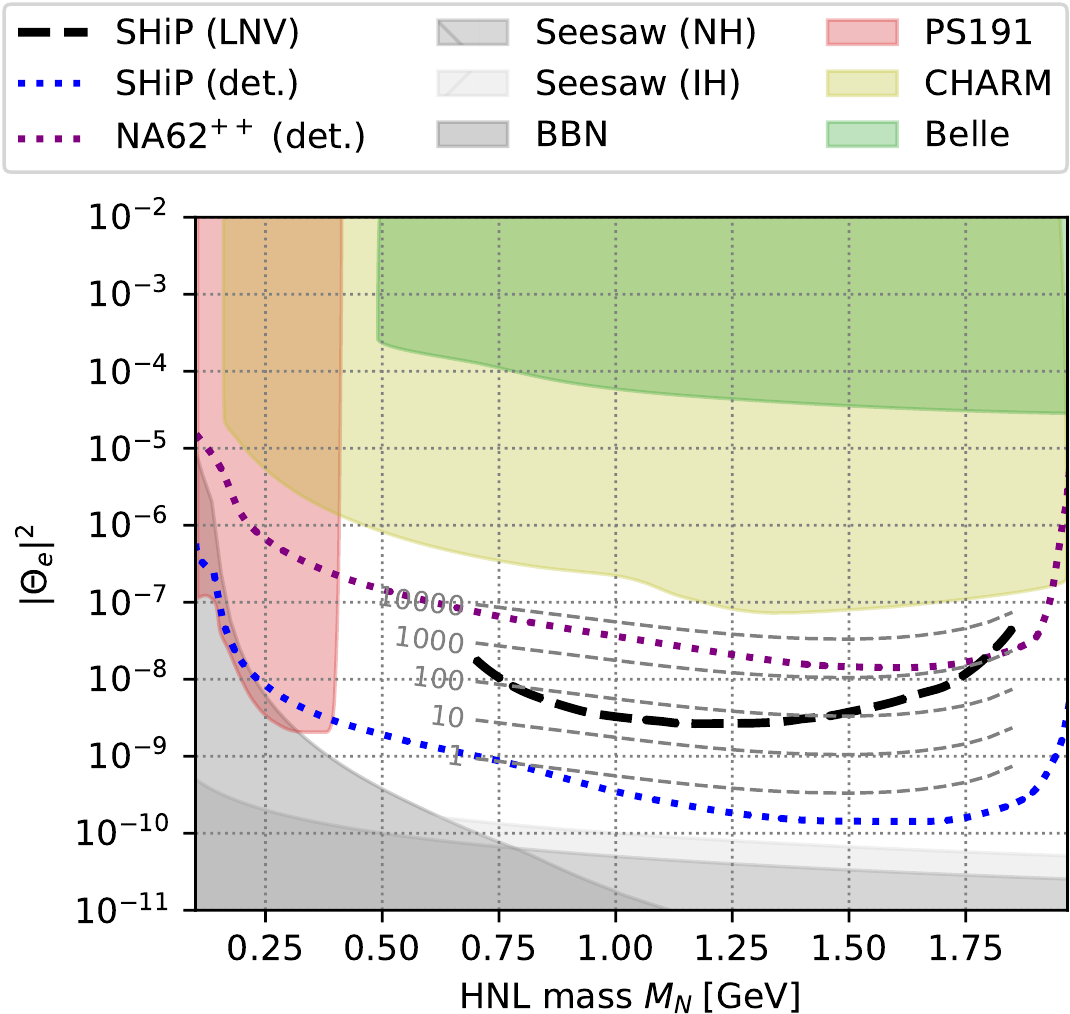}\\
        \vspace{1em}%
        (a) HNLs mixing with $\nu_e$.
    \end{minipage}%
    \begin{minipage}[b][][b]{0.5\textwidth}
        \centering
        \includegraphics[width = 0.911\linewidth]{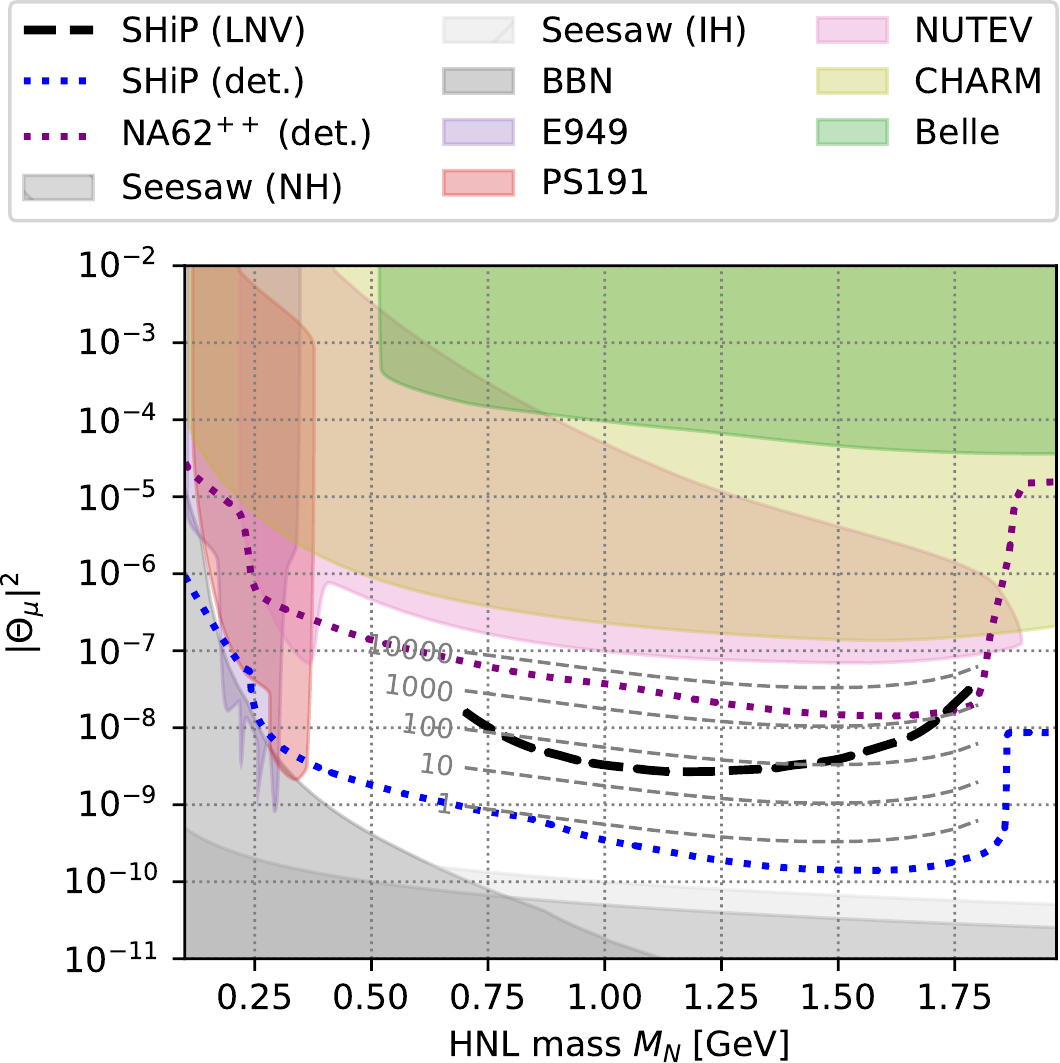}\\
        \vspace{1em}%
        (b) HNLs mixing with $\nu_{\mu}$.
    \end{minipage}
    \caption{SHiP sensitivity to lepton number violation for HNLs with pure electron (a) or muon (b) mixing. Thick dashed line is the required number of events to differentiate between the models of Dirac and Majorana-like HNLs. Gray dashed lines represent the number of fully reconstructable events at SHiP for given mass and mixing angle. The figure is taken from~\cite{Tastet:2019nqj}.}
    \label{fig:LNCsensitivity}
\end{figure}

\begin{figure}[ht]
    \centering
    \includegraphics[width = 0.52\textwidth]{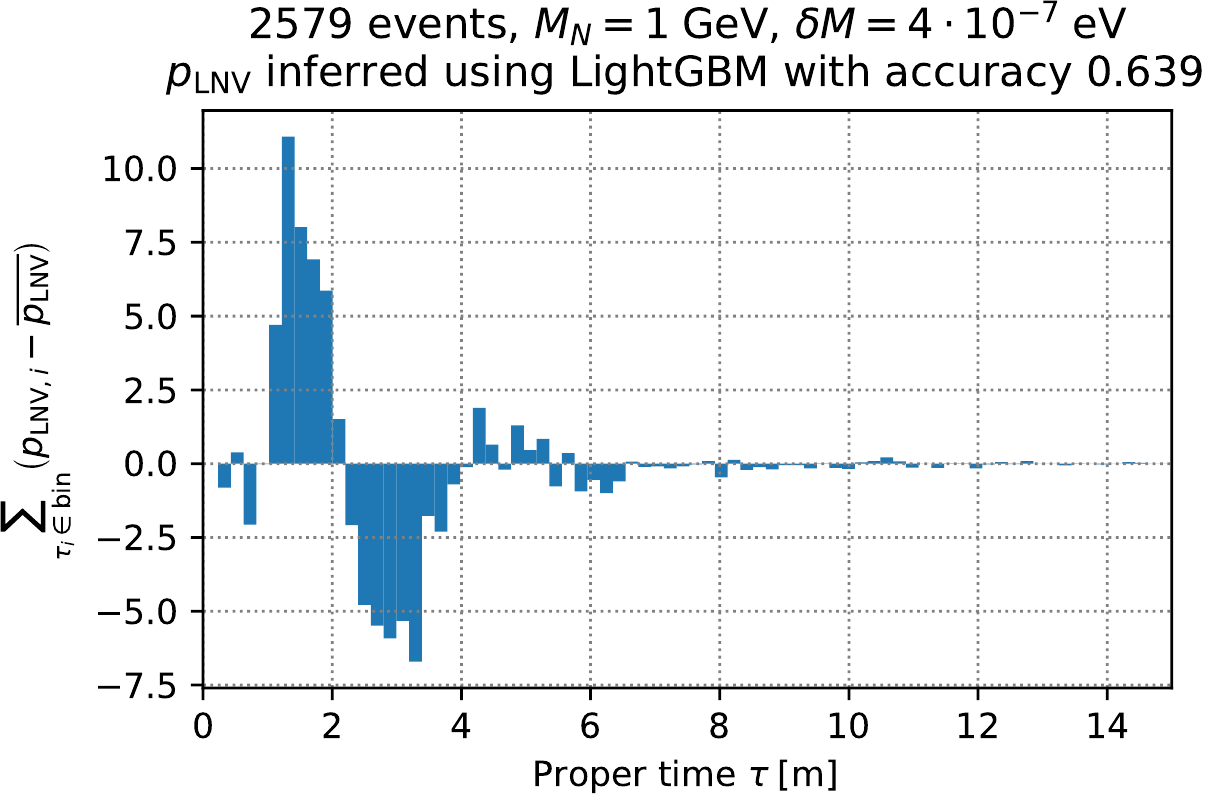}
    \caption{The oscillation pattern observed in simulated data for the model of two HNLs with average mass $M_N = 1\text{ GeV}$, mass splitting $\delta M = 4\cdot 10^{-7}\text{ eV}$ and pure muon mixing with the mixing angle $U^2_\mu = 2 \cdot 10^{-8}$. The figure is taken from~\cite{Tastet:2019nqj}.}
    \label{fig:LNCosc}
\end{figure}

The mass splitting $\delta M$ in the model of two quasi-degenerate HNLs determines the oscillation length between the two particles and therefore controls the ratio of lepton number conserving (LNC) and violating (LNV) decays. These processes may be distinguished thanks to different angular distribution of final particles in decay due to the correlation between the helicity of the HNL and the charge of the initial lepton produced together with the HNL~\cite{Tastet:2019nqj}. The required number of events to differentiate between two extreme cases of Dirac-like (only LNC decays observed) and Majorana-like (equal fractions of LNV and LNC) HNLs is shown in Fig.~\ref{fig:LNCsensitivity} for the SHiP case. In addition, SHiP may offer a unique opportunity to probe the mass splitting in the intermediate regime, when the oscillation between the two HNLs may be observed explicitly on the scale of the experiment, which corresponds to $\delta M \approx 10^{-6}/\tau$ with $\tau$ being the proper time of the HNL before decay. The oscillation pattern in a simulated data is shown in Fig.~\ref{fig:LNCosc}.

\begin{figure}[ht]
    \centering
    \includegraphics[width = 0.35\textwidth]{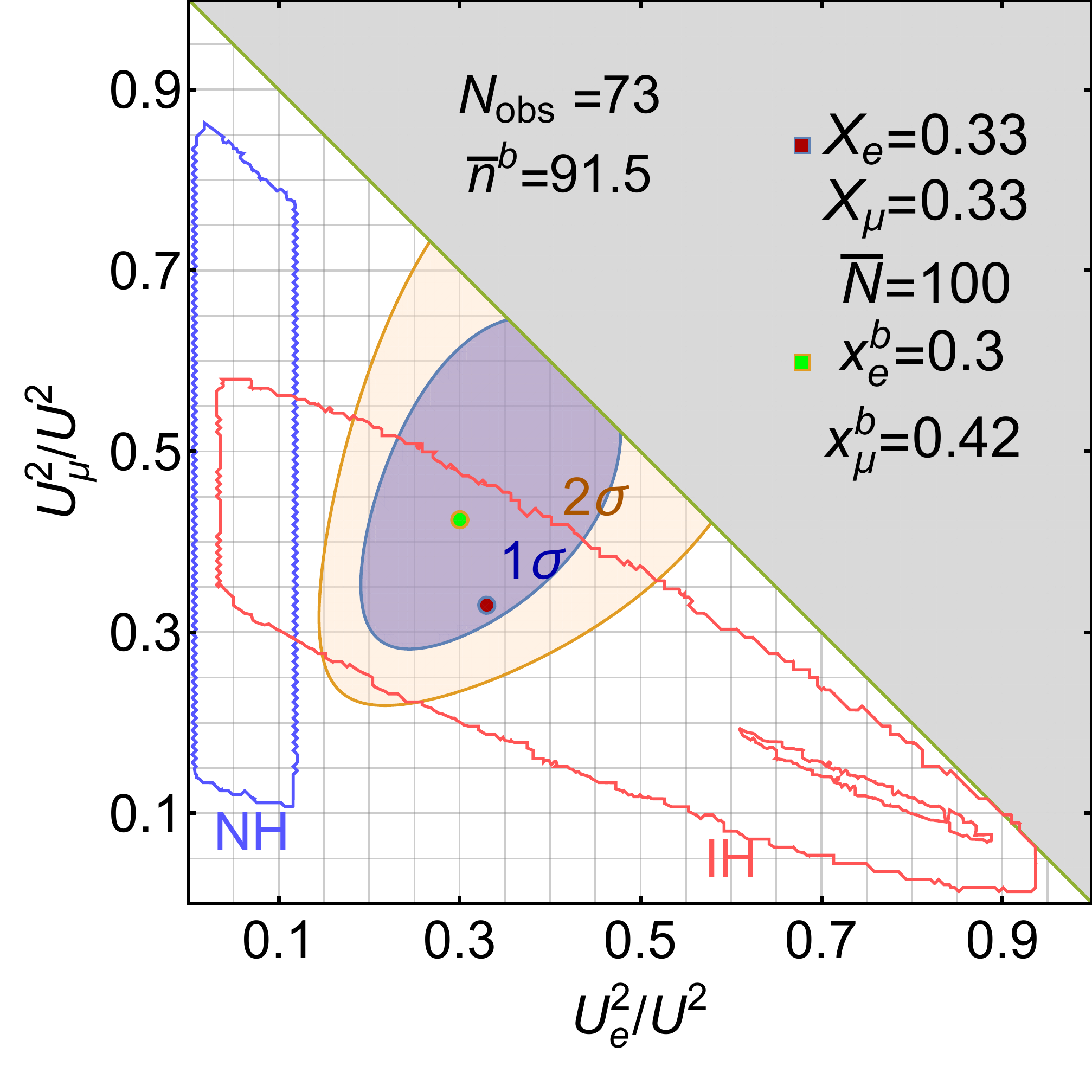}~\includegraphics[width = 0.35\textwidth]{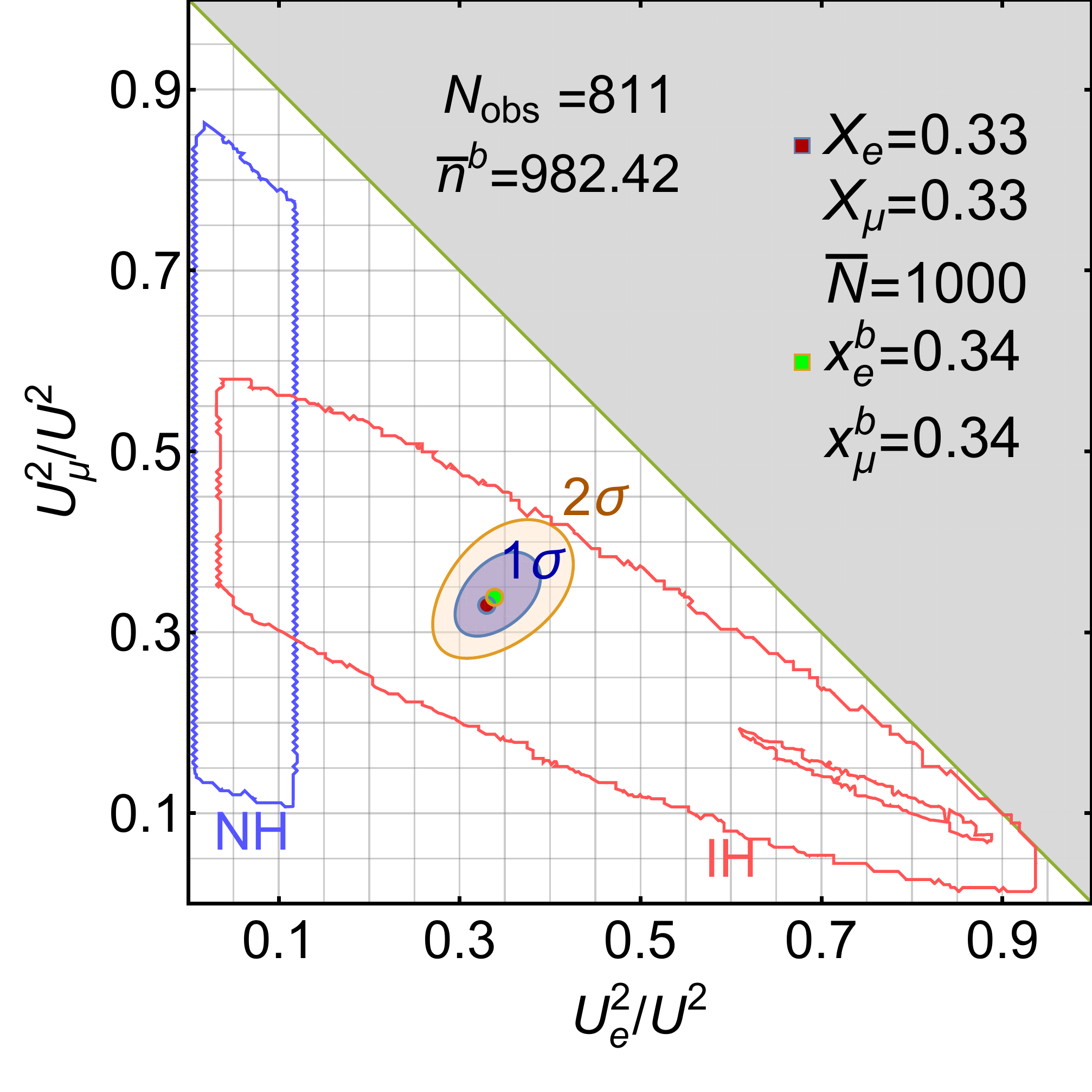}
    \caption{$1\sigma$ (blue) and $2\sigma$ (orange) confidence regions in $(U^2_e/U^2, U^2_\mu/U^2)$ plane computed using simulated data for an HNL with mass $M_N = 1\text{ GeV}$, ratio $U^2_e:U^2_\mu:U^2_\tau = 1:1:1$ (red point), and expected 100 (left) or 1000 (right) events with HNLs. Green point represents the best-fit ratio for the simulated data. $N_\text{obs}$ is the number of detected events (excluding invisible $N\to3\nu$) and $\bar n^b$ is the best-fit value for the total number of events.}
    \label{fig:mixpat}
\end{figure}

Additionally, beam-dump experiments may measure the various mixing parameters (i.e. the ratio $U^2_e:U^2_\mu:U^2_\tau$), specifically in the case where multiple (possibly quasi-degenerate) HNLs exist. This allows for testing whether the HNL is consistent with being responsible for light neutrino masses as measured by oscillation experiments and could distinguish between the  (NH) and inverted hierarchy (IH) of neutrino masses. The SHiP sensitivities at $1\sigma$ and $2\sigma$ CL in the $(x_e, x_\mu)$ plane (with $x_\alpha =U_\alpha^2/U^2$) for a simulated data corresponding to the average 100 or 1000 HNL decays are shown in Fig.~\ref{fig:mixpat}, together with the regions consistent with NH and IH from Ref.~\cite{Bondarenko:2021cpc}. Here, the two HNLs are assumed to have mixing angles that differ negligibly, which requires them to have the total mixing angle sufficiently larger than the seesaw bound. This is correct for the assumed number of events at the SHiP experiment. Detection efficiencies of all decay modes are assumed to be equal.

\textbf{Kaon-Decay Discovery:} Finally, if a new particle $N$ is discovered by studying kaon decays, e.g., $K^+\to \mu^+ N$ using the missing-mass technique, there is the possibility that $N$ is relatively short-lived and its properties can be studied by the same detector. If $N$ can decay via LNV physics, it could subsequently decay into another positively-charged muon and other particle(s).
Identifying signatures in the detector such as $K^+\to\pi^-\ell^+\ell^+$~\cite{NA482:2016sfh,NA62:2019eax,NA62:2021zxl,NA62:2022tte}, which represent a clear pattern of LNV, allows for not only the discovery of the HNL but also the determination of the nature of the neutrino mass.

\subsection{Conclusions}
\label{subsec:HNL:Conclusions}
\begin{figure*}[ht]
\includegraphics[width=\textwidth]{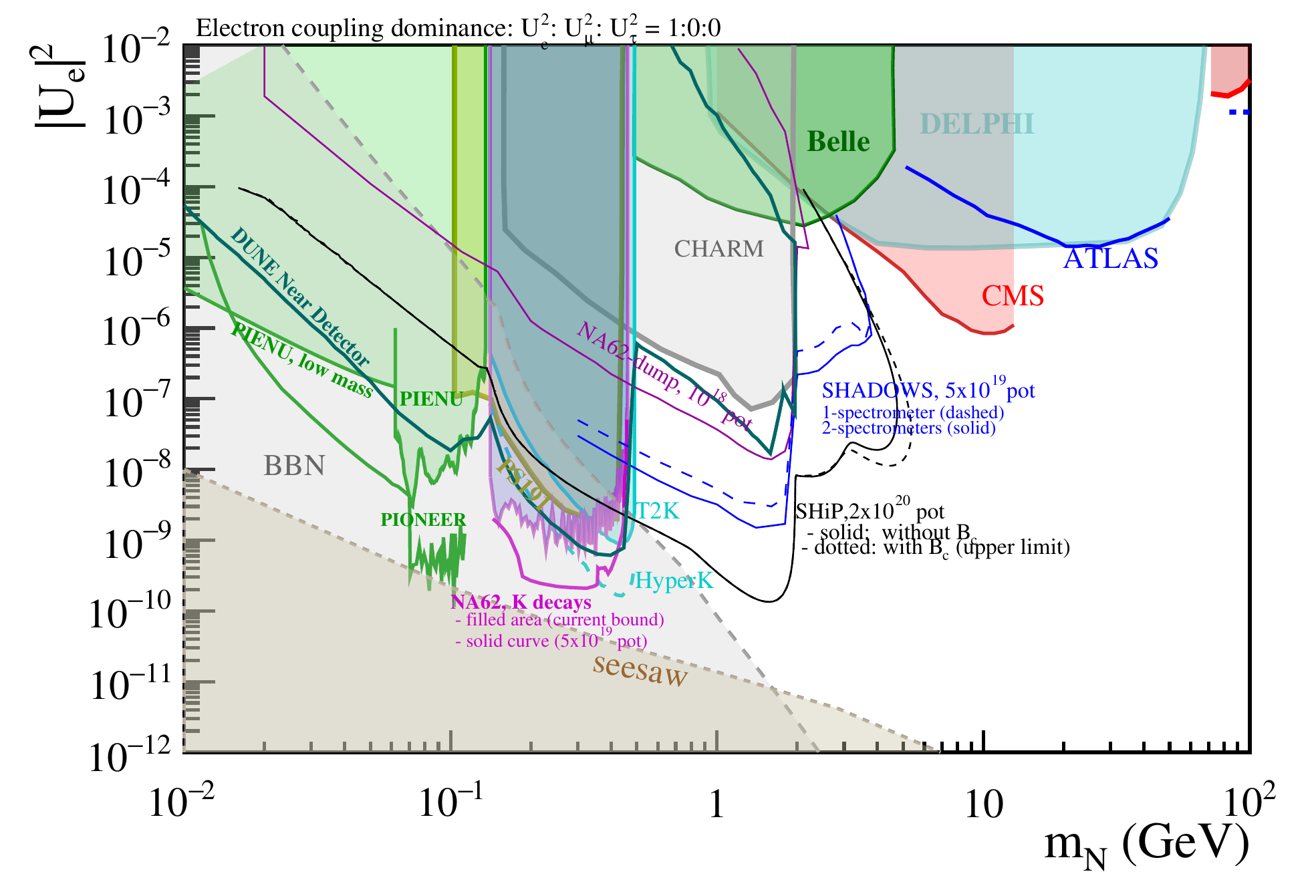}
\caption{\normalsize Heavy Neutral Leptons with coupling to the first lepton generation. Filled areas are existing bounds from:
PS191~\cite{Bernardi:1987ek}, CHARM~\cite{CHARM:1985anb}, PIENU (peak searches~\cite{PIENU:2017wbj} and bounds at low masses~\cite{PiENu:2015seu,Bryman:2019ssi,Bryman:2019bjg}),
NA62~($K_{eN}$)~\cite{NA62:2020mcv},
T2K~\cite{T2K:2019jwa},
Belle~\cite{Belle:2013ytx}; DELPHI~\cite{Abreu:1996pa}, ATLAS~\cite{ATLAS:2019kpx} and CMS~\cite{CMS:2022fut}. Colored curves are projections from:  NA62-dump~\cite{Beacham:2019nyx},
NA62 $K^+$ decays (extrapolation obtained by the Collaboration based on~\cite{NA62:2020mcv}),
PIONEER~\cite{PIONEER:2022yag},
SHADOWS~\cite{Baldini:2021hfw},
DarkQuest~\cite{Batell:2020vqn},
SHiP~\cite{SHiP:2018xqw}, DUNE near detector (projections based on methods developed in~\cite{Breitbach:2021gvv}), and Hyper-K (projections based on~\cite{T2K:2019jwa}). 
The BBN bounds are from~\cite{Boyarsky:2020dzc} and heavily depend on the model assumptions (hence they should be considered indicative). The seesaw bound is computed under the hypothesis of two HNLs mixing with active neutrinos.}
\label{fig:hnl_exclusion_electron}
\end{figure*}

\begin{figure*}[ht]
\includegraphics[width=\textwidth]{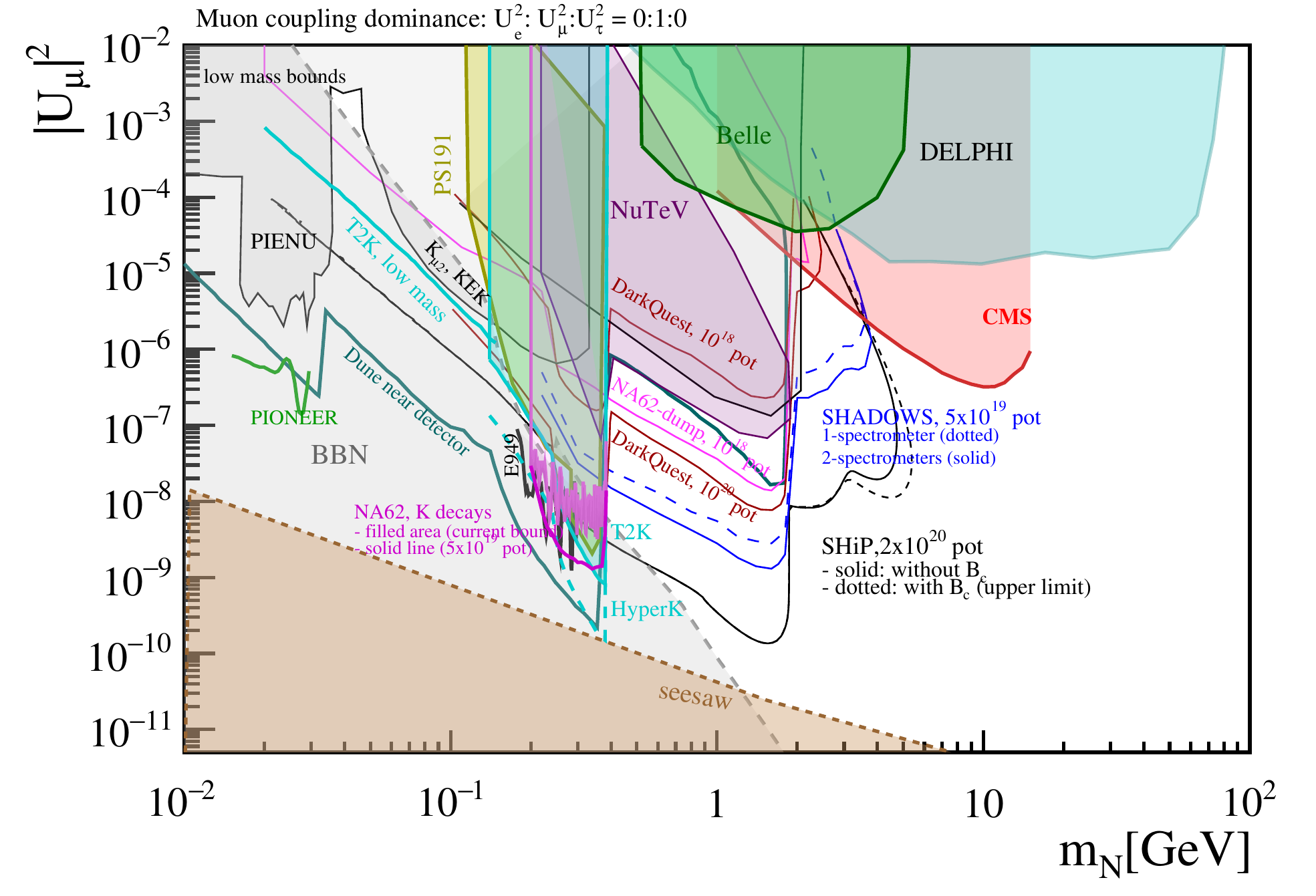}
\caption{\normalsize Heavy Neutral Leptons with coupling to the second lepton generation. Filled areas are existing bounds from:
PS191~\cite{Bernardi:1987ek}, CHARM~\cite{CHARM:1985anb}, 
NA62 ($K_{\mu N}$)~\cite{NA62:2021bji},
T2K~\cite{T2K:2019jwa}, 
E949~\cite{E949:2014gsn},
Belle~\cite{Belle:2013ytx}; DELPHI~\cite{Abreu:1996pa}, and CMS~\cite{CMS:2022fut}.
The ``low mass bounds'' label refers to a set of results obtained from $\pi$ and $K$ decays, as detailed in Ref.~\cite{Bryman:2019bjg}, namely a PIENU result~\cite{PIENU:2019usb} and $K_{\mu2}$ results at KEK~\cite{Hayano:1982wu,Yamazaki:1984de}.
Colored curves are projections from:  NA62-dump~\cite{Beacham:2019nyx},
NA62 $K^+$ decays (projections obtained by the Collaboration based on~\cite{NA62:2021bji}),
SHADOWS~\cite{Baldini:2021hfw},
DarkQuest~\cite{Batell:2020vqn},
PIONEER~~\cite{PIONEER:2022yag},
SHiP~\cite{SHiP:2018xqw}, DUNE near detector (projections based on methods developed in~\cite{Breitbach:2021gvv}), Hyper-K (projections based on~\cite{T2K:2019jwa}), T2K low mass~\cite{Arguelles:2021dqn}.
The BBN bounds are from \cite{Boyarsky:2020dzc} and heavily depend on the model assumptions (hence should be considered only indicative). The seesaw bounds are computed under the hypothesis of two HNLs mixing with active neutrinos.}
\label{fig:hnl_exclusion_muon}
\end{figure*}

\begin{figure*}[ht]
\includegraphics[width=\textwidth]{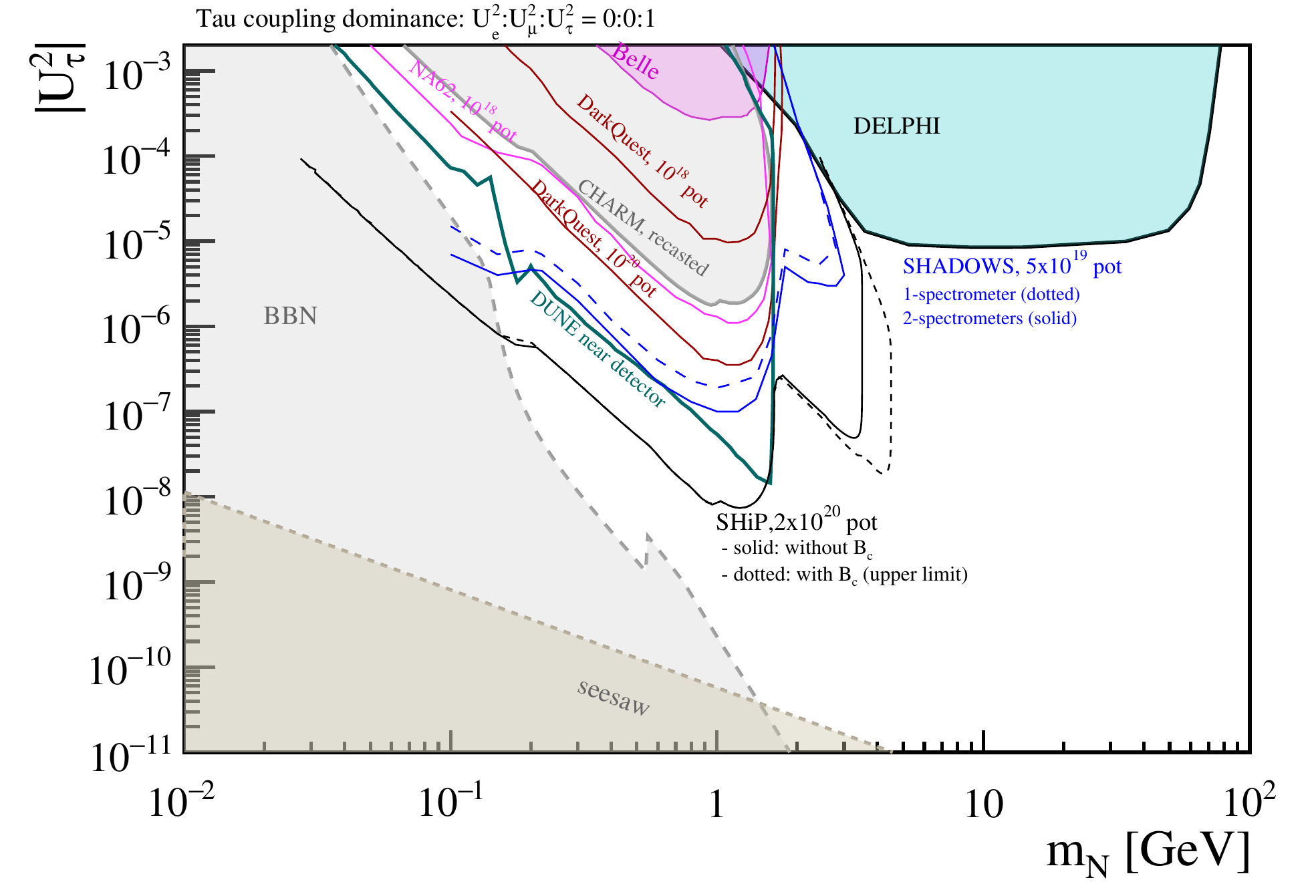}
\caption{\normalsize Heavy Neutral Leptons with coupling to the third lepton generation. Filled areas are existing bounds from:
CHARM~\cite{CHARM:1985anb} (recasted in~\cite{Boiarska:2021yho});  Belle~\cite{Belle:2013ytx}; DELPHI~\cite{Abreu:1996pa}. Colored curves are projections from:  NA62-dump~\cite{Beacham:2019nyx},
SHADOWS~\cite{Baldini:2021hfw},
DUNE (projections based on methods developed in~\cite{Breitbach:2021gvv}),
DarkQuest~\cite{Batell:2020vqn}, and
SHiP~\cite{SHiP:2018xqw}. 
The BBN bounds are from \cite{Boyarsky:2020dzc} and heavily depend on the model assumptions (hence should be considered only indicative). The seesaw bounds are computed under the hypothesis of two HNLs mixing with active neutrinos.}
\label{fig:hnl_exclusion_tau}
\end{figure*}

This section has focused on the prospects of discovering HNLs in the coming one or two decades in fixed-target environments. This broad category includes many currently-operating and next-generation experiments, each with various approaches and physics goals (many of which are orthogonal to these beyond-the-Standard-Model searches). We have broadly categorized these searches based on their experimental apparatuses, divided into searches from rare kaon decays, beam-dump setups, and searches in neutrino-beam environments. Figs.~\ref{fig:hnl_exclusion_electron},~\ref{fig:hnl_exclusion_muon}, and ~\ref{fig:hnl_exclusion_tau} summarizes the capabilities of these searches for the different coupling scenarios (electron, muon, and tau coupling dominance) in the next two decades. Here we observe not only complementarity amongst the different fixed-target probes, but also when comparing with the other types of searches discussed throughout this work. Fixed-target searches offer some of the most promising sensitivity to discovering HNLs in the near future.

Fixed-target searches additionally have more to offer -- in the hopeful event of an HNL discovery, there is the potential to study the HNL's (or HNLs') properties. This includes, but is not limited to, studying the mixing pattern(s) of the HNL(s) to determine if there is a connection to the observed light neutrino masses, as well as determining whether Lepton Number is conserved or violated, or equivalently, whether neutrinos and HNLs are Dirac or Majorana fermions. Either of these observations would revolutionize particle physics, and fixed-target environments are capable of carrying out these studies, as we discussed in Section~\ref{subsec:HNL:PostDiscovery}.

Finally, we conclude this section by including Table~\ref{tab:fixed-target-all-proposals}, a summary of the searches considered here with comparisons of the technologies, dimensions, and plans of operation. In the  ``MicroBooNE'' row, we give two sets of numbers for the distance from the protons-on-target to the detector and the number of protons-on-target collected, corresponding to the BNB and NuMI-Dump searches discussed above in Section~\ref{subsubsec:HNL:SBN}. The figures provided for the different searches allow one to compare future prospects of HNL discovery and potential follow-up studies. We note the variety of beam energies, worldwide locations, and detector technologies -- in the event of a discovery of HNLs in one experiment, confirmation in others will bolster this new-physics case.
\begin{sidewaystable}[htbp]
\caption{Summary of the fixed-target searches for Heavy Neutral Leptons considered throughout Section~\ref{sec:fixed}.}
\label{tab:fixed-target-all-proposals}
\vspace{.1cm}
\begin{center}
\begin{tabular}{ccccccccc}
\hline \hline
experiment/ & lab & beam type & detector  & detector  & detector decay & distance & N$_{\rm pot}$ & timescale \\
proposal  &  & & technology & transverse & volume &
from & \\
    &   &  &  & dimensions & length  & dump &  & \\
\hline \hline
    &  &  &  &  &  &  &  &   \\
NA62-K  & CERN & $p$, 400~GeV & spectrometer & $A = \pi r^2$, $r=1$~m & $\sim $ 80~m & $\sim$100~m & $5\cdot 10^{19}$ & by (2032--2038)\\
NA62-dump  & CERN & $p$, 400~GeV & spectrometer & $A = \pi r^2$, $r=1$~m &  $\sim $ 80~m & $\sim$100~m & $5\cdot 10^{19}$ & by (2032--2038)  \\
SHADOWS   & CERN & $p$, 400~GeV & spectrometer & 2.5$\times$2.5~m$^2$ & $\sim$~20~m & $\sim~10$~m & $5\cdot 10^{19}$ & by (2032--2038)     \\
SHiP  & CERN & $p$, 400~GeV  & spectrometer & 5$\times$10~m$^2$ & $\sim $ 50~m  & $\sim $ 45~m  & $2\cdot 10^{20}$   \\
T2K & J-PARC & $p$, 30\,GeV & composite w/ GArTPC & $\sim 3.3$\,m$^2$ & $\sim 1.7$\,m & $280$\,m & $3.8 \cdot 10^{21}$ & 2010--2021 \\
T2K-II & J-PARC & $p$, 30~GeV & composite w/ GArTPC & $\sim 3.3$\,m$^2$ & $\sim 3.6$\,m & $280$\,m & $+10 \cdot 10^{21}$ & 2022--2026 \\
Hyper-K & J-PARC & $p$, 30~GeV & composite w/ GArTPC & $\sim 3.3$\,m$^2$ & $\sim 3.6$\,m & $280$\,m & $2.70 \cdot 10^{22}$ & by 2038 \\
SBND  &  FNAL & $p$, 8~GeV & LArTPC & $16$\,m$^2$ & $5$\,m & $110$\,m & $10 \cdot 10^{20}$ & 2023--2027  \\
MicroBooNE  & FNAL & $p$, 8/120~GeV & LArTPC & $6$\,m$^2$ & $10.4$\,m & $463$\,m/$100$\,m & $1.5 \cdot 10^{21}/2.2 \cdot 10^{21}$ & 2015--2021 \\
ArgoNeuT  & FNAL & $p$, 120~GeV & LArTPC & $0.2$ m$^2$ & $0.9$ m & $318$ m & $1.25 \cdot 10^{20}$ & 2009--2010 \\
DUNE ND  & FNAL & $p$, 120~GeV & LAr/GAr TPC & ${\sim}$12 m$^2$ & ${\sim}$5 m & $574$ m & $\gtrsim 1.47 \cdot 10^{22}$ & ${\sim}$2030--2040 \\
DarkQuest  & FNAL & $p$, 120~GeV & spectrometer & 2$\times$4~m$^2$ & 20~m & 5~m & $1 \cdot 10^{18}$  &  2024-2025\\
\hline \hline
\end{tabular}
\end{center}
\end{sidewaystable}

\clearpage

\clearpage

\section{Atmospheric and Solar Searches}
\label{sec:atmos}

Solar and atmospheric neutrinos are ever-present natural sources of neutrinos that extend up to $\sim20$ MeV and $\sim$100s of GeV respectively. The Sun represents the most intense source of neutrinos outside the laboratory \emph{by far}, and supplies a mutli-flavored flux of neutrinos at the Earth's location due to the MSW effect and neutrino oscillations. In the case of atmospherics, the flux is roughly flat for $E_\nu \lesssim 1$ GeV, and then falls like a power law. Atmospheric neutrino oscillations result in a highly ``flavor mixed'' flux, much like solar neutrinos, albeit one with substantial energy-flavor correlations. Both of these natural neutrino sources are therefore an incredibly powerful resource with which to search for $e$-coupled, $\mu$-coupled, \emph{and} $\tau$-coupled HNLs. Combined with their naturally abundant fluxes, this makes both solar and atomspheric neutrinos a powerful resource with which to search for HNLs. 

The relevant phenomenology is conveniently organized by the HNL's decay length. Taken together, solar and atmospheric neutrino searches are capable of probing HNLs whose decay length can vary over orders of magnitude ranging from $10^{8}$~{\rm km} to $10's$ of meters or less. This remarkable variability in length scales stems from the fact that: {\em i}) the solar neutrino flux is sufficiently well measured that \emph{indirect} recoil measurements are possible; {\em ii}) the Sun is $\sim 1.5 \times 10^{8}$ km from the Earth; {\em iii)} decay lengths satisfying $R_\odot \gg \lambda \gtrsim R_\oplus$ can be efficiently probed by taking advantage of upscattering inside the Earth; and {\em iv}) decay lengths that are sub-detector scale, i.e.\ $\lesssim 10$ m, can leave so-called ``double bang'' signatures inside detectors which have low backgrounds. Combining all of these techniques allows one to probe a wide region of parameter space. 

In what follows we work from the longest decay lengths to the shortest ones, emphasizing mass ranges which are best probed by solar and atmospheric neutrinos. Phenomenology differs considerably when one contrasts a mass-mixing vs. dipole portal both because the decay length's dependence on the HNL mass is parametrically different, and because scattering cross sections have different behaviors in the two cases. Since both atmospherics and solar neutrinos have sub-GeV energy scales, the dipole portal is ``relevant'' despite being a dimension-5 operator because it competes with the Weak interaction which is a dimension-6 contact operator at low energies. For this section we therefore treat the conventional mass-mixing portal, and the less conventional dipole portal on equal footing. This also serves as a useful guide that facilitates an understanding of how model dependent details can affect the phenomenology. 

\subsection{Modified Recoil spectrum:} 

Independent of the HNL decay length one can search for HNLs by searching for deviations from the expected recoil signatures in $\nu X_{\rm SM}$ scattering, where $X_{\rm SM}$ is some convenient SM target. The relevant experiments can be classified as those with $X_{\rm SM}=e$ (electron recoil) and those with $X_{\rm SM}$ some nucleus (nuclear recoil). Electron recoils have $T_e \sim E_\nu$ whereas nuclear recoils have $T_A\sim  E_\nu^2/(2M_A)$ where $T$ is the kinetic energy. Electrons are therefore the natural choice for low-mass HNL searches with Borexino, and its unparalleled ability to map out the solar neutrino flux, serving as the flagship experiment. Low threshold dark matter direct detection datasets can also be repurposed to set constraints on $\nu A\rightarrow N A$ scattering. 

Famously, this kind of signature sets the neutrino floor for dark matter direct detection experiments. An observation of coherent elastic neutrino nucleus scattering (CEvNS) has not yet been achieved using solar neutrinos, and so mass-mixing portals cannot be probed with nuclear recoil data. As we will discuss below, the electron recoil spectra have been successfully leveraged for dipole portal interactions, and supply leading constraints at low masses. 

To the best of our knowledge, the first proposal of this type was made in \cite{Shoemaker:2018vii} where a neutrino dipole portal was investigated in the context of the XENON dark matter experiment. The authors of \cite{Shoemaker:2018vii} discussed possible nuclear-recoil signatures and found sensitivity at lower masses that was competitive with accelerator based experiments. Later it was realized that direct electron scattering, rather than nuclear recoil spectra, could yield more stringent results \cite{Brdar:2020quo}.

It may be possible to constrain e.g.\ $\tau$-flavored HNL mixing with solar neutrino observation. Constraints based on spectral shapes may be derived by performing a binned log-likelihood fit to a detector's binned detector spectrum. It is interesting to note that in addition to $\tau$-flavored HNL mixing, solar neutrinos may also provide competitive constraints on $\mu$-flavored mixing for $m_N \leq 2 m_e$ because below this threshold $N\rightarrow \nu e^+e^-$ becomes kinematically forbidden. This interesting line of inquiry requires further study.

\subsection{HNLs decaying en route from the Sun:} 

Indirect searches for HNLs make no assumptions about the lifetimes or decay products of an HNL and so correspond to the infinite lifetime limit in our discussion. We now turn to the largest, but finite, lifetimes of interest which is set by the distance between the Earth and the Sun. If HNLs are produced in the solar interior then they may exit the Sun, and decay in the intervening distance between the Sun and the Earth. Alternatively, they may be so long lived as to have an $O(1)$ survival probability to reach Earth and their decays may be searched for inside the volume of a neutrino detector. 

This idea was first sketched in Ref.~\cite{Toussaint1981ConstraintsOH}, in which the HNLs are produced in the Sun and then decay to $e^{+}e^{-} \nu$ outside the Sun where they can exceed the limit on the inter-planetary positron flux. More recently, Borexino has set limits on HNL mixing with electron neutrinos by searching for $e^+e^-$ pairs in their detector \cite{Borexino:2013bot}. Their null results set leading constraints on $U_{eN}$ in the few-MeV regime. More recently Hostert \& Pospelov have derived constraints on HNL mixing using KAMLAND and Super-Kamiokande data searching for anti-neutrinos via. inverse $\beta$ decay \cite{Hostert:2020oui}. It is important to emphasize that in contrast to the solar neutrino flux at Earth, HNLs produced in the Sun's interior arrise only via $U_{eN}$ mixing, and so cannot probe $U_{\mu N}$ or $U_{\tau N}$. No analgous such search strategy has been developed for dipole portal interactions, however work is ongoing to research this possible signatures \cite{Gustafson:2021yyy}.

\subsection{Upscatter in earth:}

The next relevant length scale is the radius of the Earth itself, $R_\oplus = 6563$ km. When HNL decay lengths are long relative to the dimensions of an experiment, it is rare for an HNL to decay within a fiducial volume. If HNLs are produced via upscattering inside the Earth, however, a long decay length also supplies a large volume of material on which a solar or atmospheric neutrino can upscatter. The large column density of scatters $n_\perp \sim n \lambda$ and the small probability of decay $\ell/\lambda$ (with $n$ an average density, $\lambda$ the decay length, and $\ell$ a detector length), cancel against one another making the rate of energy deposition inside detectors roughly independent of the decay length. 

This idea was first pursued in \cite{Plestid:2020vqf,Plestid:2020ssy} where limits on HNLs with masses less than 18 MeV were derived for both dipole and mass-mixing portals. In the mass mixing case decay lengths were so long, $\lambda \sim 10^6$ km, that only the $\tau$-coupled HNLs were probed in new regions of parameter space. For the dipole portal, solar neutrinos were found to provide leading limits for $e$-coupled, 
$\mu$-coupled, and $\tau$-coupled portals. The bounds in~\cite{Plestid:2020vqf,Plestid:2020ssy} rely on a semi-analytic approach that is tailored to solar neutrinos, and a conservative rate-only estimate of experimental sensitivities. The constraints on the dipole portal from solar neutrino upscattering, followed by decays at Borexino is plotted in Fig.~\ref{fig:db}. 

In~\cite{Gustafson:2021xxx} this idea was extended to atmospheric neutrinos, whose analysis is considerably more involved and demands a numerical treatment owing to the broad band nature of the atmospheric neutrino flux and the necessity to model neutrino oscillations. A full Monte Carlo implementation has been developed, and new constraints on both dipole and mass-mixing portals derived. The qualitative features mirror the solar neutrino analysis but extend to higher masses.

\subsection{Production in atmospheric showers:}

The upscattering mechanism of the previous section is highly efficient if decay lengths are hundreds, or thousands of kilometers. For decay lengths satisifying $\lambda \lesssim 5 km$, direct production in the upper atmosphere can serve as a useful tool with which to search for HNLs.  The same cosmic-ray collisions with the Earth's atmosphere which sources the atmospheric neutrino flux, may itself be a site of HNL production~\cite{Kusenko:2004qc,Coloma:2019htx,Arguelles:2019ziu,Boiarska:2021yho}. Predictions are complicated by the fact that the atmosphere is thick and diffuse such that HNLs are produced at varying heights, and for decay lengths that are commensurate with the thickness of the atmosphere may decay prior to reaching underground detectors.

\subsection{Double bang:}

Finally, at decay lengths which are smaller than a detector, one may observe both production and decay of an HNL within the same detector volume. Ultra-high energy tau neutrinos produce similar phenomenology, dubbed ``double-bang'' events at the IceCube experiment. In~\cite{Coloma:2017ppo} it was proposed that a low-energy class of double-bang events could be produced by HNLs. 

At the time it appeared that tau-mixing could be better probed via double-bang events at IceCube than any terrestrial search for GeV-scale HNLs. However since then old CHARM data was found to provide strong bounds on tau-mixing~\cite{Boiarska:2021yho} at the GeV-scale. 

\begin{figure*}[t]
\includegraphics[width=0.47\textwidth]{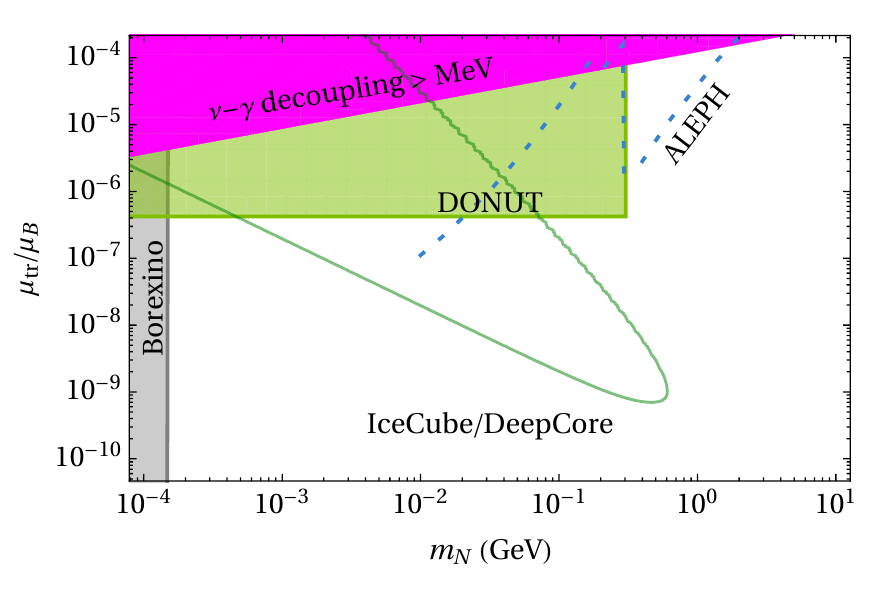}
\includegraphics[width=0.5\textwidth]{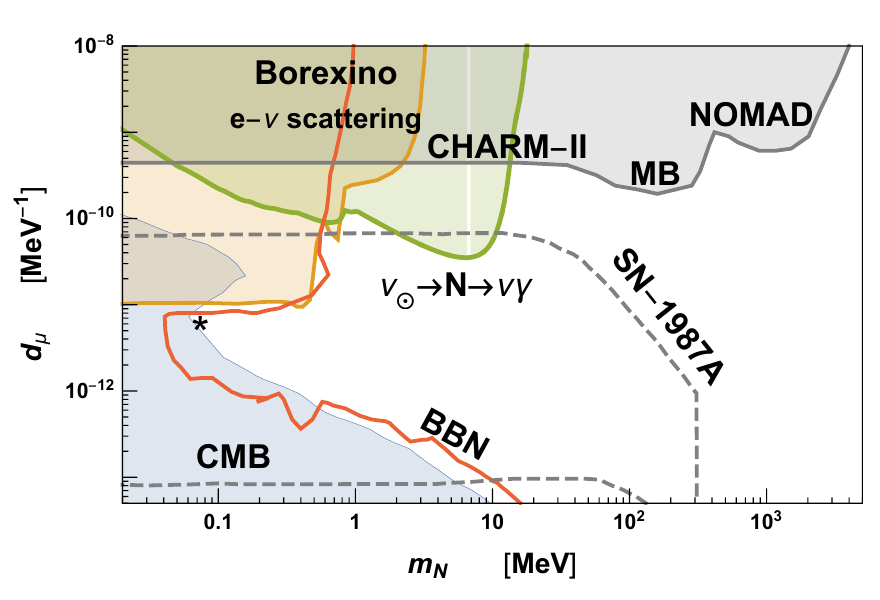}
\caption{\normalsize ({\it Left}) Constraints on HNLs with a dipole portal coupling to tau neutrinos. The green curve arises from the non-observation of low-energy double-bang events at IceCube~\cite{Coloma:2017ppo}. ({\it Right})  Constraints on HNLs with a dipole portal coupling to muon neutrinos.The green region is excluded from solar neutrinos which up-scatter to an HNL inside of the Earth, before decaying at Borexino~\cite{Plestid:2020vqf}.
}
\label{fig:db}
\end{figure*}

However, it turns out that the double bang event topology can also be used to place strong constraints on non-minimal HNLs~\cite{Coloma:2017ppo}. For example, transition dipole moments between active neutrinos and HNLs with a tau-dominant coupling can be much better probed at IceCube than terrestrial experiments. In Fig.~\ref{fig:db} we reproduce these constraints. A related study has also examined the sensitivity of Super-Kamiokande, Hyper-Kamiokande, and DUNE to double bang events~\cite{Atkinson:2021rnp}.

The naturally occurring neutrino fluxes from the Sun and atmosphere offer novel sensitivity to HNLs. On the one hand, the kinetic energy from these neutrinos can be partially converted in HNL rest mass in so-called ``up-scattering'' processes in which an incoming SM neutrino inelastically scatters to an outgoing HNL~\cite{Coloma:2017ppo,Plestid:2020vqf,Plestid:2020ssy,Atkinson:2021rnp}. On the other hand, the production source of atmospheric neutrinos itself offers additional HNL sensitivity.  \\


\section{Cosmological and Astrophysical Searches} 
\label{sec:cosmo}

Depending on their mass $M$, HNLs can affect cosmology and astrophysics in various different ways, cf.~e.g.~\cite{Drewes:2013gca}. 
Qualitatively one can distinguish constraints that HNLs must necessarily respect if they exist (Sec.~\ref{sec:ObservationalConstraints})
and additional requirements that they would need to fulfill to solve open problems in cosmology (Sec.~\ref{sec:SolvingCosmoProblems}).

\subsection{Constraints from observations}\label{sec:ObservationalConstraints}

HNLs may be produced at very high temperatures of the Early Universe and then decay away as the Universe cools and expands. Sufficiently light HNLs can directly affect the expansion of the Universe by contributing directly to the radiation density, or by adding radiation via their decay products.

\subsubsection{Big Bang Nucleosynthesis and HNLs}
The dense and hot early universe provides a natural avenue for HNL production and possible decay. However, these decays may also negatively impact the successes of standard big bang nucleosynthesis (BBN)~\cite{Dolgov:2000jw,Ruchayskiy:2012si,Gelmini:2020ekg,Boyarsky:2020dzc,Sabti:2020yrt}.

In this context, let us first discuss the case of HNLs that are heavy enough to decay into long-lived mesons such as $\pi^\pm$ and $K^{0,\pm}$ (depending on the mixing pattern, the threshold for $\pi$ production ranges from $m_N = m_\pi + m_e$ to $m_N = m_\eta$).

For small lifetimes of these particles, $\tau_N \sim (10^{-2} - 10)\,\mathrm{s}$, the main effect on BBN is the modification of the $p\leftrightarrow n$ conversion processes via the inclusion of additional meson-mediated reactions like $\pi^- + p^+ \leftrightarrow n + \pi^0/\gamma$. Since the cross-sections of these additional reactions are large compared to the weak $p\leftrightarrow n$ conversion processes of the SM, even a small fraction of mesons that gets injected into the plasma can keep the neutron-to-proton ratio $(n/p)$ close to unity $(n/p) \simeq 1$, which then also influences the subsequently produced helium-4 abundance $Y_p \simeq 2(n/p)_{T_{\text{BBN}}}/[ 1 + (n/p)_{T_{\text{BBN}}} ]$ at some characteristic BBN temperature $T_\text{BBN}$. Hence, if the neutron-to-proton ratio gets modified by the decay but cannot relax back to its SM value of $(n/p) \approx 1/7$ once the HNL decay has concluded -- and before the weak reactions freeze out --, meson injection will lead to an overproduction of helium-4 and other light elements, and thus to conflicts with observations for lifetimes already as small as $\tau_{N}= 0.02\,\mathrm{s}$~\cite{Boyarsky:2020dzc}. 
For larger lifetimes, HNLs survive until the onset of nuclear synthesis reactions, and the injected mesons also start to induce hadrodisintegration reactions like \ $\pi^- + {}^4\text{He} \rightarrow p + 3n$ (see e.g.~\cite{Kawasaki:1994sc,Jedamzik:2006xz} for more general bounds from hadrodisintegration). However, together with destroying nuclei, mesons still keep $(n/p) \simeq 1$. Therefore, as long as HNLs disappear before the decoupling of nuclear synthesis reactions, i.e.\ for $\tau_N \lesssim 10^{4}\,\mathrm{s}$, these sizable amounts of free protons and neutrons get bounded in nuclei, thus still increasing the nuclear abundances compared to their SM values~\cite{Ovchynnikov:2021zyo,Bondarenko:2021cpc}.

\begin{figure}[!h]
    \centering
    \includegraphics[width=0.33\textwidth]{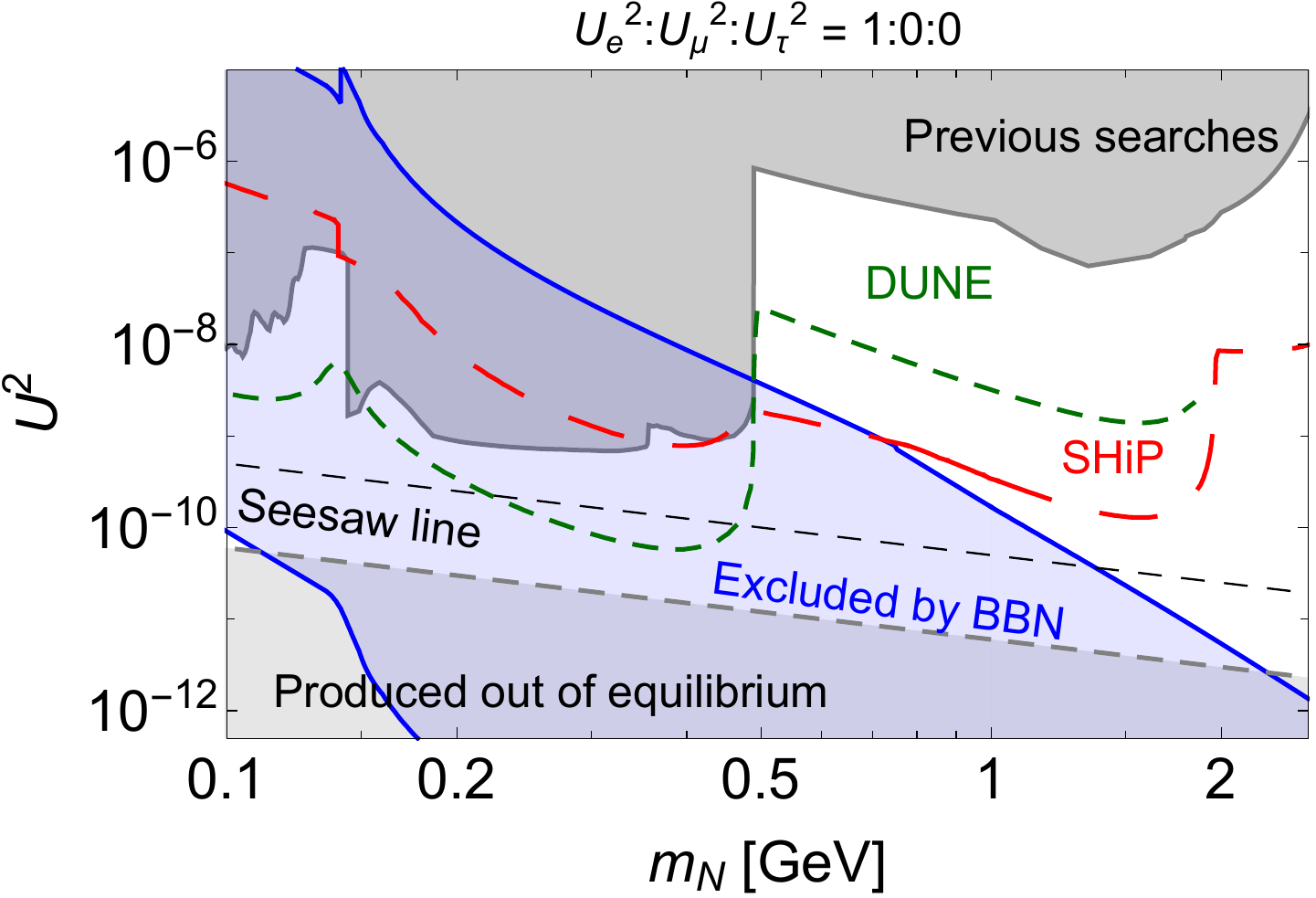} \includegraphics[width=0.33\textwidth]{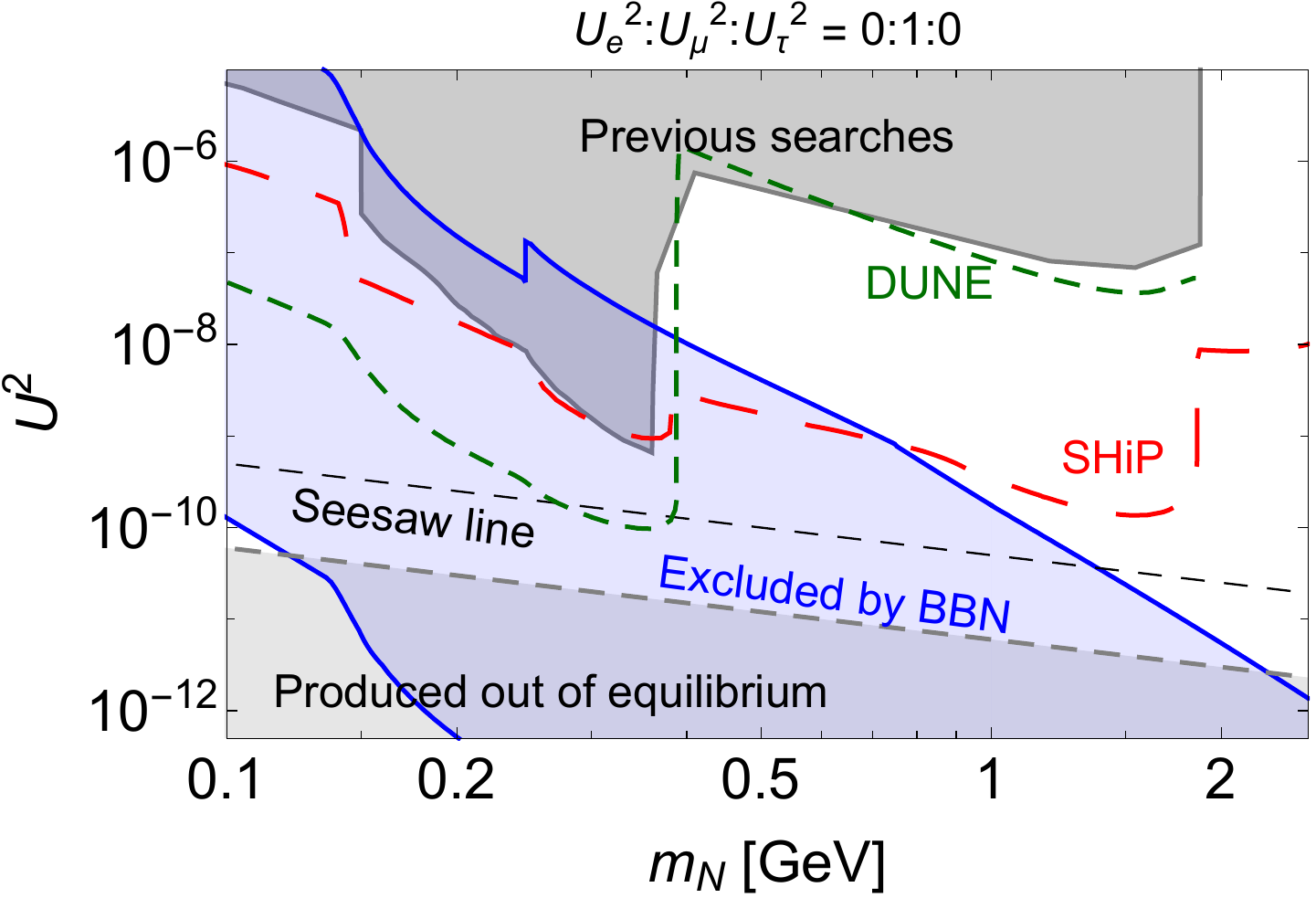} \includegraphics[width=0.33\textwidth]{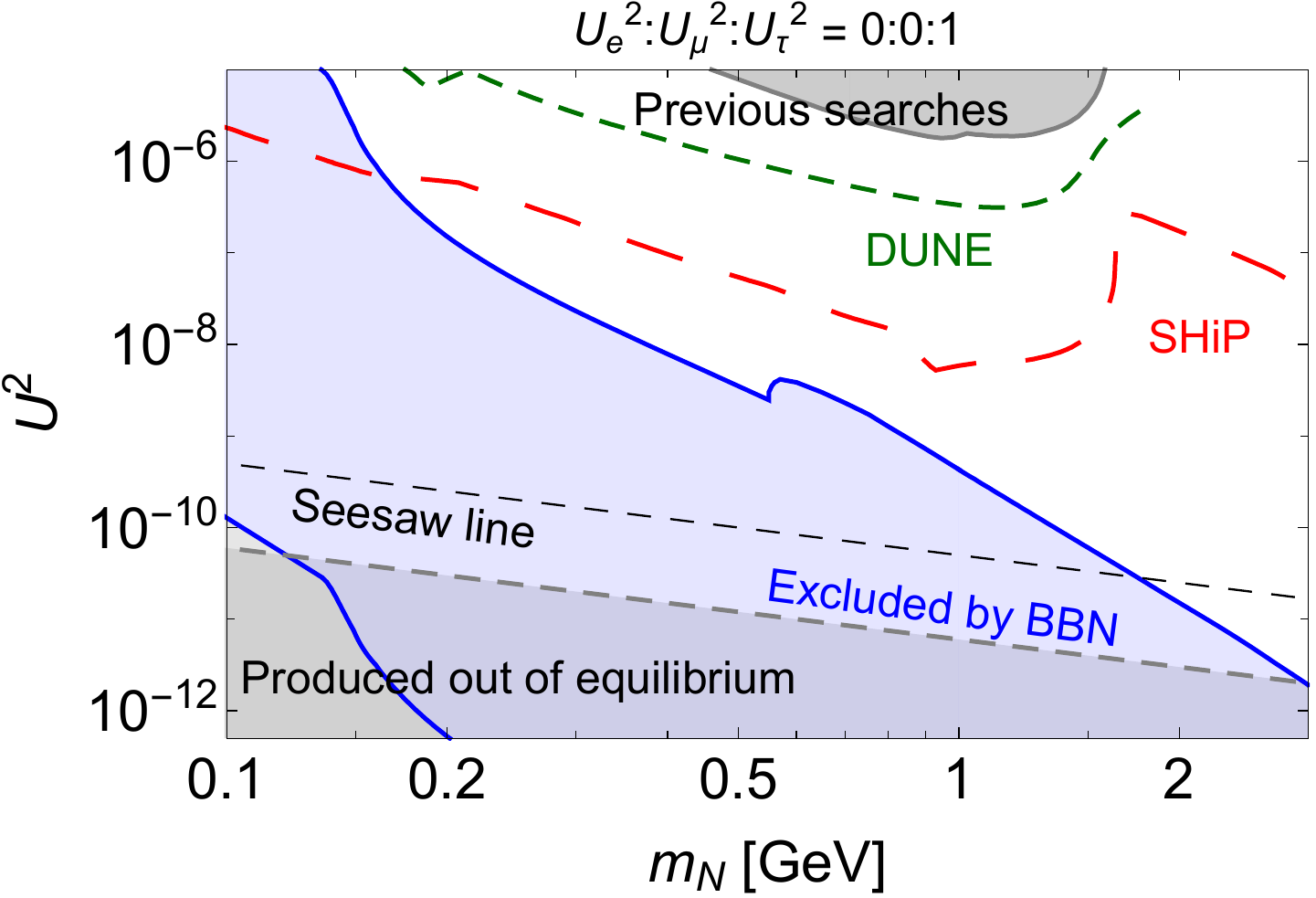}
    \caption{BBN bounds (blue) on GeV scale HNLs for the cases of the pure $e$ (left figure), $\mu$ (central figure), and $\tau$ mixing (right figure) from~\cite{Boyarsky:2020dzc,Ovchynnikov:2021zyo}. Zero lepton asymmetry in the neutrino sector is assumed. The black dashed line corresponds to the seesaw bound $U^{2}_{\text{seesaw}} = 5\cdot 10^{-11}\frac{1\text{ GeV}}{m_{N}}$. The lower bound of the blue domain, $\tau_{N} = 10^{4}\text{ s}$, is given by the applicability of the analysis, indicating the time scale at which nuclear reactions go out of equilibrium. The figures are given from~\cite{Ovchynnikov:2021zyo}. The light gray region corresponds to the parameter space in which HNLs are produced via the freeze-in mechanism.}
    \label{fig:constraints-BBN}
\end{figure}

The BBN constraints on $\mathcal{O}(\mathrm{GeV})$ HNLs in models with negligible lepton asymmetry\footnote{Effects from heavy sterile neutrinos on BBN could be modified in the presence of non-negligible lepton asymmetry~\cite{Gelmini:2020ekg}.} in the neutrino sector, i.e.~$\eta_{\nu}\ll 1$, are summarized in Fig.~\ref{fig:constraints-BBN}.

\begin{figure}[!h]
\centering
\includegraphics[width=0.7\textwidth]{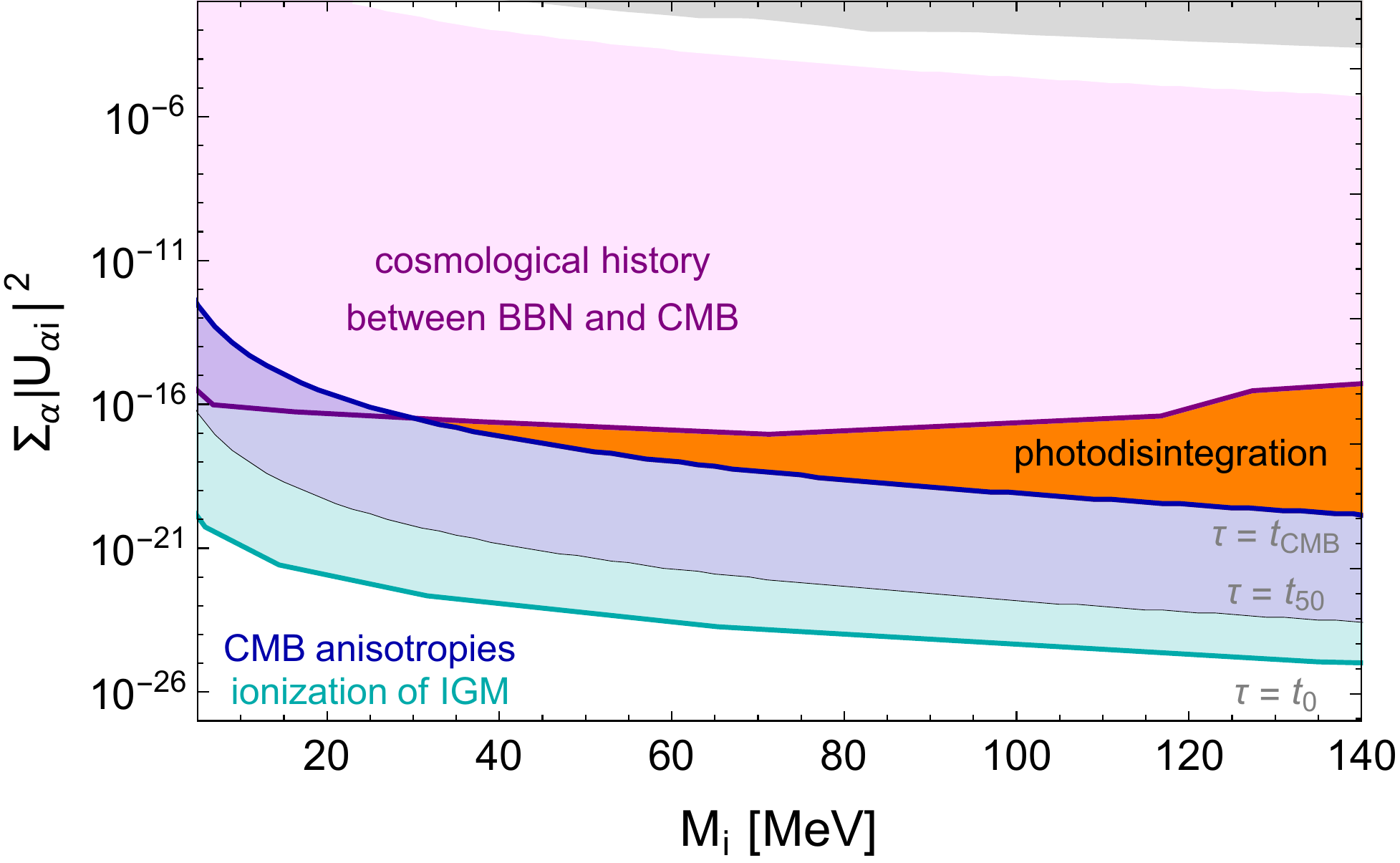}
\caption{Constraints on the mixing of sterile neutrinos with SM neutrinos. Here, the orange region is excluded by photodisintegration constraints, which complement the remaining bounds from laboratory
constraints (grey), as well as cosmological constraints from~\cite{Vincent:2014rja} (pink),~\cite{Poulin:2016anj} (blue), and~\cite{Diamanti:2013bia} (cyan). This figure has been taken from~\cite{Domcke:2020ety}, but the upper bound of the pink region has been recalculated with the lifetime bound from~\cite{Mastrototaro:2021wzl} applied to all mixing patterns.}
\label{fig:photodis}
\end{figure}

Lighter HNLs that cannot decay into mesons, mainly affect BBN via their decay into high-energy neutrinos or electromagnetic states and different effects can be dominant, namely \textit{(i)} an altered $n\leftrightarrow p$ conversion rates due to the presence of additional neutrinos (if the neutrinos are injected after neutrino decoupling), \textit{(ii}) a modification of the expansion dynamics of the Universe due to modifications to the Hubble rate or the time-temperature relation, and/or \textit{(iii)} the late-time modification of the light-element abundances after BBN via the process of photodisintegration~\cite{Kawasaki:2004qu,Poulin:2015opa,Depta:2020zbh,Domcke:2020ety}. The first two effects can cause an increase in the deuterium and helium-4 abundances for the whole range of lifetimes from $\tau_{N} \gtrsim 0.05\,\mathrm{s}-\mathcal{O}(1)\,\mathrm{s}$, with a dependence on the HNL mass $m_{N}\gtrsim 10\,\mathrm{MeV}$~\cite{Sabti:2020yrt,Ovchynnikov:2021zyo}. The latter effect instead concerns HNLs with lifetimes $> 10^4\,\mathrm{s}$, and mainly leads to deuterium underproduction and helium-3 overproduction via the dissociation of deuterium ($\gamma+\text{D}\rightarrow n + p$) and helium-4 ($\gamma + {}^4\text{He} \rightarrow D + D, \dots$), respectively. Regarding photodisintegration, the corresponding bounds can be obtained via the public code \texttt{ACROPOLIS} \cite{Depta:2020mhj,Depta:2020zbh, Hufnagel:2018bjp}, which can be used to model the process of photodisintegration for a wide range of scenarios. The corresponding BBN constraints for the type-I seesaw model with three right-handed neutrinos and Majorana masses below the pion mass, are summarized in Fig.~\ref{fig:photodis}.

\subsubsection{Cosmic Microwave Background}

Similarly to BBN, HNLs may affect CMB if surviving until temperatures $T\simeq \text{ few MeV}$. The impact of HNLs with masses $m_{N}\gtrsim \mathcal{O}(1\text{ MeV})$ and lifetimes below the seesaw bound is indirect -- via a change of the helium abundance and the effective number of ultrarelativistic degrees of freedom $N_{\text{eff}}$. The effect of these parameters on CMB that cannot be mimicked by a change of $\Lambda$CDM parameters within their error bars is the impact on the damping tail of the CMB spectrum~\cite{Zyla:2020zbs}. 

\begin{figure}[!h]
    \centering
    \includegraphics[width=\textwidth]{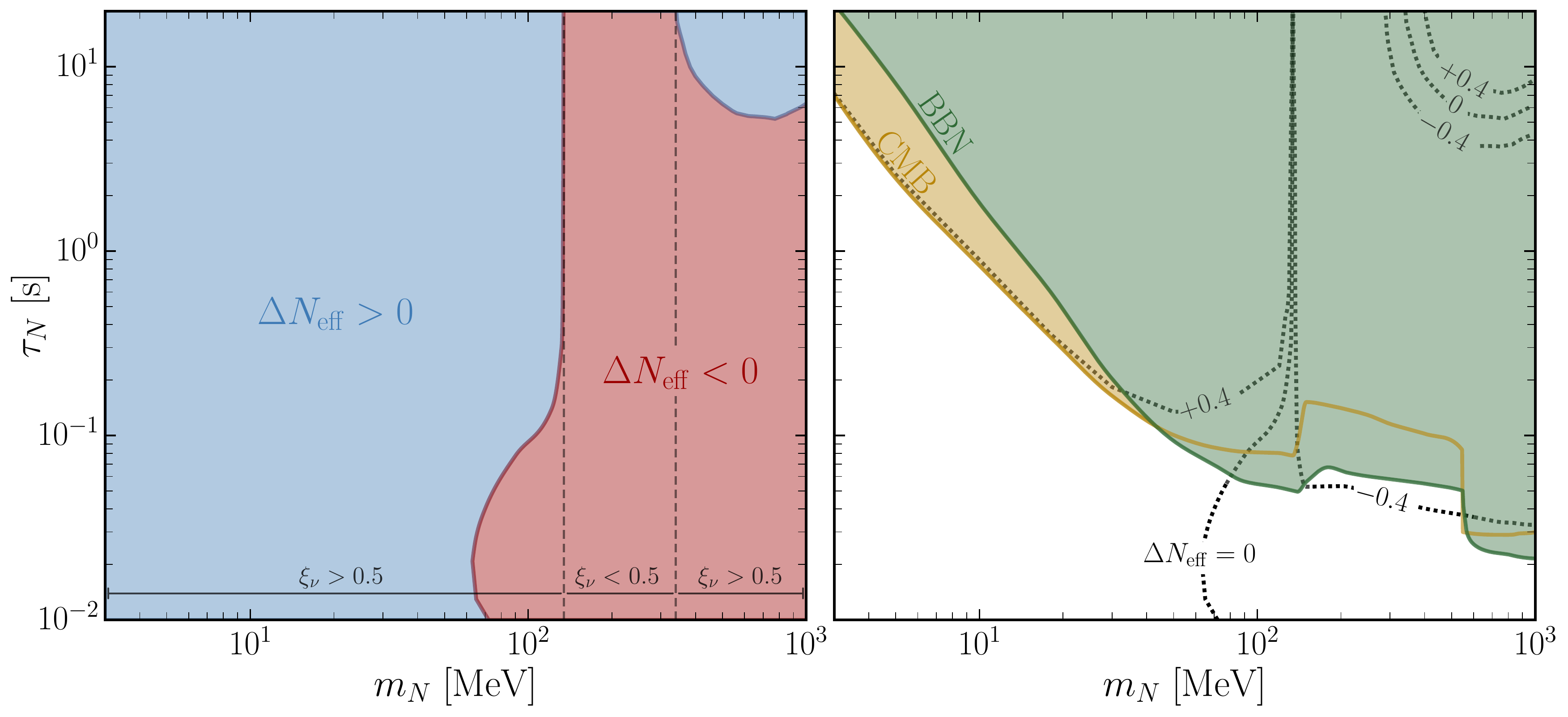}
    \caption{Parameter space of HNLs with the pure $\tau$ mixing. \textit{Left panel}: the value of $\Delta N_{\text{eff}} = N_{\text{eff}}-N_{\text{eff,SBBN}}$ as a function of the HNL mass and lifetime. \textit{Right panel}: the current CMB bounds on HNLs.}
    \label{fig:cmb-bound}
\end{figure}

The impact of short-lived HNLs on $N_{\text{eff}}$ has been recently studied in several works~\cite{Boyarsky:2021yoh,Rasmussen:2021kbf}, where it has been argued that HNLs with masses $m_{N}\gtrsim 50-70\text{ MeV}$ and lifetimes $\tau_{N}\lesssim 1\text{ s}$ decrease $N_{\text{eff}}$ even if decaying mostly into neutrinos (see Fig.~\ref{fig:cmb-bound}). 

The current CMB constraints on HNLs with the pure $\tau$ mixing from~\cite{Boyarsky:2021yoh} are shown in Fig.~\ref{fig:cmb-bound} (see also~\cite{Vincent:2014rja}); the case of the pure $e,\mu$ mixings is qualitatively similar. They are weaker than the current BBN bounds for masses $m_{N}\gtrsim 30\text{ MeV}$. Nevertheless, a number of upcoming and proposed CMB missions, such as the Simons Observatory~\cite{SimonsObservatory:2018koc} and CMB-S4~\cite{Abazajian:2019eic}, could provide a determination of $N_{\text{eff}}$ around the percent-level, which will improve the CMB bounds.

\begin{figure}
    \centering
    \includegraphics[width=0.5\textwidth]{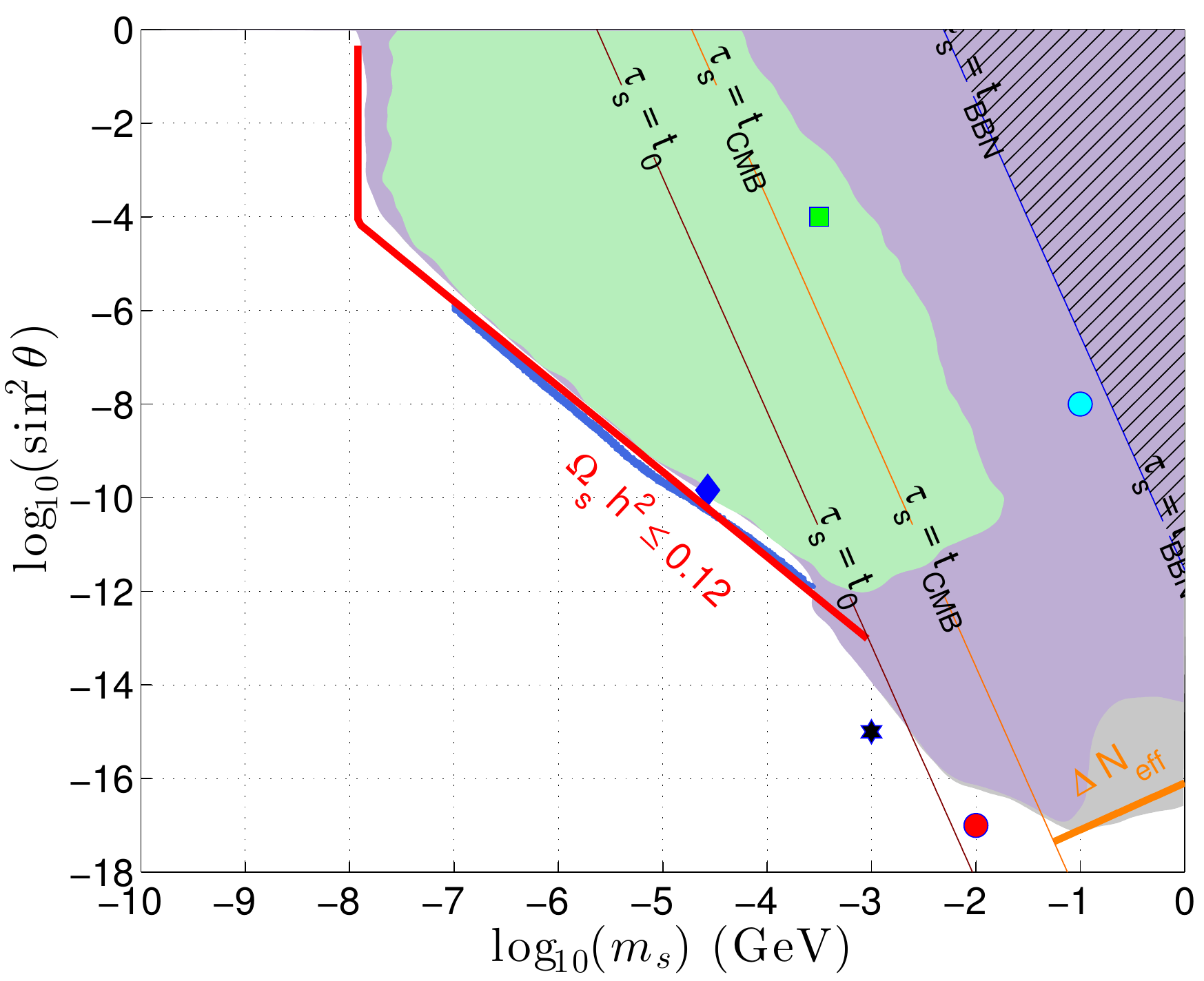}
    \caption{Cosmological HNL constraints based on production via oscillation in the Early Universe. Light green region: excluded due to effects on the current expansion rate. Purple region: Exclusions based on the effects of an early matter component that affects the sound horizon as observed in BAO and the CMB. Particles that decay before BBN (hashed region) cannot be constrained with this method. Red and orange lines respectively correspond to approximate limits on the total abundance of dark matter , and to the requirement that the number of ultrarelativistic particles in the early Universe $\Delta N_{eff} \lesssim 1$.  Figure from \cite{Vincent:2014rja}.}
    \label{fig:cosmo_expansion}
\end{figure}

More broadly, it has long been known HNLs produced in the Early Universe can lead to an overabundance of matter, directly affecting the expansion rate \cite{Sato1977,Gunn1978}. For sufficiently long-lived (lifetime $\gtrsim t_{eq}$) HNLs with a relic abundance larger than the observed dark matter abundance, this can directly affect the measured value of the Hubble parameter $H_0$. This is illustrated as a green region in Fig. \ref{fig:cosmo_expansion}. If these HNLs are abundantly produced but decay before matter-radiation equality, they do not leave an imprint on $H_0$. However, they may lead to an early period of matter domination. This can alter the sound horizon at recombination, affecting the location of the CMB peaks, as well as the BAO peak as observed in the large scale matter power spectrum \cite{Vincent:2014rja}. Such scenarios can be ruled out be CMB and LSS data, as shown in in the purple region of Fig. \ref{fig:cosmo_expansion}.

Finally, heavier HNLs decaying into charged final states after recombination can lead to extra ionization and heating \cite{Adams:1998nr} during the dark ages. Rescattered CMB photons lead to a suppression of the angular power spectrum at high multipoles, as well as increased correlation in the EE power spectrum for lower $\ell$. Including data from Planck and Lyman-$\alpha$ limits on heating on the intergalactic medium, this leads to a lifetime bound of $\tau \gtrsim 10^{26}$ s \cite{Diamanti:2013bia}. Translated to a mixing angle, for HNLs heavier than twice the electron mass, this yields \cite{Vincent:2014rja}:
\begin{equation}
\sin^2 \theta \gtrsim  \times 10^{-5} \sqrt{ \frac{192 \pi^3}{b_e}\left(\frac{\mathrm{GeV}^{-2}}{ G_F}\right)^2\left(\frac{\mathrm{eV}}{m_N}\right)^7},
\label{eq:CMBlife}
\end{equation}
where $b_e$ is the fraction of decays into $e^+e^-$ pairs. Eq. \eqref{eq:CMBlife} is valid for lifetimes longer than $\sim$ 300,000 years, corresponding to the time of recombination. 

Such exotic energy constraints are close to cosmic-variance limited and could be marginally improved by future CMB experiments such as CORE. However, a precise measurement of the baryon temperature at redshifts 20-25 to ($\delta T_b \sim 5-10$ mK) may reliably improve such constraints by around an order of magnitude \cite{Poulin:2016anj}.Energy injection before recombination could also lead to $\mu$-type distortions of the CMB spectrum. A future spectral distortion mission such as PIXIE \cite{Kogut_2011} could help set leading constraints on lifetimes between $10^{6}$ and $10^{13}$~s \cite{Poulin:2016anj}.

\subsubsection{HNLs in Astrophysical Settings}

In addition, HNLs can be produced in a variety of astrophysical environments, e.g. supernovae~\cite{Fuller:2008erj,Albertus:2015xra,Rembiasz:2018lok}.

Supernovae (SN) may be efficient $\mathcal{O}( \text{keV})$ HNL factories. This is thanks to matter-driven enhancement of mixing angle~\cite{Notzold:1987ik}, allowing it even to reach the resonant value $\theta_{res} = \pi/4$ inside particular regions of the SN. Leaving the resonance region, the produced HNLs freely escape SN, providing an additional mechanism of its cooling which may lead to shortening of the duration of the active neutrino burst (the main cooling stage). Comparing the results of observations and theoretical predictions, this might be used to impose constraints for HNL parameters.

The only supernova event occurred sufficiently close to observe the neutrino flux was the explosion of SN1987A. Three neutrino detectors have reported simultaneous detection of a series of events with electron anti-neutrinos: IMB~\cite{IMB_detector}, Kamiokande II~\cite{Hirata:1987hu} and Baksan scintillator telescope~\cite{Baksan}. They have shown that the emission duration was $\Delta t \sim 10$~s, with characteristic neutrino energy $E_\nu \sim 10$ MeV. This data allows to make an estimate on the total energy emitted by neutrinos as $E_{\bar\nu_e\text{ tot}} \sim 3 \cdot 10^{53}$ erg, assuming the neutrino flavour equipartition~\cite{Loredo:2001rx}. It fits the estimate of the gravitational energy of a neutron star as a potential remnant of SN1987a, which defines the total energy budget of the explosion. 

Considering this data, in order to maintain the observed duration of the active neutrino emission, the energy loss driven by HNLs even during short period of time must not exceed\footnote{Due to cooling, this emission will be dependent on time and this limit corresponds to the start of it.}
 \begin{equation}
     E_{N} \lesssim 10^{53}\text{ erg/s}.
 \end{equation}
The constraint was numerically justified for SN explosion in models with axions~\cite{Fischer:2016cyd} and may be applied to the case of HNLs, although there are differences of their production mechanism.
 
The HNL production inside the SN has been studied in the works~\cite{Raffelt:2011nc,Wu:2013gxa,Warren:2014qza,Warren:2016slz,Rembiasz:2018lok,Zhou:2015jha,Arguelles:2016uwb,Suliga:2019bsq,Syvolap:2019dat,Gelmini:2019deq,Mastrototaro:2019vug,Tang:2020pkp}. However, any possible constraints are subject of either theoretical uncertainties related to the SN explosion model or lack of experimental data: 
\begin{itemize}
\item[--] SN neutrinos observation did not provide the information about the temperature of the SN withing the neutrino-sphere, where neutrinos are trapped, while HNLs are mostly produced there, see~\cite{Syvolap:2019dat}.
\item[--] There was no direct observation of the SN1987a remnant yet~\cite{Esposito:2018nib,Alp:2018oek}.\footnote{However, there is a rising possibility though that it is indeed the neutron star~\cite{Page:2020gsx}.} Despite the neutron star is commonly believed to be the one, there is a probability to have a different remnant which may appear in a different explosion model~\cite{Blum:2016afe}.
\item[--] Most models are either built on toy models of SN explosion or used the results of SM-based numeric simulations (Core-collapse supernovae model~\cite{Burrows:2020qrp,Janka:2012wk}) where the back-reaction from the HNL production on the SN explosion is either omitted or simplified \cite{Suliga:2020vpz}.
\item[--] Assuming non-typical supernovae explosion with a various neutrino spectra, limits for total emitted energy might be extended (see~\cite{Suliga:2021hek}), spoiling the energy-loss argument.
\end{itemize}

To summarize, the SN explosion data does not allow to set robust bound on HNL parameters from study of SN explosion at the current status of the problem.

\subsection{Solving open problems in cosmology}\label{sec:SolvingCosmoProblems}

\subsubsection{Leptogenesis and the origin of ordinary (baryonic) matter}
\label{sub:leptogenesis}

As already mentioned in Section~\ref{sec:theory}, the observed baryon asymmetry of the Universe (BAU) is one of the strongest hints pointing at existence of physics beyond the Standard Model (SM), cf.~\cite{Canetti:2012zc}.
Leptogenesis is an attractive solution to the question of the origin of matter as it connects the observed BAU with the origin of the light neutrino masses.
In this mechanism the same HNLs that are responsible for the origin of the light neutrino masses, can produce the matter-antimatter asymmetry via their $CP$ violating decays in the early Universe~\cite{Fukugita:1986hr}.
Among the different realizations, leptogenesis in the type-I seesaw is by far the most studied one~\cite{Bodeker:2020ghk}, where the SM only needs to be extended by two or more HNLs.
Leptogenesis can also be realised with scalar~\cite{Ma:1998dx} or fermionic~\cite{Hambye:2003rt} triplets, cf.~\cite{Hambye:2012fh} for a review.

While initial calculations suggested that the mass scale of HNLs needed for leptogenesis is quite high---$M\sim 10^9$ GeV~\cite{Davidson:2002qv}, it was soon realized that the mass of HNLs can be as low as the TeV scale in ~\emph{resonant leptogenesis}~\cite{Pilaftsis:2003gt}, or even at the GeV scale in the case of \emph{leptogenesis via oscillations}~\cite{Akhmedov:1998qx,Asaka:2005pn}.
While developed independently, these mechanisms rely on a similar enhancement, and can be described by the same equations (see e.g.~\cite{Garbrecht:2018mrp}).
Together they are often known by the term \emph{low-scale leptogenesis}.

\begin{figure}[hpb]
	\centering
	\includegraphics[width=0.9\textwidth]{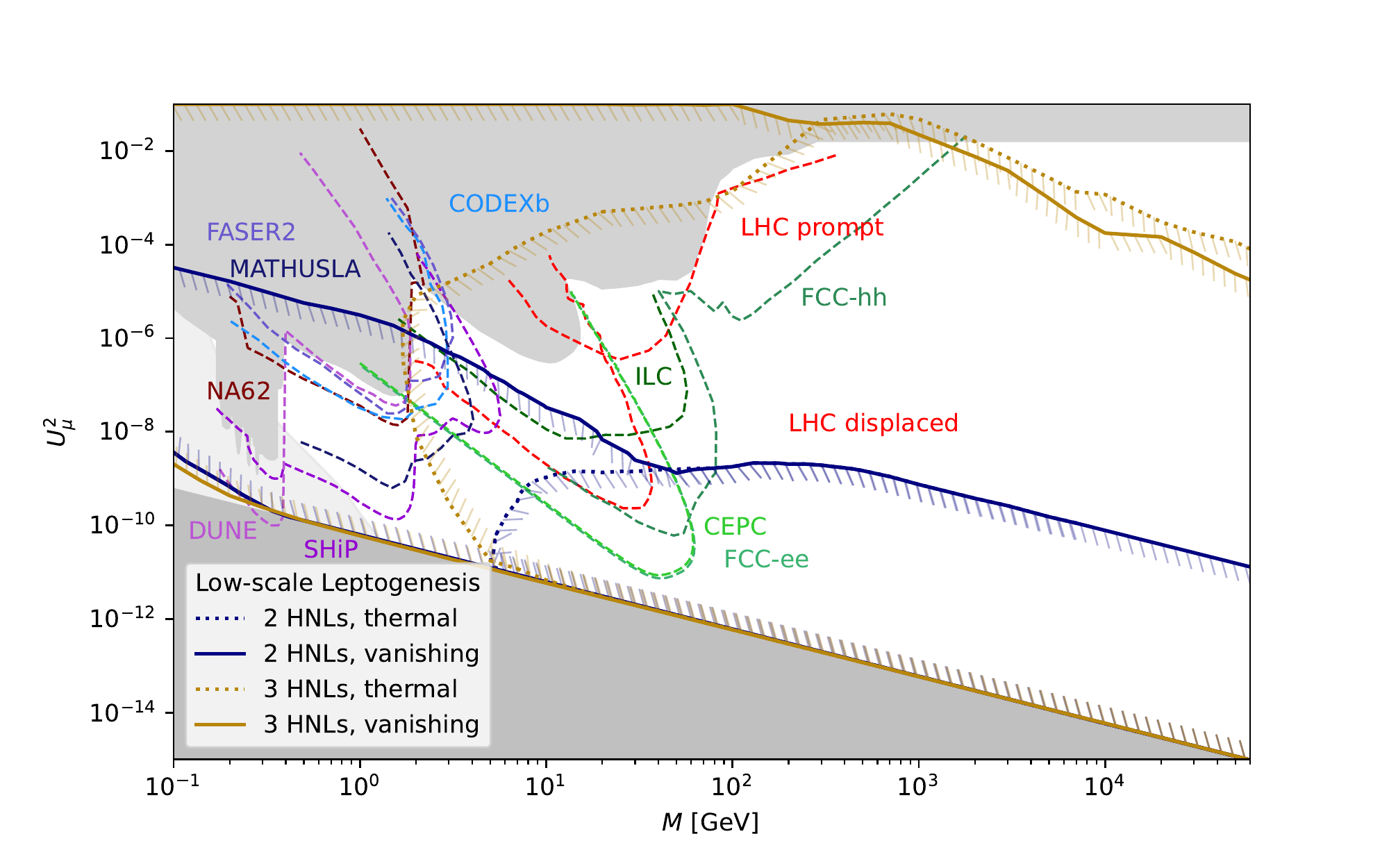}
	\caption{Comparison of the ranges of the total mixing angle $U^2$ consistent with both the seesaw mechanism and leptogenesis as a function of HNL mass $M$ for leptogenesis with 
	two \cite{Klaric:2020phc,Klaric:2021cpi} or three \cite{Drewes:2021nqr}  HNL flavours in the case of normal neutrino mass ordering (NO).
	For both vanishing and thermal initial HNL abundances.
	The shaded region indicates the region excluded by past experiments~\cite{CHARM:1985nku,Abela:1981nf,Yamazaki:1984sj,E949:2014gsn,Bernardi:1987ek,NuTeV:1999kej,Vaitaitis:2000vc,CMS:2018iaf,DELPHI:1996qcc,ATLAS:2019kpx,CMS:2022fut} (gray),
	complemented by the updated BBN bounds (light gray) from~\cite{Sabti:2020yrt,Boyarsky:2020dzc} and the lower bound from the seesaw mechanism (darker gray).
    See~\cite{Klaric:2020phc,Klaric:2021cpi,Drewes:2021nqr} for details.
    The various colored lines indicate existing~\cite{Izaguirre:2015pga,Das:2017gke,Pascoli:2018heg,Drewes:2019fou,Drewes:2018gkc} and future~\cite{Ballett:2019bgd,FASER:2018eoc,SHiP:2018xqw,Gorbunov:2020rjx,Curtin:2018mvb,Aielli:2019ivi,Antusch:2017pkq,Antusch:2016ejd,Verhaaren:2022ify} experiments that will be able to probe the low-scale leptogenesis parameter space.}
	\label{fig:leptogenesis_23HNLs}
\end{figure}

Due to the discovery potential associated with such light HNLs, this mechanism has received significant attention in the past decade.
Below we discuss the constraints on the HNL properties of the imposed by the baryogenesis.
In the case with two HNLs, their masses should be nearly degenerate, and we denote the common mass as $M$.
Phenomenologically, one is interested in  the maximal size of the HNL mixing angles $U^2$ as a function of $M$~\cite{Shaposhnikov:2008pf,Canetti:2012kh,Drewes:2016gmt,Hernandez:2016kel,Drewes:2016jae,Antusch:2017pkq,Abada:2018oly,Eijima:2018qke,Klaric:2020phc,Klaric:2021cpi}.
The most recent results~\cite{Klaric:2020phc,Klaric:2021cpi, Drewes:2021nqr}, accounting for the significant theoretical progress~\cite{Shuve:2014zua,Ghiglieri:2016xye,Eijima:2017anv,Ghiglieri:2017csp,Ghiglieri:2018wbs} (cf.~also \cite{Biondini:2017rpb,Garbrecht:2018mrp}), are shown in Fig.~\ref{fig:leptogenesis_23HNLs}.

In addition to the total mixing angles $U^2$ and masses $M$ various properties of the HNLs (e.g. their mass splittings, mixing with individual SM flavours, and the amount of LNV) can be probed in direct searches (cf.~Sec.~\ref{sec:collider} and \ref{sec:fixed}) or indirect probes (cf.~e.g.~\cite{Chrzaszcz:2019inj}) to constrain the viable parameter space. 
In particular in the minimal model with two HNLs the dimensionality of the parameter space is small enough
that the combined information from direct search experiments, neutrino oscillation data, and neutrinoless double $\beta$-decay provides a powerful probe of leptogenesis~\cite{Hernandez:2016kel,Drewes:2016jae,Drewes:2016lqo,Antusch:2017pkq,Eijima:2018qke,Bondarenko:2021cpc}, allowing for a full testability at least in part of the parameter space~\cite{Hernandez:2016kel,Drewes:2016jae}, cf. Section~\ref{sub:type-i_seesaw}. Future energy and intensity frontier experiments hence have the potential to unveil the common origin of neutrino masses and matter in the universe. 

\clearpage
\subsubsection{Dark matter}

Cosmological and astrophysical measurements indicate that about 25\% of the total energy budget of the Universe does not interact with known particles or emit any light.
This phenomenon is known as dark matter \cite{Peebles:2017bzw}.
The nature of dark matter remains an unsolved puzzle.
The central hypothesis of today is that dark matter (DM for short) is made of particles.
The only SM particles that could play role of DM are neutrinos.
This would require the sum of neutrino masses to be $\sum_i m_i \sim 11.5$ eV~\cite{Lesgourgues:2006nd} (clearly violating laboratory \cite{KATRIN:2021uub} and cosmological \cite{Planck:2015fie} constraints).
Moreover, the neutrinos would be relativistic during cosmic structure formation, leading to clustering properties inconsistent with observation \cite{White:1983fcs}.
Finally, neutrino DM would violate the 'Tremaine-Gunn bound'~\cite{Tremaine:1979we} on the phase space density of DM.
Summarizing drawback of neutrino dark matter, one can say that they are ``too light'' and
``too abundant''.
HNLs cure all of these drawbacks, see \emph{e.g.}~\cite{Boyarsky:2018tvu} for a review, and provide a testable dark matter candidate with reach astrophysical phenomenology.

\emph{HNL DM production models.}
First, the HNLs' feeble interactions imply a smaller abundance than SM neutrinos.
This is true for DM produced via mixing with neutrinos~\cite{Dodelson:1993je,Shi:1998km, Dolgov:2000ew,Abazajian:2001vt,Abazajian:2001nj,Asaka:2005an,Asaka:2005pn,Asaka:2006nq} in the minimal models like $\nu$MSM (Sec.~\ref{sec:nuMSM}).
Another production mechanism that does not require the introduction of new particles is provided by Einstein-Cartan theory~\cite{Shaposhnikov:2020aen}. Einstein-Cartan theory is a description of gravity which is obtained by gauging the Poincar{\'e} group~\cite{Kibble:1961ba}, see e.g.,~\cite{Hehl:1976kj} for a review. This theory is equivalent to general relativity augmented by a universal four-fermion interaction.
Dark matter $N$ is produced  via the annihilation of the SM particles $X + \bar{X} \to N + \bar{N}$ at very high temperatures. The momentum distribution of DM produced in this way has a unique form and differs from both non-resonantly and resonantly produced sterile neutrinos.
In this mechanism, the mixing of sterile neutrinos is not related to their abundance. Moreover, the mass of $N$ can vary in a wide range of values, not being limited by the keV range and thus avoiding many of the structure formation constraints (see below).

Beyond the minimal HNL models, the particle content of the theory is broader, and thus more production mechanisms exist. 
HNL DM can be produced in decays of heavier particles: neutral scalars \cite{Shaposhnikov:2006xi,Petraki:2007gq,Boyanovsky:2008nc,Matsui:2015maa}, charged scalars \cite{Frigerio:2014ifa,Drewes:2015eoa}, an additional Higgs doublet \cite{Adulpravitchai:2015mna,Drewes:2015eoa}, vector bosons \cite{Shuve:2014doa} or fermions~\cite{Abada:2014zra}.
HNLs can also be charged with respect to new gauge interactions (as in left-right symmetric models) \cite{Pati:1974yy,Mohapatra:1974hk,Senjanovic:1975rk,Wyler:1982dd}, see \cite{Dror:2020jzy} for the recent summary.
In these cases, the abundance is generically ``too high''.
However, owing to the presence of extra species, these models can produce extra relativistic species (``generate entropy'') thus bringing the total HNL abundance to the dark matter level \cite{Asaka:2006ek,Bezrukov:2009th,Bezrukov:2012as,Nemevsek:2012cd,King:2012wg}. 

\emph{Probing HNL DM.}
HNL DM particles can be constrained in a number of ways.
Although none of them are specific for HNL DM, their combination and cross-correlation with the results of ground-based searches may provide a convincing argument for the detection of the HNL DM.

The phase-space density analysis (similar to that of \cite{Tremaine:1979we}) leads a lower limit on the HNL DM mass (as on any other fermionic DM).
There are model-independent and model-dependent bounds (see \cite{Boyarsky:2008ju} for a general discussion).
The recent updates, based on the improved Jeans analysis of dwarf galaxies can be found in \cite{Alvey:2020xsk}.

\emph{Structure formation constraints.}
HNL DM is produced relativistic both in the minimal and non-minimal models.
Therefore, it free-streams out of overdense regions, erasers (or at least modifies) the power spectrum of primordial inhomogeneities beyond the free-streaming horizon scale~\cite{Boyarsky:2008xj} and affects the formation of small-scale structures.
Observables in which this can be visible include the Lyman-$\alpha$ forest, weak lensing constraints  on the matter power spectrum, see \emph{e.g.}\ \cite{Stafford:2021uvk} or \cite{Amorisco:2021iim} for the recent results, cosmic void properties\cite{Tikhonov:2009jq,Reed:2014cta,Yang:2014upa}, the number of high-redshift galaxies~\cite{Bozek:2015bdo,Nierenberg:2013lqa,Menci:2016eui,Menci:2017nsr,Menci:2018lis,Rudakovskyi:2021jyf},  counts of various objects on comparably small scales (galaxy halos, Milky Way satellites)~\cite{Klypin:2014ira,Papastergis:2011xe,Horiuchi:2013noa,Bozek:2015bdo,Horiuchi:2015qri,Lovell:2016fec,Lovell:2016nkp}.

The widely used tool is the Lyman-$\alpha$ data~\cite{Seljak:2006qw,Viel:2007mv,Boyarsky:2008xj,Boyarsky:2008mt, Viel:2013fqw,Garzilli:2015iwa,Baur:2015jsy,Armengaud:2017nkf,Baur:2017stq,Irsic:2017ixq,Irsic:2017yje,Murgia:2017lwo,Garny:2018byk}. The most crucial result of this analysis is that at small scales the data show a cut-off as predicted by realistic
DM models~\cite{Viel:2013fqw, Garzilli:2015iwa, Boera:2018vzq,Garzilli:2018jqh}. However, current data do not allow to identify the nature of this signal. Indeed, \emph{astrophysical effects} can strongly  intervene at scales of interest~\cite{Viel:2013fqw,Garzilli:2015iwa,Garzilli:2018jqh}.
Most prominently, gas pressure prevents small lines from forming and thermal broadening also smooths lines~\cite{Gnedin:1997td,Theuns:1999mz}. To derive conclusions about the nature of DM this degeneracy must be resolved~\cite{Garzilli:2015iwa, Garzilli:2018jqh}.
Significant improvement in the data quality, better modeling of intergalactic media physics at higher numerical resolution and
the use of novel statistical observables~\cite{Garzilli:2015bha} will allow to
detect this effect, remove the degeneracy and obtain a definite result about the nature of DM.

Gravitational lensing (and microlensing) is a promising way to probe the presence of substructures directly (c.f.~\cite{Vegetti:2009cz,Vegetti:2012mc,Hezaveh:2014aoa,Xu:2014dda,Li:2015xpc, Li:2016afu,Kamada:2016vsc,Mahdi:2016las,Fedorova:2016gpa,Birrer:2017rpp,Vegetti:2018dly,Gilman:2019nap,Gilman:2019vca}).
 The detection of low mass halos within the main lensing halo \cite[see e.g.][]{Koopmans:2005nr} or along the line-of-sight \cite[see e.g.][]{Despali:2017ksx} relies on the detailed information contained within the multiple lensed images.
 A large sample of lenses together with a robust way to predict the average distribution of sub-halos is required in order to make a statistically robust prediction.

 A promising way to the detection of dark substructures relies on \textit{gaps in the stellar streams}. Stellar streams are thin bands of stars lying on trajectories of dissolved dwarf galaxies or globular clusters in the Milky Way halo \cite{Eyre:2009tf,Koposov:2009hn}.
 Dark subhalos can perturb these streams, leading to the \emph{gaps} and other features in the streams.
 Several works have attempted to use these gaps to infer the number of dark satellites (and thus, constrain DM) \cite{Banik:2018pjp,Banik:2019smi,Enzi:2020ieg,Hermans:2020skz,Lovell:2021vmq}.
However, gaps can also be created by many astrophysical reasons, including time-varying bar potential  or giant molecular clouds \cite{Erkal:2014tda,Bonaca:2018fek,Amorisco:2016evb,Read:2017lvq,DOnghia:2009xhq,Balbinot:2017upo} making robust constrains challenging. The ambiguity will somewhat be reduced for distant streams, potentially observable with the next generation of photometric surveys.

Future potentially interesting observations include 21cm \cite[see e.g.][]{Carucci:2015bra,Neutsch:2022hmv}. The recent EDGES result demonstrated a lot of potential power in these observables for the structure formation constraints~\cite{Leo:2019gwh,Boyarsky:2019fgp} as well as uncertainties, associated with unknown reionization history.

The wealth of new cosmological data expected in the coming years as well as methodological breakthroughs will eventually allow to distinguish  ``nuisance'' astrophysical processes from (potential) effects of DM free-streaming and thus to infer the nature of DM from cosmological  observations. Joint analyses \cite[see e.g.][]{Enzi:2020ieg,Lovell:2021vmq,Nadler:2021dft} will help to reduce systematic uncertainties of improper astrophysical modeling in each specific 

\emph{DM decay line.}
Weak-like interactions of HNLs together with their mass allow them to decay \cite{Bondarenko:2018ptm}.
The most interesting channel from a point of view of DM detection is the loop mediated radiative process $N\to \nu\gamma$~\cite{Pal:1981rm,Barger:1995ty}, which predicts a sharp photon emission line  from DM regions \cite{Dolgov:2002wy,Abazajian:2001vt,denHerder:2009sxr}.
The phase-space constraints limit HNL DM mass to be above $\mathcal{O}(0.1)$~keV which means that this line can be detected in X-ray or $\gamma$-ray telescopes~\cite{Boyarsky:2012rt}.
The searches for DM decay lines have been extensively done in the last 20 years, with a generic bound for the lifetime being $10^{27}-10^{28}$~seconds, see~\cite{Boyarsky:2012rt} for review.

In 2014 an unidentified feature  at $3.5$ keV
in the X-ray spectra of galaxy clusters~\cite{Bulbul:2014sua,Boyarsky:2014jta} as well as Andromeda~\cite{Boyarsky:2014jta} and the Milky Way galaxies~\cite{Boyarsky:2014ska} was reported.
The DM interpretation of this signal has been the subject of an active discussion within the community ever since, see~\cite{Boyarsky:2018tvu} for a summary.
The central point of the discussion is related to the background modelling \cite{Abazajian:2020unr,Boyarsky:2020hqb,Dessert:2020hro,Bhargava:2020fxr} and statistical interpretation of the data \cite{Boyarsky:2018ktr,Dessert:2018qih}.
Similar uncertainties exist in observations with the NuStar \cite{Neronov:2016wdd,Perez:2016tcq,Roach:2019ctw}, for which the $3.5$ keV lies at the edge of the sensitivity interval, making a quantitative estimate of the errors difficult.
The preliminary analysis with \emph{Hitomi} data \cite{Hitomi:2016mun} demonstrates the power of high-resolution X-ray instruments.
It has also excluded astrophysical origin of the 3.5~keV line~\cite{Riemer-Sorensen:2014yda,Jeltema:2014qfa,Carlson:2014lla,Phillips:2015wla}. Depending on cosmology and theoretical model, the
presumed 3.5 keV X-ray signal could correspond to a sterile neutrino with a mixing testable in upcoming laboratory experiments~\cite{Gelmini:2019esj,Gelmini:2019wfp}. 
The final word in this discussion will be put by the XRISM instrument
\cite{XRISMScienceTeam:2020rvx} that will either detect the line.
Additionally, microcalorimeters on sounding rockets~\cite{XQC:2015mwy} looking into the direction of Galactic Centre may confirm the origin of the line.

\emph{In summary:} models with HNLs can accommodate dark matter particles in a number of ways. The range of masses of HNL DM is from sub-keV range to MeV or even GeV (depending on the model/production mechanism).
These models have many observational signatures that allow to probe them with existing and forthcoming astronomical and cosmological data.


\subsubsection{Hubble Parameter}

Measurement of the Hubble constant $H_0$, 
which describes the expansion rate of the Universe, supplies vital information about cosmology as well as potential insights about new physics~\cite{Suyu:2012ax}. Discrepancy between the measurements of Hubble constant in the local universe (e.g. Cepheids, Type-Ia supernovae) of $H_0 = 74.03 \pm 1.42$~km s$^{-1}$ Mpc$^{-1}$~\cite{Riess:2019cxk} and Planck satellite measurements of CMB and BAO data of $H_0 = 67.66 \pm 0.42$~km s$^{-1}$ Mpc$^{-1}$~\cite{Planck:2018vyg} is at the $\sim 5 \sigma$ level~\cite{Wong:2019kwg}.

Consider a minimal scenario with a heavy sterile neutrino that primarily couples to the Standard Model through a mixing $\sin \theta$ with an active neutrino $\nu_{a}$ ($a = e, \mu, \tau$). Ref. \cite{Gelmini:2019deq} demonstrated that such sterile neutrino with mass $m_s \simeq \mathcal{O}(30)$~MeV decaying just before BBN, predominantly into active neutrinos, can increase the effective number of relativistic neutrino species $\Delta N_{\rm eff}$ and thus alleviate the $H_0$ discrepancy. Intriguingly, the necessary masses and couplings of the sterile
neutrino, assuming it mixes primarily with $\nu_{\tau}$ and/or $\nu_{\mu}$ neutrinos, are within reach of the Super-Kamiokande as well as upcoming laboratory experiments such as NA62 and DUNE. More so, improved upcoming measurements of $\Delta N_{\rm eff}$ by CMB-S4~\cite{CMB-S4:2016ple} will further explore this scenario.
As demonstrated by Ref.~\cite{Gelmini:2020ekg}, in the presence of non-negligible lepton asymmetry, decays of even heavier sterile neutrinos with mass $m_s \simeq 150 - 450$~MeV can assist with alleviating the Hubble tension.

\newpage
\section{Executive Summary}

Heavy Neutral Leptons are right-handed neutrino partners to the 
Standard Model active neutrinos. Their existence
can provide elegant solutions to present open questions in fundamental
physics such as the origin of neutrino masses, the nature of dark matter and the observed matter antimatter asymmetry in the Universe.
These HNLs, named as such because they are significantly heavier than the 
Standard Model active neutrinos, are (quasi) sterile and are produced through mixing 
with the active neutrinos.
The allowed mass range for these for putative particles is 
unknown and spans any value between a fraction of an eV up to the 
GUT scale. Hence HNLs are searched for with a large number
of complementary experimental approaches, from nuclear decays to
the high energy frontier experiments. In this report we give a 
survey of existing and new opportunities for the hunt for HNLs
in the coming decades, covering the keV to TeV mass ranges. Moreover, in addition to examining HNLs which only interact via neutrino mass mixing, we have also surveyed the phenomenological consequences of non-minimal HNLs which have additional interactions. 

A novel approach to searching for HNLs involves exploiting energy-momentum conservation in nuclear reactions in which an electron neutrino or electron antineutrino is involved, such as
beta decay and electron capture processes.
Proposed new experiments to probe these processes in the next decade
will provide valuable handles to make a direct search for heavy neutral leptons in the full keV HNL mass range, with 
coupling sensitivities that will improve the 
present experimental reach by several orders of magnitude, in 
particular when including the envisaged potential upgrades 
of these experiments.

Present or future planned or upgraded short baseline experiments have a window to improve the sensitivity for HNL searches  in the mass region of 1-10 MeV.
The challenge for these experiments will be to have good handles 
on the background control, and ensure dedicated triggers for
HNL decays in flight, which would allow to cover
a substantial extension of the present search region in that mass 
range. We strongly recommend the reactor experiment community to 
study and invest in this particular opportunity.

The prospects for discovering HNLs in the coming one or two decades in fixed-target experiment environments have been examined. This broad category includes many currently-operating and next-generation experiments, each with various approaches and physics goals (many of which are orthogonal to these beyond-the-Standard-Model searches). These can be  broadly categorized based on their experimental 
equipment deployed, and
can be used to divide these  into searches from rare kaon decays, beam-dump setups, and searches in neutrino-beam environments. Up to date summary figures  show the capabilities of these searches for the different coupling scenarios (electron, muon, and tau coupling dominance) in the next two decades.  Complementarity among the
different fixed-target probes is evident, but also when comparing with the other types of searches discussed e.g. for colliders
as discussed below. Fixed-target searches offer some of the most promising sensitivity to discovering HNLs 
in the tens of MeV to few GeV range in the near future.

The LHC and possible future high energy colliders will offer excellent opportunities to search for heavy neutral leptons. With the 
full high luminosity event statistics the CMS and ATLAS experiments 
can potentially reach values of couplings in the minimal 
HNL model
on $|V_{e N}|^2$ and $|V_{\mu N}|^2$ down to or below $10^{-7}-10^{-8}$, in
the mass region $m_N $ of 5-20 GeV. LHCb will extend the range for lower mass values. One of the issues hindering the reach to 
smaller couplings at small mass hypotheses is the decreasing acceptance of the 
central detectors due to the  correspondingly increasing HNL 
lifetimes. 
Current proposals for new experiments at the LHC, made to overcome this limitation, are grouped in transverse and forward type 
of detectors. The transverse 
detector proposals encompass the MATHUSLA, CODEX-b, AL3X, ANIBUS and MAPP-LLP
experiments. These are typically experiments optimized for searches for observing new weakly 
interacting neutral particle decays, and placed at distances of tens to more than a 
hundred meters away from the new particle production point.
Forward detectors, such as FASER, SND@LHC, the Forward Physics 
Facility and FACET, are located along the  
direction of the LHC beam line and are mostly sensitive to 
the production of new neutral particles 
originating in decays of mesons.

These additional detectors will cover an important part of the HNL 
parameter space, mostly 
for masses $m_N$ less than 5 GeV, and will be complementary to  experiments at high intensity fixed target experiments.
The sensitivities  will be reaching values  of
$|V_{e N}|^2$ and $|V_{\mu N}|^2$ roughly 
down to $10^{-8}-10^{-9}$, possibly even lower values, 
in a mass region between 100 MeV and 5 GeV. This constitutes a large
newly explored region of the  HNL parameter space.


In a more distant future a facility like
the FCC project could be realized. 
In terms of searches for HNLs, the FCC in its different complementary stages, can probe very large areas of the parameter space towards
in the tens of GeV mass region, due to the copiously produced 
heavy bosons (Z's at the FCC-ee Z-factory and W's at a high luminosity FCC-hh hadron collider)
that are large sources of 
neutrinos, and covering regions
that are not constrained by astrophysics or cosmology, and are complementary to beam dump and neutrino facilities. Heavy new neutrinos with masses larger than 
10~TeV can be searched for at the high energy frontier at the 
FCC-hh.

In the event of an HNL discovery, it will be important 
to study the HNL's (or HNLs') properties. This includes, but is not limited to, studying the mixing pattern(s) and flavor structure of the HNL(s) to determine if there is a connection to the observed light neutrino masses, as well as determining whether Lepton Number is conserved or violated, or equivalently, whether neutrinos and HNLs are Dirac or Majorana fermions. Either of these observations would revolutionize particle physics. In particular fixed target experiments 
would be in excellent position for such measurements if the HNLs
happen to live in their covered parameter space, but 
present and future collider detectors will prepare for 
this too.

In addition to the terrestrial bounds discussed above, this report also surveyed the landscape of constraints arising from solar, atmospheric, astrophysical, and cosmological considerations. The solar and atmospheric neutrino fluxes are large sources of naturally occurring neutrinos which can be utilized for HNL searches. At present, solar neutrino up-scattering searches provide strong constraints on minimal HNLs, while atmospheric neutrino up-scattering provides novel sensitivity to non-minimal HNLs (such as those interacting with a transition magnetic moment). Likewise, the presence of HNLs in the early universe can be strongly constrained given that they can disrupt the success of big bang nucleosynthesis.  These cosmological constraints provide complementary sensitivity to HNLs, reaching lower mixing angles than any existing terrestrial constraint. At the same time, to the extent that terrestrial and cosmological sensitivities overlap, there is the possibility of detecting HNLs which could require modifications to cosmology. 

For most of the future options and proposal given in this report
- both for the near and more distant future - first estimates on the
sensitivity for HNL discoveries have been made, and demonstrate 
the potential HNL parameter space coverage.
Certainly further studies e.g on detector optimization
are strongly desirable and needed for this important physics target.
Such studies are encouraged to go beyond the simplest version of the HNL models, covering non-minimal scenarios.

We make the following recommendations:

{\bf Recommendation 1:} We very strongly encourage the present experimental developing program
to pursue new ideas and make proposals for
HNL sensitive experiments at the current existing accelerator  facilities, as well as 
continue to explore the hunt for HNLs with 
already existing detectors and/or upgrades.
Some facilities  were perhaps not designed for 
BSM particle hunt studies per se, but thanks to
their high intensity proton source and the newly planned
near detectors that will have excellent resolutions and efficiencies, these will become very competitive and one should exploit this superb  opportunity  to
``upgrade'' the searches for such new particles to become a key part of these experiment's baseline physics program.

{\bf Recommendation 2:} We very strongly 
recommend that future collider facilities take 
into account from the start the strong
interest and need for searches for long 
lived particles in their infrastructure plans.
Detector designs should from the start take
searches for HNLs and LLPs in general in
their baseline physics targets.

{\bf Recommendation 3:} In order to facilitate apples-to-apples comparisons, and for simplicity, we encourage experimental analyses to examine sensitivity to the electron-, muon-, and tau-HNL mixing angles separately one at a time. Of course many other flavor assumptions are possible, and possibly even more realistic.  We
encourage analyses to examine scenarios beyond single flavor dominance.
For example, a discussion of such new benchmarks is ongoing in the context of the FIPS workshop series~\cite{Agrawal:2021dbo}.

{\bf Recommendation 4:} The keV mass scale
can be covered using nuclear process and 
the proposed experiments are very important to be conducted. We should also make sure to capitalize on 
the existing and planned (upgraded) reactor
experiments to cover the low MeV mass range.

{\bf Recommendation 5:} In the aftermath of HNL discovery, the most immediate question will be the experimental determination of HNL properties. One would like to extract the data-preferred mixing angles and HNL mass, determine the nature of the quantum statistical nature of the 
new particles and perhaps stress-test the assumption that the detection is consistent with minimal HNL couplings.

\clearpage
\bibliographystyle{apsrev}
\bibliography{bibliography}

\end{document}